\documentclass[traditabstract]{aa} 
\usepackage[pdftex]{graphicx}
\usepackage{amsmath}
\usepackage{subfig}

\def\la{\lower.5ex\hbox{$\; \buildrel < \over \sim \;$}}
\def\ga{\lower.5ex\hbox{$\; \buildrel > \over \sim \;$}}

\begin{document}
   \title{From the Circumnuclear Disk in the Galactic Center to thick, obscuring tori of AGNs}
   \subtitle{Modelling the molecular emission of a parsec-scale torus as found in NGC~1068}

   \author{B.~Vollmer\inst{1}, R.I.~Davies\inst{2}, P.~Gratier\inst{3}, Th.~Liz\'ee\inst{1}, M.~Imanishi\inst{4,5}, J.~F.~Gallimore\inst{6}, C.~M.~V.~Impellizzeri\inst{7,8}, S.~Garc\'ia-Burillo\inst{9}, F.~Le~Petit\inst{10}}

   \institute{Universit\'e de Strasbourg, CNRS, Observatoire astronomique de Strasbourg, UMR 7550, F-67000 Strasbourg, France \and 
        Max-Planck-Institut f\"ur extraterrestrische Physik, Postfach 1312, D-85741, Garching, Germany \and
        Laboratoire d’Astrophysique de Bordeaux, Univ. Bordeaux, CNRS, B18N, all\'ee Geoffroy Saint-Hilaire, 33615 Pessac, France \and
        National Astronomical Observatory of Japan, National Institutes of Natural Sciences (NINS), 2-21-1 Osawa, Mitaka, Tokyo 181-8588, Japan \and
        Department of Astronomy, School of Science, The Graduate University for Advanced Studies, SOKENDAI, Mitaka, Tokyo 181-8588, Japan \and
        Department of Physics and Astronomy, Bucknell University, Lewisburg, PA 17837, USA \and
        Leiden Observatory—Allegro, Leiden University, P.O. Box 9513, 2300 RA Leiden, The Netherlands \and
        Joint ALMA Observatory, Alonso de Cordova 3107, Vitacura 763-0355, Santiago de Chile, Chile \and
        Observatorio  Astron\'omico  Nacional  (OAN-IGN)-Observatorio  de Madrid, Alfonso XII, 3, 28014 Madrid, Spain \and
        LERMA, Observatoire de Paris, PSL Research University, CNRS, Sorbonne Universit\'es, 92190 Meudon, France
         }

   \date{Received ; accepted }

 
  \abstract
{
The high accretion rates needed to fuel the central black hole in a
galaxy can be achieved via viscous torques in thick disks and rings, which can be resolved by millimetre interferometry within the inner
$\sim 20$~pc of the active galaxy NGC~1068 at comparable scales and sensitivity to single dish observations of the Circumnuclear Disk (CND)
in the Galactic Center. To interpret observations of these regions and determine the physical
properties of their gas distribution, we present a modelling effort that includes (i) a simple dynamical simulations involving partially 
inelastic collisions between disk gas clouds, (ii) an analytical model of a turbulent clumpy gas disk calibrated by the dynamical model and
observations, (iii) local turbulent and cosmic ray gas heating and cooling via H$_2$O, H$_2$, and CO emission, and (iv) determination of 
the molecular abundances. We also consider photodissociation
regions (PDR) where gas is directly illuminated by the central engine. We compare the resulting model datacubes of the CO, HCN, HCO$^+$, and CS
brightness temperatures to available observations. In both cases the kinematics can be explained by one or two clouds colliding with
a pre-existing ring, in a prograde sense for the CND and retrograde for NGC~1068.
And, with only dense disk clouds, the line fluxes can be reproduced to within a factor of about two.
To avoid self-absorption of the intercloud medium, turbulent heating at the largest scales, comparable to the disk height, has to be decreased
by a factor of 50--200. Our models indicate that turbulent mechanical energy input is the
dominant gas heating mechanism within the thick gas disks. Turbulence is maintained by the gain of potential energy via radial gas
accretion, which is itself enhanced by the collision of the infalling cloud. In NGC~1068, we cannot exclude that intercloud gas contributes
significantly to the molecular line emission. In this object, while the bulk of the AGN X-ray radiation is absorbed in
a layer of Compton-thick gas inside the dust sublimation radius, the optical/UV radiation may enhance the molecular line emission from
photodissociation regions by $\sim50$\,\% at the inner edge of the gas ring. Infrared pumping may also increase the HCN(3-2) line flux throughout the
gas ring by about a factor of two. Our models support the scenario of infalling gas clouds onto pre-existing gas rings in galactic centers, 
and is viable and consistent
with available observations of the CND in the Galactic Center and the dense gas distribution within the inner $20$~pc of NGC~1068.
}


   \authorrunning{Vollmer et al.}
   \titlerunning{Modelling the molecular emission of the parsec-scale torus of NGC~1068}

   \maketitle
%

\section{Introduction\label{sec:introduction}}

Active galactic nuclei (AGN) can outshine the optical light of the whole galaxy which surrounds them. This huge luminosity and
thus this huge amount of energy is gained through the conversion of gravitational potential energy to electromagnetic radiation. 
The ingredients for an AGN are (i) a supermassive black hole with a mass exceeding about $10^6$~M$_{\odot}$, (ii) interstellar gas
that surrounds the central black hole, and (iii) gas infall onto the black hole.
As the material falls in toward the black hole, angular momentum will cause it to form an accretion disk. During infall the disk 
gas is heated to temperatures between $1$ and $4 \times 10^{5}$~K (e.g., Bonning et al. 2007) giving rise to intense optical and
UV emission. Moreover, the geometrically thin accretion disk is surrounded by a hot corona giving rise to strong X-ray emission.
The typical size of the accretion disk is about $10$ lightdays or $\sim 0.01$~pc (e.g., Kokubo 2018).
Further out, at distances of a few parsec to a few tens of parsecs a massive gas disk is present in AGNs 
(Sani et al. 2012, Hicks et al. 2013, Combes et al. 2019).
The estimated disk masses are of the order of $10$\,\% of the dynamical mass, exceeding $10^6$~M$_{\odot}$ (Garcia-Burillo et al. 2021).
Because of the uncertain conversion factor between luminosity and mass, the mass estimates are quite uncertain.
Most frequently, these massive gas disks are geometrically thick, i.e. they have a high gas velocity dispersion (Davies et al. 2007,
Hicks et al. 2009, Sani et al. 2012). 

The Galactic Center is not an AGN although it hosts a supermassive black hole ($M_{\rm BH}=4.3 \times 10^6$~M$_{\odot}$, Gillessen et al. 2009)
and a gas disk (Circumnuclear Disk or CND) of several $10^4$~M$_{\odot}$ located between $1$ and $5$~pc (e.g., G\"usten et al. 1987, 
Christopher et al. 2005, Etxaluze et al. 2011). The reason for the quiescence of the Galactic Center is a negligible mass 
accretion rate onto the black hole.

With a distance of $8.2$~kpc (Gravity Collaboration 2019), the gas of the CND can be observed with a resolution of $\sim 0.1$~pc.
The CND in the Galactic Center is made of gas clouds with sizes of $\sim 0.1$~pc, masses
of a few $10$~M$_{\odot}$, an area filling factor of $\Phi_{\rm A} \sim 0.1$, and a volume filling factor of $\Phi_{\rm V} \sim 0.01$
(Jackson et al. 1993; however, Requena-Torres et al. 2012 suggest a much higher volume filling factor of $\Phi_{\rm V} \sim 0.2$). 
Since the densities derived from the HCN lines (Jackson et al. 1993) are close to those of selfgravitating clouds with a thermal sound 
speed of $\sim 1$~km\,s$^{-1}$, Vollmer \& Duschl (2001a) and Vollmer et al. (2004) proposed a model of a collisional disk made of stable
gas clouds. This model was extended to AGN gas tori in Vollmer et al. (2008). The small linewidths of the selfgravitating gas
clouds are in stark contrast to the observed large linewidths in the CND (Montero-Casta\~no et al. 2009). This means that the
clouds are not selfgravitating and thus short-lived. Strong gas turbulence can create the observed large velocity dispersion.

Vollmer \& Davies (2013) developed an analytical model for turbulent clumpy gas disks where the energy driving turbulence is supplied
by external infall or the gain of potential energy by radial gas accretion within the disk.
The gas disk is assumed to be stationary ($\partial \Sigma/\partial t=0$) and the external mass accretion rate to be close to the 
mass accretion rate within the disk (the external mass accretion rate feeds the disk at its outer edge).
Within the model, the disk is characterized by the disk mass accretion rate $\dot{M}$ and the Toomre $Q$ parameter which is
used as a measure of the gas content of the disk.
It is suggested that the velocity dispersion of the gas disk at a $10$-pc scale or torus is increased through adiabatic compression by the infalling gas.
The turbulent and collisional model were applied to the CND in the Galactic Center and thick obscuring tori in AGNs.
Despite the different gas masses (CND: $10^{4}$~M$_{\odot}$; AGN torus: $10^{6}$~M$_{\odot}$) and 
velocity dispersions (CND: $20$~km\,s$^{-1}$; AGN torus: $50$~km\,s$^{-1}$), both models share disk gas clouds of similar masses and sizes.

With the advent of ALMA, it is now possible to observe the centers of other galaxies with a decent resolution ($\la 1$~pc) in the millimeter regime.
NGC~1068 is the archetypical type~2 AGN with a massive black hole of $8 \times 10^6$ (Lodato \& Bertin 2003), 
where the massive thick gas disk is seen edge-on and thus entirely obscurs the central engine.

Imanishi et al. (2016, 2018) observed the center of NGC~1068 in the HCN(3-2) line with a resolution of $0.1''$-$0.2''$ ($7$-$14$~pc) and $0.06''$ 
($4$~pc), respectively. Garcia-Burillo et al. (2016) observed the same target in the CO(6-5) line with a $0.06''$ ($4$~pc) resolution.
These authors detected a massive gas disk of a one to a few $10^5$~M$_{\odot}$ and a size of $\la 10$~pc.
Subsequent higher resolution observations (Garcia-Burillo et 2019, Impellizzeri et al. 2019, Imanishi et al. 2020) with beamsizes
of the order of $20$~mas ($1.4$~pc) clearly resolved the massive gas disk and gave access to its detailed kinematics.
Compared to its appearance in the HCN(3-2), HCO$^+$(3-2), and HCO$^+$(4-3) lines, 
the molecular disk is bigger and more lopsided in the lower-J CO transitions (Garcia-Burillo et 2019, Impellizzeri et al. 2019). 
Moreover, Impellizzeri et al. (2019) revealed a counter-rotating inner disk at radii smaller than $\sim 1.2$~pc
(see also Imanishi et al. 2020). This is surprising, because counter-rotating gas disks are supposed to be unstable against
shear instability (Quach et al. 2015). 

Gallimore et al. (2016), Garcia-Burillo et al. (2019), and Impellizzeri et al. (2019) found signs of a molecular outflow in the inner region of 
the massive gas disk ($R \la 3$~pc). Furthermore, the torus is connected to the 200pc-size gas ring through a network of gas 
lanes whose kinematics are accounted for by a 3D outflow geometry  (Garcia-Burillo et al. 2019). The latter authors argued that about half of the mass of
the massive gas disk is outflowing. Gas compression by the outflow or by radiation pressure may be considered as sources of mechanical heating of the 
molecular gas at the inner edge of the massive disk.

In this article we continue the comparison of Vollmer \& Davies (2013) between the CND in the galactic center and the massive gas disk in the 
center of NGC~1068. In the following, the term CND is exclusively used for the $1$-$5$~pc gas disk or ring in the Galactic Center.
We produced CND-like and NGC~1068-like dynamical models and calculated their CO, HCN, and HCO$^+$ emission for 
multiple transitions. Whereas traditional models derive the gas properties from molecular line ratios, our forward modelling aims at 
reproducing qualitatively and, to a certain extent, quantitatively the observed molecular brightness temperatures. 

The article is structured in the following way: the model is described in Sect.~\ref{sec:model}, its application to the CND in the
Galactic Center in Sect.~\ref{sec:cnd} and to the inner $20$~pc of NGC~1068 in Sect.~\ref{sec:n1068}. 
The cosmic ray (CR) ionization fractions, a more quantitative comparison between the model and observations, the role of the continuum emission,
and the relevance of the different gas heating mechanisms are discussed in Sect.~\ref{sec:discussion}. Finally, we give our
conclusions in Sect.~\ref{sec:conclusions}.

\section{The model\label{sec:model}}

The inner $\sim 50$~pc of galaxies are complex environments involving a dense nuclear disk and star cluster and a central massive black hole.
Misalignments between the ionization cones of AGN, and hence the obscuring gas structures, and the disks of the host galaxies are frequently observed
(e.g., Fischer et al. 2013). Indeed, the  nuclear  gas  angular  momentum content can be significantly different from
that of the galactic disk because of large-scale gas fragmentation or secondary stellar bars (Hopkins et al. 2012).
Therefore, gas disks or rings with arbitrary inclinations and position angles are expected at scales of $\sim 10$~pc.
These rotating gas structures are perturbed by infalling gas from larger distances with a different angular momentum.
The simplest scenario is that of a gas cloud falling onto a pre-existing gas disk or ring.
Even in such a simple configuration there are more than ten free parameters and parameter space is vast.
Therefore, we do not aim at reproducing details of the available observations of the Galactic Center and NGC~1068.
Our work has a broader scope. We want to find out if such a simple scenario can explain the main
characteristics of the observations in terms of gas distribution and kinematics. In a second step, the combination
of the large-scale dynamical model with the small-scale analytical model of turbulent clumpy accretion disks
allows us to calculate the molecular line emission of the largest turbulent disk gas clouds.
It is then possible to investigate if our models can reproduce the characteristics of the available observations
in terms of gas density, temperature, and chemistry, clump area filling factor, and brightness temperatures of the molecular lines.
In this respect, our models are not expected to exactly reproduce the available observations, but to serve as a guideline
to interpret them. With this work we want to make a step forward in the understanding of the large-scale gas dynamics and 
the physics in terms of turbulence, heating, cooling, and chemistry in the inner $\sim 30$~pc around a central supermassive black hole.

Our modelling of thick gas disks located at distances between $1$ and $10$~pc from the central black hole
comprises five components: (i) a sticky-particles dynamical code for the large-scale gas dynamics, where the particles 
represent gas clouds, (ii) an analytical model of a turbulent clumpy accretion disk from which the
properties (density, size, velocity dispersion) of the largest turbulent gas cloud are derived, (iii) an analytical model for the calculation
of the disk cloud temperature via the equilibrium between gas heating and cooling, (iv) a model for the gas chemistry,
and (v) the calculation of the molecular line emission from the disk gas clouds.
The dynamical model is used to follow the pc-scale evolution of the gas distribution within a distance of $\sim 30$~pc
around the central black hole. The dynamical simulations of a collisional gas disk can be regarded as an approximation of a 
simulation of a turbulent gas disk in terms of large-scale gas density, velocity dispersion, and gas viscosity (Vollmer \& Davies 2013).
The analytical accretion disk model, which is used to calculate the disk cloud properties,
is linked to the dynamical model via the total gas mass and the particle velocity dispersion: the analytical model
assumes a Toomre $Q$ parameter and a gas mass of the central gas disk or ring, which is consistent with the gas properties 
of the timestep of interest of the dynamical model and observations.
Furthermore, we made sure that the disk mass accretion rate of the dynamical model is also consistent with that of
the analytical model. Given the size, density, and velocity dispersion of a disk gas cloud, its temperature was calculated via the 
equilibrium between turbulent mechanical and cosmic ray heating and radiative cooling. The molecular abundances
were determined for each disk gas cloud independently by the two-phase gas-grain code {\tt Nautilus}
(Hersant et al. 2009). The cloud molecular gas density, surface density, temperature, and velocity dispersion 
were then used to calculate the CO, HCN, HCO$^{+}$, CS, and CN line emission and 
datacubes were produced from the simulations. These datacubes are directly compared to the available observations.

\subsection{The dynamical model \label{sec:dynmodel}}

Following Vollmer et al. (2002), the ISM is simulated as a collisional component, i.e. as discrete particles that possess a mass and a 
radius and can have partially inelastic collisions. In contrast to smoothed 
particle hydrodynamics (SPH), which is a quasi-continuous approach where the particles cannot penetrate each other, our approach 
allows a finite penetration length between particles, which is given by the mass-radius relation of the particles.
During the disk evolution, the cloud particles can have partially 
inelastic collisions, the outcome of which (coalescence, mass exchange, or fragmentation) is simplified following
the geometrical prescriptions of Wiegel (1994). 

We  follow  the  orbits of these  clouds  in  the  three  dimensional gravitational potential. The radial distribution of the total enclosed 
mass $M(R)$ is given by 
\begin{equation}
\label{eq:massdist}
M(R)=M_{\rm BH}+M_0\,R^{\frac{5}{4}} \ ,
\end{equation}
where $M_{\rm BH}$ is the mass of the central black hole, and $M_0$ describes the mass distribution of the stellar content.
We used $M_{\rm BH}=4 \times 10^{6}$~M$_{\odot}$ and $M_0=1.6 \times 10^6$~M$_{\odot}$pc$^{-5/4}$ (Genzel et al. 2010) for the CND in the
Galactic Center and $M_{\rm BH}=10^{7}$~M$_{\odot}$ (Greenhill et al. 1996) and  $M_0=0.8 \times 10^6$~M$_{\odot}$pc$^{-5/4}$ for NGC~1068.

The integration of the ordinary differential equation is done with the Burlisch-Stoer method (Stoer  \&  Burlisch 1980) using  
a Richardson extrapolation and Stoermer's rule. Vollmer \& Duschl (2002) chose the error level to match the theoretical collision rates.
The global timestep\footnote{In addition, the integrator divided this timestep at least into three sub-timesteps of about $30$~yr.} 
is typically around $50$-$100$~yr. For a velocity of $100$~km\,s$^{-1}$ this corresponds to $\sim 0.01$~pc.
During each cloud-cloud collision the overlapping parts of the clouds are calculated. 
Let $b$ be impact parameter and $r_1$ and $r_2$ the radii of the larger and smaller clouds. 
If $r_1+r_2 > b > r_1-r_2$ the collision can result into fragmentation (high-speed encounter) or
mass exchange. If $b < r_1-r_2$ mass exchange or coalescence (low speed encounter) can occur. In this way a cloud mass distribution is naturally
produced. The cloud masses and velocities resulting from a cloud-cloud collision are calculated by assuming mass and momentum conservation.

As in Vollmer \& Duschl (2002) we allowed for additional energy dissipation by lowering the final gas cloud velocities
by a constant fraction $\xi$: $\vec{v}_{\rm end}= \sqrt{\xi} \times \vec{v}_{\rm ini}$ where $\vec{v}_{\rm ini}$ and
$\vec{v}_{\rm end}$ are the velocity vectors of the initial and resulting gas clouds after the collision.
The energy dissipation rate of a collisional disk is
\begin{equation}
\frac{\Delta E}{\Delta A \Delta t}= \eta \frac{\Sigma\,v_{\rm turb}^2}{t_{\rm coll}}\ ,
\end{equation}
where $t_{\rm coll}=4\,r_{\rm cl}/(3\,\Phi_{\rm V} v_{\rm turb})$ (Vollmer et al. 2008) is the collision timescale, 
$\Phi_{\rm V}$ the cloud volume filling factor, and $\Sigma$ the mean gas surface density of the disk, that of a turbulent gas disk is
\begin{equation}
\frac{\Delta E}{\Delta A \Delta t}= \Phi_{\rm A} \frac{\Sigma\,v_{\rm turb}^3}{l_{\rm driv}}\ ,
\end{equation}
where the turbulent driving length equals the disk height $l_{\rm driv}=H$ (Vollmer \& Davies 2013). 
Both expressions are equivalent if the dissipated energy fraction per collision is
\begin{equation}
\eta=\Phi_{\rm A} \frac{v_{\rm turb}^2 4\,r_{\rm cl}/(3\,\Phi_{\rm V})}{v_{\rm ini}^2 H}\ .
\end{equation}

\begin{figure}[!ht]
  \centering
  \resizebox{\hsize}{!}{\includegraphics{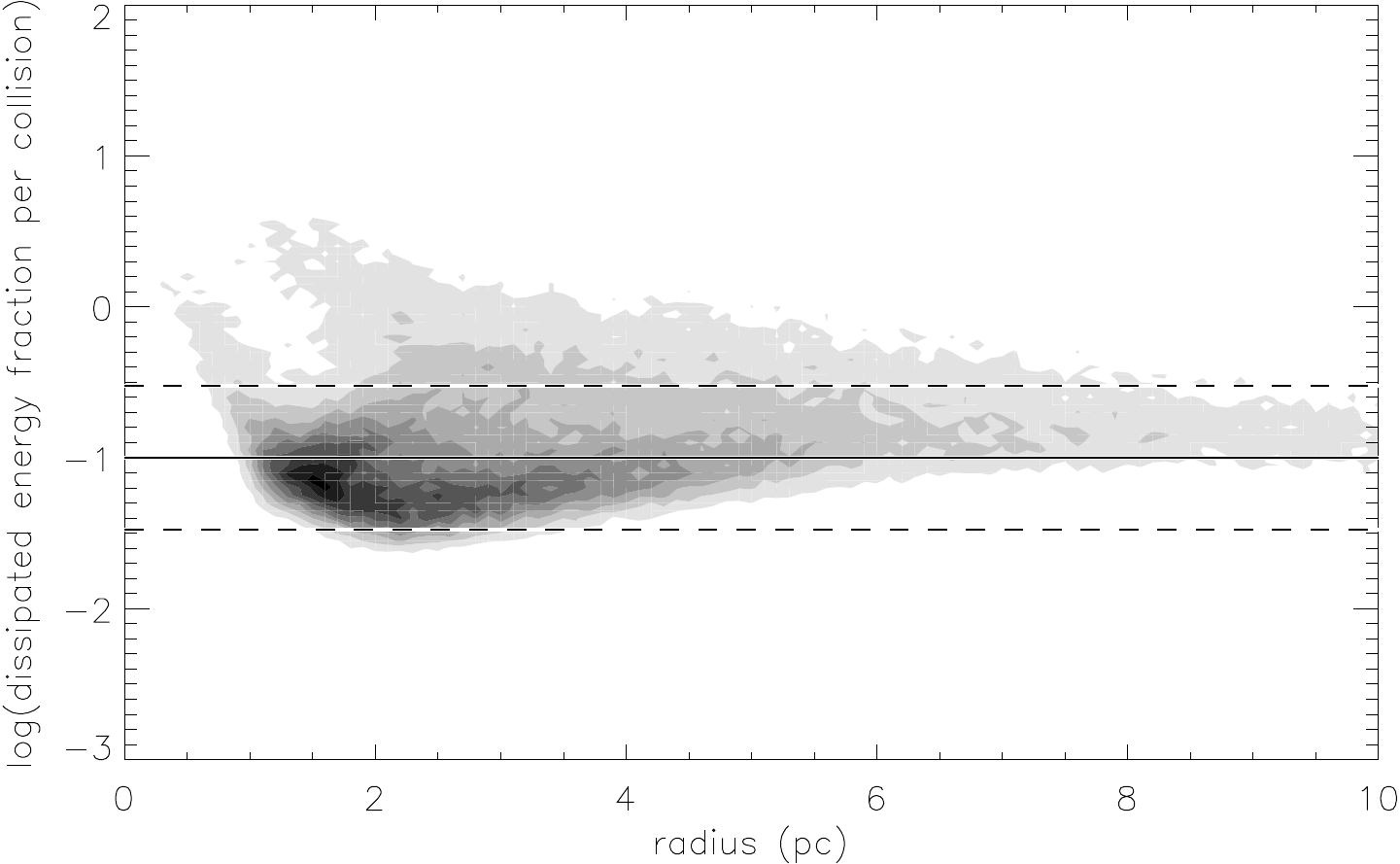}}
  \caption{Fraction of dissipated energy per cloud-cloud collision as a function of the distance to the central black hole.
    The dissipation energy fraction is shown for all clouds during the NGC~1068 model evolution presented in Fig.~\ref{fig:sim_N1068COLLdata2pot_diss1a_light1_hcn32b}.
  \label{fig:dissipationeta}}
\end{figure}

For simplicity we used the cloud radii from the analytical model if $r_{\rm cl} \leq 0.05$~pc and $r_{\rm cl}=0.05$~pc elsewhere.
In this way our simulated disk is broadly consistent with $\eta=0.1$ and thus $\xi=(1-\eta)=0.9$ for the NGC~1068-like simulations 
(Fig.~\ref{fig:dissipationeta}) .

By construction, the dynamical model does not include a wind or outflow. Such a wind is 
observed in NGC~1068 in the line emission of ionized gas (the narrow-line region) at 
scales of 10 to 100~pc (e.g., May  \& Steiner 2017), which has an hourglass structure.
Most importantly, its kinematics are consistent with an outflow within a hollow-cone 
structure (Das et al. 2006; Miyauchi \& Kishimoto 2020).
At smaller scales, the morphology of the point-source subtracted ALMA band~6 continuum 
emission (Impellizzeri et al. 2019) and the polar dust emission from MIR interferometric 
observations (Lopez-Gonzaga et al. 2014) are also consistent with an outflow. 
Gallimore et al. (2016), Garcia-Burillo et al. (2019), and Impellizzeri et al. (2019) 
interpreted the high radial velocities and large linewidths of the molecular line emission
along the minor axis of the molecular torus as a molecular outflow in the inner region of the massive gas disk 
($R \la 3$~pc). Furthermore, the torus is connected to the 200pc-size gas ring through a 
network of gas lanes whose kinematics are accounted for by a 3D outflow geometry 
(Garcia-Burillo et al. 2019). Radiation pressure, the radio jet, and the high-velocity 
ionized outflow certainly help to launch the molecular outflow. A magnetocentrifugal 
molecular and dusty wind starting at the inner edge of the massive gas disk is also viable 
(Vollmer et al. 2018). For a discussion of the outflow scenario we refer to Sect.~\ref{sec:outflow}.
 
The cloud particle masses of our dynamical simulations range between $1$ and $90$~M$_{\odot}$. The mean mass of the clouds is $\sim 20$~M$_{\odot}$.
For the simulations corresponding to the CND in the Galactic Center the number of gas clouds was $N \sim 3200$,
and the total gas mass is $M_{\rm gas,tot}=7.5 \times 10^{4}$~M$_{\odot}$.
For the simulations of the gas distribution around the central black hole as in NGC~1068 the number of gas clouds was $N \sim 1.6 \times 10^4$
and the total gas mass is $M_{\rm gas,tot}=3.1 \times 10^{5}$~M$_{\odot}$.

For all simulations the pre-existing gas ring was located in the $x$-$y$ plane. A massive approximately spherical gas cloud
was located in the same plane with initial conditions that lead to a prograde orbit and apocentric distances of $10$, $15$, and $20$~pc and
pericentric distances of $2$, $3$, and $4$~pc for the CND and $3$, $4$, and $5$~pc for NGC~1068. In this way nine models
were calculated. The infalling gas cloud was then rotated around the $x$-axis from $\beta=20^{\circ}$ to $340^{\circ}$ in steps of $20^{\circ}$
leading to a total number of $162$ models for the CND. Because we focused on counter-rotating gas infall in NGC~1068,
the infalling cloud was rotated around the $x$-axis from $\beta=120^{\circ}$ to $240^{\circ}$ in steps of $20^{\circ}$
leading to a total number of $63$ models for NGC~1068.
For each model timestep a datacube was produced with the following projections in steps of $10^{\circ}$:
CND: $0^{\circ} \leq i \leq 350^{\circ}$, $0^{\circ} \leq PA \leq 350^{\circ}$, $0^{\circ} \leq az \leq 350^{\circ}$;
NGC~1068: $40^{\circ} \leq i \leq 140^{\circ}$, $220^{\circ} \leq PA \leq 320^{\circ}$, $0^{\circ} \leq az \leq 350^{\circ}$,
where $i$ is the inclination angle, $PA$ the position angle, and $az$ the azimuthal angle within the $x$-$y$ plane.
We had to restrict the number of projections of the NGC~1068 model because of the long computation times.

The model datacubes were directly compared to the IRAM 30m CS(2-1) observations of G\"usten et al. presented in Vollmer \& Zylka (2003)
for the CND and to the ALMA CO(2-1) observations of Garcia-Burillo et al. (2019) for NGC~1068.
The best-fit models were determined by searching for the minimum reduced $\chi^2$.
For the CND, we additionally required the western edge to be closer to the observer (e.g., Liu et al. 2012).  
Since the environment of the CND is complex, it is not possible to reproduce the gas distribution with a single
infalling gas cloud and therefore, the $\chi^2$ are quite high.
Our timestep of interest has $\chi^2=161$. We considered timesteps with up to a 10\,\% higher $\chi^2$ as acceptable.
To better reproduce the gas distribution of the inner $20$~pc, we decided to add a second acceptable timestep to our final CND model.
Among the acceptable timesteps we selected the timestep with the best-fit kinematics by setting 
all voxels of the model and observed datacubes to unity for the $\chi^2$ calculations.
The reduced $\chi^2$ of the combined model ($\chi^2=148$) is smaller than those of all single timesteps.
The properties of the two models and the combined model are presented in Table~\ref{tab:bestfit}.

\begin{figure*}[!ht]
  \centering
  \resizebox{\hsize}{!}{\includegraphics{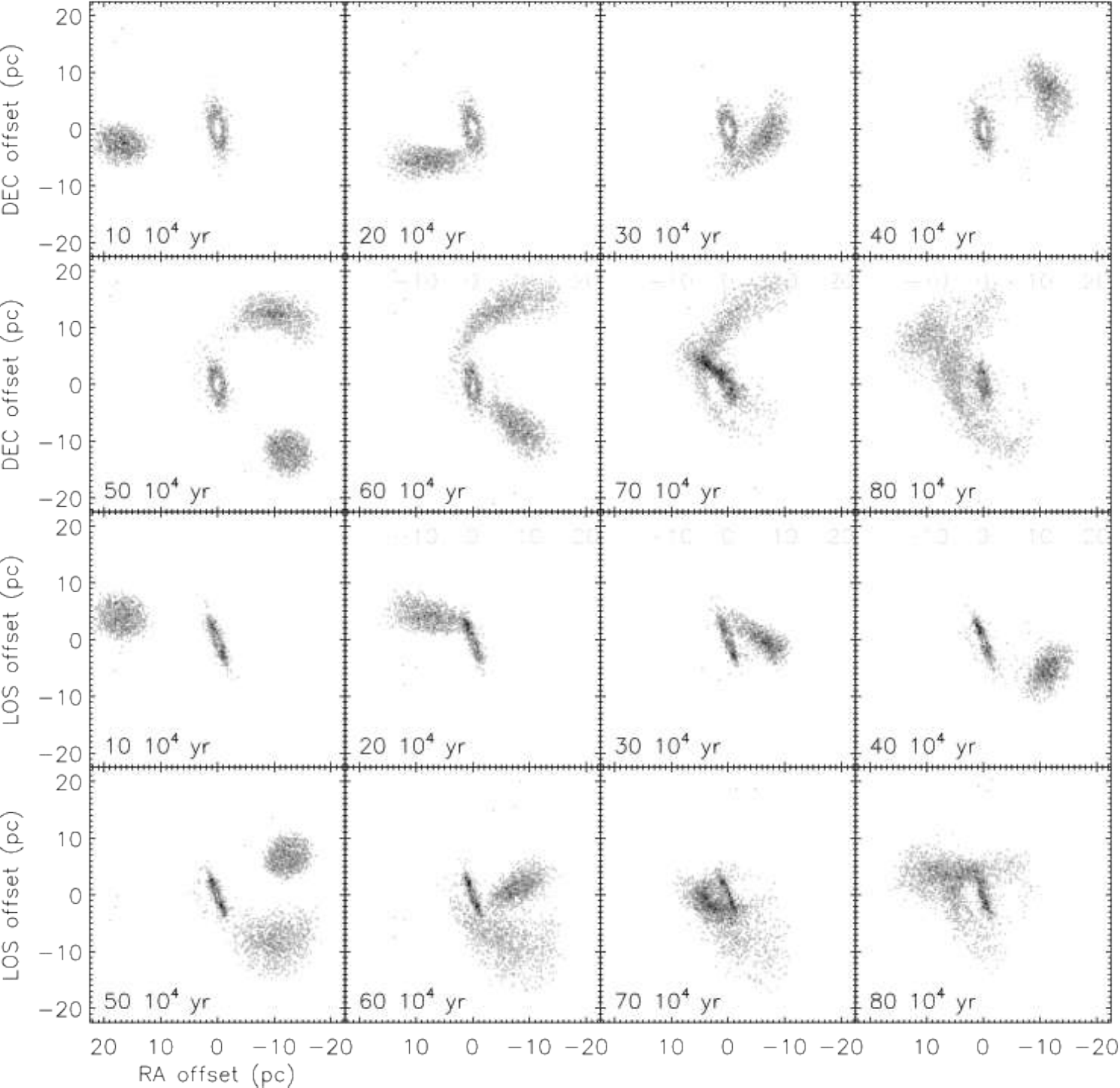}}
  \caption{Evolution of the CND-like model. The times since the beginning of the simulations are indicated.
    Upper half: RA-DEC projection. Lower half: RA-LOS (line-of-sight) projection.
  \label{fig:sim_CNDCOLLdata2small1}}
\end{figure*}

The evolution of the CND-like is shown in Fig.~\ref{fig:sim_CNDCOLLdata2small1}.
The impact of the first infalling cloud onto the CND occurs at about $t=0.23$~Myr, that of the second infalling cloud at
about $t=0.65$~Myr. The timestep, which has the minimum $\chi^2$ and which we compare to observations, i.e. the time of interest, is $t=0.8$~Myr.
At this timestep the gas mass of the CND within a radius of $4$~pc is $1.2 \times 10^4$~M$_{\odot}$.

\begin{figure*}[!ht]
  \centering
  \resizebox{\hsize}{!}{\includegraphics{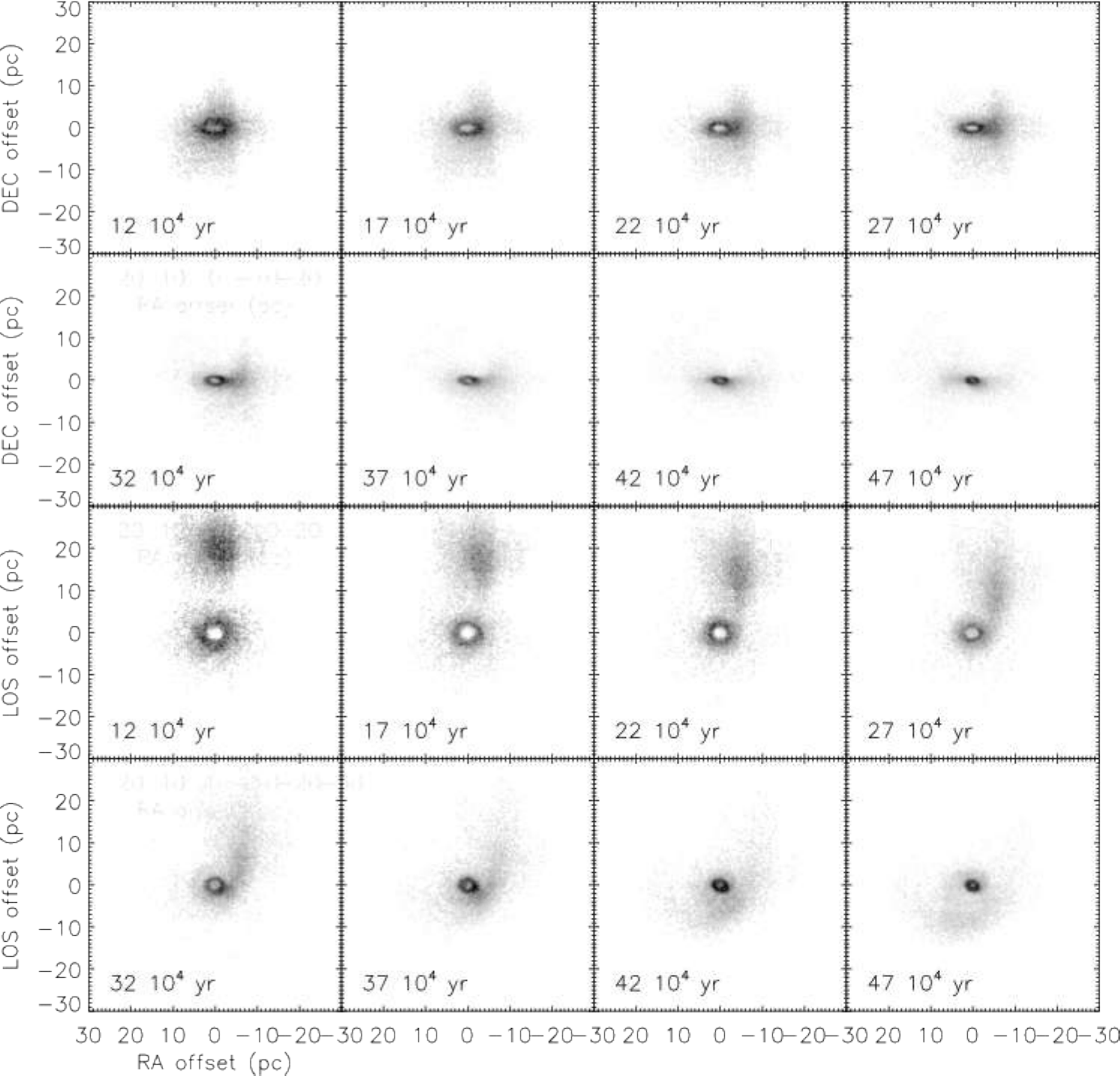}}
  \caption{Evolution of the NGC~1068-like model. The times since the beginning of the simulations are indicated.
    Upper half: RA-DEC projection. Lower half: RA-LOS (line-of-sight) projection.
  \label{fig:sim_N1068COLLdata2pot_diss1a_light1_hcn32b}}
\end{figure*}

\begin{figure*}[!ht]
  \centering
  \resizebox{\hsize}{!}{\includegraphics{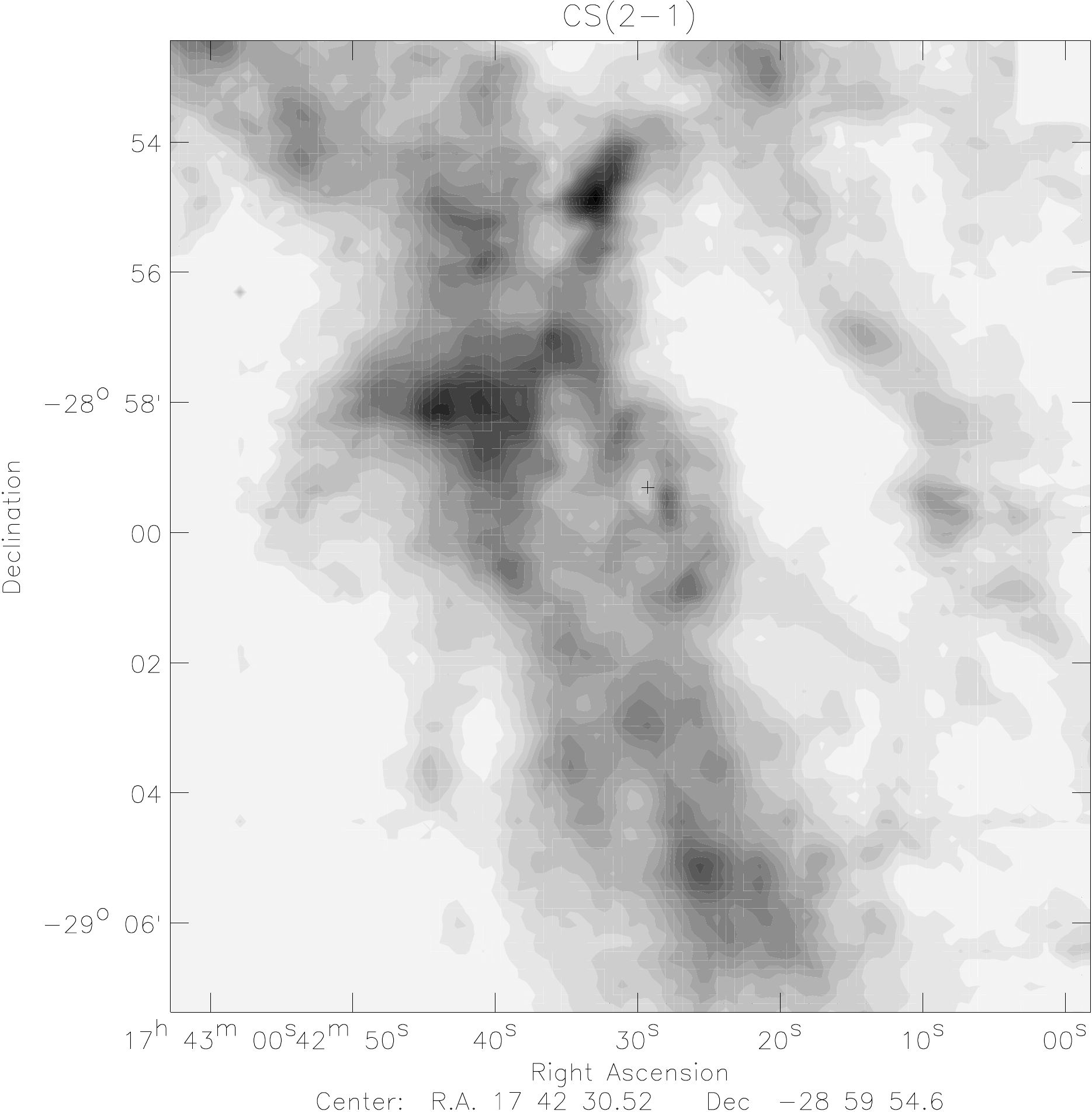}\includegraphics{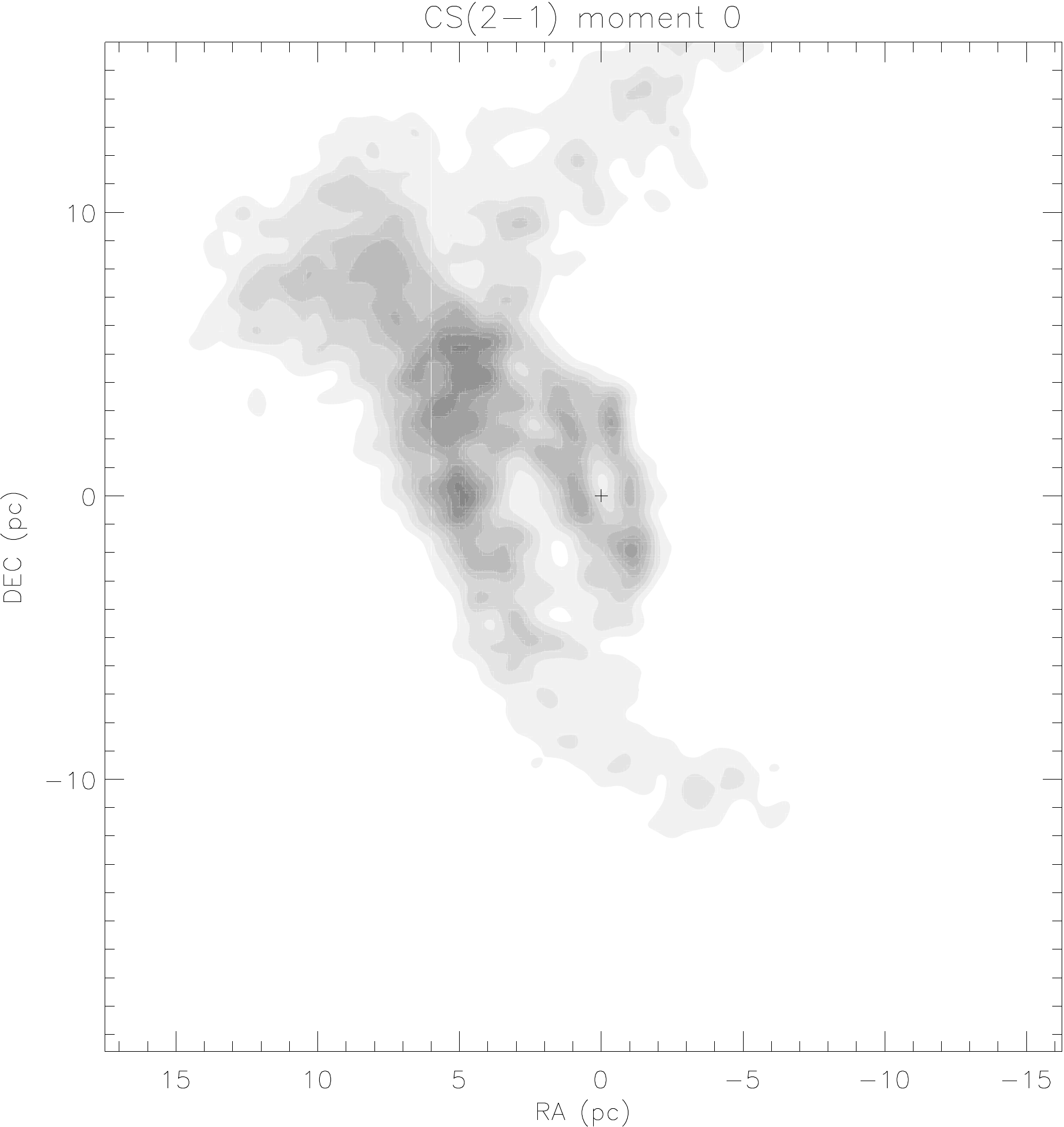}}
  \caption{CS(2-1) emission distribution. Left panel: IRAM 30m observations of the CND in the Galactic Center (G\"usten et al.).
    Right panel: best-fit combined model (see Table~\ref{tab:bestfit}). The levels are $13$ to $260$~K\,km\,s$^{-1}$ in
    steps of $13$~K\,km\,s$^{-1}$. Both images have the same physical size.
  \label{fig:cndiram}}
\end{figure*}

The evolution of the NGC~1068-like model is shown in  Fig.~\ref{fig:sim_N1068COLLdata2pot_diss1a_light1_hcn32b}.
The impact of the infalling cloud occurs at about $t=0.33$~Myr, the time of interest, where the reduced $\chi^2$ is minimum, is $t=0.47$~Myr.
At this timestep the gas mass located between $1.5$ and $7$~pc is $1.4 \times 10^5$~M$_{\odot}$.
The CS(2-1) emission distribution of the combined CND-like model is presented in Fig.~\ref{fig:cndiram},
the CO(2-1) emission distribution of the NGC~1068-like best-fit model in panel (c) of Fig.~\ref{fig:plottingvollcnd_hcn32radex_new_garciaburillo_co21hr}.
\begin{table*}
      \caption{Best-fit dynamical models}
         \label{tab:bestfit}
      \[
       \begin{tabular}{lccccccc}
        \hline
         apocenter & pericenter & $\beta$ & $t$ & $i$ & $PA$ & $az$ & $\chi^2$ \\
         \hline
         CND & & & & & & &  \\
          20~pc & 4.0~pc & $300^{\circ}$ & 0.80~Myr & $70^{\circ}$ & $10^{\circ}$ & $70^{\circ}$ & 161 \\ 
          20~pc & 3.0~pc & $40^{\circ}$ & 0.35~Myr &  $80^{\circ}$ & $10^{\circ}$ & $160^{\circ}$ & 175 \\
          combined & & & 0.80~Myr & $70^{\circ}$ & $10^{\circ}$  & $60^{\circ}$ & 148 \\
         NGC~1068  & & & & & & &  \\
         20~pc & 3.0~pc & $160^{\circ}$ & 0.47~Myr & $60^{\circ}$ & $270^{\circ}$ & $270^{\circ}$ & 106 \\
        \hline
        \end{tabular}
      \]
\end{table*}

We inspected the $150$ model snapshots with the smallest $\chi^2$ by eye. All CND-like model snapshots with $i > 50^{\circ}$ where the
western edge is the near side of the CND are prograde encounters between the gas ring and the infalling gas cloud 
($0^{\circ} \leq \beta \leq 60^{\circ}$ or $320^{\circ} \leq \beta \leq 360^{\circ}$). 
Furthermore, retrograde encounters generally lead to much higher mass accretion rates
than prograde encounters. The higher mass accretion rates lead to centrally concentrated gas distributions, which are not
observed in the Galactic Center. We thus conclude that the gas clouds interacting with the CND are most probably on 
prograde orbits with respect to the CND.
On the other hand, only a retrograde encounter can explain the observed kinematics in the inner $20$~pc of NGC~1068.

\subsection{The analytical model of a turbulent clumpy gas disk \label{sec:anamodel}}

We assume a turbulent clumpy gas disk where the energy to drive turbulence is supplied by external infall or the gain of potential
energy by radial gas accretion within the disk. The analytical model is fully described in Vollmer \& Davies (2013).
An overview of the model is given in Appendix~\ref{sec:amod}. Within the model, the disk is characterized by the
disk mass accretion rate $\dot{M}$ and the Toomre $Q$ parameter, which is used as a measure of the gas content of the disk.
Both parameters are assumed to be constant within the disk.
The size of the largest turbulent gas clouds is determined by the size of a continuous (C-type) shock propagating in dense molecular clouds 
with a low ionization fraction at a given velocity dispersion. We used the expressions derived by Vollmer \& Davies (2013) for the expected volume and
area filling factors, mass, density, column density, and velocity dispersion of the disk clouds. The latter is based on 
scaling relations of intermittent turbulence whose open parameters are estimated for the Circumnuclear Disk
in the Galactic Center. The turbulent clumpy gas disks of the CND and NGC~1068 are calibrated in terms of the radial profiles of the gas 
surface density and velocity dispersion (Fig.~\ref{fig:sigmadisk}) by the dynamical model and observations.

\subsection{The cloud temperature \label{sec:cloudtemp}}

For the determination of the temperature of the turbulent disk gas clouds we follow the model of Vollmer et al. (2017),
which is based on the calculations of Neufeld \& Kaufman (1993) and Neufeld et al. (1995).  
Their model for the radiative cooling of molecular gas includes a detailed treatment of the interstellar 
chemistry that determines the abundances of important coolant molecules, and a detailed treatment of the 
excitation of the species H$_2$, CO, H$_2$O, HCl, O$_2$, C, O, and their isotopic variants where important.
For simplicity, we only take the main cooling agents, CO, H$_2$, and H$_2$O, into account.

For the calculation of the thermal balance within molecular clouds one needs 
to consider processes affecting the gas and the dust in addition to the radiative gas cooling.

We assume gas heating via turbulence, X-rays, and cosmic rays:
\begin{equation}
\label{eq:heating}
\Gamma_{\rm g} = \Gamma_{\rm turb} + \Gamma_{\rm X} + \Gamma_{\rm CR}\ .
\end{equation}
Photoelectric heating by UV photons within photodissociation regions is neglected (for a discussion of photodissociation regions
in NGC~1068 see Sect.~\ref{sec:pdr}).
The turbulent heating is
\begin {equation}
\Gamma_{\rm turb} = 0.3\, \rho\, \frac{|\vec{v}_{\rm turb}^{\rm cl}|^3}{l_{\rm cl}}\ ,
\label{eq:turbheat}
\end{equation}
(e.g., Mac Low 1999). 
Following Maloney et al. (1996) the X-ray heating rate at the cloud center is
\begin{multline}
\label{eq:xheat}
\Gamma_{\rm X} = 7 \times 10^{-22} \\ \big(\frac{L_{\rm X}}{10^{44}~{\rm erg\,s}^{-1}}\big) \big(\frac{R}{100~{\rm pc}}\big)^{-1} \big(\frac{N_{\rm H}}{10^{22}~{\rm cm}^{-2}} \big) n_{\rm H}~{\rm erg\,cm^{-3}s^{-1}}\ ,
\end{multline}
where $L_{\rm X}$ is the X-ray luminosity between 1 and 100~keV, and $N_{\rm H}=n_{\rm H} r_{\rm cl}$ is half the column density of the cloud.
Following Nelson \& Langer et al. (1997) we adopted the cosmic ray heating rate
\begin{equation}
\label{eq:crheat}
\Gamma_{\rm CR} = 6 \times 10^{-28} n_{\rm H} \big(\frac{\zeta_{\rm CR}}{2 \times 10^{-17}~{\rm s}^{-1}}\big)~{\rm erg\,cm^{-3}s^{-1}}\ ,
\end{equation}
where $\zeta_{\rm CR}$ is the cosmic ray ionization rate.

\subsection{Chemical network \label{sec:network}}

Chemical modelling was carried out using the {\tt Nautilus} gas-grain code
presented in detail in Hersant et al. (2009), Semenov et al. (2010), and Ruaud et al. (2015).
This code computes the abundances of chemical species (atoms and
molecules) as a function of time by solving the rate equations for a
network of reactions.
For gas-phase reactions, we used the kida.uva.2014 network
(Wakelam et al. 2015\footnote{the network is available online on the
KIDA website {\rm http://kida.astrophy.u-bordeaux.fr}}) comprising $489$
species and $6992$ reactions. For grain surface reactions, we used the
desorption, diffusion, activation barrier energies along with a set of
grain surface reactions, all from the KIDA database. Both, thermal and
non-thermal desorption processes are taken into account, the latter
consisting mainly of CR-induced desorption following the formalism
presented by Hasegawa \& Herbst (1993).

The model parameters are time, density, gas temperature, grain
temperature, UV flux, cosmic ray ionization rate, and the elemental
abundances of the elements C, O, N, and S (C/H$=1.7 \times 10^{-4}$, O/H$=2.4 \times 10^{-4}$, N/H$=6.2 \times 10^{-5}$, S/H$=10^{-6}$). 
At the beginning of the calculation all hydrogen is in molecular form.

Grids of models were obtained by varying the disk cloud lifetime (20 log spaced steps
between $10^3$ and $10^8$~yr), the cloud density (20 log spaced steps
between $10^3$ and $10^9$~cm$^{-3}$), and cloud gas temperatures (20 log
spaced steps between $10$ and $1000$~K). For each type of cloud, the CO, HCN, and HCO$^+$ abundances
were interpolated on the grid given the lifetime, density, and temperature of the cloud.

\subsection{The X-ray dominated region \label{sec:XDR}}

In the vicinity of the central engine, the interstellar medium is exposed to a tremendously strong UV and X-ray radiation field.
Both can ionize the gas and dissociate the molecules comprised within the gas. About $50$\,\% and $10$\,\% of the bolometric
luminosity are emitted in the UV and X-rays. Close to the central engine the dust is sublimated if it is heated to temperatures 
in excess of $\sim$1500~K. In NGC~1068 this is the case at a distance of $0.25$~pc (Gravity Collaboration 2020).
Inside this radius, the hydrogen can be still in molecular form if its column density is high enough to permit self-shielding.
Outside the dust sublimation radius the UV emission is absorbed by dust with an optical depth of
$\tau_{\rm UV} \sim N_{\rm H}/(10^{21}~{\rm cm}^{-2})$. The regions, which are directly illuminated by the central UV
emission, are called photodissociation regions (see, e.g., Tielens \& Hollenbach 1985). 
These regions generally include a hot ($T > 100$~K) atomic region near the surface with an extinction $A_{\rm V}<3$, a warm ($T \sim 100$~K) 
partially dissociated region at about $A_{\rm V} \sim 3-5$, and a cooler ($T < 100$~K) interior region at $A_{\rm V} \sim 10$ where
oxygen is still photodissociated to atomic form. Only the X-ray emission can penetrate into region of higher $A_{\rm V}$.
In these regions the X-rays dominate the heating of the gas and its chemistry through X-ray ionization and dissociation.
Following Maloney et al. (1996), the energy deposition rate per particle at the center of a gas cloud is
\begin{multline}
\label{eq:hx}
H_{\rm X}=7 \times 10^{-22} \\ (\frac{L_{\rm X}}{10^{44}~{\rm ergs\,s}^{-1}}) (\frac{R}{100~{\rm pc}})^{-2} (\frac{N_{\rm H}}{10^{22}~{\rm cm}^{-2}})^{-1}~{\rm ergs\,s}^{-1}\ ,
\end{multline}
where $L_{\rm X}$ is the X-ray luminosity and $N_{\rm H}$ the hydrogen column density.
In the low-ionization limit, the X-ray ionization rate is
\begin{multline}
\label{eq:zetax}
{\zeta_{\rm X}=1.4 \times 10^{-11}} \\ (\frac{L_{\rm X}}{10^{44}~{\rm ergs\,s}^{-1}}) (\frac{R}{100~{\rm pc}})^{-2} (\frac{N_{\rm H}}{10^{22}~{\rm cm}^{-2}})^{-1}~{\rm s}^{-1}\ .
\end{multline}
For the chemical network we replace the cosmic ray ionization rate by the X-ray ionization rate whenever the latter exceeds the former.

\begin{figure*}[!ht]
  \centering
  \resizebox{\hsize}{!}{\includegraphics{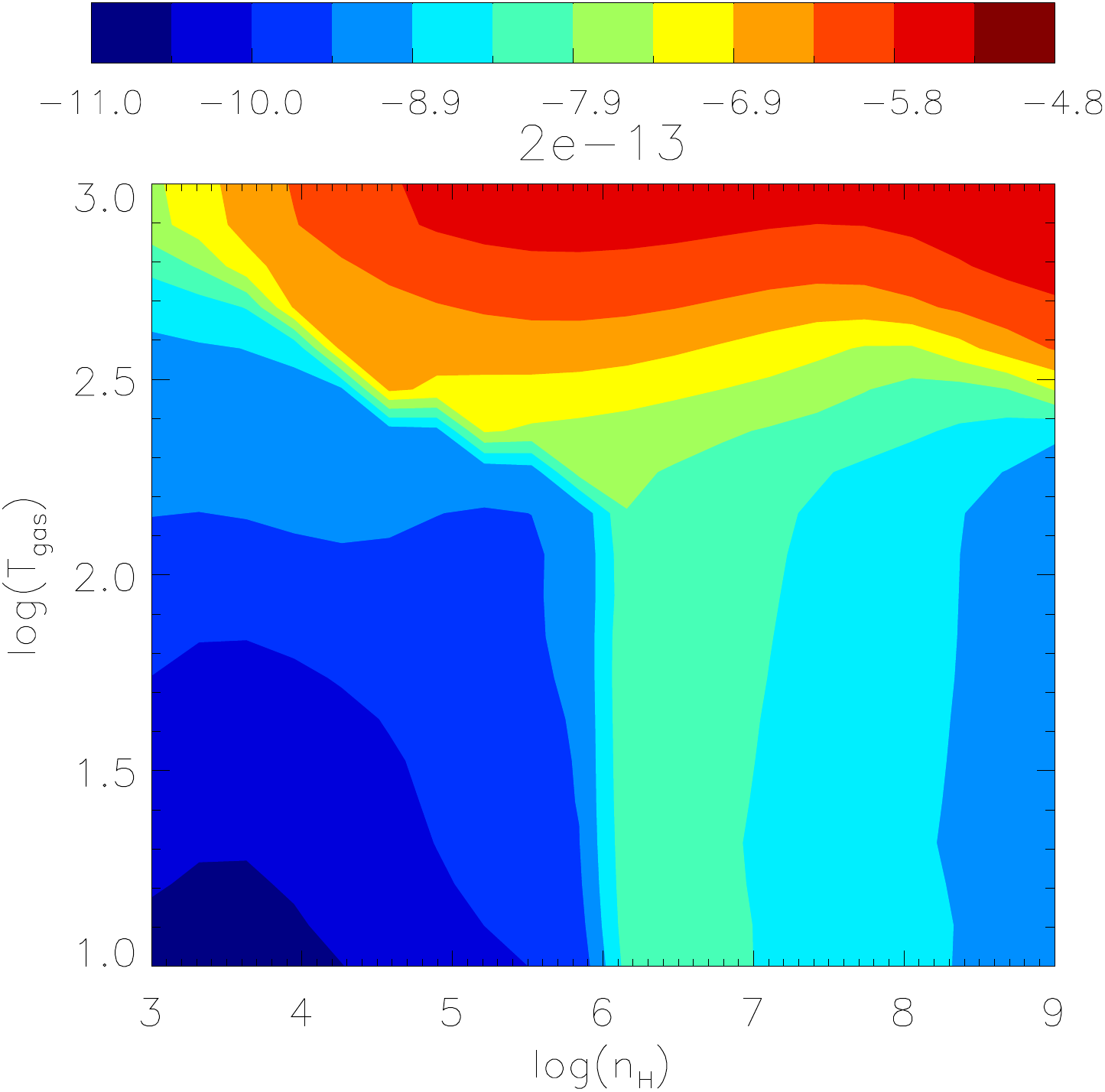}\includegraphics{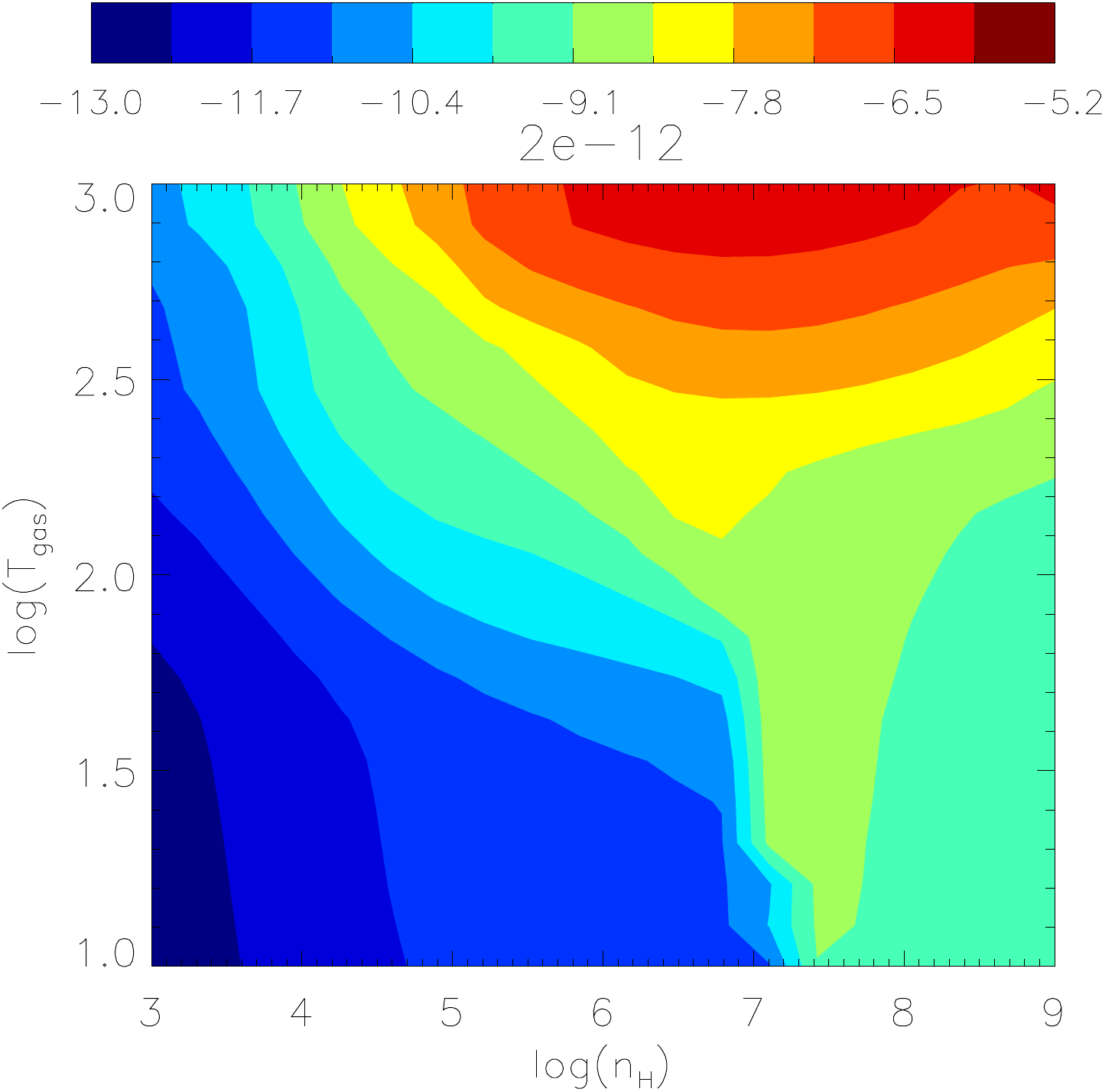}\includegraphics{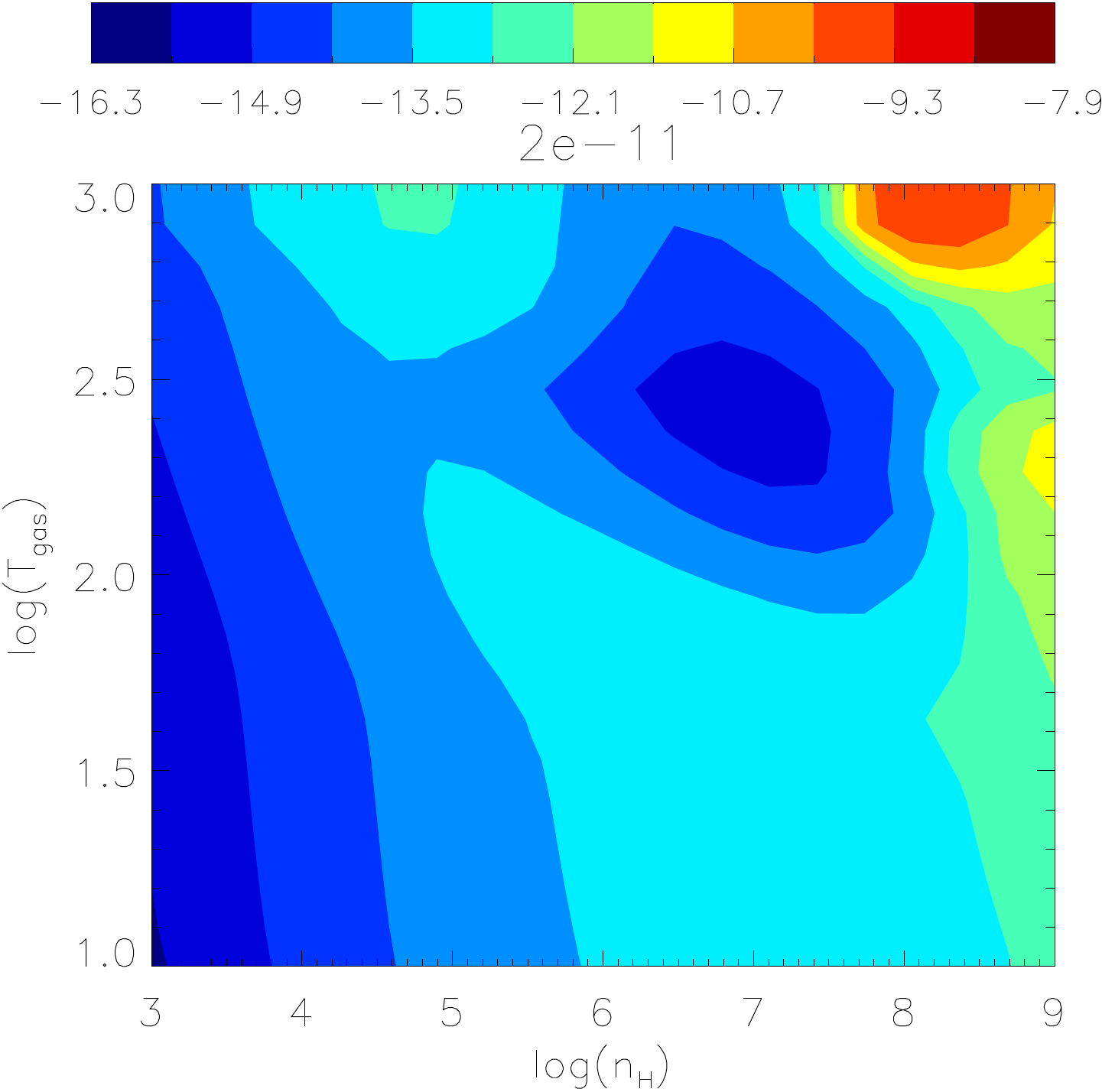}}
  \put(-500,0){\large (a)}
  \put(-320,0){\large (b)}
  \put(-150,0){\large (c)}
  \caption{HCN abundance $x_{\rm HCN}=n_{\rm HCN}/n_{\rm H}$ as a function of density and temperature for different cosmic ray ionization fractions.
    Panel (a) $\zeta_{\rm CR}=2 \times 10^{-13}$~s$^{-1}$, (b) $2 \times 10^{-12}$~s$^{-1}$, and (c) $2 \times 10^{-11}$~s$^{-1}$.
  \label{fig:chemistryhcn}}
\end{figure*}

For an X-ray luminosity of $L_{\rm X} \sim 5 \times 10^{43}$~ergs\,s$^{-1}$ in NGC~1068 (Bauer et al. 2015, Marinucci et al. 2016),
a distance of $3$~pc, and a column density of $N_{\rm H}=10^{24}~{\rm cm}^{-2}$, we obtain $\zeta_{\rm T}=8 \times 10^{-11}$~s$^{-1}$.
This is about a factor $4 \times 10^4$ higher than the cosmic ray ionization rate in the Circumnuclear disk in the Galactic Center
(Yusef-Zadeh et al. 2013, Harada et al. 2015). The Nautilus model HCN abundances as a function of the gas density and temperature are presented in Fig.~\ref{fig:chemistryhcn}
for ionization rates between $2 \times 10^{-15}$~s$^{-1}$ and $6 \times 10^{-11}$~s$^{-1}$.
It becomes clear that HCN abundances in excess of $x_{\rm HCN}=10^{-8}$ are absent for ionization rates that are higher than a few $10^{-12}$~s$^{-1}$.
The reason for this absence is the destruction of H$_2$ molecules through photodissociation, which blocks the chemical reactions
leading to the formation of the HCN molecule.
Thus, the observed high HCN abundances (Imanishi et al. 2020) cannot be reproduced with models including an
ionization rate which is expected if the gas is directly illuminated by the X-ray emission of the central engine. 
The same result is found for the HCO$^+$ and CO abundances.
Since an X-ray ionization rate of $\zeta_{\rm T} \ga 8 \times 10^{-11}$~s$^{-1}$ is expected, 
we conclude that the bulk of the central X-ray emission must be absorbed in a gas layer of column density $N_{\rm H} \ga 10^{25}$~cm$^{-2}$,
a Compton-thick gas layer. The same conclusion was reached by Burtscher et al. (2016) who stated that deviations from the Galactic $N_{\rm H}/A_{\rm V}$ 
can be simply explained by dust-free neutral gas within the broad-line region in some sources.

If the absorbing gas has a radial extent of $\sim 0.2$~pc, these column densities ($N_{\rm H}=\int_{R_1}^{R_2} \rho\,{\rm d}R$) 
can be reached in the midplane of the molecular gas distribution if the inner wall of a thick disk or ring is located 
at $R_1 \la 2.6$~pc for $Q=10$, $R_1 \la 1.5$~pc for $Q=30$, and $R_1 \la 0.7$~pc for $Q=100$.
Alternatively, the Compton-thick gas layer can be located inside the dust sublimation radius and therefore be dust-free.
The latter possibility is in agreement with the conclusions of Gravity Collaboration (2020).
These authors stated that large column densities ($\sim 10^{25}$~cm$^{-2}$) are found in the X-rays,
which may largely originate from dust-free plasma inside the dust sublimation radius at distances smaller than $\sim 0.25$~pc.
In the following we will assume that the X-ray emission is entirely absorbed by the dense gas inside or outside
the dust sublimation radius and that the X-ray emission does not play a role for the heating and ionization of the molecular gas.

\subsection{Molecular line emission \label{sec:molemission}}

We employ the emission line modelling used by Vollmer et al. (2017).
For simplicity we consider only a single collider (H$_2$).
We consider two-level molecular systems in which the level populations are determined by
a balance of collisions with H$_2$, spontaneous decay and line photon absorption, and stimulated emission with
$\tau > 1$. The molecular abundances were calculated using the chemical network (see Sect.~\ref{sec:network}).
For simplicity, we neglected the hyperfine structure of HCN. 
A more detailed description of our modelling is given in Appendix~\ref{sec:molli}.

\subsection{Model datacubes \label{sec:datacubes}}

A brightness temperature and a cloud surface area were assigned to each gas cloud particle of the dynamical simulations according to the
assumed gas density, temperature, velocity dispersion, and size obtained from the analytical model (Sects.~\ref{sec:anamodel}
to \ref{sec:molemission}). A first datacube with a voxel size of $0.01$~pc in the two spatial axes and a voxel size
of $10$ or $20$~km\,s$^{-1}$ according to the available observations was established and the gas clouds were placed into this
datacube according to their projected two-dimensional position and their radial velocity. Within this first datacube the vast
majority of the gas clouds is spatially resolved. A Gaussian line profile was applied to the clouds according to the analytical model.
The emission from gas clouds, which were hidden (spatially and by radial velocity) by other gas clouds whose optical depth exceeds unity,
was removed from the datacube. Each of the velocity channels was then convolved to the spatial resolution of the corresponding
observations and the datacubes were clipped with the rms of the available molecular line observations (see Table~\ref{tab:obstab} for NGC~1068).
Moment maps and position-velocity (pv) diagrams were calculated from the model datacubes.
\begin{table*}
      \caption{ALMA molecular line observations of NGC~1068}
         \label{tab:obstab}
      \[
       \begin{tabular}{lcccl}
        \hline
        Line & resolution & sensitivity & channel width & reference \\
         & & (mJy/beam) (K)& (km\,s$^{-1}$) & \\
        \hline  
        HCN(3-2) & $0.020'' \times 0.019''$ & $0.13$ $(5.9)$ & $10$ & Impellizzeri et al. (2019) \\
        HCN(3-2) & $0.02'' \times 0.02''$ & $0.2$ $(8.6)$ & $10$ & Imanishi et al. (2020) \\ 
        H$^{13}$CN(3-2) & $0.03'' \times 0.03''$ & $0.2$ $(4.0)$ & $10$ & Imanishi et al. (2020) \\
        HCO$^+$(3-2) & $0.02'' \times 0.02''$ & $0.2$ $(8.5)$ & $10$ & Imanishi et al. (2020) \\
        H$^{13}$CO$^+$(3-2) &  $0.03'' \times 0.03''$ & $0.2$ $(4.0)$ & $10$ & Imanishi et al. (2020) \\
        HCO$^+$(4-3) & $0.04'' \times 0.03''$ & $0.28$ $(2.2)$ & $20$ & Garcia-Burillo et al. (2019) \\
        CO(2-1) & $0.04'' \times 0.03''$ & $0.14$ $(2.7)$ & $20$ & Garcia-Burillo et al. (2019) \\
        CO(3-2) & $0.04'' \times 0.03''$ & $0.23$ $(2.0)$ & $20$ & Garcia-Burillo et al. (2019) \\
        CO(6-5) & $0.084'' \times 0.064''$ & $5.3$ $(2.5)$ & $10$ & Gallimore et al. (2016) \\
        HCN(4-3) & $0.15'' \times 0.11''$ & $1.1$ $(0.7)$  & $10$ & this work \\
        CN(3-2) & $0.16'' \times 0.15''$ & $0.5$ $(0.2)$ & $15$ & this work \\
        CS(7-6) & $0.15'' \times 0.11''$ & $0.4$ $(0.3)$ & $15$ & this work \\
	\hline
        \end{tabular}
      \]
\end{table*}

\section{A CND-like model \label{sec:cnd}}

\begin{figure}[!ht]
  \centering
  \resizebox{\hsize}{!}{\includegraphics{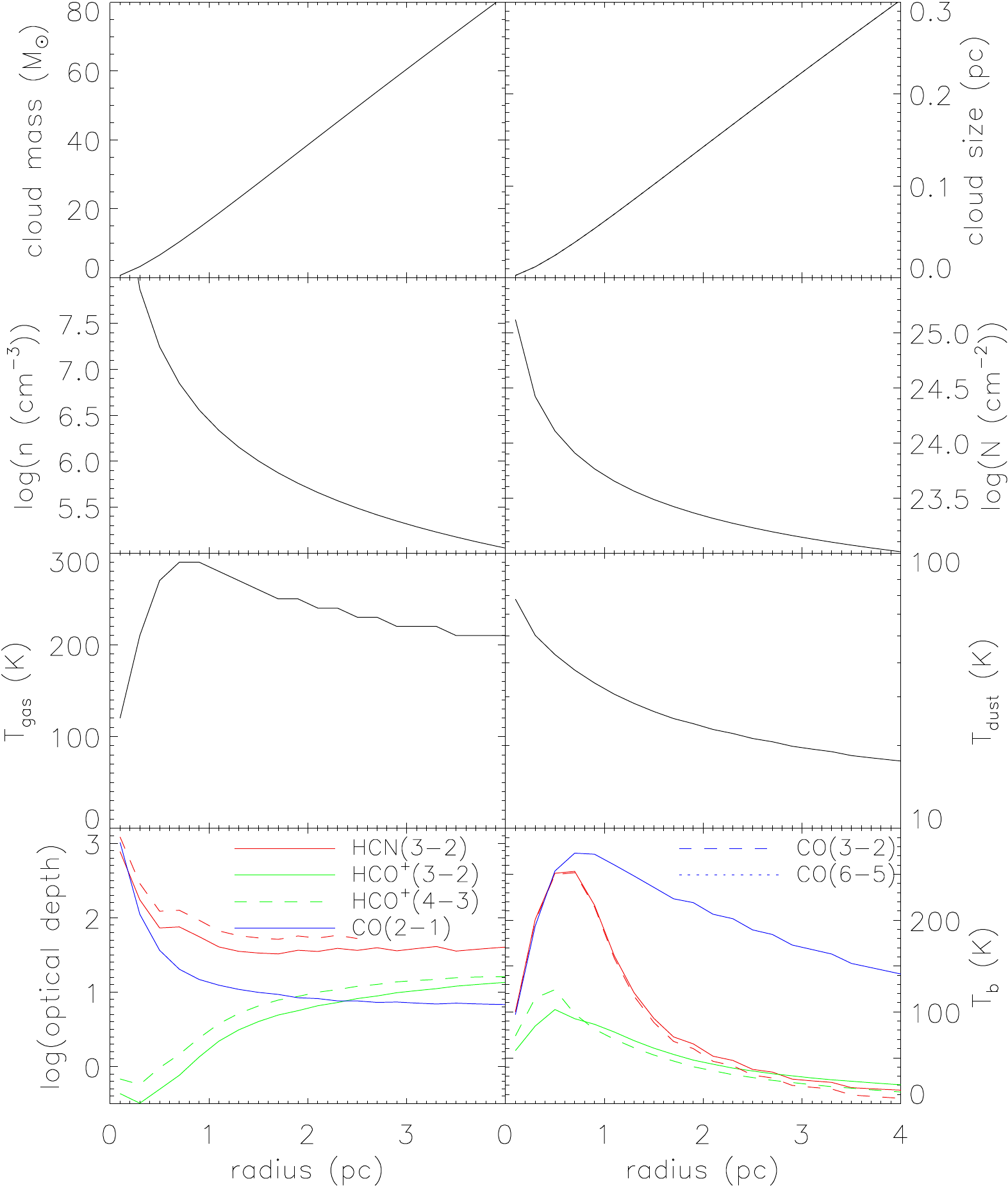}}
  \caption{Characteristics of individual model disk gas clouds across the CND at the time of interest. 
    From upper left to lower right: cloud mass, size, density, H$_2$ column density, gas and dust temperatures,
    optical depth, and brightness temperature.
  \label{fig:profiles_CND_plotall}}
\end{figure}

The underlying dynamical simulation is shown in Fig.~\ref{fig:sim_CNDCOLLdata2small1}.
We used a cosmic ray ionization rate  of $\zeta_{\rm CR}=10^{-15}$~s$^{-1}\,(R/1~{\rm pc})$, which is consistent with
the results of Yusef-Zadeh et al. (2013) and  Harada et al. (2015).
The radial dependence was introduced because the model with a constant cosmic ray ionization rate of $\zeta_{\rm CR}=2 \times 10^{-15}$~s$^{-1}$
lead to a too extended HCO$^+$ emission distribution. Thus, a centrally concentrated source of 
cosmic ray particles is assumed, most probably the colliding winds of the massive stars located within the inner half parsec of the Galaxy
(Paumard et al. 2006).
It was assumed that the relevant dust temperature for the molecular line emission is set by the heating through dust-gas collisions.
These dust temperatures correspond to the cold dust component discovered by Etxaluze et al. (2011).
In this way we neglected the molecular line emission from the photodissociation regions at the inner edge of the 
CND which is irradiated by the UV emission of early type stars located within the central cavity (Genzel et al. 2010).
The radial distribution of the disk cloud mass, size, density, H$_2$ column density, and gas and dust temperatures
are presented in Fig.~\ref{fig:profiles_CND_plotall}. At the inner edge of the CND ($R \sim 1$~pc)
the characteristics of the clouds are $M_{\rm cl} \sim 20$~M$_{\odot}$, $l_{\rm cl} \la 0.1$~pc,
$n_{\rm H_2} \sim 2 \times 10^6$~cm$^{-3}$, $N_{\rm H_2} \sim 4 \times 10^{23}$~cm$^{-2}$, $T_{\rm gas} \sim 300$~K,
and $T_{\rm dust} \sim 30$~K. The volume filling factor of the clouds is $\Phi_{\rm V}=0.035$. 

The resulting radial profiles of the CO(2-1), CO(3-2), CO(6-5), HCN(3-2), HCO$^+$(3-2), and HCO$^+$(4-3) optical depths and
brightness temperatures are shown in the two lower panels of Fig.~\ref{fig:profiles_CND_plotall}.
For $R > 1$~pc all lines are optically thick ($\tau_{\nu} > 1$). Whereas the optical depths of the CO and HCN lines decrease with increasing
radius, the opposite trend is observed for the HCO$^+$ lines. The HCO$^+$ optical depth is small for $R<1.5$~pc.
This is due to the low HCO$^+$ abundance, which is $x_{\rm HCO+} \sim 10^{-8}$ compared to $x_{\rm HCN} \sim 10^{-6}$ at $R=1$~pc.
The HCN brightness temperature profile show a
narrow peak around $R \sim 0.6$~pc with a peak brightness temperature of about $250$~K. The characteristics of these inner disk clouds
are similar to those of hot cores. Their high HCN brightness temperatures are due to their high HCN abundances, up to
$x_{\rm HCN} \sim 10^{-6}$, as observed in Galactic hot cores (Boonman et al. 2001, Rolffs et al. 2011).
The HCN brightness temperatures strongly decrease for radii larger than $\sim 1$~pc. 
The HCO$^+$ brightness temperatures decrease monotonically with increasing radius from $\sim 100$~K in
the center to $\sim 20$~K at $R=4$~pc.
The CO brightness temperature has a peak of $\sim 280$~K at $\sim 0.7$~pc and decreases monotonically
with increasing radius. This decrease is much shallower than that of the HCN lines.

As noted by Mills et al. (2013), the HCN line becomes a maser for temperatures higher than $T \sim 100$~K and column densities in excess 
of $N_{\rm HCN}=10^{16}$~cm$^{-2}$. These conditions are fulfilled by the disk clouds at $R \ga 2$~pc. The reason for the high
column density is the hot core chemistry which produces very high HCN abundances ($x_{\rm HCN} \sim 10^{-7}$-$10^{-6}$).
To avoid HCN maser emission, our model clouds should have an about five times lower HCN abundance. This can be achieved by about $20$\,\% 
lower gas temperatures ($T_{\rm gas} \sim 220$~K instead of $T_{\rm gas} \sim 260$~K) or an about two times longer turbulent lifetime of the clouds.
Most importantly, the resulting HCN brightness temperatures did not significantly change when these modifications were applied. 

\subsection*{CO(6-5)}

\begin{figure}[!ht]
  \centering
  \resizebox{\hsize}{!}{\includegraphics{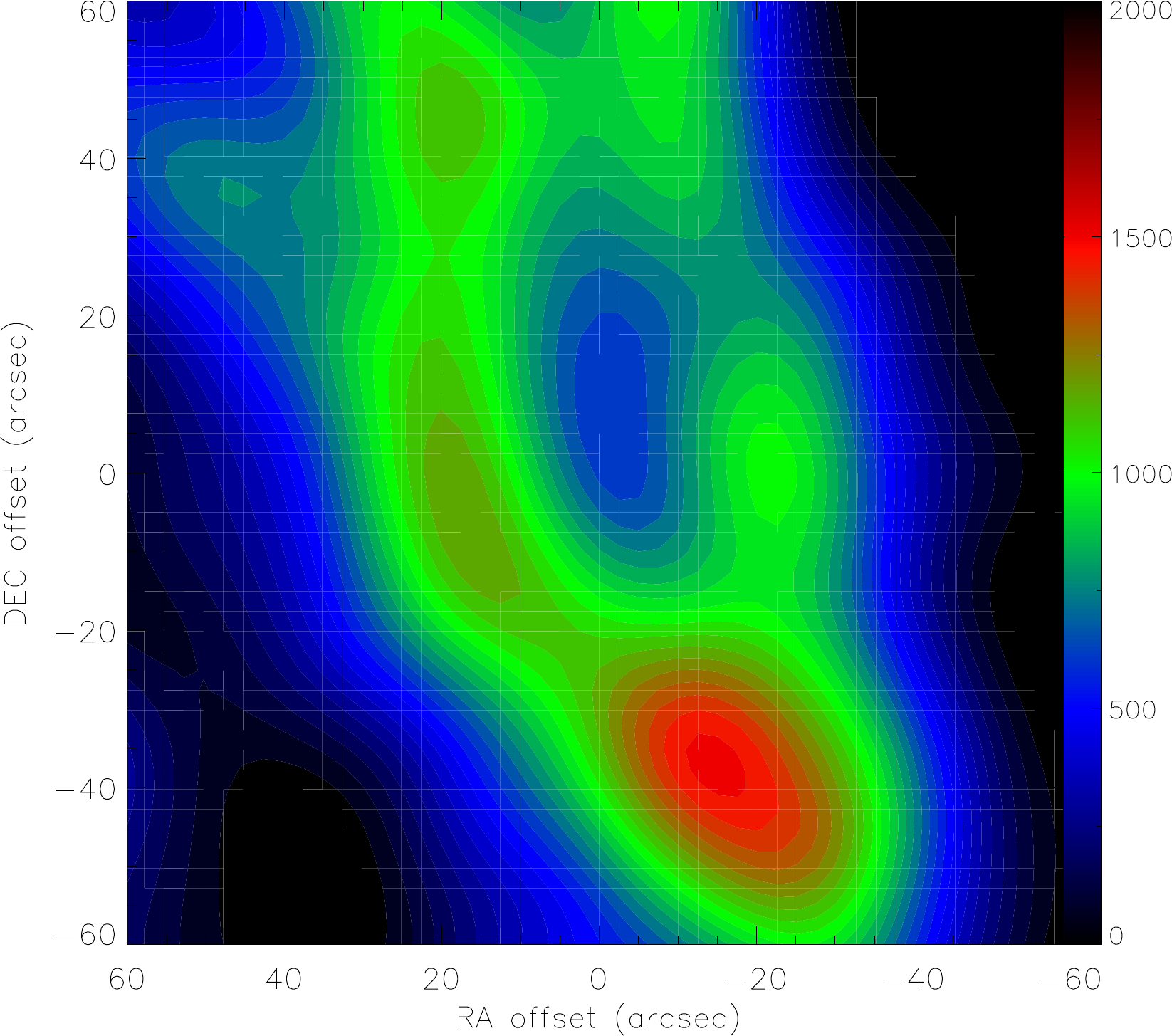}}
  \resizebox{\hsize}{!}{\includegraphics{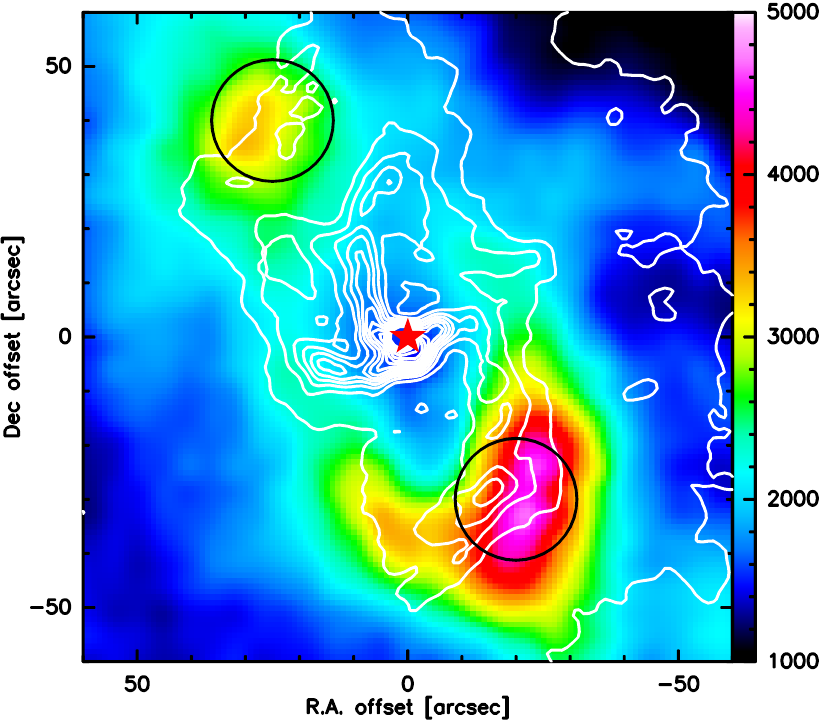}}
  \caption{CO(6-5) emission distribution. Upper panel: model; lower panel: APEX/CHAMP$^+$ observations integrated between
    $-150$ and $150$ ~km\,s$^{-1}$ (Requena-Torres et al. 2012). 
    The spatial resolution is $9.5''$. The flux is given in units of K\,km\,s$^{-1}$, in $T_{\rm mb}$.
  \label{fig:cndchamp}}
\end{figure}

\begin{figure}[!ht]
  \centering
  \resizebox{\hsize}{!}{\includegraphics{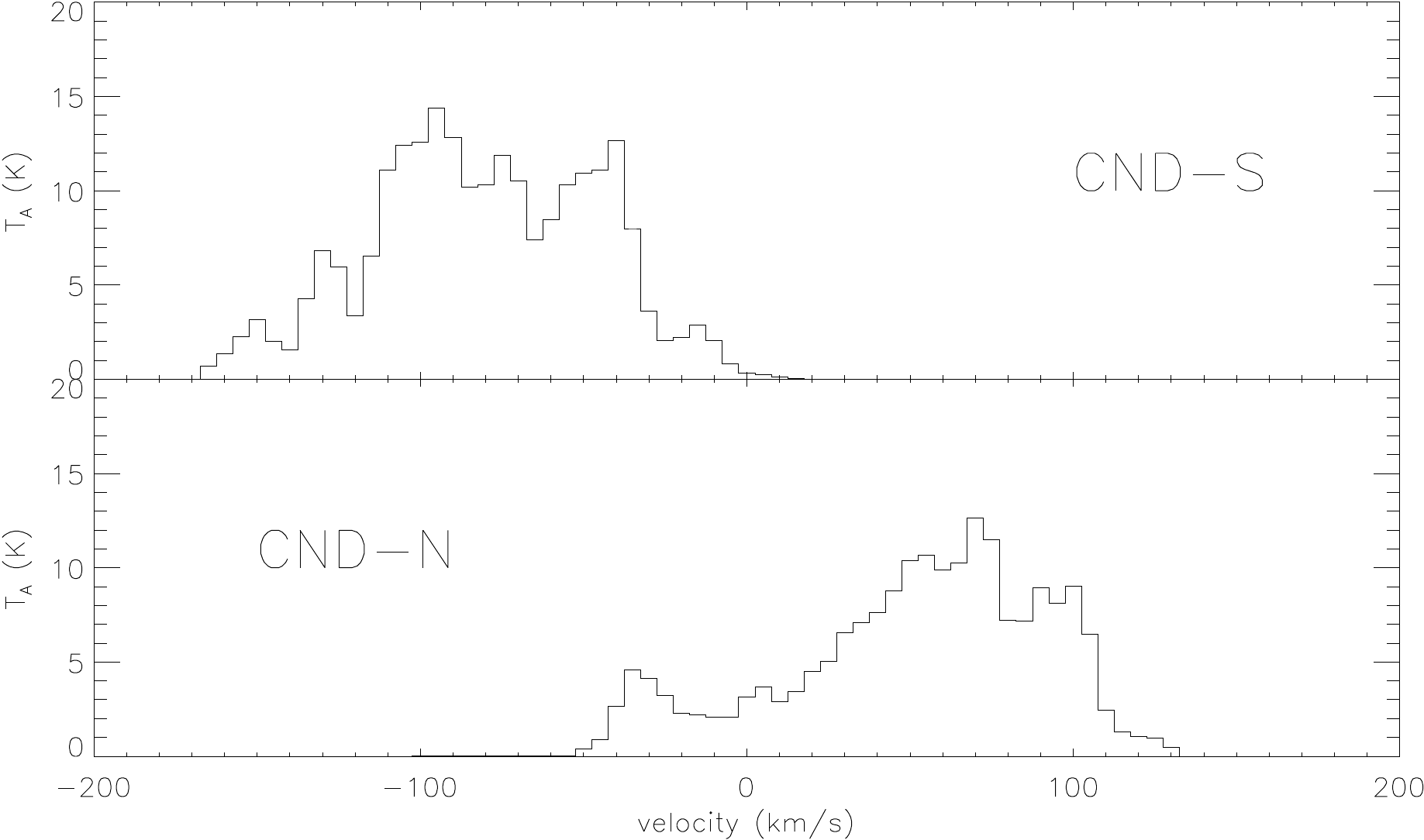}}
  \resizebox{\hsize}{!}{\includegraphics{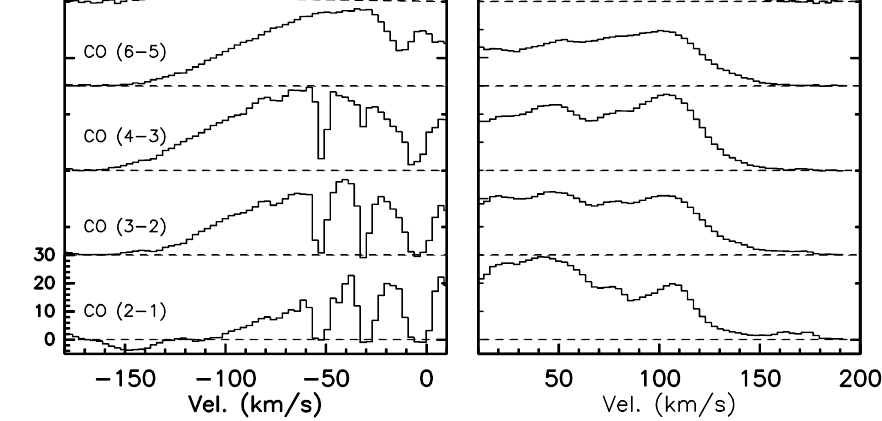}}
  \caption{CO spectra. Upper panel: CO(6-5) model spectra of the southern and northern lobes.
    Lower panel: CO spectra toward the southern (left) and northern (right) lobe of the CND (from Requena-Torres et al. 2012). 
    All spectra are convolved to the same angular resolution of $22.5''$. The flux density is given in units of K, in $T_{\rm mb}$.
  \label{fig:cndcoNS}}
\end{figure}

The comparison of the CO(6-5) moment~0 map with the observations of Requena-Torres et al. (2012) 
is presented in Fig.~\ref{fig:cndchamp}. The observed CO(6-5) map shows an inclined ring structure with two
prominent lobes. The southern lobe has an about two times higher flux than the northern lobe. 
The model CO(6-5) flux of the southern lobe is $1060$~K\,km\,s$^{-1}$ compared to the observed value of $1780$~K\,km\,s$^{-1}$,
that of the northern lobe (from $60$ to $150$~km\,s$^{-1}$) is $490$~K\,km\,s$^{-1}$ compared to the observed value of $790$~K\,km\,s$^{-1}$.
Thus, the flux ratio between the lobes is well reproduced by the model but the model fluxes are about $60$\,\% smaller than the
observed fluxes.
The inspection of the spectra of the northern and southern lobes (Fig.~\ref{fig:cndcoNS})
shows that the observed linewidths are comparable to those of the model. However, the relative emission at low velocities 
($\la -80$~km\,s$^{-1}$) is overestimated by the model by about a factor of two. The difference between the observed and model fluxes
in the lobes is mainly due to a difference in the CO(6-5) brightness temperatures, which is linked to the disk cloud densities, sizes,
and temperatures.

\subsection*{HCN(4-3) and CS(7-6)}

\begin{figure}[!ht]
  \centering
  \resizebox{\hsize}{!}{\includegraphics{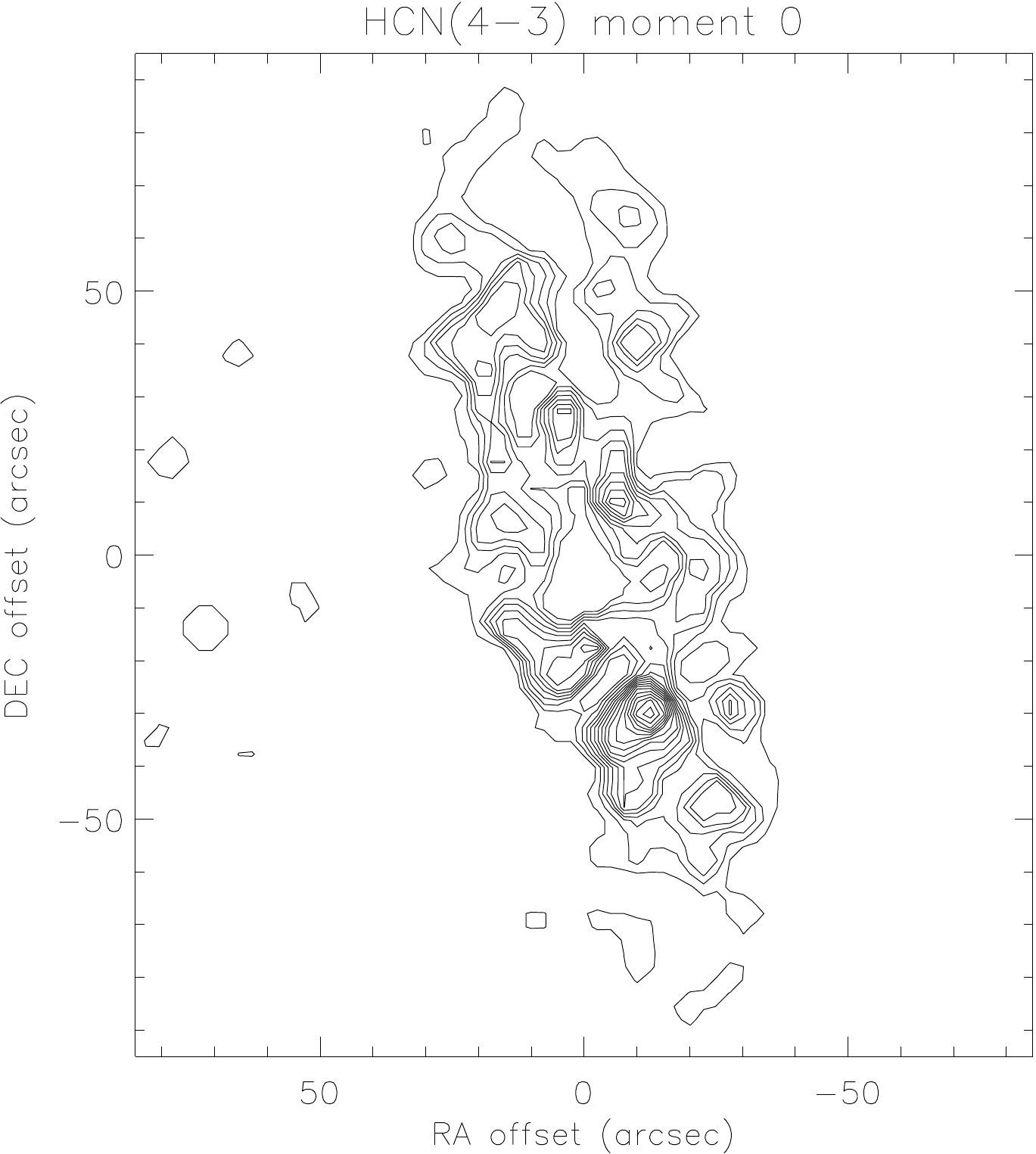}}
  \resizebox{\hsize}{!}{\includegraphics{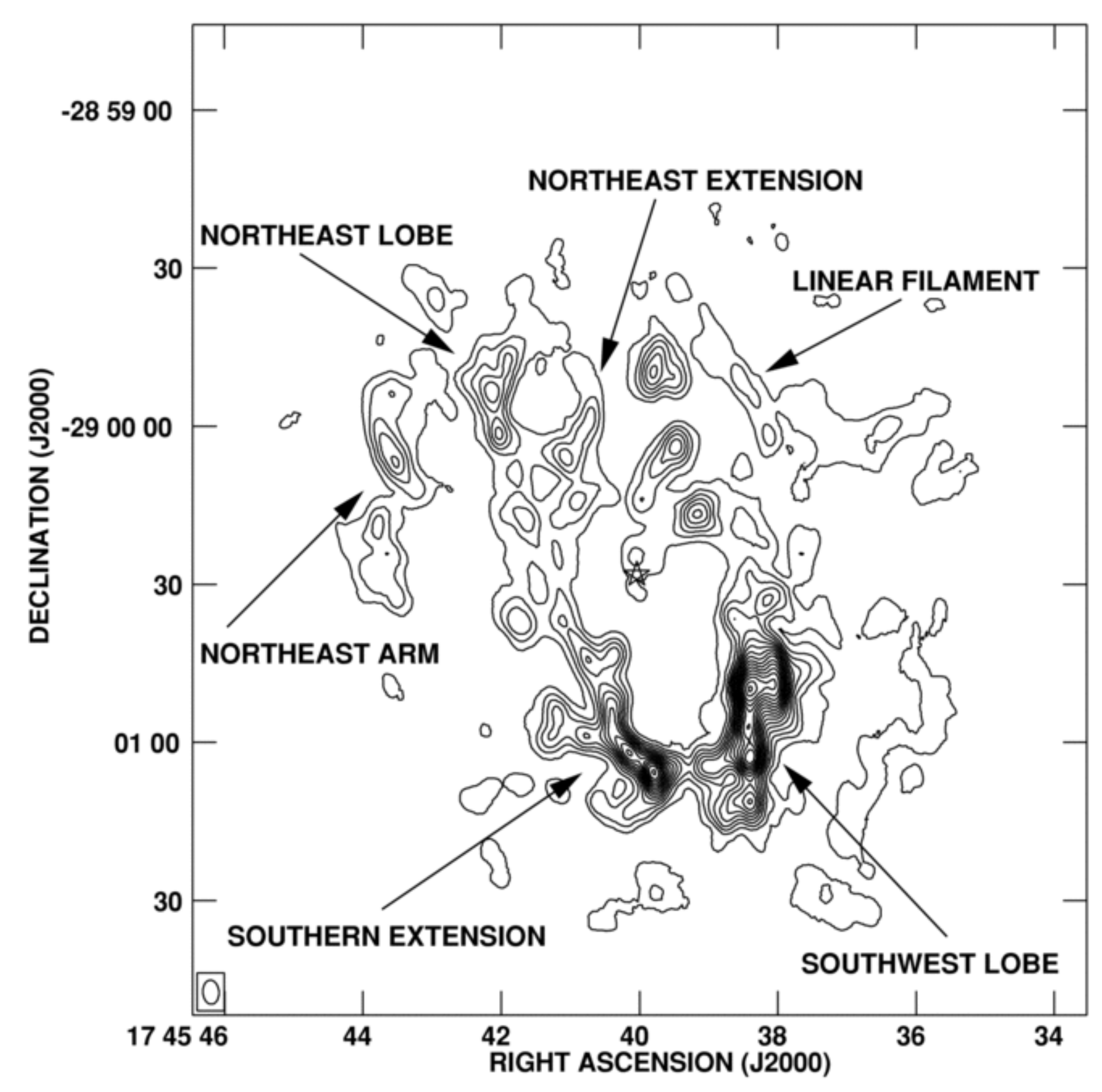}}
  \caption{HCN(4-3) integrated intensity map. Upper panel: model. 
    The contour levels are in steps of $53$~K\,km\,s$^{-1}$ from $27$ to $875$~K\,km\,s$^{-1}$
    The spatial resolution is $4.6'' \times 3.0''$.
    Lower panel: observations from Montero-Casta\~{n}o et al. (2009). The rms is $5.9$~K\,km\,s$^{-1}$.
    The contour levels are in steps of $35$~K\,km\,s$^{-1}$ from $18$ to $550$~K\,km\,s$^{-1}$, but for the highest contour level
    at $574$~K\,km\,s$^{-1}$. Sgr A$^*$ is marked with a star.
  \label{fig:cndsma}}
\end{figure}

Whereas the CO abundance in dense gas clouds is never significantly different from the canonical value of $x_{\rm CO} \sim 10^{-4}$
(e.g., Lacy et al. 1994), the HCN and CS abundances
depend on the gas chemistry. The comparison between of the model HCN(4-3) and CS(7-6) maps thus tells us if
our model is not only able to reproduce the gas densities, surface densities, velocity dispersions, and temperatures but also 
the gas chemistry. The comparison of the HCN(4-3) moment~0 map with the observations of Montero-Casta\~{n}o et al. (2009) is 
shown in Fig.~\ref{fig:cndsma}. {\tt Nautilus} yields HCN abundances of $x_{\rm HCN} \sim 10^{-6}$ at a radius of $1$~pc and
$x_{\rm HCN} \sim 10^{-7}$ at $2$~pc. The HCN abundance is thus significantly enhanced compared to typical values in dense
gas clouds ($x_{\rm HCN} \sim 10^{-8}$; e.g. Hirota \& Yamamoto 1998) due to the high density and temperatures that lead to hot core physics.
The HCO$^+$ abundance of several $10^{-8}$ is much lower than the HCN abundance in this region.
The model gas ring is about $30$\,\% larger and somewhat more 
inclined than the observed CND.
The bulk of the observed HCN(4-3) emission is confined to radii between $1$ ($25''$) and $2$~pc ($50''$) and the southern
lobe, which shows an arc structure, is more prominent than the northern lobe. The maximum model HCN(4-3) integrated intensity 
is about $50$\,\% higher than the maximum observed integrated intensity.
The observed clumpy structure of the CND is well reproduced by the model. 
Moreover, the observed arc structure of the southern lobe is much less prominent in the model.
The observed faint HCN(4-3) emission east and west of the inner ring is not present in the model.

\begin{figure}[!ht]
  \centering
  \resizebox{\hsize}{!}{\includegraphics{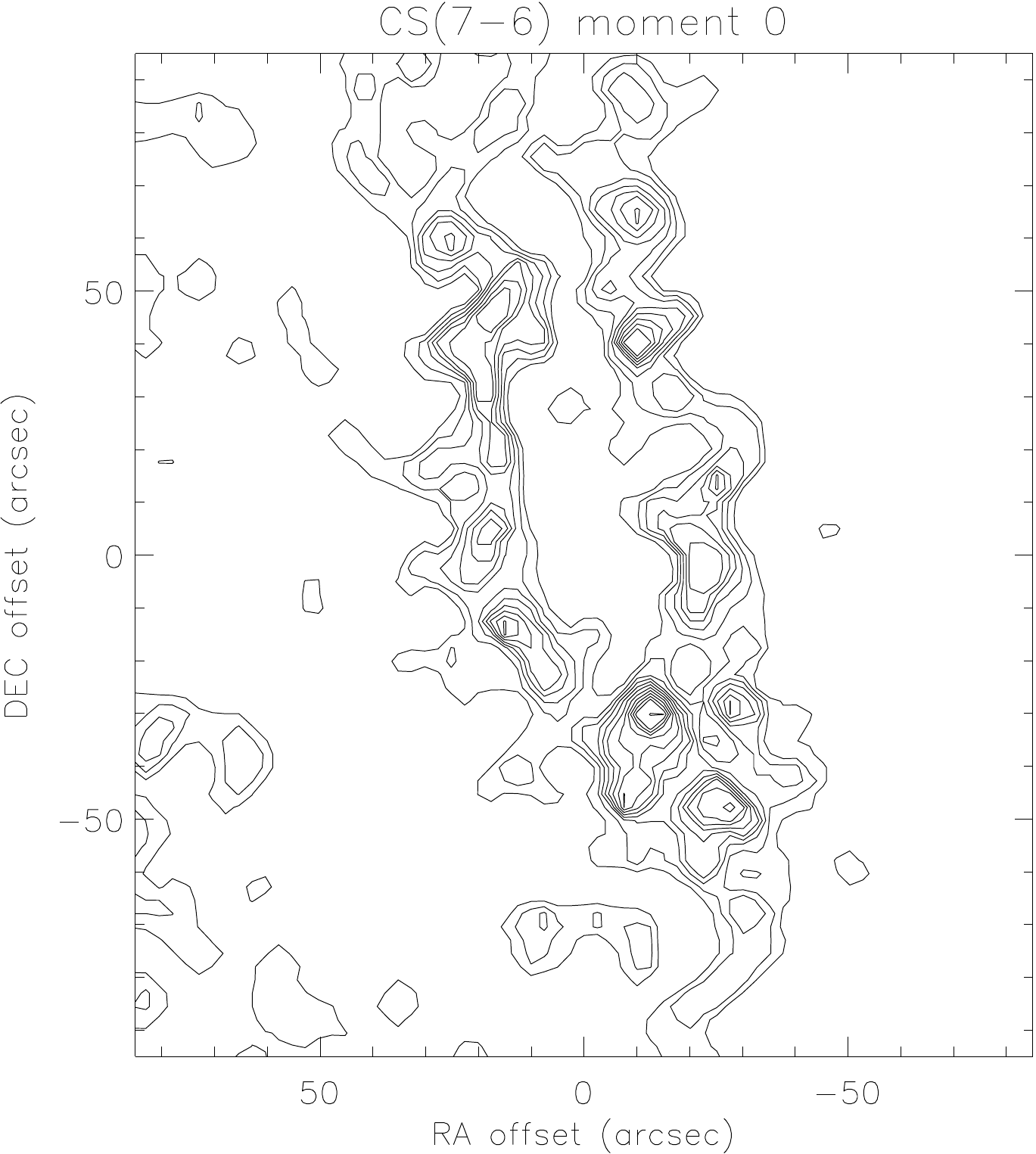}}
  \resizebox{\hsize}{!}{\includegraphics{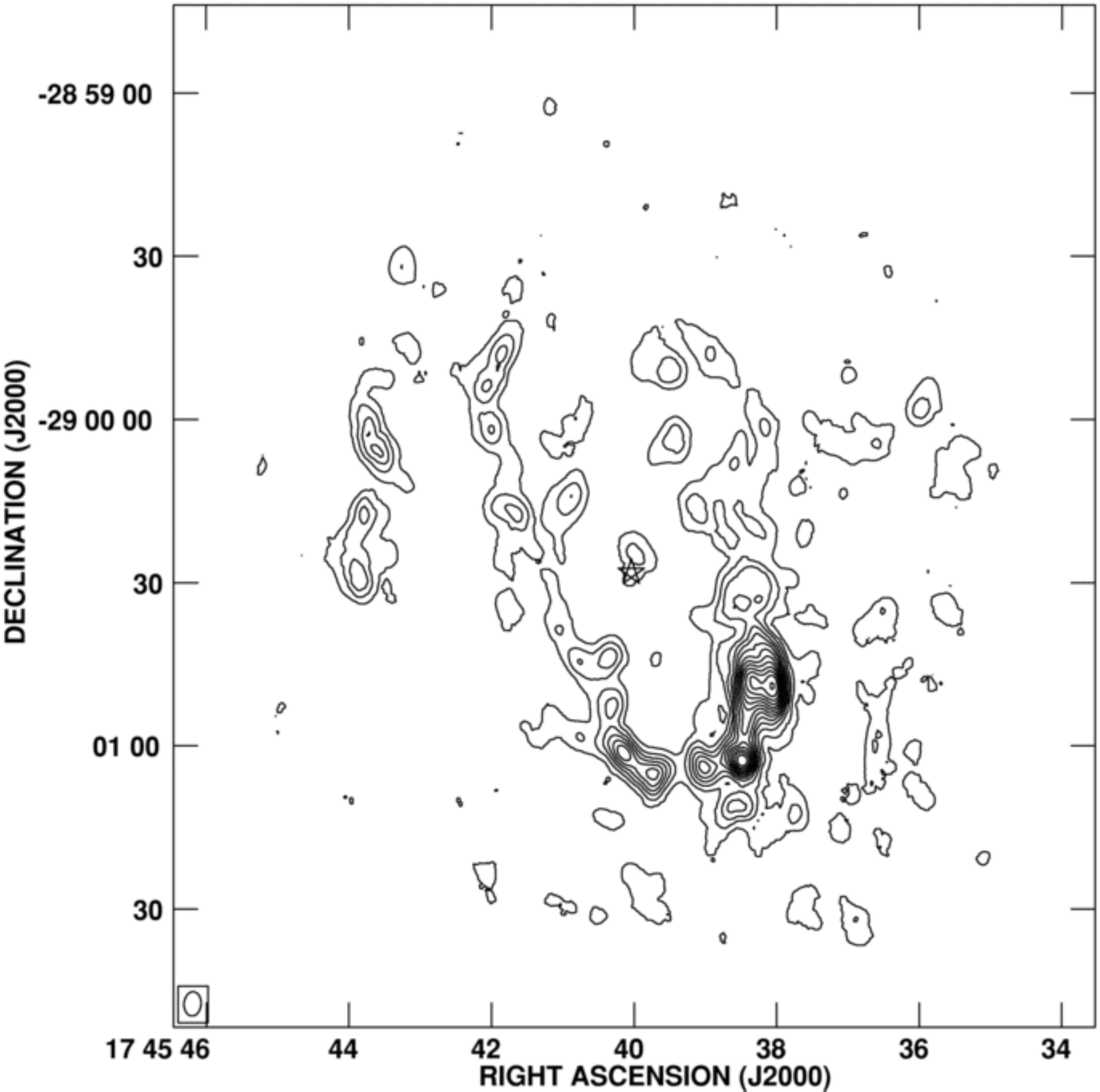}}
  \caption{CS(7-6) integrated intensity map. Upper panel: model. The contour levels are in steps of $22$~K\,km\,s$^{-1}$ from 
    $11$ to $275$~K\,km\,s$^{-1}$. The spatial resolution is $4.5'' \times 3.1''$.
    Lower panel: observations from Montero-Casta\~{n}o et al. (2009). The rms is $2.5$~K\,km\,s$^{-1}$.
    The contour levels are in steps of $15$~K\,km\,s$^{-1}$ from $7$ to $170$~K\,km\,s$^{-1}$, but for the highest contour level
    at $190$~K\,km\,s$^{-1}$. Sgr A$^*$ is marked with a star.
  \label{fig:cndsmacs}}
\end{figure}

The comparison of the CS(7-6) moment~0 map with the observations of Montero-Casta\~{n}o et al. (2009)
is shown in Fig.~\ref{fig:cndsmacs}. The CS abundance increases from $x_{\rm CS}=6 \time 10^{-10}$ at a radius of $1$~pc
to  $x_{\rm CS}=1.5 \time 10^{-9}$ at $2$~pc.
The maximum model CS(7-6) integrated intensity is about $35$\,\% higher than the maximum observed integrated intensity.
The observed north-south asymmetry is not present in the model. 
The southern and northern extensions beyond offsets of $\pm 50''$ in the model moment~0 map are not observed. 
As for the HCN(4-3) moment~0 map, the observed clumpy structure of the CND
is well reproduced by the model. Hsieh et al. (2021), who derived CS column densities from ALMA multi-transition CS observations,
found $N_{\rm CS} =10^{15}$-$10^{16}$~cm$^{-2}$. Our model CS column densities of $N_{\rm CS} \sim 1$-$3 \times 10^{15}$~cm$^{-2}$ are situated
at the lower end but are comparable to the observed distribution. Our assumed sulfur abundance of S/H$=10^{-6}$ is therefore justified.

\subsection*{HCN(3-2), HCN(4-3), HCO$^+$(3-2), and HCO$^+$(4-3)}

\begin{figure*}[!ht]
  \centering
  \resizebox{\hsize}{!}{\includegraphics{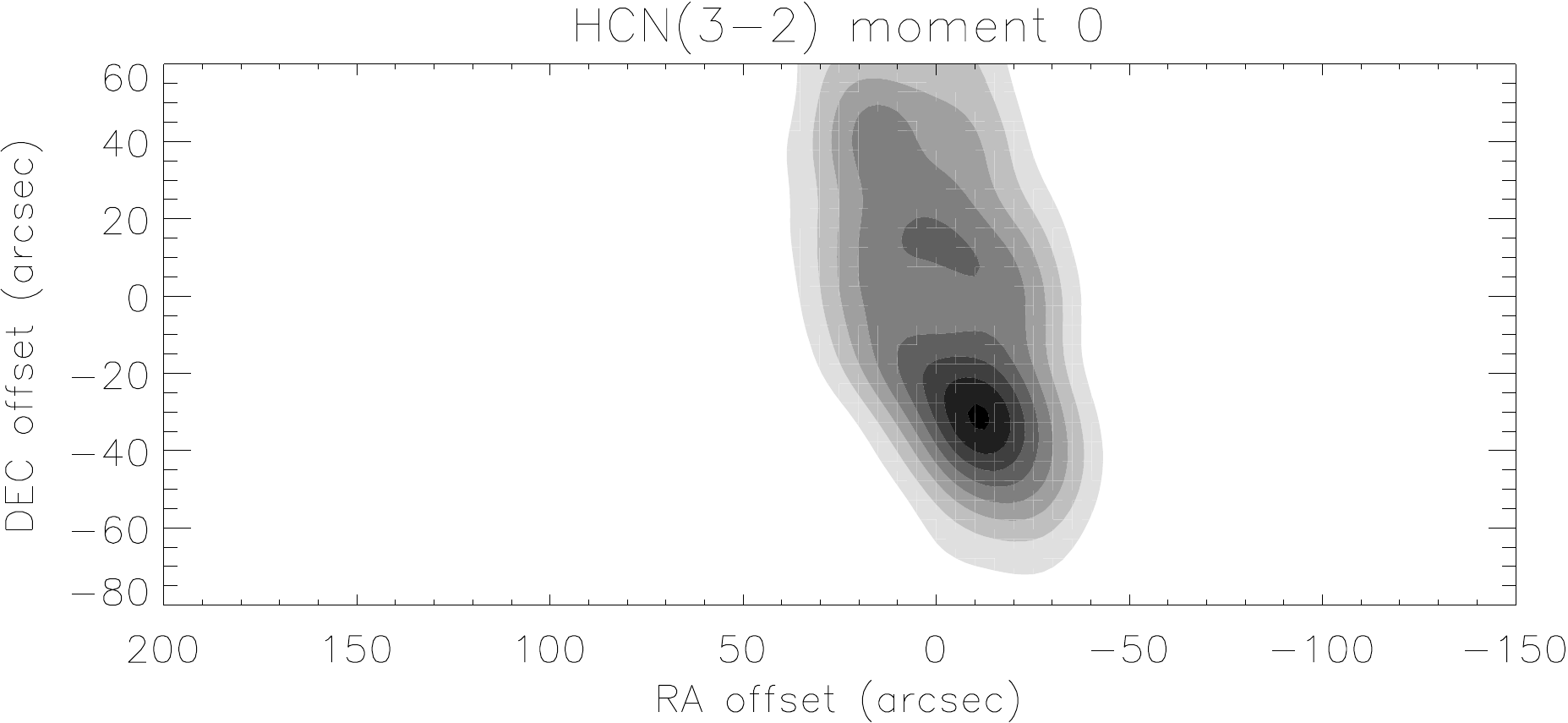}\includegraphics{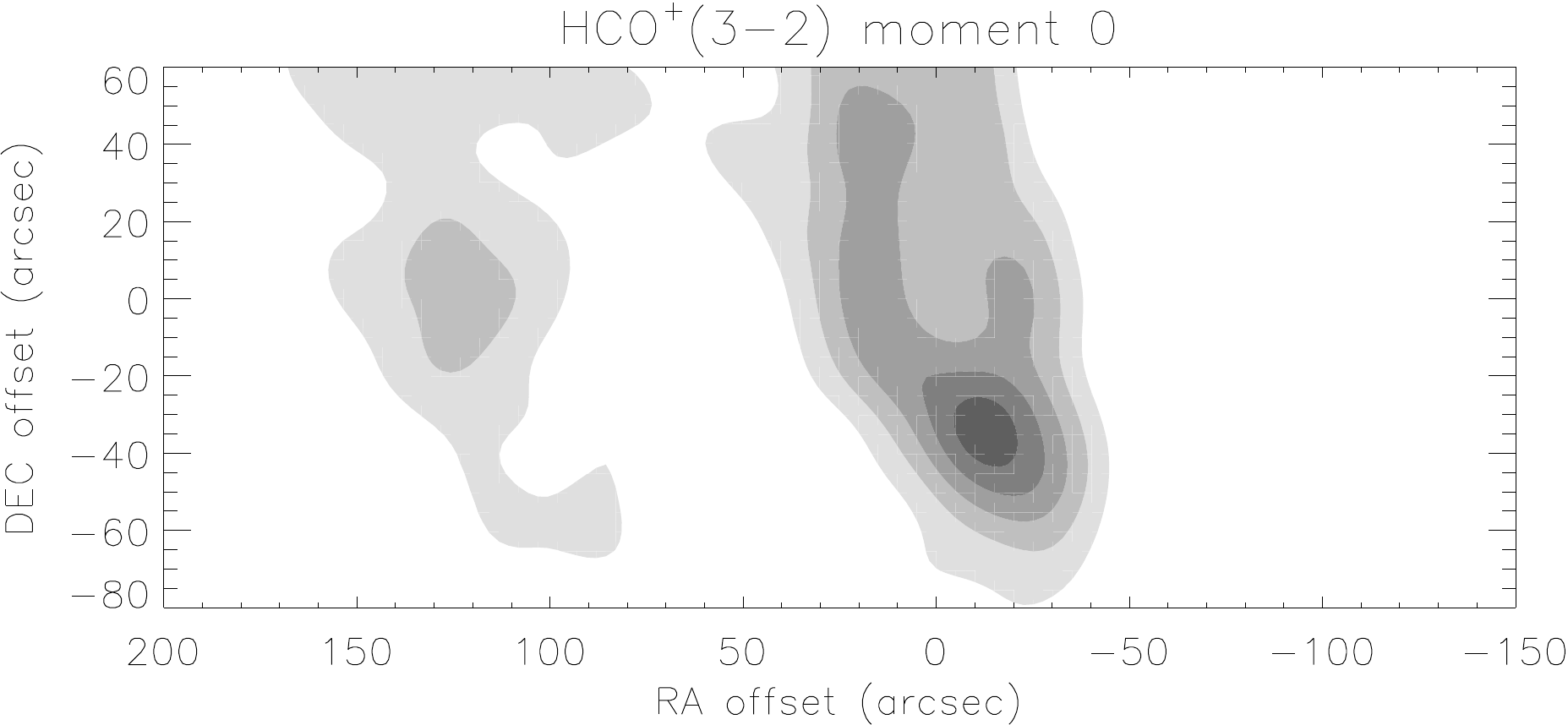}\put(-1000,200){\Huge (a)}\put(-470,200){\Huge (b)}}
  \resizebox{\hsize}{!}{\includegraphics{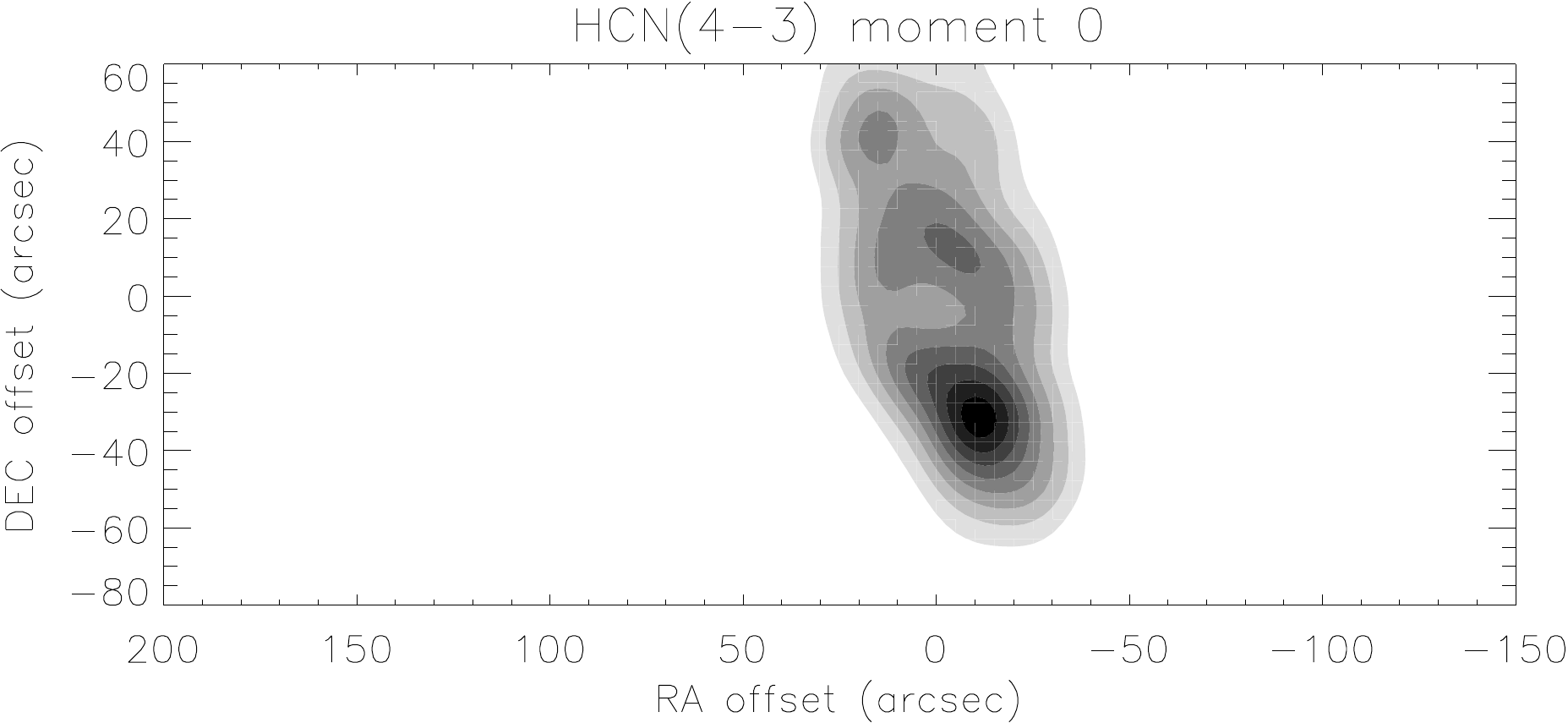}\includegraphics{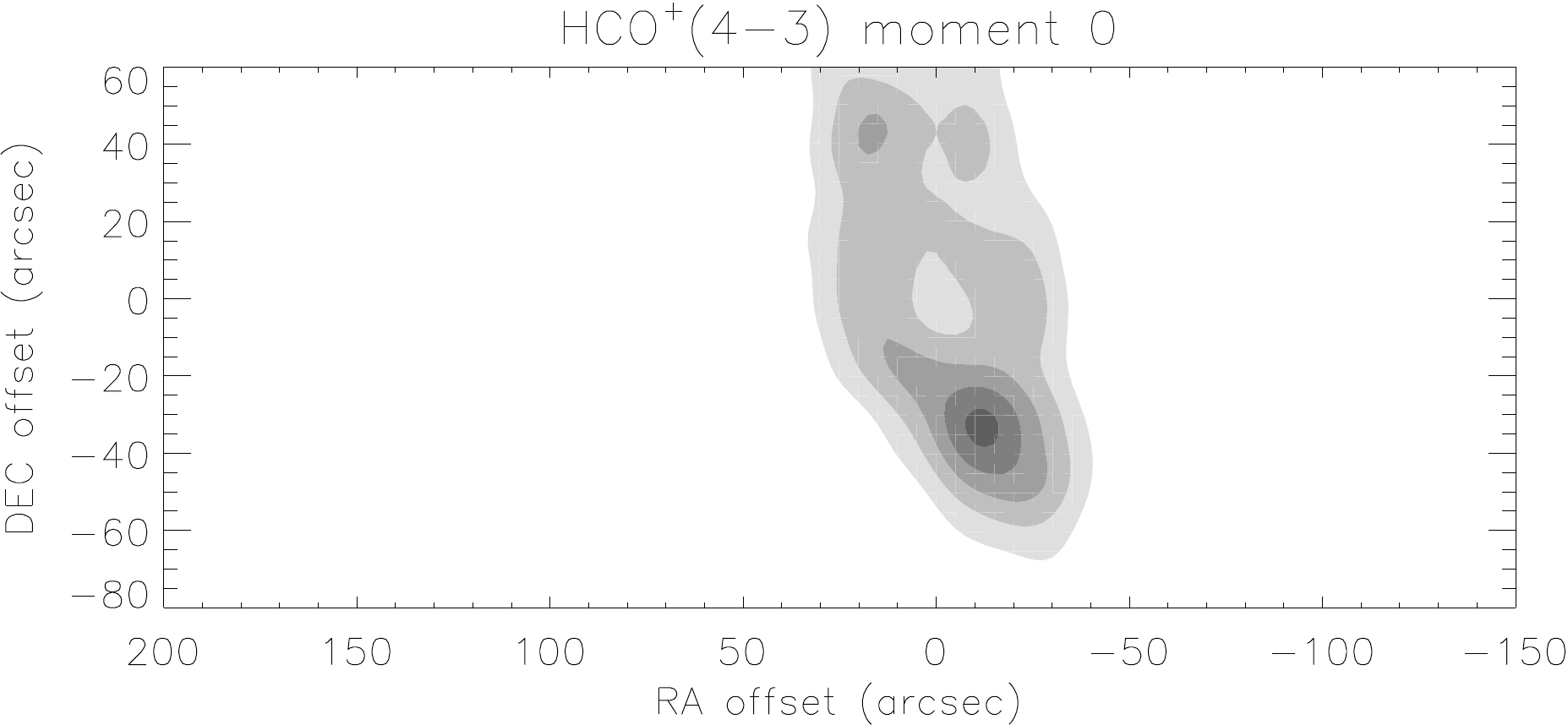}\put(-1000,200){\Huge (c)}\put(-470,200){\Huge (d)}}
  \resizebox{\hsize}{!}{\includegraphics{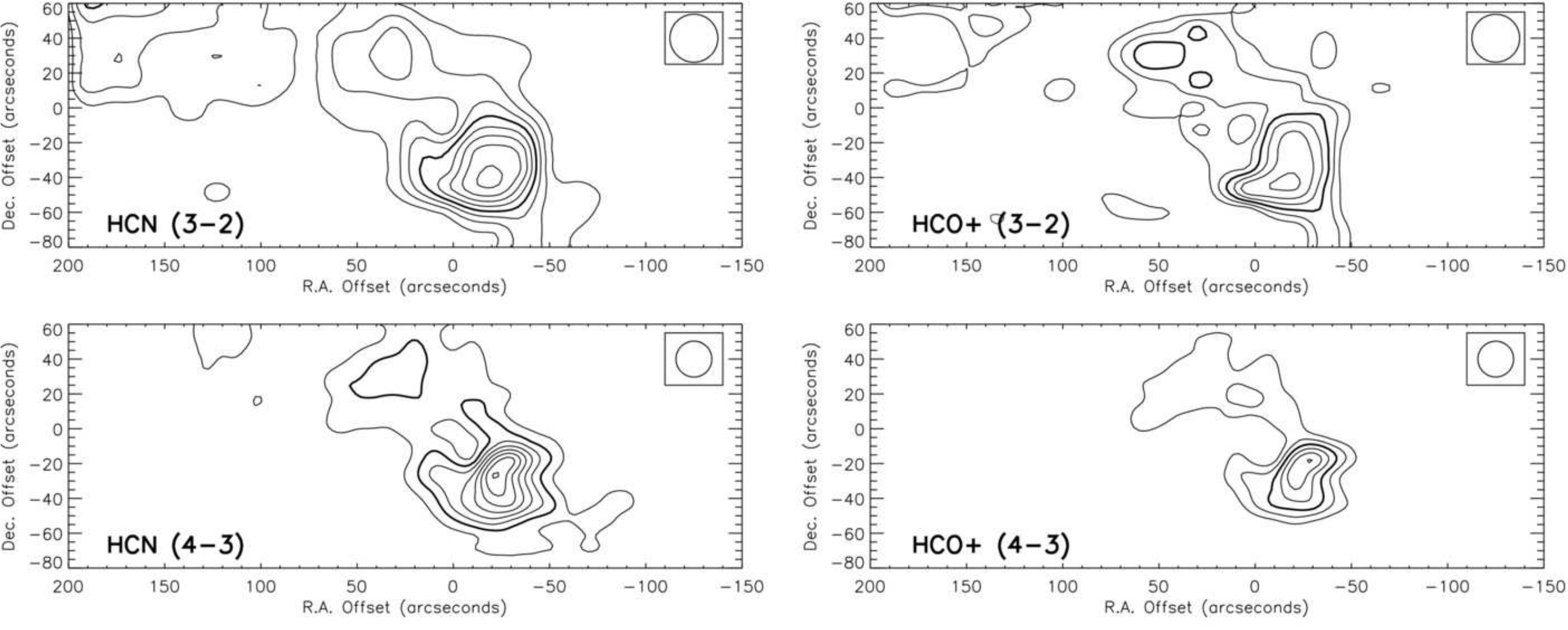}\put(-720,280){\huge (e)}\put(-330,280){\huge (f)}\put(-720,130){\huge (g)}\put(-330,130){\huge (h)}}
   
  \caption{HCN(3-2), HCN(4-3), HCO$^+$(3-2), and HCO$^+$(4-3) moment~0 maps. Panels (a)-(d): model.
    Contour levels are from $100$ to $500$~K\,km\,s$^{-1}$, spaced by $50$~K\,km\,s$^{-1}$.
    Panels (e)-(h): observations (Mills et al. 2013). The spatial resolutions are: HCN(3-2): $23.6''$,
    HCN(4-3): $17.7''$, HCO$^+$(3-2): $23.5''$, HCO$^+$(4-3): $17.6''$.
    Contours are linearly spaced. (e) HCN(3-2), contours from $75$ to $265$~K\,km\,s$^{-1}$, spaced by $27$~K\,km\,s$^{-1}$. 
    (f) HCO$^+$(3-2), contours from $58$ to $140$~K\,km\,s$^{-1}$, spaced by $17$~K\,km\,s$^{-1}$. (g) 
    HCN(4-3), contours from $60$ to $262$~K\,km\,s$^{-1}$, spaced by $29$~K\,km\,s$^{-1}$. (h) 
    HC0$^+$(4-3), contours from $58$ to $140$~K\,km\,s$^{-1}$, spaced by $17$~K\,km\,s$^{-1}$.
    \label{fig:cndapex}}
\end{figure*}

Mills et al. (2013) observed the CND with APEX in the HCN(3-2), HCN(4-3), HCO$^+$(3-2), and HCO$^+$(4-3) lines (lower four
panels of Fig.~\ref{fig:cndapex}).
They detected strong line emission from the southern lobe and rather faint line emission from the northern lobe.
The observed mean HCN/HCO$^+$ line ratio is about $1.7$-$1.8$.
The corresponding model maps are presented in the four upper panels of Fig.~\ref{fig:cndapex}.
The model HCN and HCO$^+$ surface brightnesses of the two lobe are about a factor of two higher than the observed 
surface brightnesses.
The model HCN/HCO$^+$ line ratio is about $1.3$-$1.4$ there, about $25$\,\% lower than the observed value.

\begin{figure*}[!ht]
  \centering
  \resizebox{\hsize}{!}{\includegraphics{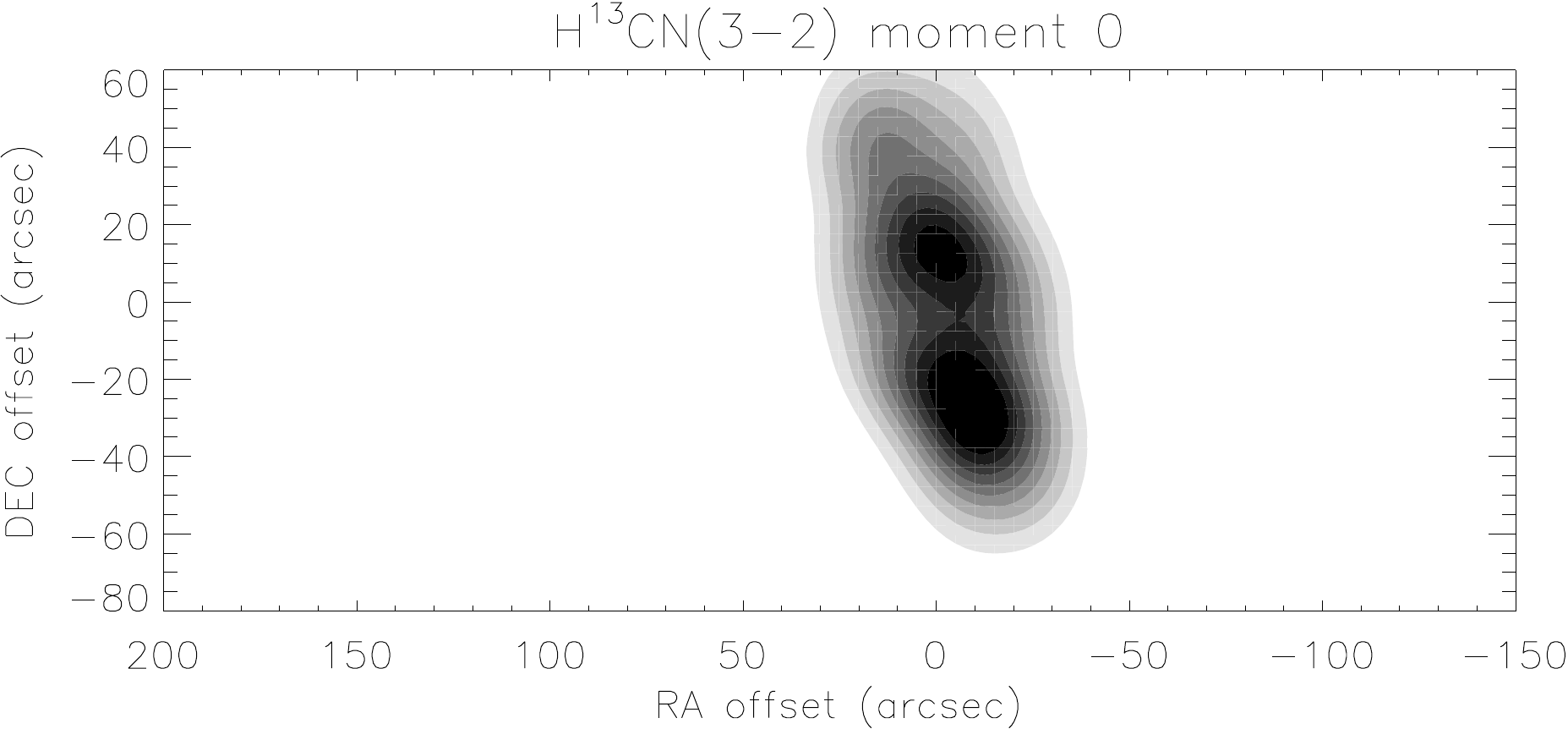}\includegraphics{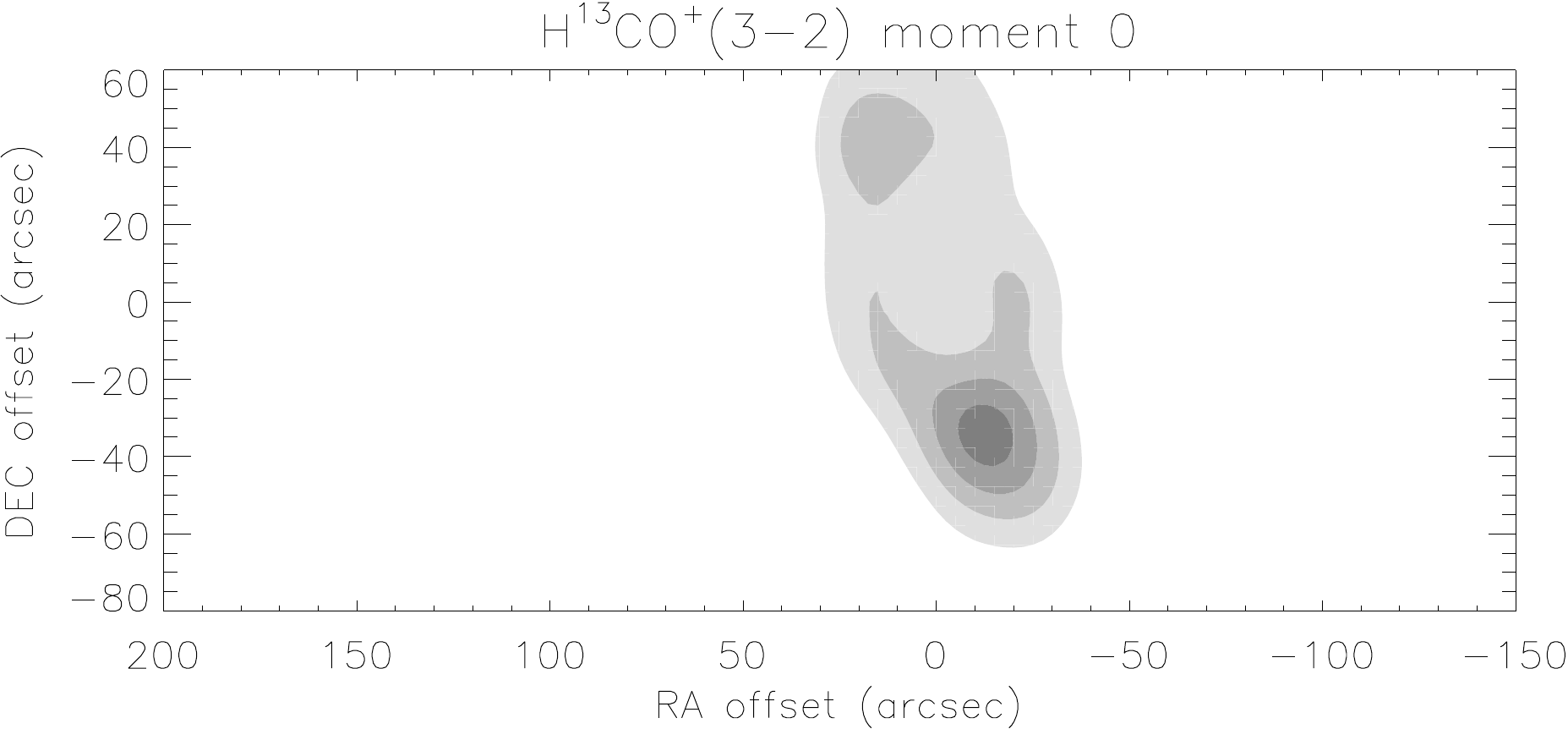}\put(-1000,200){\Huge (a)}\put(-470,200){\Huge (b)}}
  \resizebox{\hsize}{!}{\includegraphics{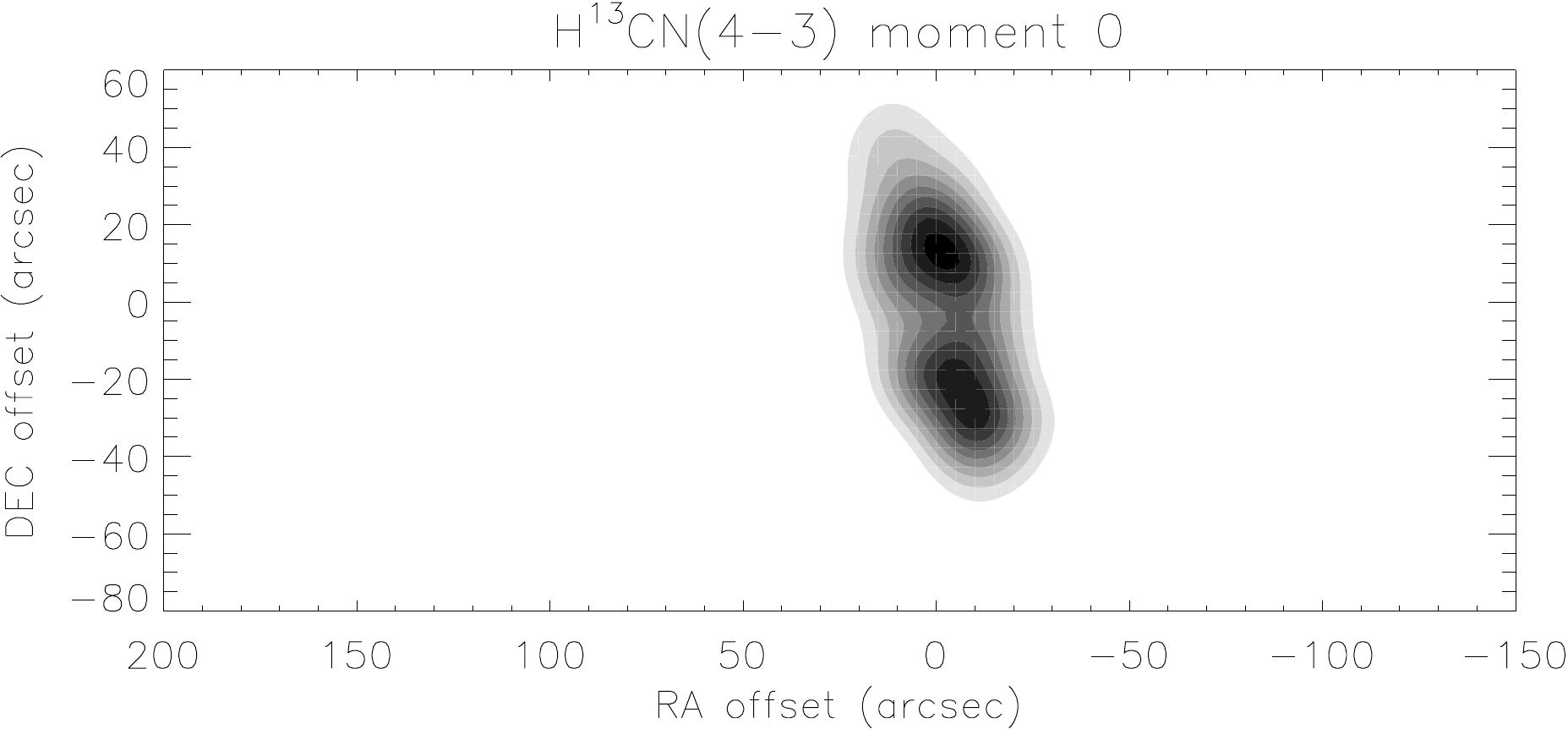}\includegraphics{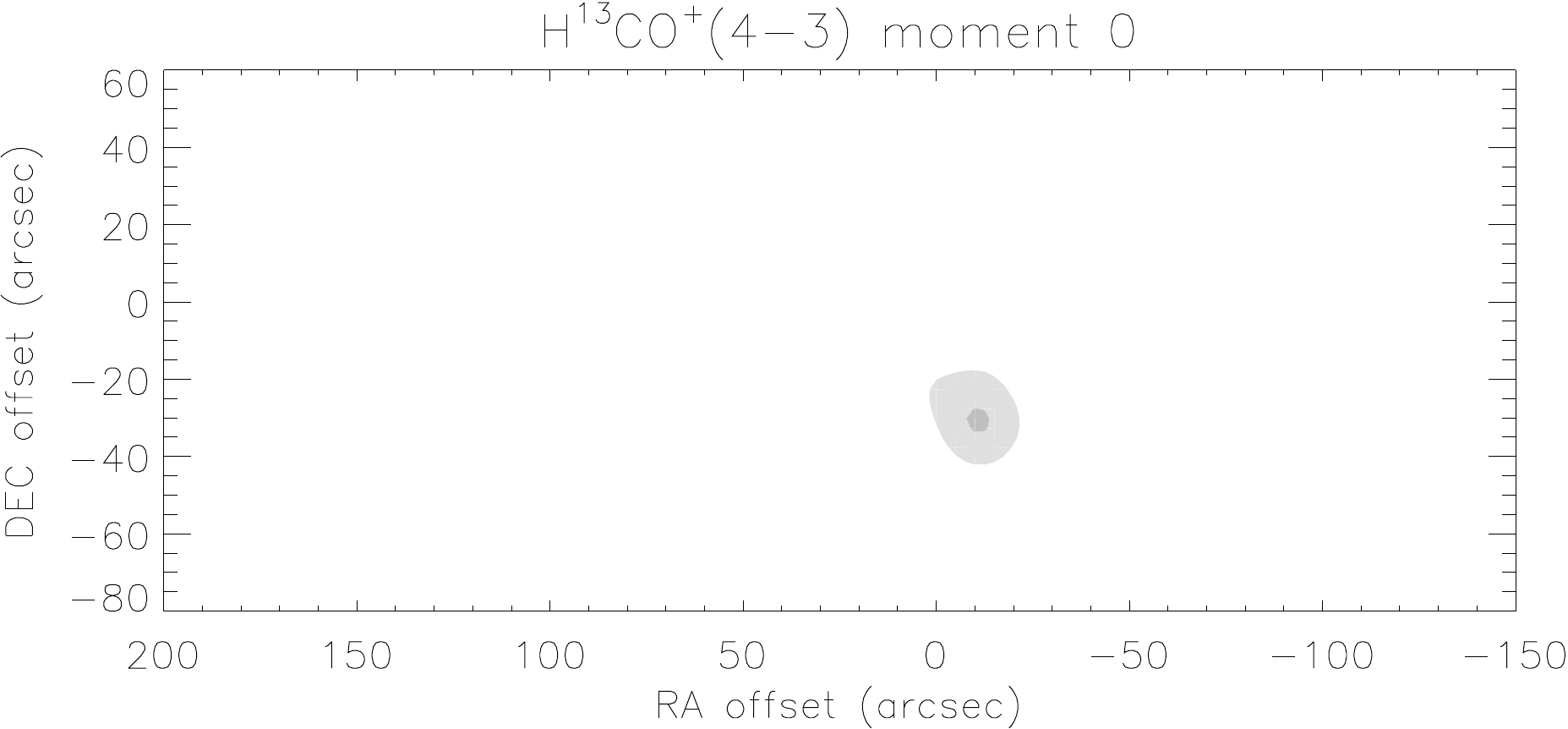}\put(-1000,200){\Huge (c)}\put(-470,200){\Huge (d)}}
  \caption{Model H$^{13}$CN(3-2) (a), H$^{13}$CN(4-3) (b), H$^{13}$CO$^+$(3-2) (c), and H$^{13}$CO$^+$(4-3) (d) moment~0 maps. 
    The contour levels are those of Fig.~\ref{fig:cndapex} divided by a factor $4$.
    \label{fig:cndapex13}}
\end{figure*}

Mills et al. (2013) observed four positions in the CND in the H$^{13}$CN(3-2), H$^{13}$CN(4-3), HC$^{13}$O$^+$(3-2), and H$^{13}$CO$^+$(4-3) lines.
Within our model, we set the carbon isotope ratio [$^{12}$C]/[$^{13}$C]$=30$, which is close the the value of $25$ adopted by
Mills et al. (2013) to calculate the isotopologue emission. The resulting moment~0 model maps are presented in Fig.~\ref{fig:cndapex13}.
The resulting model line ratios together with the observed line ratios are presented in Table~\ref{tab:lineratioscnd}.
The model line ratios are about a factor of two higher than the observed line ratios.
We conclude that our model reproduces the observed brightness temperatures and line ratios within a factor two.
This represents an additional verification of our model gas chemistry.
\begin{table*}
      \caption{Line ratios observed in the Galactic Center CND.}
         \label{tab:lineratioscnd}
      \[
       \begin{tabular}{lcccc}
        \hline
         & H$^{13}$CN(3-2)/HCN(3-2) &  H$^{13}$CN(4-3)/HCN(4-3) &  H$^{13}$CO$^+$(3-2)/HCO$^+$(3-2) & H$^{13}$CO$^+$(4-3)/HCO$^+$(4-3) \\
         & &  & & \\
        \hline
        observed & $0.12 \pm 0.04$ & $0.10 \pm 0.03$ & $0.06 \pm 0.03$ & $0.05 \pm 0.01$ \\
        model & $0.25 \pm 0.08$ & $0.21 \pm 0.09$ & $0.16 \pm 0.03$ & $0.10 \pm 0.02$ \\
	\hline
        \end{tabular}
      \]
\end{table*}

\subsection{Discussion}

Whereas the observed flux ratio between the northern and southern lobes is well reproduced in the CO line, 
it is significantly overestimated by the model in the HCN, HCO$^+$, and CS lines.
Thus, the conditions in terms of density, temperature, and molecule abundances of the gas in the northern lobe seem to be different 
from those of the southern lobe. Since we used a symmetric radial profile of the model brightness temperatures, we could not model such an effect.
Our disk cloud densities are in broad agreement with those derived by Mills et al. (2013). These authors derived an about $10$ times
higher gas density from the HCN emission than from the HCO$^+$ emission. We suggest that this difference at least partly stems from the
fact that within the $\sim 1$~pc APEX beam the HCN emission mainly comes from the inner edge of the CND, whereas the HCO$^+$
emission has a significant contribution from gas at larger radii. The model disk cloud
density at a radius of $1$~pc is $n_{\rm H_2} \sim 2 \times 10^6$~cm$^{-3}$, which is within the error bars of the density quoted by Mills et al. (2013). 
The model cloud density at a radius of $1.7$~pc is about three times lower, $n_{\rm H_2} \sim 7 \times 10^5$~cm$^{-3}$.

The somewhat smaller model HCN/HCO$^+$ line ratio compared to the observations is most probably caused by the assumed cosmic ray ionization rate.
The chosen radially dependent cosmic ray ionization rate represents the best compromise between the HCN surface brightness, which
increases with increasing $\zeta_{\rm CR}$, and the HCN/HCO$^+$ line ratio, which decreases with increasing $\zeta_{\rm CR}$.
As an alternative we calculated a model with a constant cosmic ray ionization rate of $\zeta_{\rm CR}=2 \times 10^{-15}$~s$^{-1}$.
This modification did not significantly change the HCN emission but lead to an extended low surface density HCO$^+$ envelope emitted by gas
located at radii between $3$ and $5$~pc. Such an envelope is not observed.  

We conclude that our forward modelling can reproduce the CO(6-5), HCN(3-2), H$^{13}$CN(3-2), HCN(4-3), H$^{13}$CN(4-3), HCO$^+$(3-2), 
H$^{13}$CO$^+$(3-2), HCO$^+$(4-3), and H$^{13}$CO$^+$(4-3)  line emission distributions 
within a factor of two. Therefore, our model is able to reproduce in a satisfactory way the gas cloud sizes, densities, temperatures, 
velocity dispersions, and chemistry in the central $10$~pc around the supermassive black hole of the Galaxy.
Moreover, the assumed carbon isotope ratio of [$^{12}$C]/[$^{13}$C]$=30$ is justified. 
The available molecular line observations of the CND can be
reasonably reproduced by two massive infalling gas clouds on prograde orbits, which interact with a pre-existing molecular gas disk.

\subsection{The intercloud gas \label{sec:intercloudCND}}

The dense clouds within the CND might not be the only sources of molecular line emission. According to our analytical model
(Sect.~\ref{sec:anamodel}), the gas is of turbulent and clumpy nature. This means that turbulent eddies at scales 
between the driving length and the cloud size with different densities and temperature coexist. A consistent
treatment of the scales as attempted by Vollmer et al. (2017) is beyond the scope of this article. For simplicity, we
divided the ISM located in the CND into clouds and intercloud gas. The available interferometric molecular line observations of the 
CND indicate that the clumps are prominent and well-resolved. This implies that neither the brightness temperature of the
intercloud gas dominates the overall emission nor the intercloud gas is optically thick. 
In the latter case, the intercloud gas, which has a high area filling factor, is expected to hide a significant fraction of the clouds via self-absorption
in the molecular lines. The condition that the optical thickness of the intercloud medium should be below unity has important implications
for its properties. 

To calculate the mass fraction between of the dense clouds, we
use the density probability distribution function of Padoan et al. (1997) for overdensities $x$:
\begin{equation}
p(x){\rm d}x=\frac{1}{x \sqrt{2\pi \sigma^2}}{\rm exp}\big(-\frac{({\rm ln}\,x+\sigma^2/2)^2}{2 \sigma^2}\big) {\rm d}x
\end{equation}
where the standard deviation, $\sigma$, is given by
\begin{equation}
\sigma^2 \simeq {\rm ln}\big(1+({\cal{M}}/2)^2\big)
\end{equation}
and ${\cal{M}}=v_{\rm turb}/c_{\rm s}$ is the Mach number with the sound speed $c_{\rm s}$.
The mass fraction of gas with overdensities exceeding $x$ is then
\begin{equation}
\frac{\Delta M}{M} = \frac{1}{2} \big(1+{\rm erf}(\frac{\sigma^2-2{\rm ln}(x)}{2^{\frac{3}{2}}\sigma})\big) \ .
\label{eq:pdfgasfrac}
\end{equation}
For ${\cal{M}} = 20$ and $x=30$ we find $\frac{\Delta M}{M} \sim 0.3$. Thus, about $70$\,\% of the gas mass is expected to 
have lower gas densities than those of the dense clouds.

\begin{figure}[!ht]
  \centering
  \resizebox{\hsize}{!}{\includegraphics{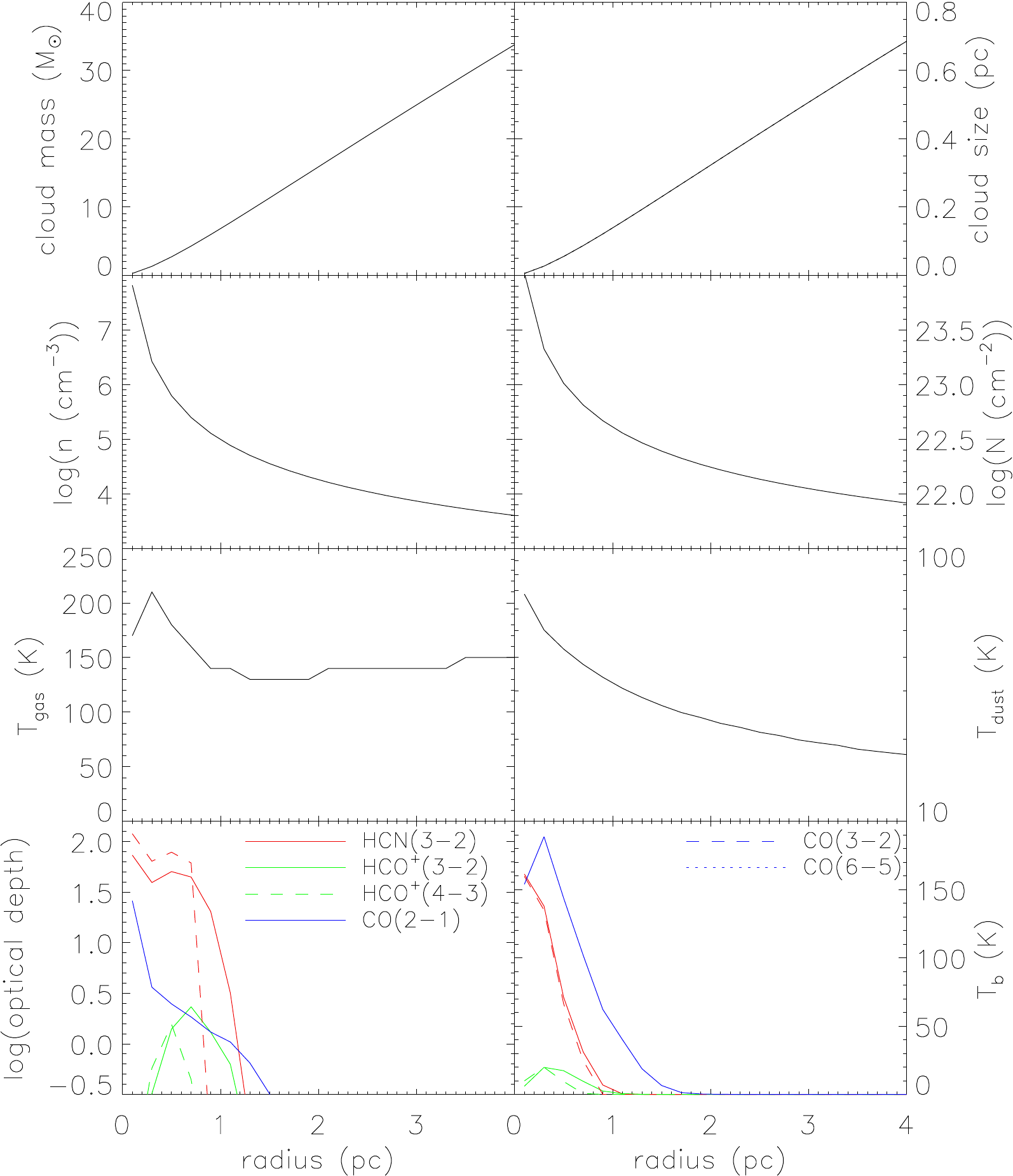}}
  \caption{Characteristics of individual model intercloud gas across the CND at the time of interest. From upper left to lower right: 
    cloud mass, size, density, H$_2$ column density, gas and dust temperatures, optical depth, and brightness temperature.
  \label{fig:profiles_CND_plotall_diff}}
\end{figure}

For the intercloud gas we assumed a size of $l_{\rm driv}=H$ and area and volume filling factors of unity.
It turned out that the CO, HCN and HCO$^+$ optical depths exceed unity for the turbulent heating 
presented in Eq.~\ref{eq:turbheat}. The molecular lines emitted by the intercloud gas only become optically
thin if the heating efficiency is decreased by a factor of $50$. We therefore used 
\begin{equation}
\Gamma_{\rm turb} = 6 \times 10^{-3}\, \rho\, \frac{|\vec{v}_{\rm turb}^{\rm cl}|^3}{l_{\rm cl}}
\end{equation}
for the intercloud medium.
The properties of these largest turbulent eddies are shown in Fig.~\ref{fig:profiles_CND_plotall_diff}.
Compared to the gas clouds (Fig.~\ref{fig:profiles_CND_plotall}) these eddies are larger, have an about $30$
times lower density, an about $10$ times lower column density, and an about two times lower gas temperature
due to the low turbulent heating efficiency. 
Most importantly, the optical depth of these eddies is below unity for distances greater than $\sim 1$~pc
from the central black hole. Therefore, it cannot be detected in the CO, HCN, and HCO$^+$ lines.

ISM turbulence is thought to be intermittent (see, e.g. Lazarian 2006): turbulence self-similarity is not exactly true 
even along the inertial range, i.e. at length scales greater than the dissipation length. 
Instead the turbulent fluctuations become increasingly sparse in time and space at smaller scales.
Pan \& Padoan (2009) found that in compressible 3D hydrodynamic simulations the  turbulent  dissipation  is  characterized 
by strongly intermittent fluctuations. A significant fraction of the kinetic energy is viscously dissipated in the finest, 
most intermittent structures. Dense filaments are sites of strong dissipation, but filaments of high dissipation rate are also found
at the interface of low- and high-density regions. As a result, the dissipation rate and the gas density are practically uncorrelated
in their simulations (Fig.~3 of Pan \& Padoan 2009). Due to the intermittent fluctuations in the turbulent heating rate, a significant 
mass fraction of a molecular cloud is not heated by turbulence and the mass-averaged cloud temperature is decreased by 
a factor of two to three.  

In the presence of magnetic fields ambipolar diffusion can represent an
important contribution to the local intermittent turbulent heating if the ratio between the ambipolar diffusion and
the turbulent crossing times is between about one and hundred (Li et al. 2012, Momferratos et al. 2014).
The latter condition is approximately fulfilled only by the clouds. 
Stone et al. (1998) found that the dissipation time of MHD turbulence is of the order of the flow crossing time or smaller, 
even in the presence of strong magnetic fields. Weak magnetic fields are amplified and tangled by the turbulence,
while strong fields remain well ordered. Moreover, these authors found that the density contrasts are larger for strong fields 
at fixed turbulent Mach number. This might indicate a lower turbulent dissipation rate at larger scales.

If we want the intercloud medium to be transparent for the molecular line emission, the area filling factor of optically thick
portions has to be small. The intermittent heating of low-density gas certainly helps to decrease the area filling factor of 
optically thick gas at these densities. However, it might only be the fact that low-density, high-dissipation regions are found close to
high density regions which ensures the low area filling factor of low-density, optically thick gas. 
Moreover, we can only speculate that a strong uniform magnetic field, as it is observed in the CND by Hsieh et al. (2018), helps to
suppress the turbulent heating rate within the inertial range $l_{\rm cl} \la l \la H$.

\section{An NGC~1068-like model \label{sec:n1068}}

Since we are able to reproduce the observed line emission distributions in terms of
area filling factor, linewidth, and brightness temperature of the Circumnuclear Disk in the Galactic Center 
within a factor two,
we are confident that the same model can be applied to the distribution of the molecular gas in the central
$10$~pc of NGC~1068, which is about $1800$ times farther away than the Galactic Center.
With a distance of $14.4$~Mpc, $1$~pc corresponds to $0.014''$. 

\begin{figure}[!ht]
  \centering
  \resizebox{\hsize}{!}{\includegraphics{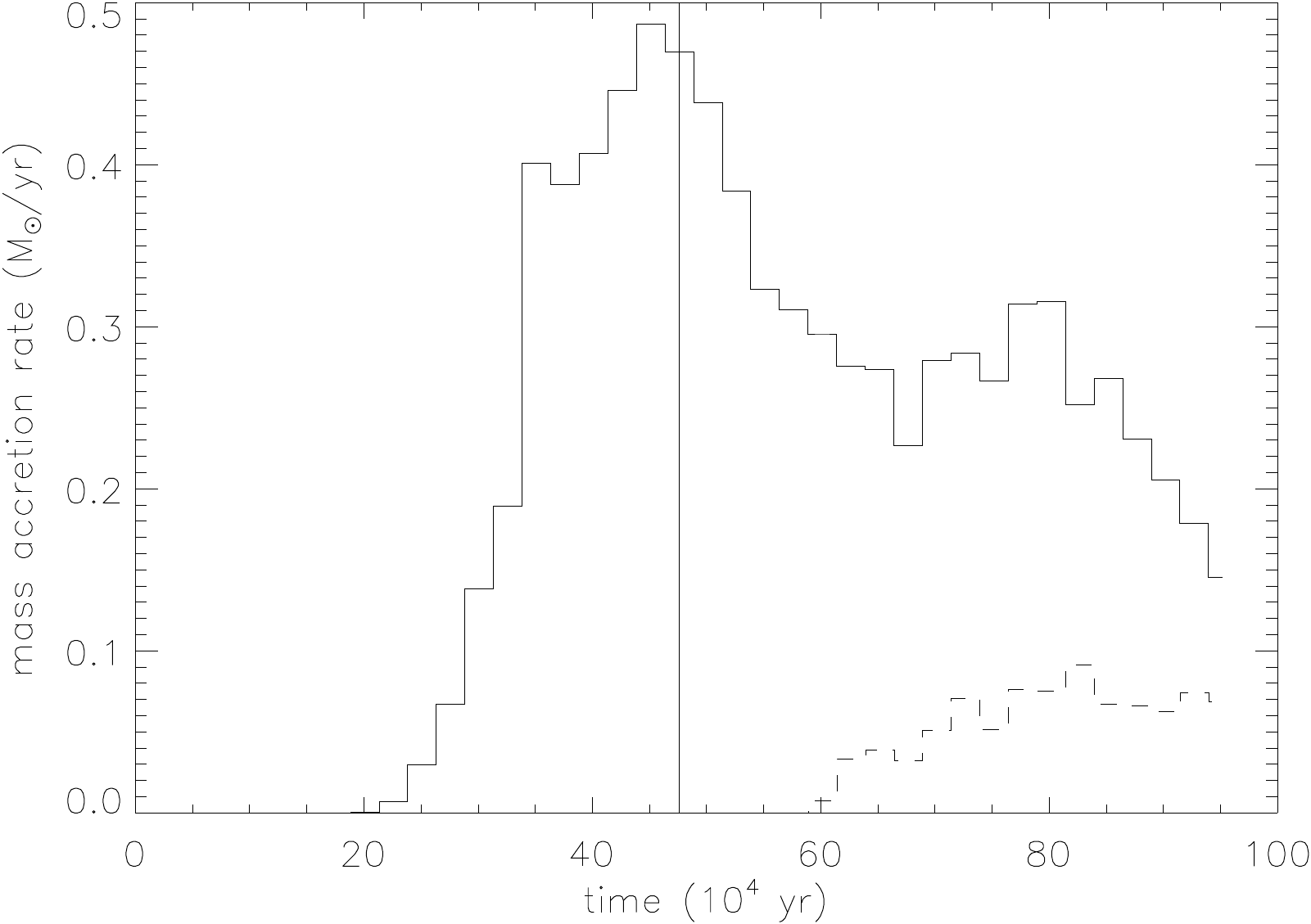}}
  \caption{Model mass accretion rate measured at $R=0.25$~pc. Solid line: preferred model with $\xi=0.9$, dashed line: preferred model without the massive
    infalling gas cloud, vertical line: time of interest.
  \label{fig:massaccretionrates}}
\end{figure}

The underlying dynamical simulation is shown in Fig.~\ref{fig:sim_N1068COLLdata2pot_diss1a_light1_hcn32b}.
During the evolution of the simulation all gas cloud particles whose distances to the central black hole are smaller than 
$0.25$~pc, which is close to the dust sublimation radius (Gravity Collaboration 2020), are
removed from the simulation and counted as accreted.
The so defined mass accretion rate is presented in Fig.~\ref{fig:massaccretionrates}. In addition, we show the mass accretion rate of
a simulation where the infalling cloud has been removed, i.e. a simulation of an unperturbed accreting disk.

In the case of an isolated gas ring the inner edge of the ring approaches the center and reaches a radius of $0.25$~pc at $t=0.6$~Myr.
The mass accretion rate then increases to $\dot{M}=0.07$~M$_{\odot}$yr$^{-1}$ at $80$~Myr and stays constant to $95$~Myr.
In the case where the infalling cloud hits the gas disk, the inner edge of the ring
reaches a radius of $0.25$~pc at $t=0.2$~Myr. The mass accretion rate rapidly increases to 
$\dot{M}=0.5$~M$_{\odot}$yr$^{-1}$ within $0.2$~Myr. It then decreases more slowly to $\dot{M}=0.3$~M$_{\odot}$yr$^{-1}$
at $t=0.6$~Myr and $\dot{M}=0.2$~M$_{\odot}$yr$^{-1}$ at $t=0.9$~Myr. 
At the time of interest ($t=0.47$~Myr corresponding to the minimum $\chi^2$; see Sect.~\ref{sec:model}) 
the mass accretion rate is close to its maximum ($\dot{M} \sim 0.5$~M$_{\odot}$yr$^{-1}$).

For the central region of NGC~1068 we assume $\gamma=300$~cm$^{-\frac{3}{2}}$s$^{\frac{1}{2}}$, $v_{\rm A}=1.5$~km\,s$^{-1}$, and
$\zeta_{\rm CR}=2 \times 10^{-13}$~s$^{-1}$ (Table~\ref{tab:model}).
This ionization rate leads to molecular line emission distributions from the dense disk clouds, which are comparable to the available observations.
We refer to Sect.~\ref{sec:CRion} for a discussion on the different ionization fractions.
In Sect.~\ref{sec:XDR} we argue that the X-ray emission is entirely absorbed by dense gas located inside the molecular emitting region at
($R \la 0.3$~pc). Therefore, X-ray heating and ionization can be ignored.
The molecular gas is heated by cosmic rays and turbulent mechanical energy injection. In all presented models the mechanical energy injection
is by far the dominant heating mechanism. The effects of optical and UV emission on the gas are treated separately in Sect.~\ref{sec:pdr}.
For all comparisons with observations the centroid coordinates of the central continuum source S1 are 
$02^h42^m40.709444''$, $-00^{\circ}00'47.9446''$, J2000.0 (Gallimore et al. 2004).

The face-on view of the cloud distribution at the time of interest is shown in Fig.~\ref{fig:sim_N1068COLLdata2pot_diss1a_light1_hcn32b}.
A very dense gas disk with a radius of $\sim 1.5$~pc might correspond to the maser disk observed in NGC~1068 (Greenhill et al. 1996, 
Gallimore et al. 1996). A rather symmetric distribution of clouds with a lower cloud density is observed at radii
between $1.5$ and $5$~pc. As in Sect.~\ref{sec:datacubes} brightness temperature and a cloud surface area were assigned to each gas 
cloud particle of the dynamical simulations according to the assumed gas density, temperature, velocity dispersion, and size obtained 
from the analytical model. Strictly speaking, our analytical model, which assumes a symmetric disk with fully developed turbulence, is only 
applicable within this region. To take into account an intercloud medium (see Sect.~\ref{sec:intercloudN1068}) we assigned half of the 
total gas mass to the dense clouds and the other half to the intercloud medium.

\subsection{Continuum emission and line absorption}

We assume that the millimeter continuum emission is free-free emission produced by thermal electrons (Gallimore et al. 2004)
whenever a gas cloud is directly illuminated by the AGN, and thermal dust emission otherwise.
Within the dynamical models all gas clouds are optically thick with respect to the UV and optical emission of the AGN. 
We therefore identified the gas clouds in our simulation, which are not hidden by other clouds.
In Vollmer et al. (2018) we suggested that there is a transition between the thick outer gas disk and the thin inner maser disk
at a radius of $\sim 1.5$~pc where a magnetocentrifugal wind is launched. We expect that the gas clouds become disrupted within
this region leading to larger volume and area filling factors. To mimic this effect, we increased the cloud sizes for radii smaller than $2$~pc
to $0.01$~pc for the calculation of the continuum emission. This also includes the location of the observed molecular maser disk at $R \la 1.5$~pc.
 
It is assumed that the ionization front has a constant surface density where the cloud becomes optically thick
to the optical emission of the AGN. The free-free emission is proportional to the emission measure $EM=\int n_{\rm ion}^2 {\rm d}l$.
We adopted the relation between the emission measure and the gas density $n$ found by Vollmer \& Duschl 
(2001; panels (a) and (c) of their Fig.~5): $EM \propto n^{4/3}$. A 3D datacube with the three position axes with the
emission measure of all model clouds was created and a 2D map was created which was then convolved to the 
beamsize of the Impellizzeri et al. (2019) observations and normalized to their maximum flux density
(see, e.g., lower left panel of Fig.~\ref{fig:plottingvollcnd_hco32radex_newQ=30}).
The 3D datacube was used to calculate molecular line absorption from clouds with optical depths higher than unity via
\begin{equation}
\Delta T_{\rm b}(v)=T_{\rm ex}(v) (1-{\rm e}^{-\tau(v)}) + T_{\rm bg} {\rm e}^{-\tau(v)} - T_{\rm bg}\ ,
\end{equation}
where $v$ is the radial velocity. The excitation temperatures $T_{\rm ex}$ comes from our molecular line emission calculations 
(Sect.~\ref{sec:mollinem}) and the background temperature $T_{\rm bg}$ from the normalized 3D datacube.

\subsection{Molecular line emission \label{sec:mollinem}}

As for the CND, we first present the results for the distribution of the dense disk clouds (Sect.~\ref{sec:denseclouds}).
The model moment~0, moment~1, pv diagrams, and spectra along the major axis 
are compared to the HCN(3-2), HCO$^+$(3-2), HCO$^+$(4-3), CO(2-1), CO(3-2), and CO(6-5) observations. We then discuss the influence of
a putative intercloud medium (Sect.~\ref{sec:intercloudN1068}). 
In addition, we separately evaluated the influence of the photodissociation regions close to the central engine (Sect.~\ref{sec:pdr}).

HCN has a large dipole moment and therefore does not trace dense gas if there is another excitation mechanism
that is faster than the H$_2$ collisions and independent of gas density. One such excitation path is through a vibrationally excited
state, to  which  molecules  can  be  pumped  by  infrared  radiation (Carroll \& Goldsmith 1981).
The first vibrationally excited state of HCN is its bending state ($v_2=1$) $1024$~K above the ground  with an emitting wavelength of $\lambda=14~\mu$m
(Sakamoto et al. 2010). Following Sakamoto et al. (2010), we define an equivalent gas density 
\begin{equation}
n_{\rm equiv}=\exp(-T_0/T_{\rm vib})\,A_{\rm vib}/\gamma_{J,J-1}\ ,
\end{equation}
where $T_0=1024$~K corresponds to the energy gap between the two vibrational levels $v=0$ and $1$,
$A_{\rm vib}=3.7$~s$^{-1}$ is the Einstein coefficient for the vibrational transition, and $\gamma_{J,J-1}$ is the collisional rate coefficient. 
Following Vollmer et al. (2017), HCN IR-pumping is implemented in the model by replacing the cloud density $n_{\rm cl}$ by $n_{\rm equiv}$ if 
$n_{\rm equiv} > n_{\rm cl}$ in the HCN emission calculations.

\subsubsection{The dense clouds \label{sec:denseclouds}}

\begin{figure}[!ht]
  \centering
  \resizebox{\hsize}{!}{\includegraphics{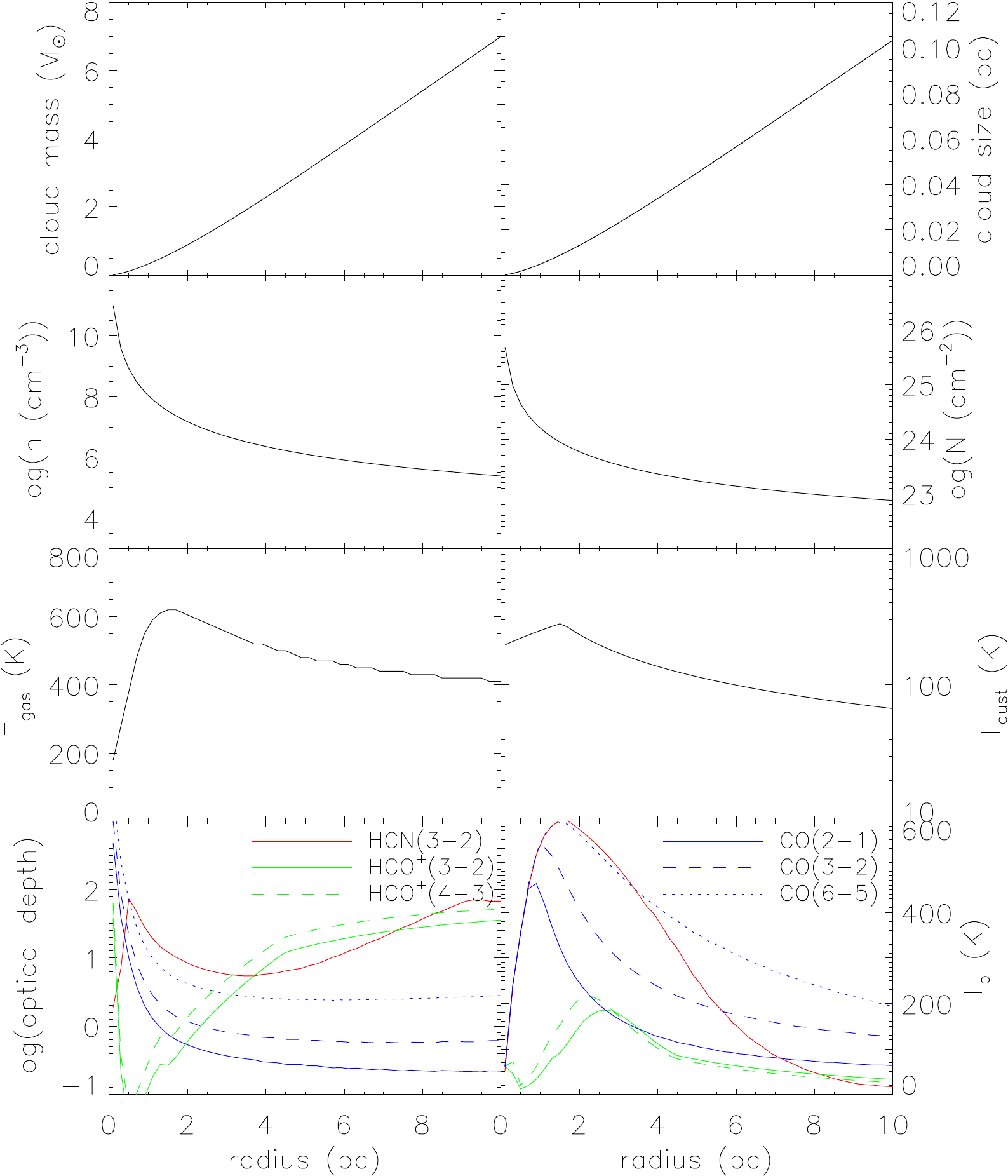}}
  \caption{NGC~1068 $Q=30$ analytical model of a turbulent clumpy gas disk (Sect.~\ref{sec:anamodel}).
  Characteristics of individual model disk gas clouds across the disk at the time of interest. From upper left to lower right: 
  cloud mass, size, density, H$_2$ column density, gas and dust temperatures, optical depth, and brightness temperature.
  IR pumping is included for the HCN and HCO$^+$ lines.
  \label{fig:profiles_n1068_plotall1}}
\end{figure}

We used a cosmic ray ionization rate  of $\zeta_{\rm CR}=2 \times 10^{-13}$~s$^{-1}$, which lead to HCO$^+$ brightness 
temperatures closest to the available observations.
As stated in Sect.~\ref{sec:XDR}, we assumed that the X-ray emission is absorbed within the Compton-thick, fully ionized, and dust-free gas
within the dust sublimation radius. 
The radial distribution of the disk cloud mass, size, density, H$_2$ column density, and gas and dust temperatures
of the $Q=30$ model are presented in Fig.~\ref{fig:profiles_n1068_plotall1}.
At a radius of $R \sim 2$~pc the characteristics of the clouds are $M_{\rm cl} \sim 1$~M$_{\odot}$, $l_{\rm cl} \sim 0.02$~pc,
$n_{\rm H_2} \sim  10^7$~cm$^{-3}$, $N_{\rm H_2} \sim 3 \times 10^{23}$~cm$^{-2}$, and $T_{\rm gas} \sim 600$~K.
We adopted the radial profile of the dust temperature of the disk midplane from the 3D radiative transfer calculations of Vollmer et al. (2018) for all models:
\begin{equation}
\begin{array}{lll}
T_{\rm dust}= & 190+60 \times R & {\rm for\ R \le 1.6\ pc} \\
 & 400 \times R^{-0.78} & {\rm for\ R > 1.6\ pc} \ .
\end{array}
\end{equation}

The resulting radial profiles of the CO(2-1), CO(3-2), CO(6-5), HCN(3-2), HCO$^+$(3-2), and HCO$^+$(4-3) optical depths and
brightness temperatures are shown in the two lower panels of Fig.~\ref{fig:profiles_n1068_plotall1}.
The optical depths of the CO lines decrease monotonically with radius. The CO(2-1) is optically thin for $R > 1$~pc.
The HCN(3-2) line is optically thick everywhere. 
As for the CND in the Galactic Center, the small HCO$^+$ optical depth for $R \la 4$~pc is due to the low HCO$^+$ abundance: 
at $R=2$~pc the model yields $x_{\rm HCO+} \sim 4 \times 10^{-8}$ whereas $x_{\rm HCN} \sim 2 \times 10^{-6}$ due to the 
hot core physics.
All brightness temperature radial profiles show a central depression.
Whereas the maximum of the HCN and CO lines is $T_{\rm b} \sim 500$-$600$~K, the maximum brightness temperature of the  
HCO$^+$ lines is about $\sim 200$~K. The maxima are reached at a radius of $R=1$-$3$~pc.

\subsubsection*{A quantitative comparison between the model and observations}

We anticipate our conclusion that the model can reproduce the available molecular line observations within a factor of two
(Table~\ref{tab:fluxes}).
\begin{table}
      \caption{Integrated intensities in Jy\,km\,s$^{-1}$ of the observations and the dense disk cloud model.}
         \label{tab:fluxes}
      \[
       \begin{tabular}{lcc}
        \hline
        line & $I_{\rm obs}$ & $I_{\rm model}$ \\
        \hline
        CO(2-1) & 1.6 & 1.2 \\
        CO(3-2) & 5.3 & 7.2 \\
        CO(6-5) & 31.3 & 55.4 \\
        HCN(3-2) & 2.5 & 2.7 \\
        HCN(4-3) & 5.3 & 5.4 \\
        HCO$^+$(3-2) & 0.8 & 0.6 \\
        HCO$^+$(4-3) & 3.0 & 1.0$^{(a)}$ \\
        CS(7-6) & 0.35 & 0.09$^{(b)}$ \\
        \hline
        \end{tabular}
      \]
      \begin{list}{}{}
      \item[$^{\rm (a)}$ The intercloud gas model (Sect.~\ref{sec:intercloudN1068}) yields]
      \item[\ \ \ \ $I_{\rm model}=2.0$~Jy\,km\,s$^{-1}$.]
      \item[$^{\rm (b)}$ The intercloud gas model (Sect.~\ref{sec:intercloudN1068}) yields]
      \item[\ \ \ \ $I_{\rm model}=0.26$~Jy\,km\,s$^{-1}$.]
      \item[IR pumping is assumed for the HCN and HCO$^+$ lines.]
      \end{list}
\end{table}
The observed emission of all lines is reproduced by the model within $40$\,\%, except that of the CO(6-5) line,
which is overestimated by about a factor of two by the model. 
The reason for this might be an overestimation of the gas densities or temperatures. 
We suggest that the small integrated HCO$^+$(4-3) intensity indicates the presence of an intercloud medium (Sect.~\ref{sec:intercloudN1068}).
In the following, all observed emission distributions were rotated counter-clockwise by $20^{\circ}$.

\subsubsection*{CO(2-1) and CO(3-2)}

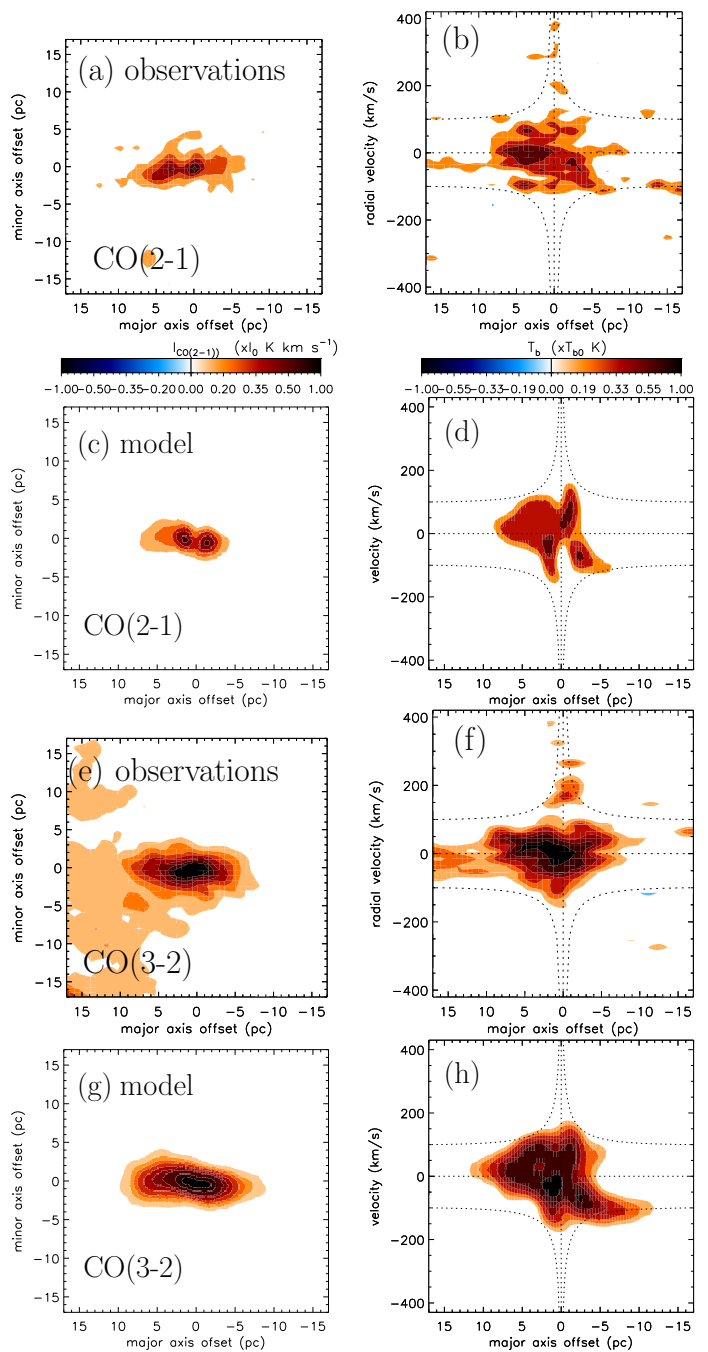
\begin{figure}[!ht]
  \centering
  \resizebox{\hsize}{!}{\includegraphics{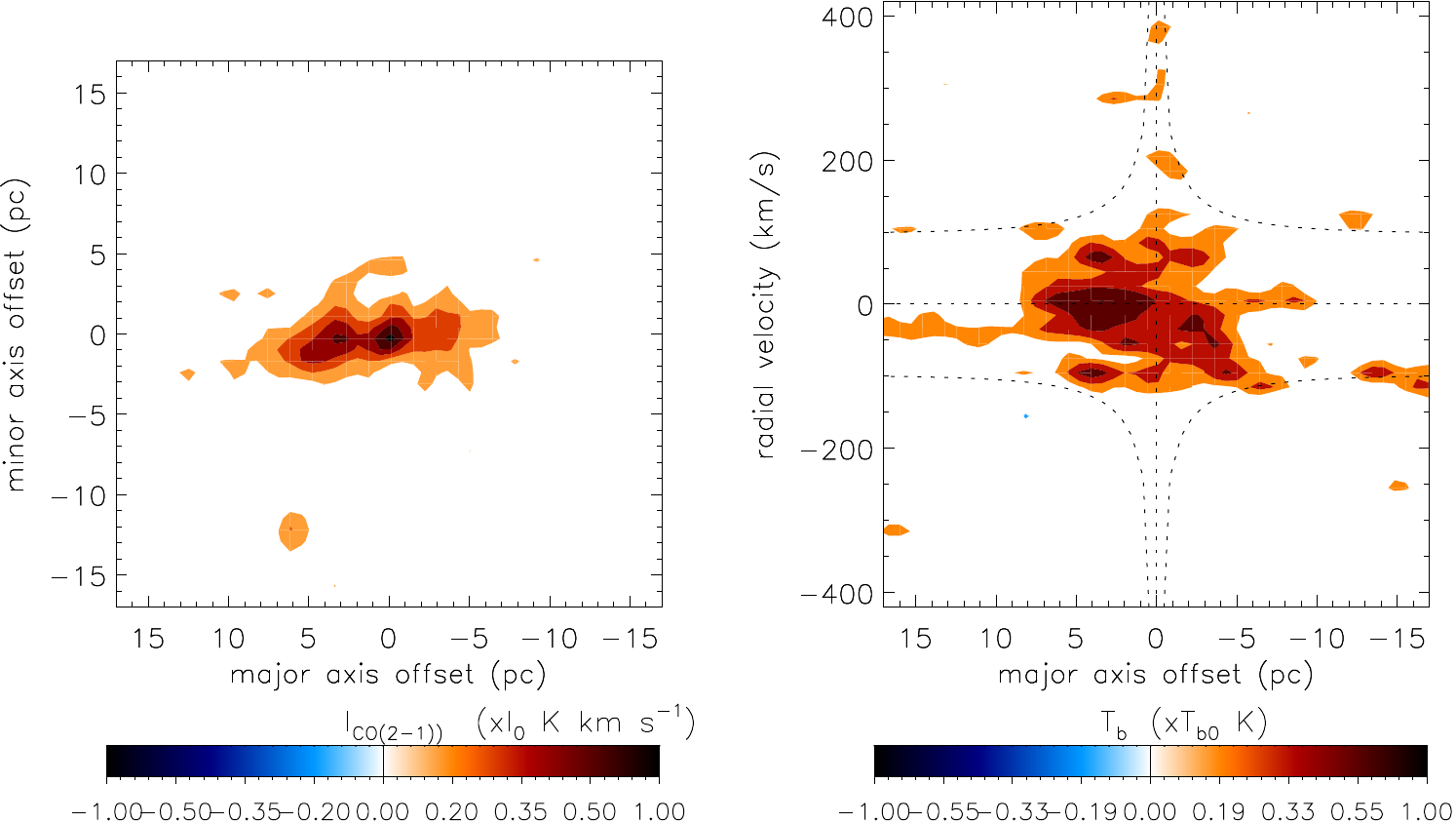}\put(-360,200){\huge observations}\put(-380,80){\huge CO(2-1)}\put(-390,200){\huge (a)}\put(-160,220){\huge (b)}}
  \resizebox{\hsize}{!}{\includegraphics{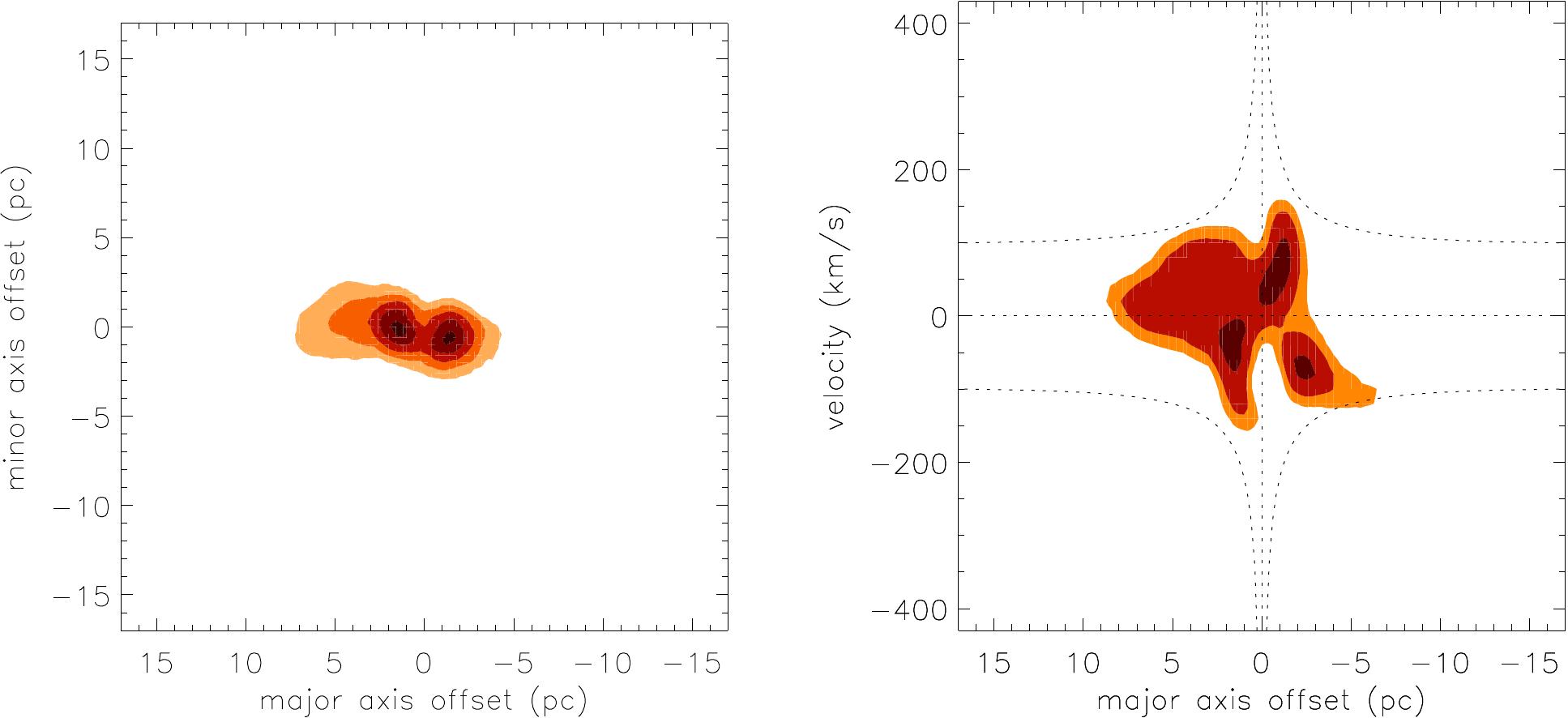}\put(-470,210){\Huge model}\put(-500,60){\Huge CO(2-1)}\put(-505,210){\Huge (c)}\put(-205,220){\Huge (d)}}
  \resizebox{\hsize}{!}{\includegraphics{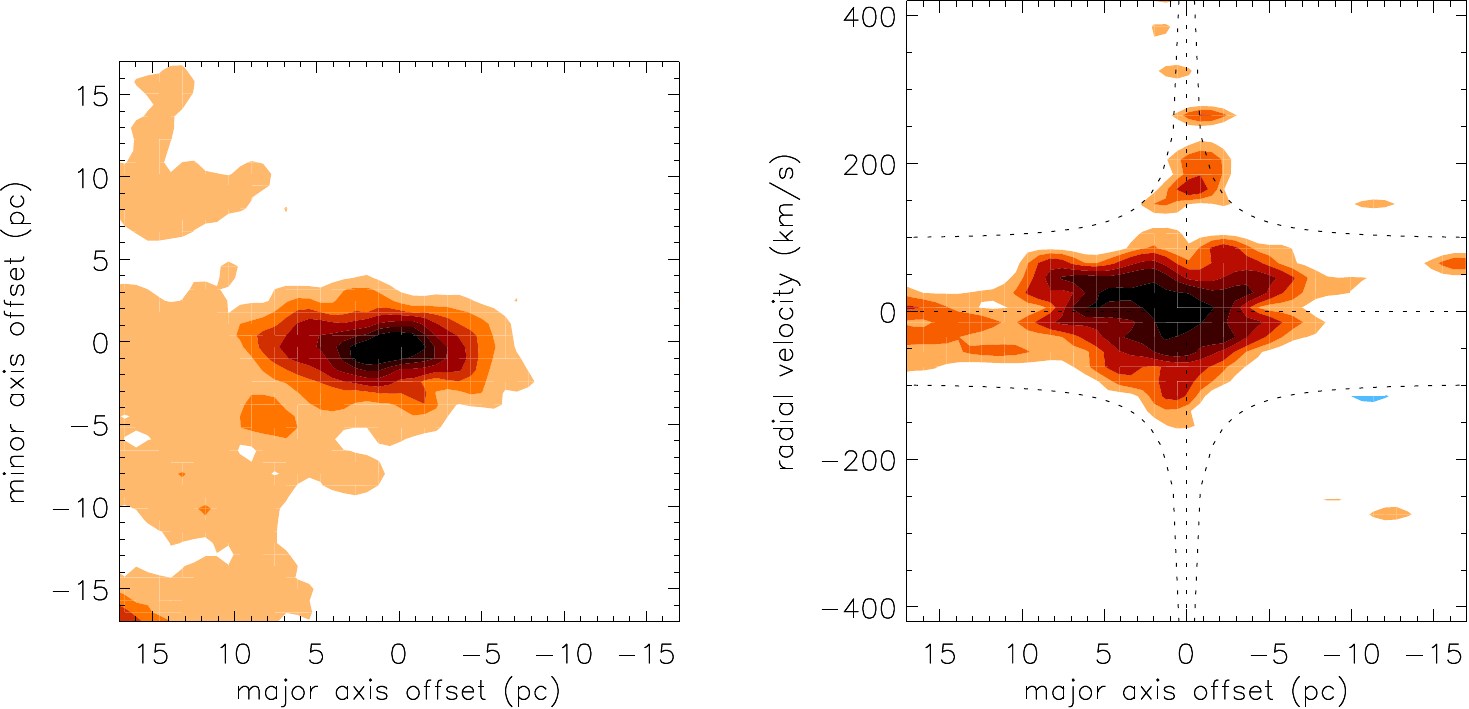}\put(-360,160){\huge observations}\put(-380,40){\huge CO(3-2)}\put(-390,160){\huge (e)}\put(-150,180){\huge (f)}}
  \resizebox{\hsize}{!}{\includegraphics{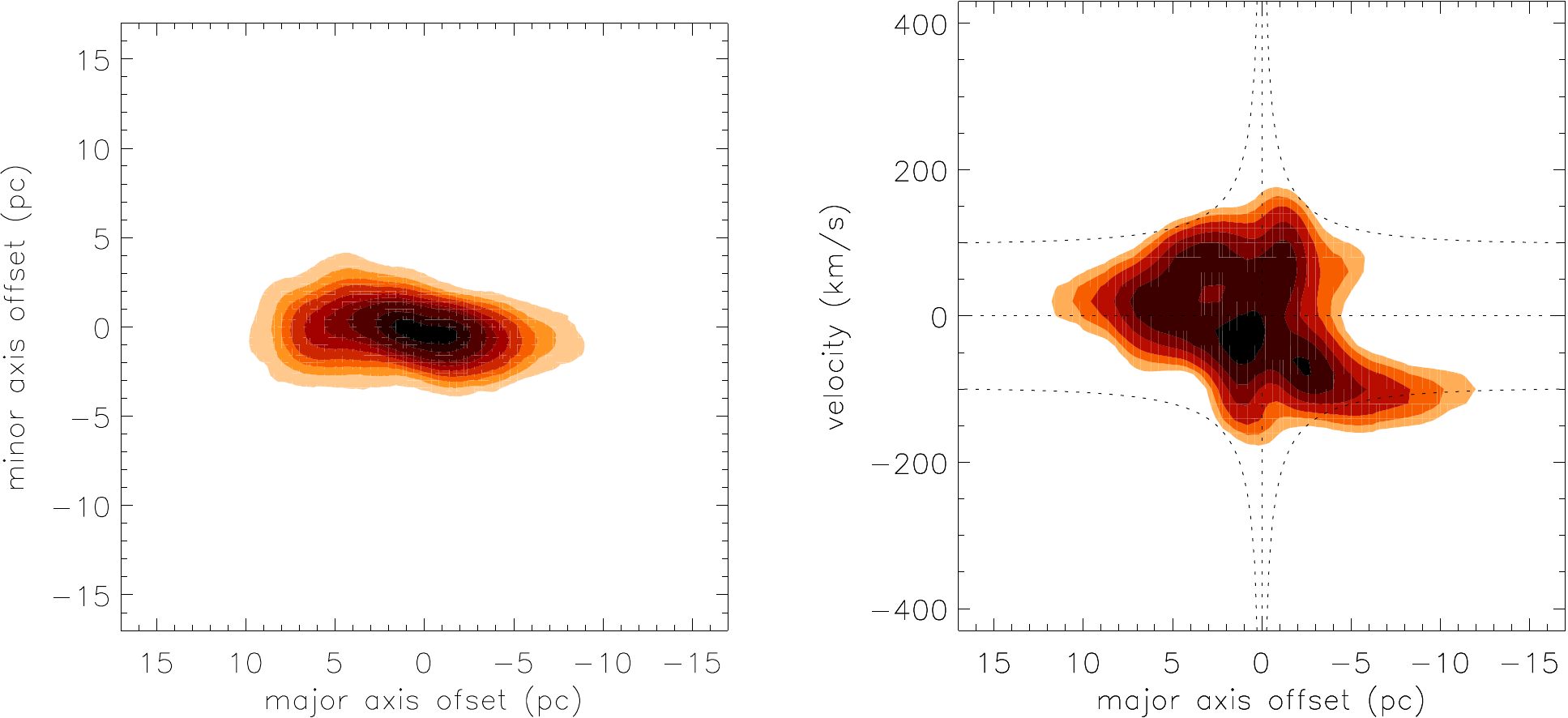}\put(-470,210){\Huge model}\put(-500,60){\Huge CO(3-2)}\put(-505,210){\Huge (g)}\put(-205,220){\Huge (h)}}
  \caption{NGC~1068. 
    Panel (a) observed CO(2-1) moment~0 map ($I_0=7923$~K\,km\,s$^{-1}$), (b) pv diagram (contours are $T_{\rm b0}=37.8$~K) from 
    Garcia-Burillo et al. (2019), (c) dense disk cloud model CO(2-1) moment~0 maps, (d) pv diagrams along the major axis,
    (e) observed CO(3-2) moment~0 map ($I_0=7923$~K\,km\,s$^{-1}$), (f) pv diagram 
    ($T_{\rm b0}=37.8$~K) from Garcia-Burillo et al. (2019), and
    (g) dense disk cloud model CO(3-2) moment~0 maps, (h) pv diagrams  along the major axis.
    The model levels are the same as those of the observed maps.
  \label{fig:plottingvollcnd_hcn32radex_new_garciaburillo_co21hr}}
\end{figure}

Garcia-Burillo et al. (2019) observed NGC~1068 in the CO(2-1) and CO(3-2) line with a resolution of $40$~mas ($2.8$~pc;
Fig.~\ref{fig:plottingvollcnd_hcn32radex_new_garciaburillo_co21hr}).
The CO(2-1) and CO(3-2) emission distributions are extended ($0.12''$
or $8.4$~pc to the east and $0.1''$ or $7.0$~pc to the west). The pv diagram shows a velocity gradient in the outer disk
with positive velocities to the east and negative velocities to the west. This gradient is more prominent in the CO(2-1) line than 
in the CO(3-2) line. A velocity gradient in the opposite direction is seen in the inner disk. This velocity gradient
is more prominent in the CO(3-2) line than in the CO(2-1) line. The identification of this apparent counter-rotation signature 
in the major axis pv diagram is explained by Garcia-Burillo et al. (2019) as due to the presence of outflowing gas inside the torus 
superposed to rotation. The highest brightness temperatures of both lines are found at velocities close to the systemic velocity.

The corresponding model CO(2-1) and CO(3-2) moment~0 maps and pv diagrams are presented in 
Fig.~\ref{fig:plottingvollcnd_hcn32radex_new_garciaburillo_co21hr}.
The extent of the model CO(2-1) and CO(3-2) emission distributions is well comparable to those of the observations. 
The linewidths of the observed CO(2-1) emission of the pv diagram are somewhat smaller but comparable to those of the model. 
Especially the observed western extension at negative velocities is present in the model CO(2-1) diagram.
As observed, the model CO(3-2) emission distribution is more centrally concentrated than the model CO(2-1) emission distribution.
The observed CO(3-2) linewidths are about $30$\,\% smaller than the model linewidths. Moreover, the western extension at negative velocities
present in the model CO(3-2) pv diagram is absent in the observed CO(3-2) data.

\subsubsection*{CO(6-5)}

\begin{figure}[!ht]
  \centering
  \resizebox{\hsize}{!}{\includegraphics{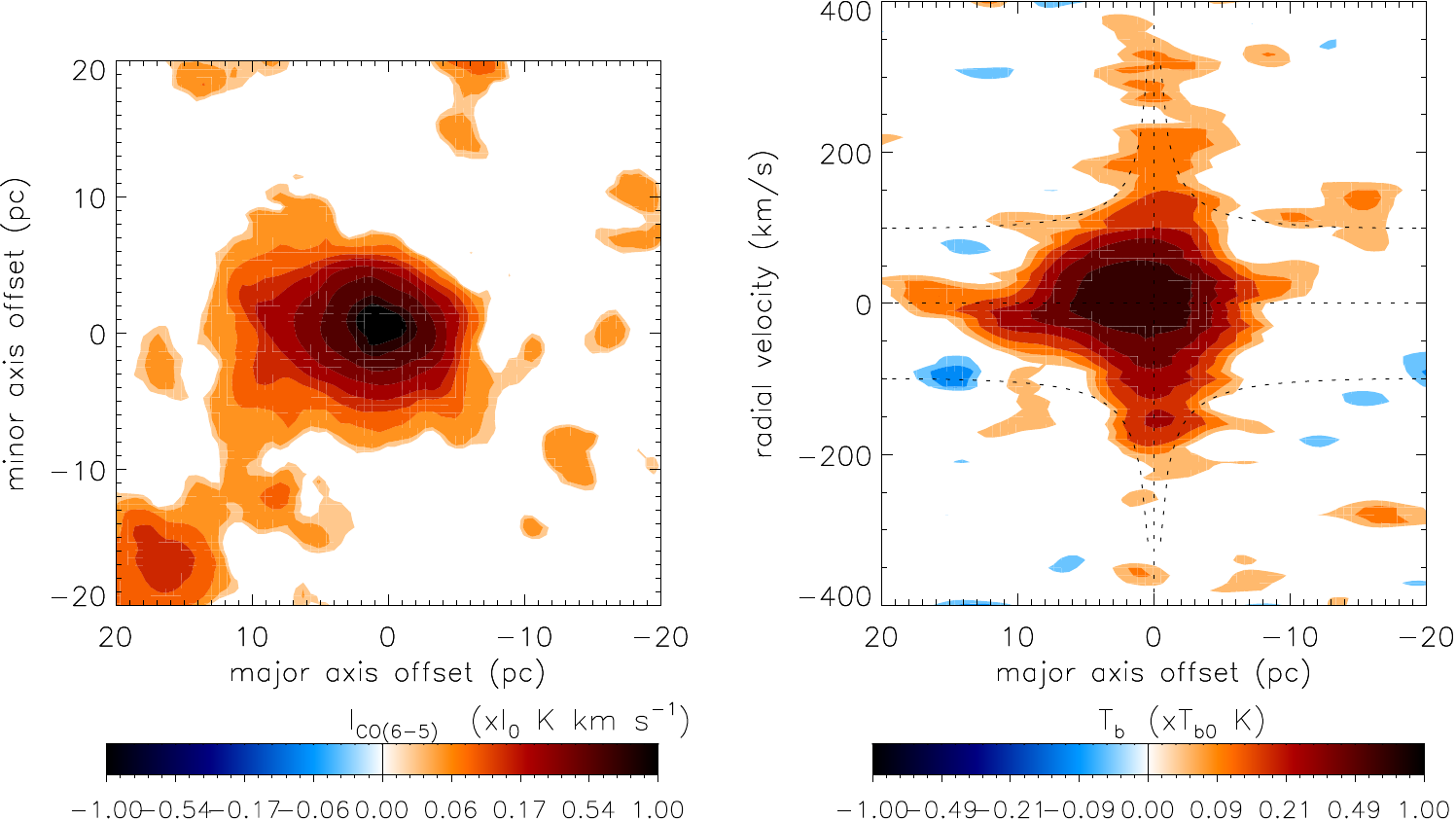}\put(-360,200){\huge observations}\put(-380,80){\huge CO(6-5)}\put(-390,200){\huge (a)}\put(-160,220){\huge (b)}}
  \resizebox{\hsize}{!}{\includegraphics{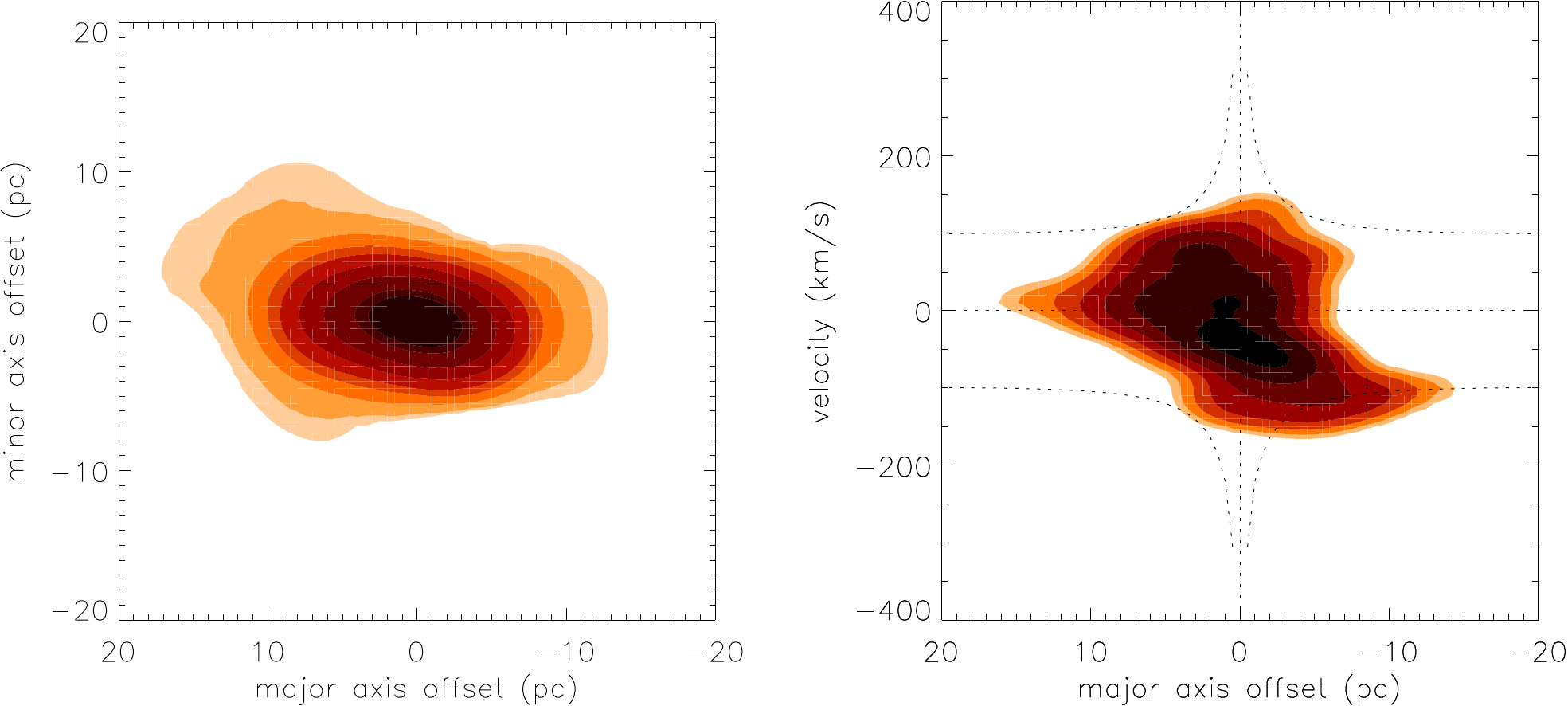}\put(-510,50){\Huge CO(6-5)}\put(-470,210){\Huge model}\put(-510,210){\Huge (c)}\put(-210,220){\Huge (d)}}
  \caption{NGC~1068. Panel (a) observed CO(6-5) moment~0 map (Gallimore et al. 2016)
    ($I_0=6657$~K\,km\,s$^{-1}$), (b) CO(6-5) pv diagram along the major axis ($T_{\rm b0}=38.0$~K),
    (c) dense disk cloud model CO(6-5) moment~0 maps, and (d) pv diagrams along the major axis.
    The model levels are the same as those of the observed maps.
  \label{fig:plottingvollcnd_hcn32radex_new_gallimore1}}
\end{figure}
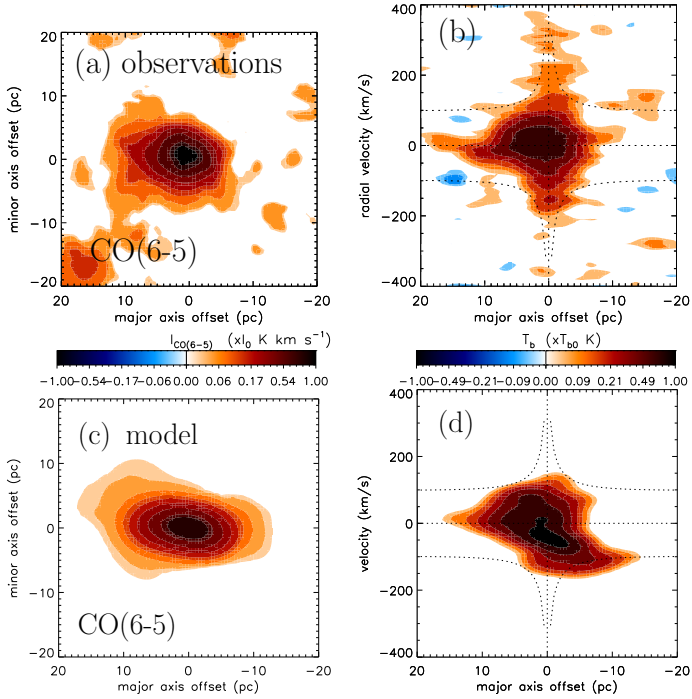

Gallimore et al. (2016) observed the center of NGC~1068 in the CO(6-5) line with a resolution of $80$~mas ($5.6$~pc).
The moment~0 map and the pv diagram are presented in the upper panels of Fig.~\ref{fig:plottingvollcnd_hcn32radex_new_gallimore1}.
The pv diagram revealed CO(6-5) emission at very high velocities close to the central black hole
($v > 400$~km\,s$^{-1}$). The highest CO(6-5) brightness temperatures are found in the
center at the systemic velocity.
The corresponding model CO(6-5) moment~0 map and pv diagram are shown in Fig.~\ref{fig:plottingvollcnd_hcn32radex_new_gallimore1}.
The extent of the observed CO(6-5) distribution and the linewidth of the high brightness temperature
emission (pv diagram) are reproduced by the model. As for the CO(3-2) emission, the western extension at negative velocities
present in the model CO(6-5) pv diagram is absent in the observed CO(6-5) diagram. The observed high-velocity emission
is absent in the model pv diagram.

\subsubsection*{HCN(3-2) \label{sec:hcn32}}

\begin{figure*}[!ht]
  \centering
  \resizebox{\hsize}{!}{\includegraphics{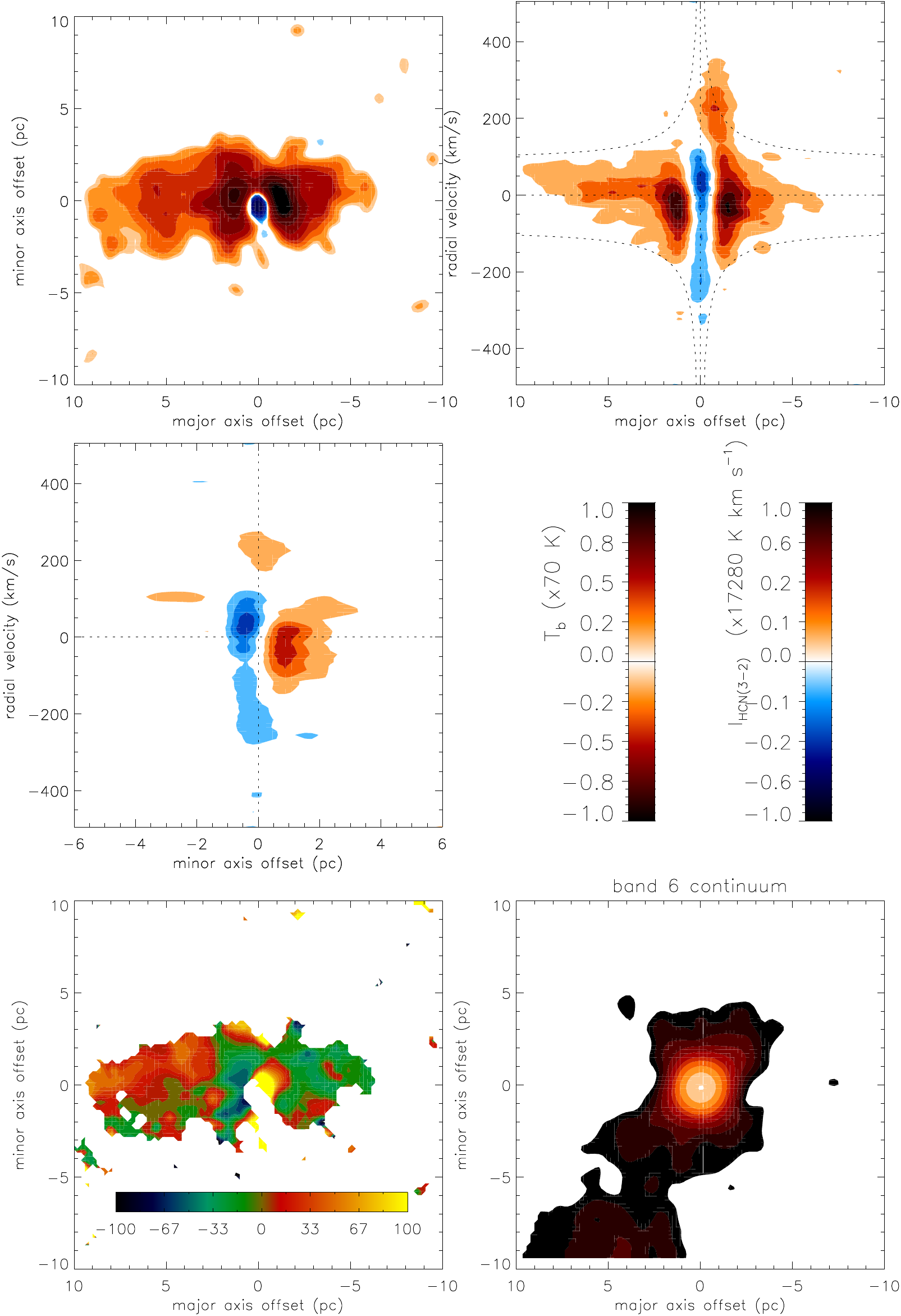}\includegraphics{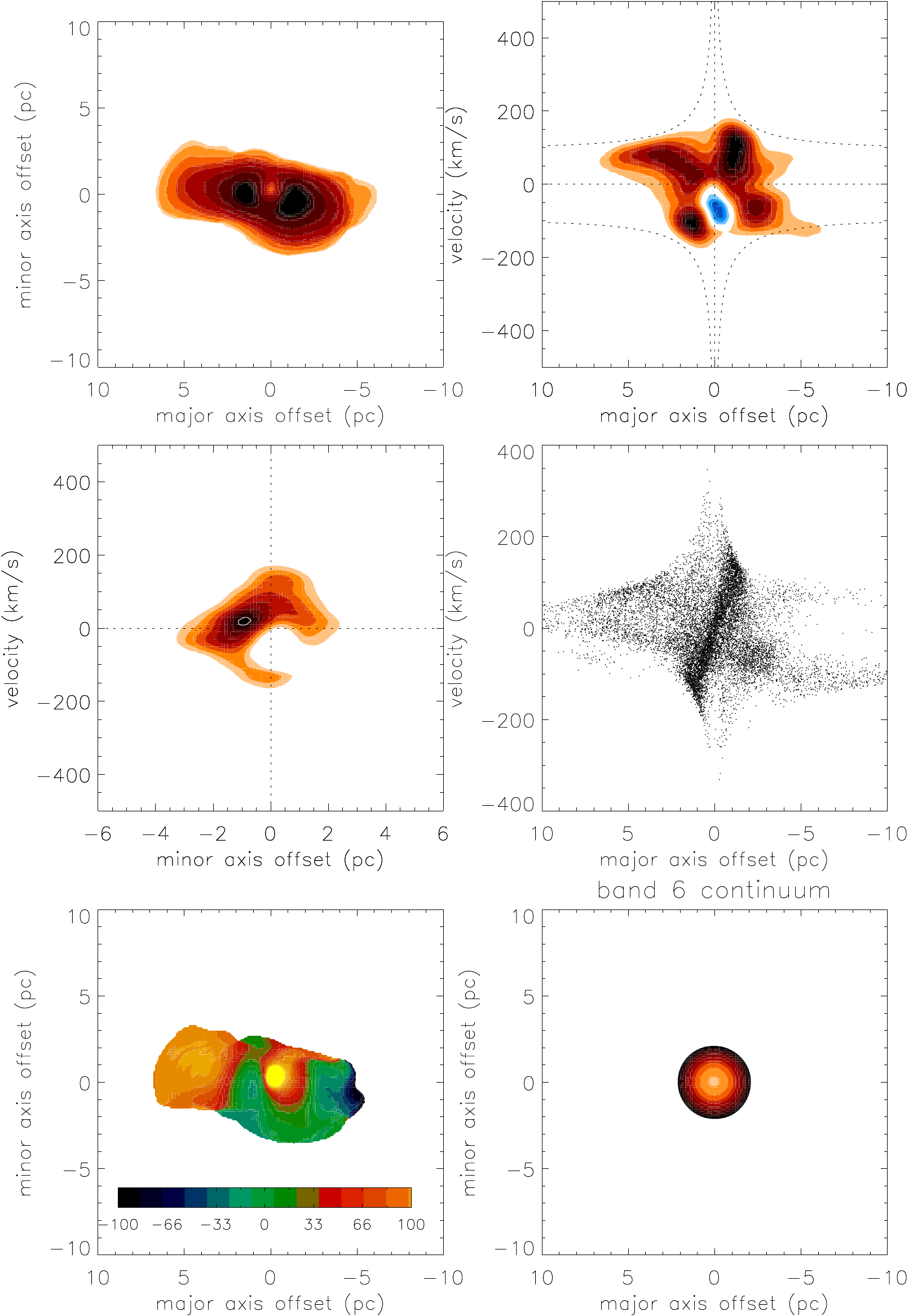}}
  \put(-485,360){\large observations}
  \put(-490,280){\large HCN(3-2)}
  \put(-215,360){\large model}
  \put(-500,360){\large (a)}\put(-370,360){\large (b)}
  \put(-500,235){\large (c)}
  \put(-230,280){\large HCN(3-2)}
  \put(-500,105){\large (d)}\put(-370,105){\large (e)}
  \put(-230,360){\large (f)}\put(-100,360){\large (g)}
  \put(-230,235){\large (h)}\put(-100,235){\large (i)}
  \put(-230,105){\large (j)}\put(-100,105){\large (k)}
  \caption{NGC~1068. 
    Left half: high resolution HCN(3-2) observations (Impellizzeri et al. 2019).
    Panel (a) moment~0 map ($I_0=17300$~K\,km\,s$^{-1}$), 
    (b) pv diagram along the major axis, (c) pv diagram along the minor axis ($T_{\rm b0}=70$~K), 
    (d) moment~1 map (in km\,s$^{-1}$), (e) continuum map (contour levels are $(0.019,0.037,0.075,0.15,0.3,0.6,1,2,4,8) \times 43$~K).
    Right half: high resolution HCN(3-2) emission of the dense disk cloud model with IR pumping. Panel (f) moment~0 map, (g) pv diagram along the major axis,
   (h) pv diagram along the minor axis, (i) unsmoothed pv diagram along the major axis, 
   (j) moment~1 map (in km\,s$^{-1}$), (k) continuum map.
   The model levels are the same as those of the observed maps. The additional white contour is at $80$~K.
  \label{fig:plottingvollcnd_hcn32radex_newQ=30}}
\end{figure*}
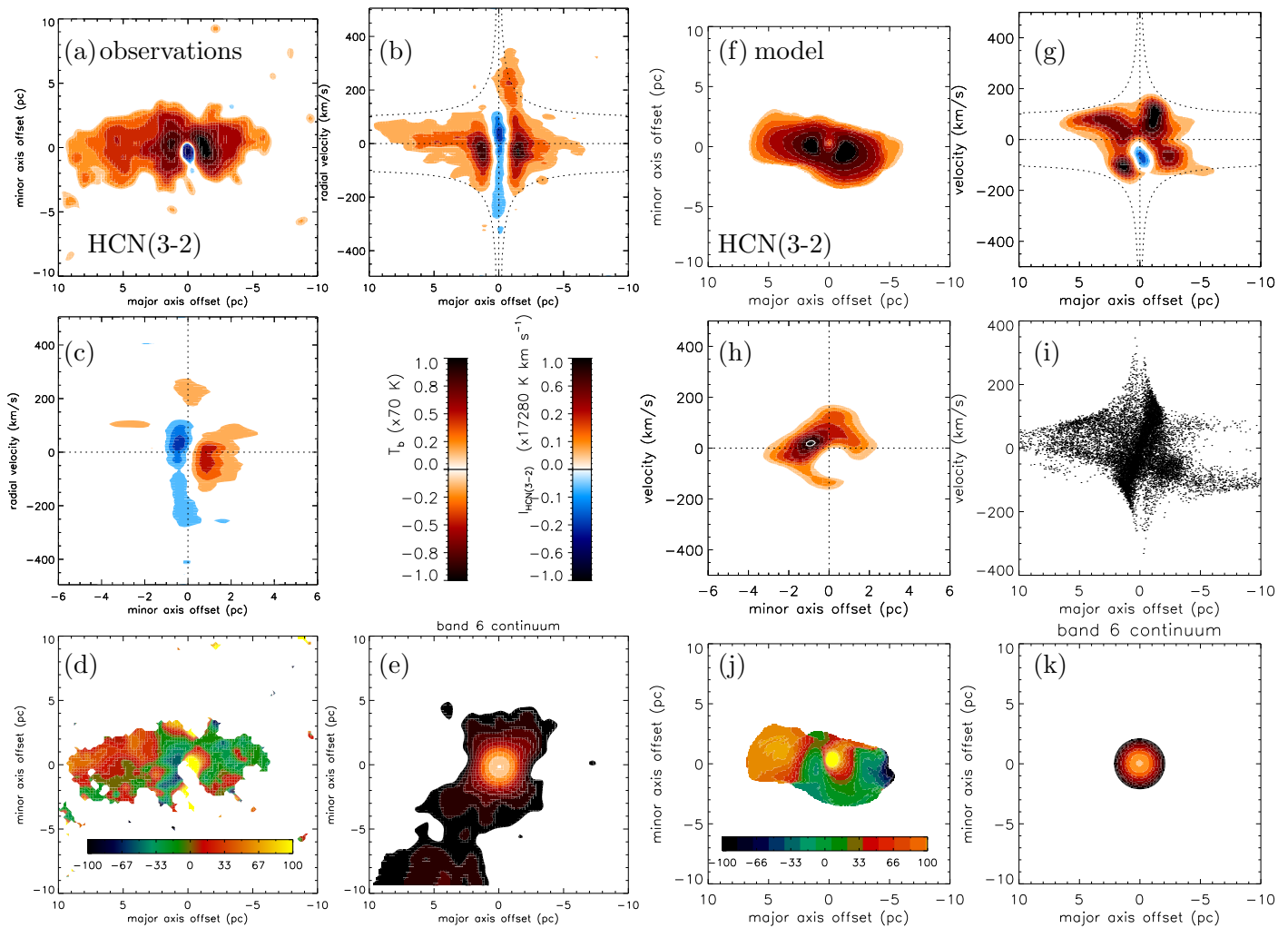

Imanishi et al. (2018, 2020) and Impellizzeri et al. (2019) observed NGC~1068 in the HCN(3-2) line with ALMA.
Whereas the Imanishi et al. (2018) data have a spatial resolution of $40$-$70$~mas ($2.8$-$4.9$~pc), the data of Impellizzeri et al. (2019) and
Imanishi et al. (2020) have a resolution of $20$~mas ($\sim 1.4$~pc).
The comparison between the model and the high resolution HCN(3-2) datacube is presented in Fig.~\ref{fig:plottingvollcnd_hcn32radex_newQ=30}
together with the observations by Impellizzeri et al. (2019).
The main characteristics of the observations are: (i) HCN(3-2) emission is detected in absorption against the continuum in the central resolution element,
(ii) the bulk of the emission comes from a $\la 3$~pc region, with (iii) the western side being 
significantly brighter than the eastern side, (iv) emission at radii $\ga 3$~pc is blueshifted in the west and redshifted in the east. 
On the other hand, emission at radii $\la 2$~pc is redshifted in the west and blueshifted in the east and coincides with the maser disk. 
The dense molecular gas in the 
gas disk appears counter-rotating between the innermost and outer parts, and (iv) a redshifted high-velocity ($170$-$370$~km\,s$^{-1}$) 
emission component is present at the innermost western side of the gas disk.

The observed east-west asymmetry of the moment~0 map is present but less prominent in the model.
Qualitatively, points (i), (ii) and (iv) are reproduced by the model. The observed western high-velocity emission
is partly seen in absorption in the model.
In particular, the observed counter-rotation between the inner and outer gas disk in the model pv diagram 
and the velocity field (moment~1 map) is present in the model. The model and observed linewidths are comparable.

On the other hand, the model has the following shortcomings:
The model moment~0 map does not show a central absorption.
Whereas the maxima of the emission lines are slightly negative and close to the systemic velocity in the observations,
they are located at $\sim \pm 70$~km\,s$^{-1}$ in the model disk.
The model pv diagram along the minor axis shows HCN in emission on the eastern side, whereas it is observed
on the western side.  

\subsubsection*{HCO$^+$(3-2)}

\begin{figure*}[!ht]
  \centering
  \resizebox{\hsize}{!}{\includegraphics{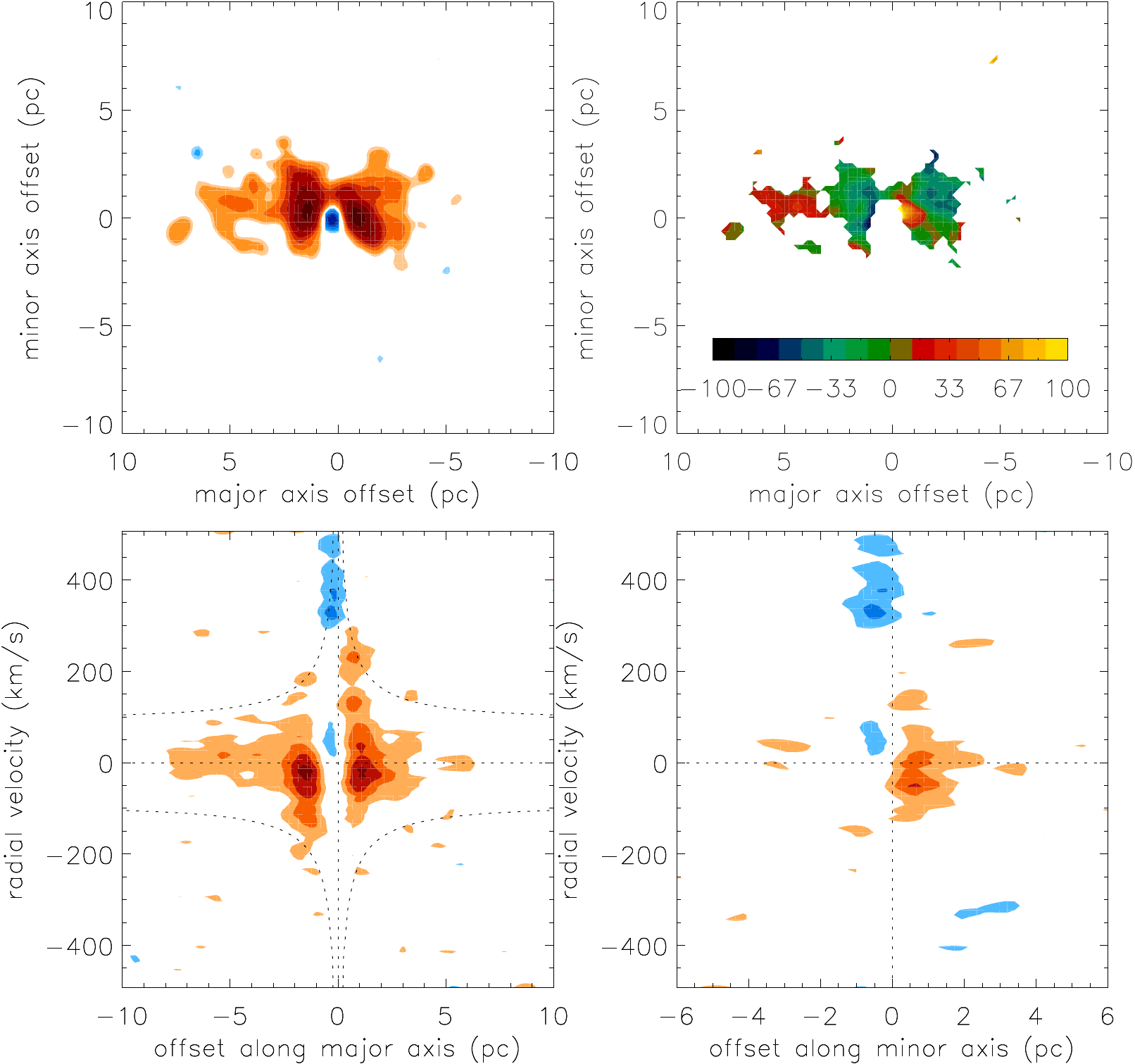}\includegraphics{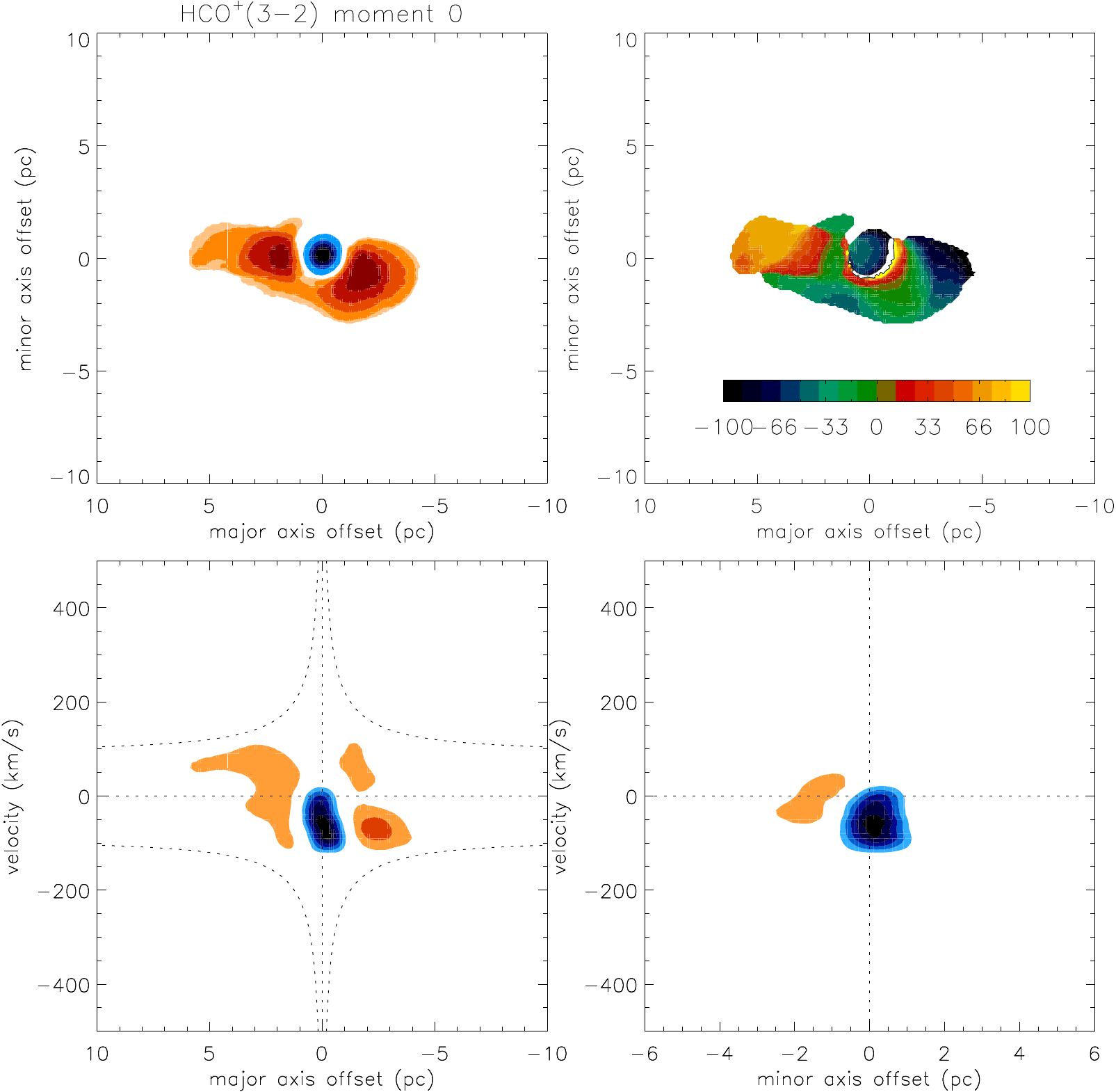}}
  \put(-475,230){\large observations}
  \put(-490,155){\large HCO$^+$(3-2)}
  \put(-210,230){\large model}
   \put(-490,230){\large (a)}\put(-360,230){\large (b)}
   \put(-490,105){\large (c)}\put(-360,105){\large (d)}
   \put(-230,155){\large HCO$^+$(3-2)}
  \put(-230,230){\large (e)}\put(-100,230){\large (f)}
  \put(-230,105){\large (g)}\put(-100,105){\large (h)}
  \caption{NGC~1068. 
    Left half: high resolution HCO$^+$(3-2) observations (Imanishi et al. 2020).
    Panel (a) moment~0 map ($I_0=18420 $~K\,km\,s$^{-1}$), (b)
    moment~1 map (in km\,s$^{-1}$), (c) pv diagram along the major axis, and (d) pv diagram along the minor axis ($T_{\rm b0}=75$~K).
    Right half: dense disk cloud model. Panel (e) moment~0 map, (f) moment~1 map (in km\,s$^{-1}$), (g) pv diagram along the major axis,
   and (h) pv diagram along the minor axis. The model levels are the same as those of the observed maps.
  \label{fig:plottingvollcnd_hco32radex_newQ=30}}
\end{figure*}

Imanishi et al. (2018, 2020) observed NGC~1068 in the HCO$^+$(3-2) line with ALMA at a resolution of $50$-$70$~mas ($3.5$-$4.9$~pc) and 
$20$~mas ($1.4$~pc), respectively. 
The comparison between the model high resolution HCO$^+$(3-2) datacube is presented in Fig.~\ref{fig:plottingvollcnd_hco32radex_newQ=30}
together with the observations by Imanishi et al. (2020).  
The observed counter-rotation between the inner and outer parts of the 
gas disk is also present in the model HCO$^+$(3-2) pv diagram and velocity field (moment~1 map).
In contrast to the observations, the model HCO$^+$(3-2) pv diagram shows a significantly deeper absorption line in the 
central resolution element than the model HCN(3-2) pv diagram. As for the HCN(3-2) line, 
the HCO$^+$ emission of the model pv diagram along the minor axis is located on the opposite side of
the observed HCO$^+$ emission. 

\subsubsection*{HCO$^+$(4-3)}

\begin{figure}[!ht]
  \centering
  \resizebox{\hsize}{!}{\includegraphics{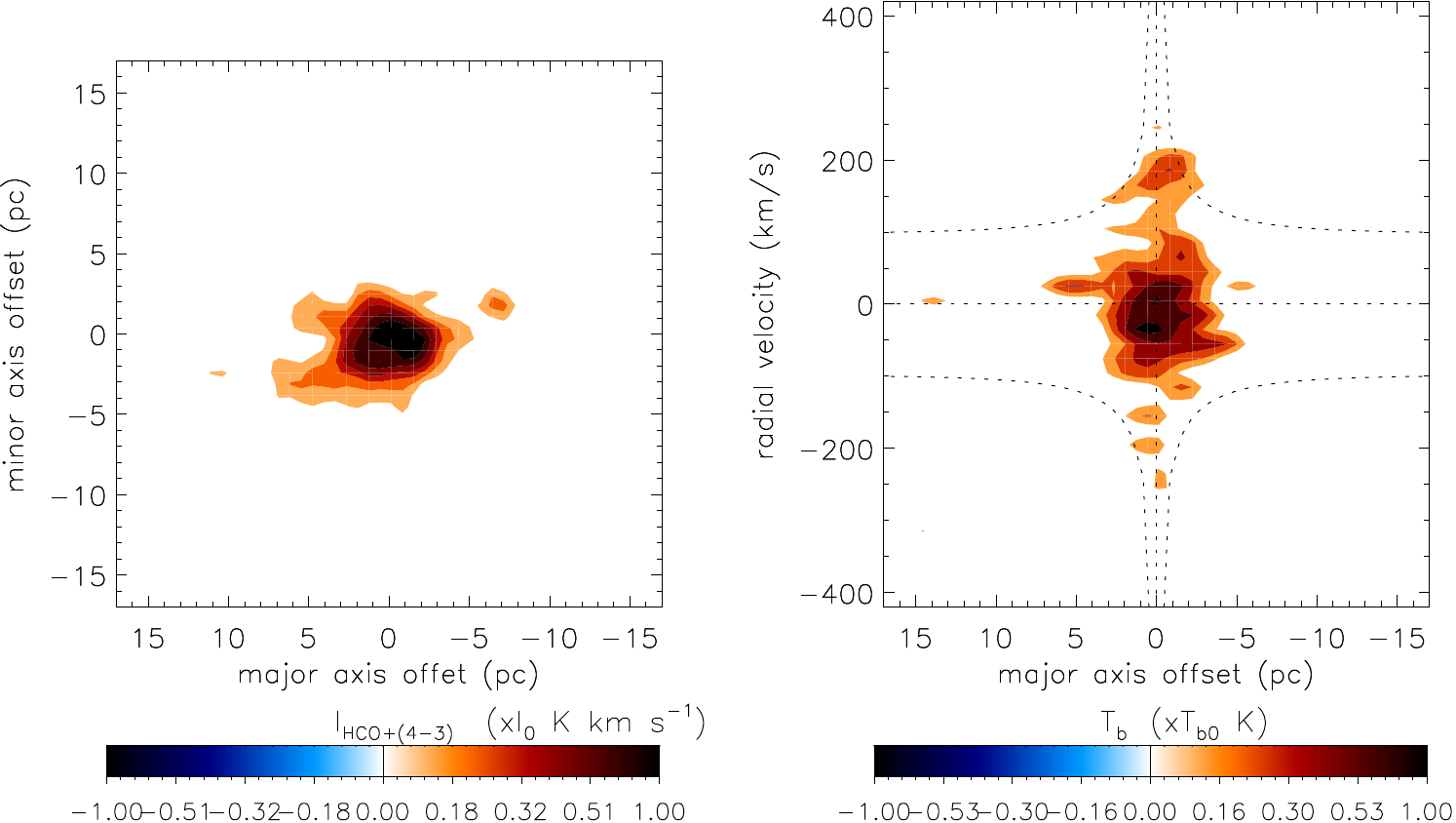}\put(-360,200){\huge observations}\put(-380,80){\huge HCO$^+$(4-3)}\put(-390,200){\huge (a)}\put(-160,210){\huge (b)}}
  \resizebox{\hsize}{!}{\includegraphics{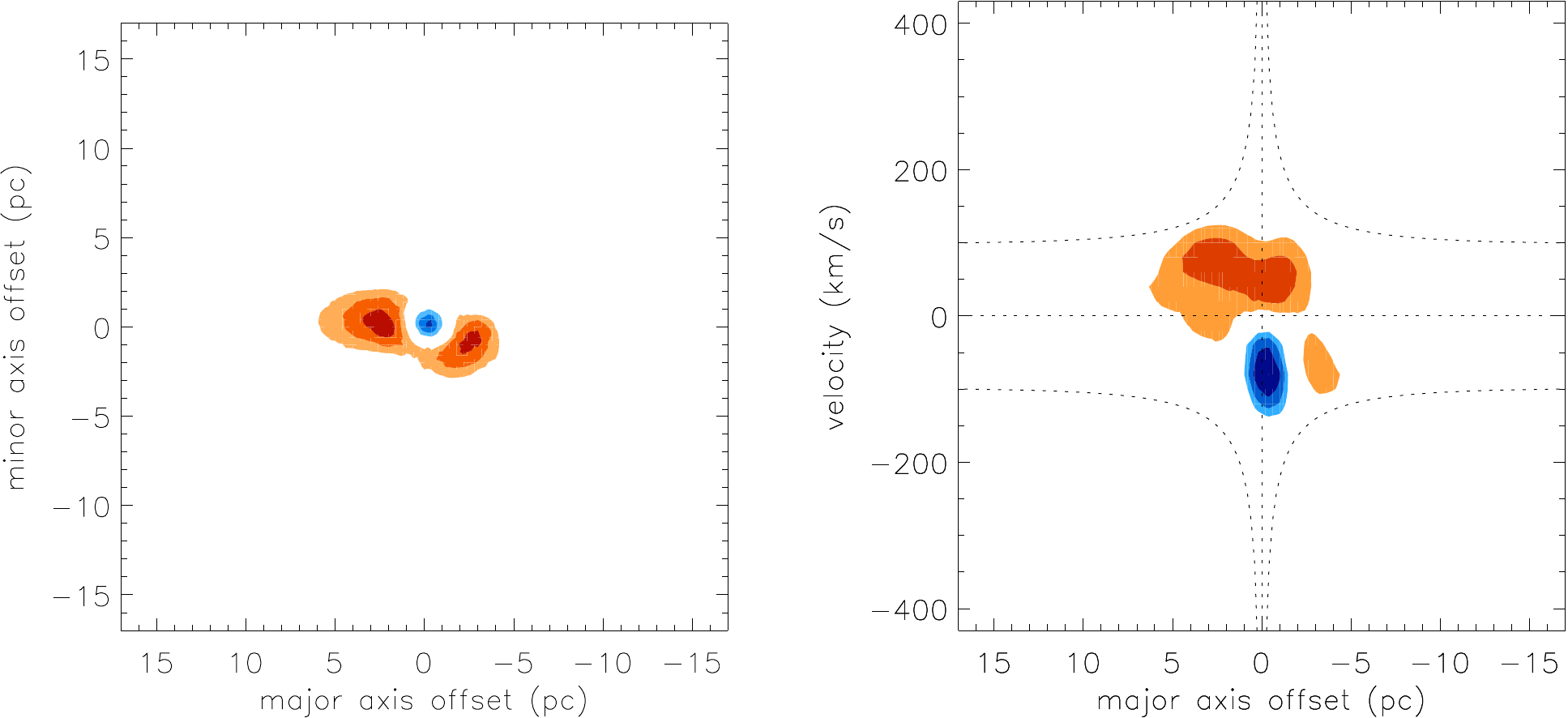}\put(-500,50){\Huge HCO$^+$(4-3)}\put(-465,210){\Huge model}\put(-500,210){\Huge (c)}\put(-200,210){\Huge (d)}}
  \caption{NGC~1068. Panel (a) observed HCO$^+$(4-3) moment~0 map ($I_0=7217$~K\,km\,s$^{-1}$), (b) pv diagram ($T_{\rm b0}=47.5$~K) 
    from Garcia-Burillo et al. (2019), (c) dense disk cloud model HCO$^+$(4-3) moment~0 maps, and (d) pv diagrams along the major axis.
    The model levels are the same as those of the observed maps.
  \label{fig:plottingvollcnd_hcn32radex_new_garciaburillo_hco43hr}}
\end{figure}

Garcia-Burillo et al. (2019) observed NGC~1068 in the HCO$^+$(4-3) line with a resolution of $40$~mas or $2.8$~pc 
(left side of Fig.~\ref{fig:plottingvollcnd_hcn32radex_new_garciaburillo_hco43hr}).
The emission distribution is concentrated within the inner $\pm 50$~mas or $3.5$~pc. 
The corresponding model HCO$^+$(4-3) moment~0 map and pv diagram are presented in Fig.~\ref{fig:plottingvollcnd_hcn32radex_new_garciaburillo_hco43hr}.
The observed HCO$^+$(4-3) emission is about three times stronger than the model emission. 
Contrary to the observations, most of the model HCO$^+$(4-3) line is seen in absorption at negative velocities.
We suggest that the small integrated HCO$^+$(4-3) intensity indicates the presence of an intercloud medium (Sect.~\ref{sec:intercloudN1068}).

\subsubsection*{HCN, HCO$^+$, H$^{13}$CN, and H$^{13}$CO$^+$ spectra along the major axis}

To make a more quantitative comparison between the model and the observations, we present the observed HCN(3-2) spectra 
(Impellizzeri et al. 2019) and HCO$^+$(3-2), H$^{13}$CN(3-2), and H$^{14}$CO$^+$(3-2) spectra (Imanishi et al. 2020) from resolution elements 
along the major axis in Appendix~\ref{sec:spectramaj}. The model resembles the observed HCN(3-2) and HCO$^+$(3-2) spectra only qualitatively.
Given the tiny amount of H$^{13}$CN(3-2) emission and the absence of H$^{13}$CO$^{+}$(3-2) emission in the model spectra, 
our assumed isotope ratio $^{12}C/^{13}C = 30$ is justified.

\subsubsection{Infrared pumping \label{sec:pumping}}

\begin{figure*}[!ht] 
  \centering
  \resizebox{\hsize}{!}{\includegraphics{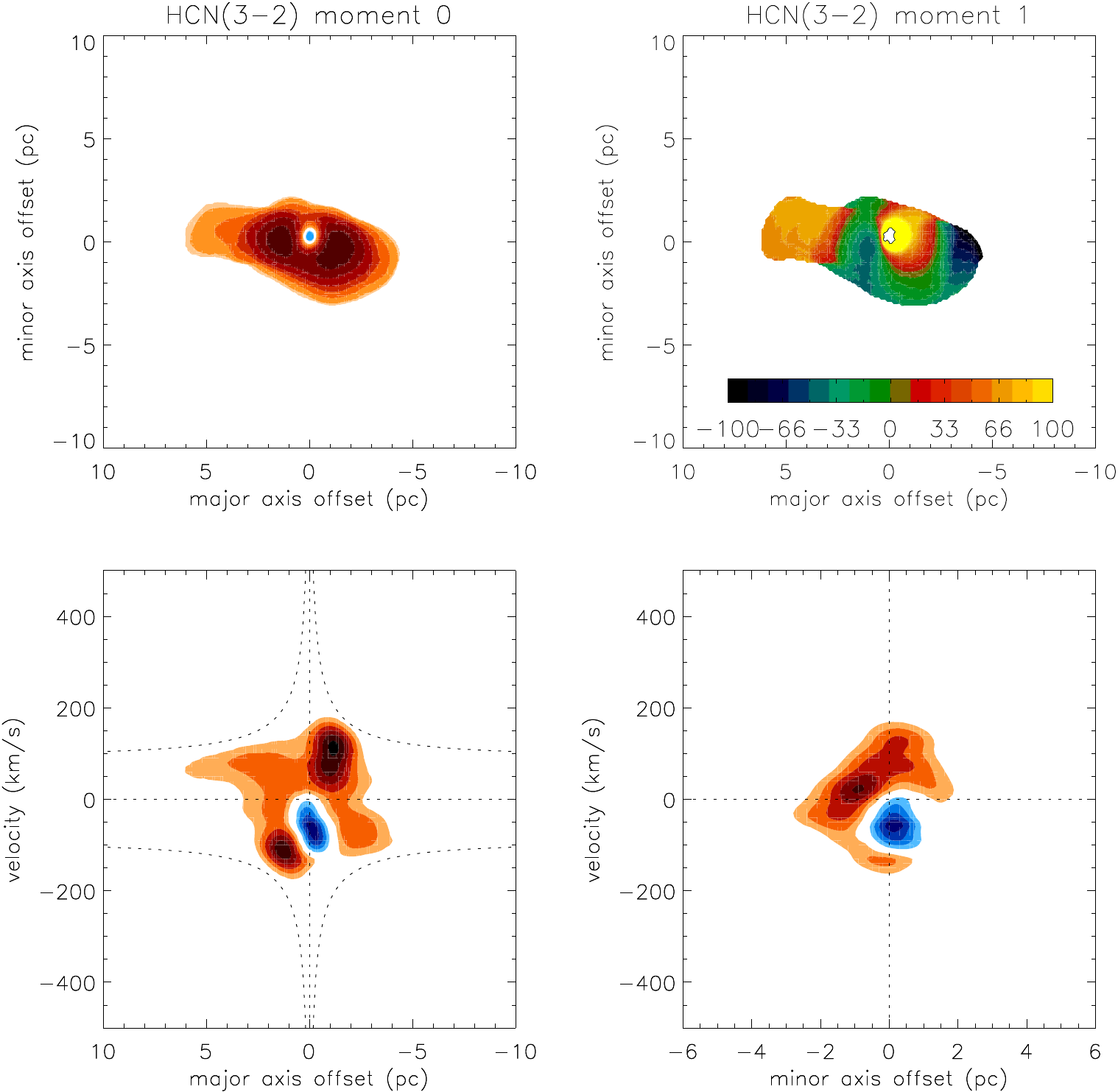}\includegraphics{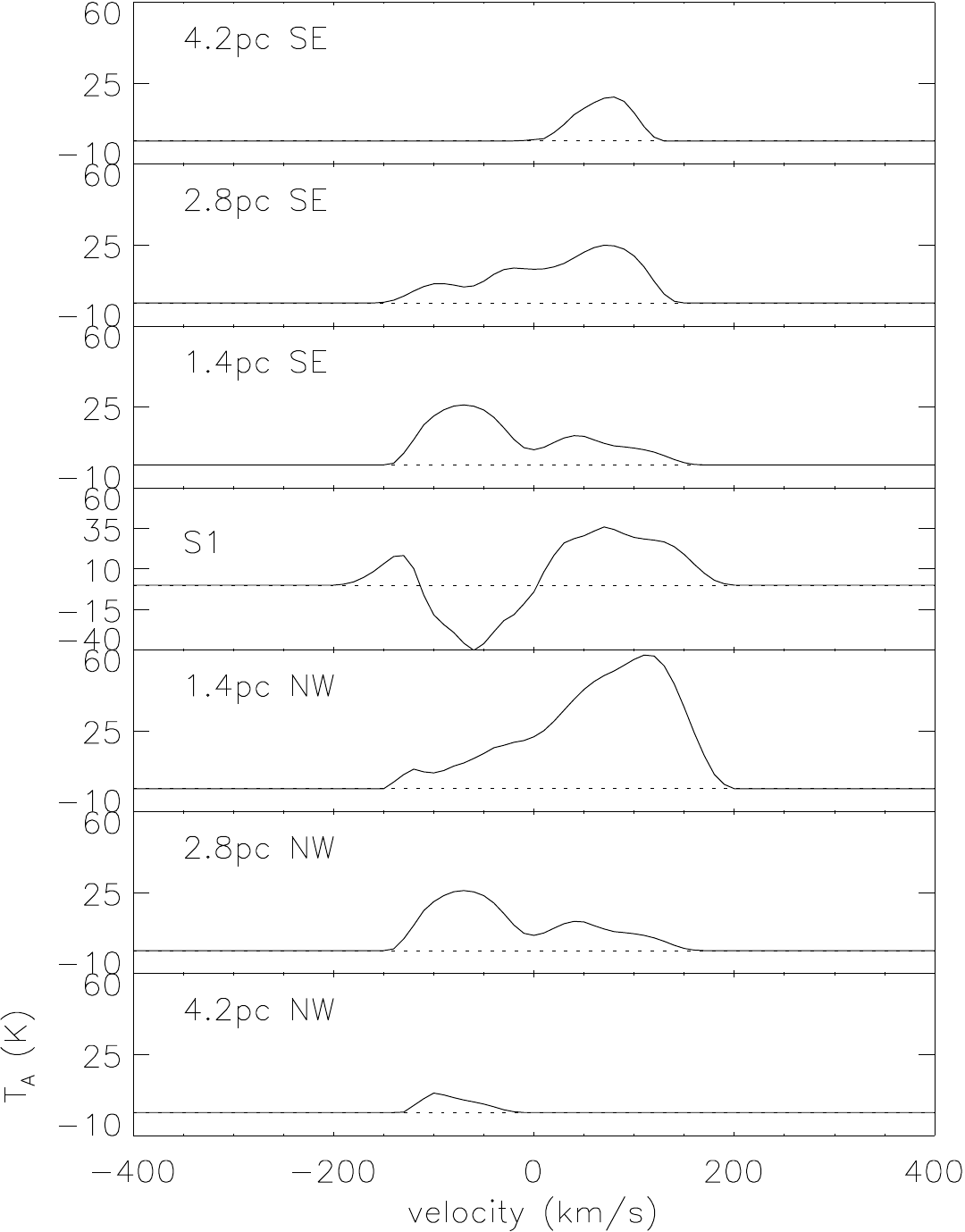}}
  \put(-490,185){\large without IR-pumping}
  \put(-110,269){\large without IR-pumping}
   \put(-490,270){\large (a)}\put(-330,270){\large (b)}
   \put(-490,125){\large (c)}\put(-330,125){\large (d)}
  \caption{IR pumping.  Left half: high resolution HCN(3-2) emission of the dense disk cloud model. Panel (a) moment~0 map,
    (b) moment~1 map, (c) pv diagram along the major axis, and (d) pv diagram along the minor axis.
    The model levels are the same as those of the observed maps (Fig.~\ref{fig:plottingvollcnd_hcn32radex_newQ=30}).
    Right half: model high-resolution HCN(3-2) spectra along the major axis. 
  \label{fig:plottingvollcnd_hcn32radex_newQ=30pump}}
\end{figure*}

The results for the HCN(3-2) emission in the absence of IR pumping (see Sect.~\ref{sec:mollinem}) of the $Q=30$ model are presented in 
Fig.~\ref{fig:plottingvollcnd_hcn32radex_newQ=30pump}.
The influence of IR-pumping is best appreciated in the pv diagrams and spectra: 
the brightness temperatures decrease by about a factor of two and thus the area of regions with detected HCN(3-2) emission decreases. 
The total integrated intensity is $1.2$~Jy\,km\,s$^{-1}$ instead of $2.7$~Jy\,km\,s$^{-1}$ with IR pumping.
We conclude that IR-pumping increases the HCN(3-2) emission by about a factor of two. This is consistent with the small amount of 
HCN-VIB J=3-2 emission detected by Imanishi et al. (2020).

\subsubsection{The intercloud gas \label{sec:intercloudN1068}}

As for the CND (Sect.~\ref{sec:intercloudCND}), the contribution of the intercloud gas to the molecular line emission has to be evaluated.
Assuming a lognormal probability density function for the gas density (Eq.~\ref{eq:pdfgasfrac}) with ${\cal{M}} = 35$ and $x=50$ we find $\frac{\Delta M}{M} \sim 0.3$. 
Thus, as in the CND about $70$\,\% of the gas mass is expected to have lower gas densities than that of the dense clouds.
Since the mean gas density is proportional to $Q^{-1}$, the mean gas density of the thick gas disk in NGC~1068 is expected to
be about $10$ times higher than in the CND. The probability that the intercloud gas in the thick disk of NGC~1068 is optically
thick  leading to important self-absorption in the molecular lines is thus higher than in the CND.

\begin{figure}[!ht]
  \centering
  \resizebox{\hsize}{!}{\includegraphics{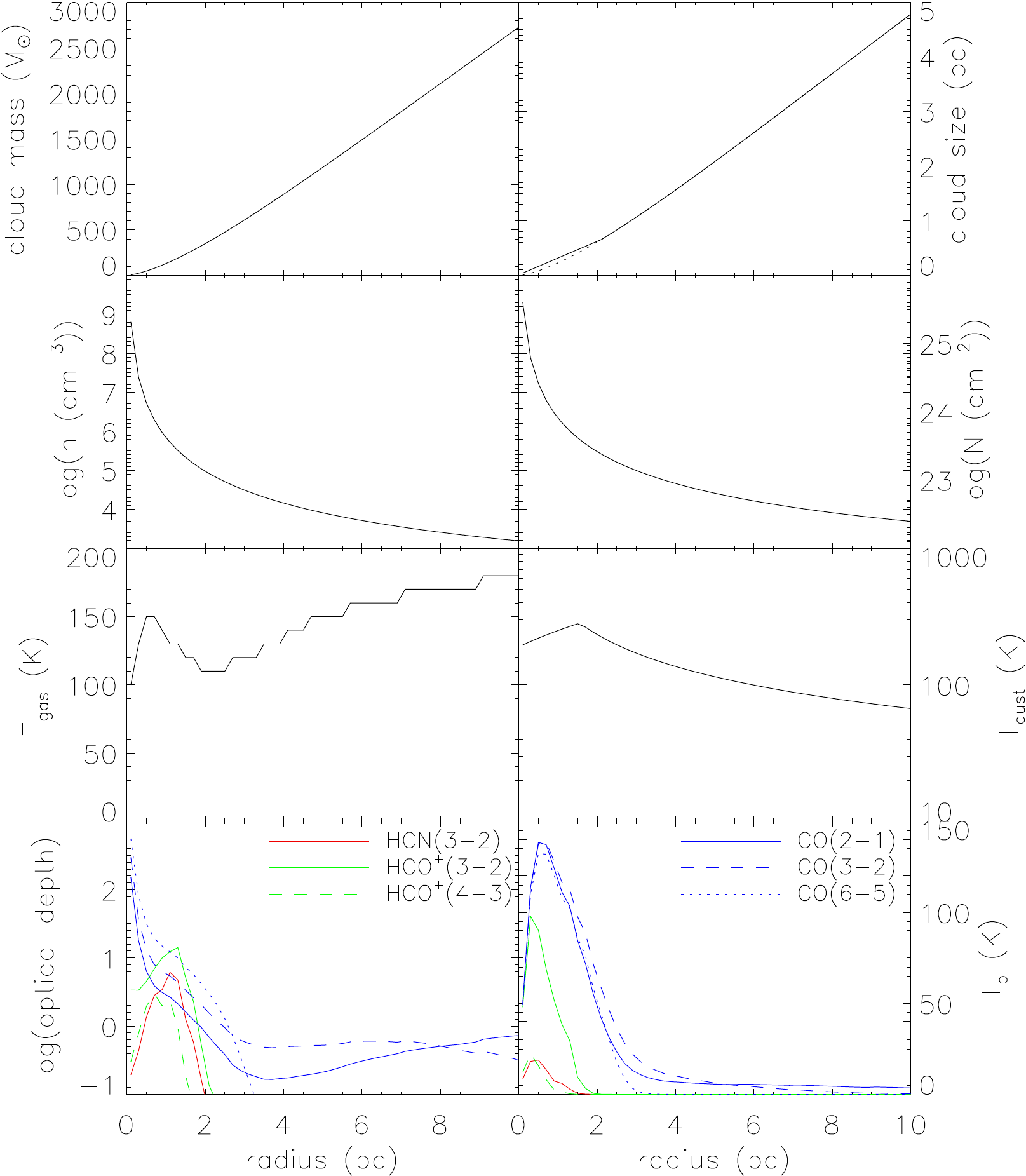}}
  \put(-220,280){\large Q=30 model}
  \caption{NGC~1068 intercloud gas model. Characteristics of individual model intercloud gas across the disk at the time of interest.
    $Q=30$, $\zeta_{\rm CR}=2 \times 10^{-13}$~s$^{-1}$, and full turbulent heating.
    Gas cloud characteristics. From upper left to lower right: cloud mass, size, density, H$_2$ column density, 
    gas and dust temperatures, optical depth, and brightness temperature.
  \label{fig:profiles_n1068_plotall_diffQ=10}}
\end{figure}

The NIR interferometric observations of the Gravity Collaboration (2019) revealed an extinction of $A_{\rm V} \sim 100$ in front
of the elongated central emission corresponding to a gas column density of $\sim 10^{23}$~cm$^{-2}$. 
The same column density was measured towards the position of the AGN from the CO(2-1) and CO(3-2) intensities by Garcia-Burillo et al. (2019).
This gas has to have a high area filling factor because a partial coverage of the extended NIR emission by optically thick clouds would not lead 
to the observed reddening. It is remarkable that the model of continuous thick stratified gas disk of Vollmer et al. (2018) reproduces the 
observationally derived column density in front of the central source.
Based on the available observations it cannot be excluded that the intercloud gas dominates the molecular
line emission. In order to obtain an optically thin intercloud gas, the Toomre $Q$ parameter has to be set to $Q=100$ and the dissipation 
rate at the driving length must be decreased by at least a factor of $1/200$ compared to Eq.~\ref{eq:turbheat} 
(left panel of Fig.~\ref{fig:profiles_n1068_plotall_diffQ=10}). This is four times less than the factor required for the CND.
As for the CND, we can only speculate that a strong uniform magnetic field, as it is observed by Lopez-Rodriguez et al. (2020), helps to
suppress the turbulent heating rate within the inertial range $l_{\rm cl} \la l \la H$. Indeed, these authors determined the ratio
between the turbulent and the uniform magnetic field strengths to be $0.06$.

On the other hand, for an optically thick intercloud gas with brightness temperatures high enough to be detected by the available observation setups,
we had to assume a $Q=30$ disk with full heating (Eq.~\ref{eq:turbheat}) and diffuse clouds of size $H/2$ and density of three times the mean density. 
As for the dense clouds we assumed a CR ionisation rate of $\zeta_{\rm CR}=2 \times 10^{-13}$~s$^{-1}$. 
The resulting properties of the diffuse clouds are presented in 
the right panel of  Fig.~\ref{fig:profiles_n1068_plotall_diffQ=10}. For radii $\la 2$~pc all lines are optically thick. 
The maximum of the molecular brightness temperatures is located at a radius of $\sim 0.5$~pc. 

\begin{figure*}[!ht]
  \centering
  \resizebox{\hsize}{!}{\includegraphics{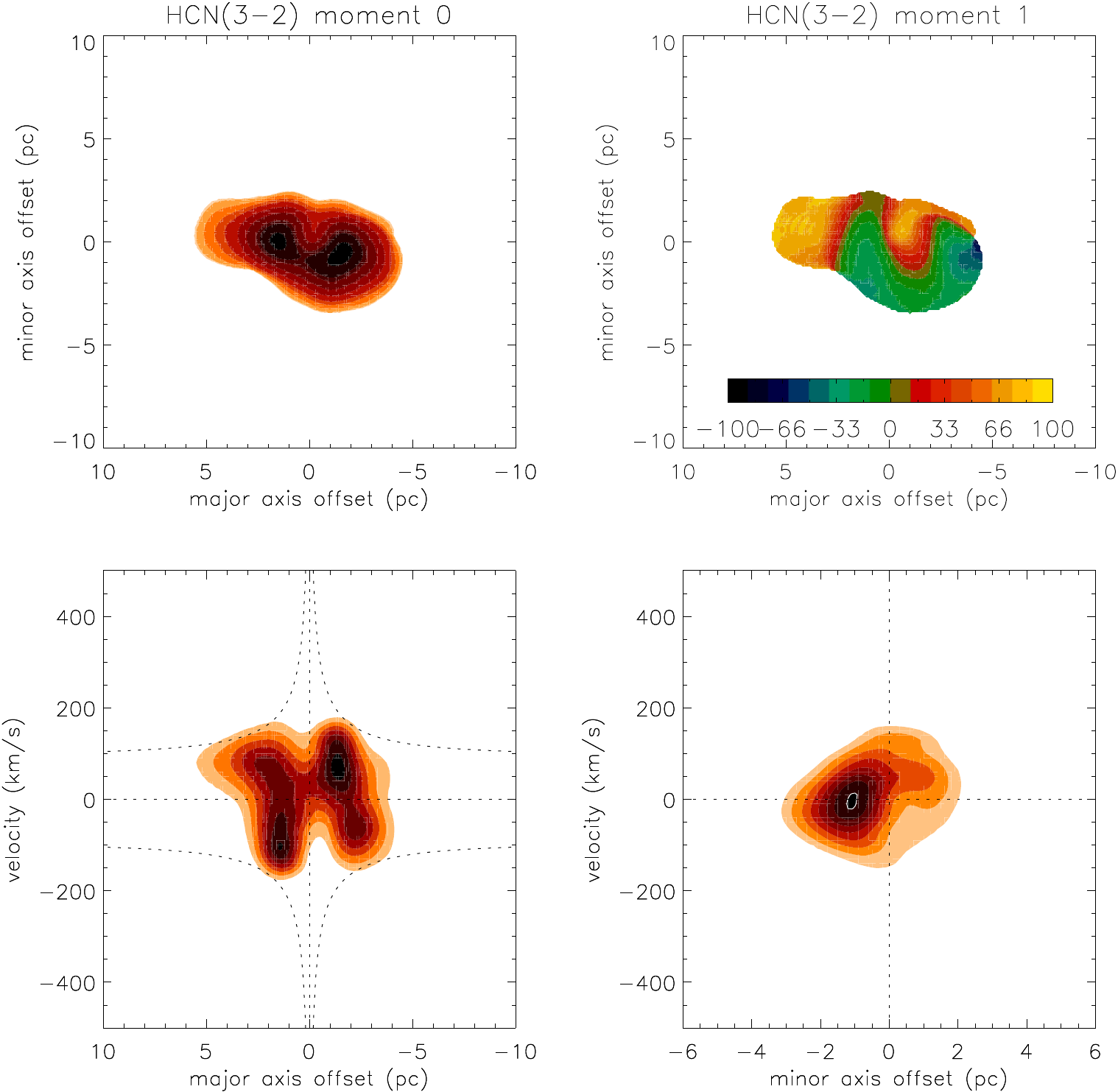}\includegraphics{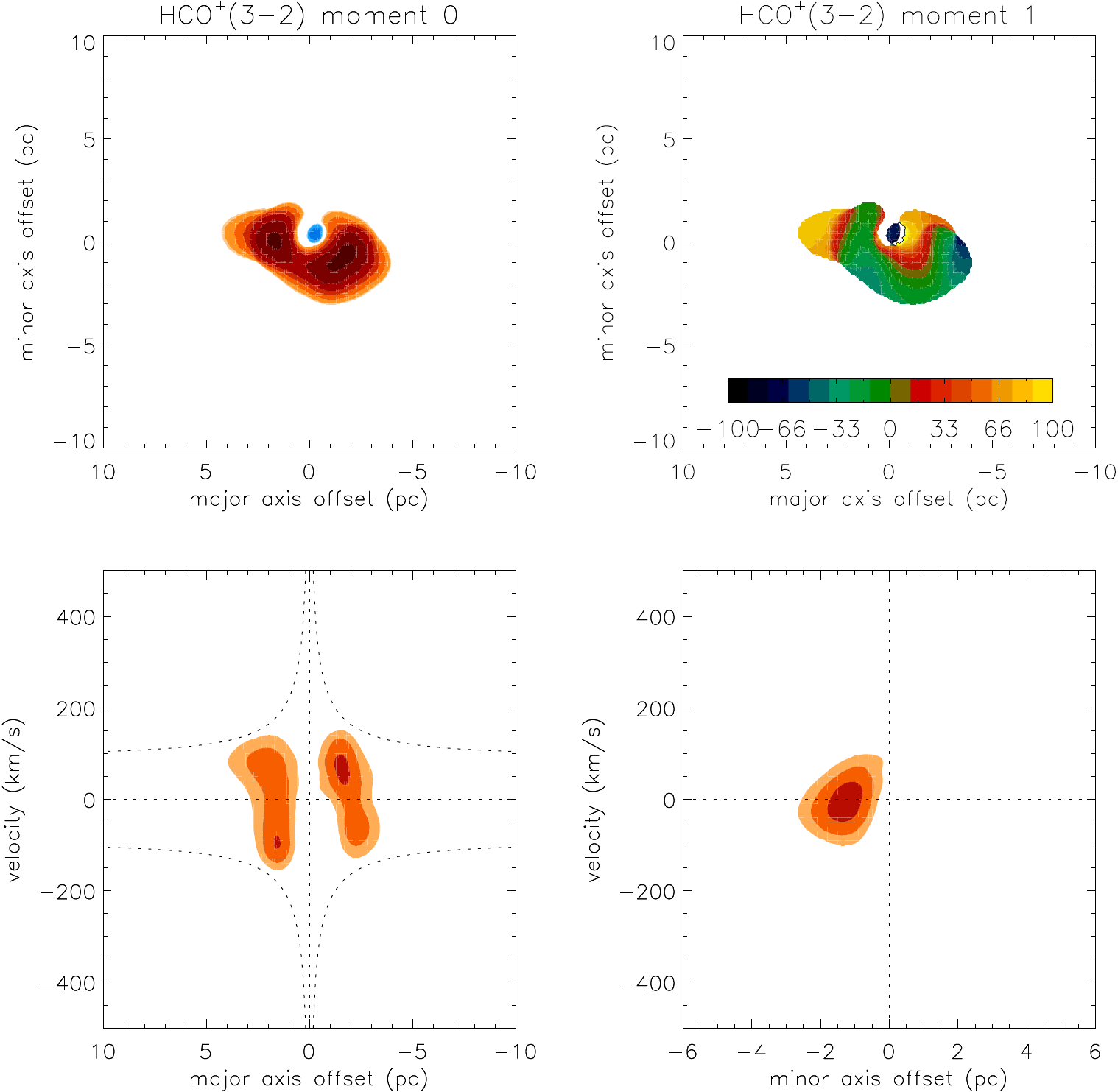}}
   \put(-490,230){\large (a)}\put(-355,230){\large (b)}
   \put(-490,105){\large (c)}\put(-355,105){\large (d)}
  \put(-230,230){\large (e)}\put(-95,230){\large (f)}
  \put(-230,105){\large (g)}\put(-95,105){\large (h)}
  \caption{NGC~1068 intercloud gas model. Left half: HCN(3-2); right half: HCO$^+$(3-2). Panels (a) and (e) model moment~0 maps,
    (b) and (f) model moment~1 maps, (c) and (g)
    pv diagrams along the major axis, (d) and (h) pv diagrams along the minor axis. 
    The model levels are the same as those of the observed maps (Figs.~\ref{fig:plottingvollcnd_hcn32radex_newQ=30} and 
    \ref{fig:plottingvollcnd_hco32radex_newQ=30}), except for the white contour at $80$~K.
  \label{fig:plottingvollcnd_hcn32radex_newQ=30diffvturbfullheat33Q=10_3}}
\end{figure*}

The resulting high-resolution model HCN(3-2) and HCO$^+$(3-2) observations are presented in 
Fig.~\ref{fig:plottingvollcnd_hcn32radex_newQ=30diffvturbfullheat33Q=10_3}. 
The HCN(3-2) emission of  the intercloud gas is comparable to that of the dense disk clouds (Fig.~\ref{fig:plottingvollcnd_hcn32radex_newQ=30}).
As expected from the surface brightness profiles,
the line emission is concentrated within the inner $\sim 3$~pc. The extent of the model HCN intercloud emission is thus smaller than that of the
observations (lower panels of Fig.~\ref{fig:plottingvollcnd_hcn32radex_newQ=30}).
The model HCN linewidth is smaller but comparable to that of the observations. The highest observed velocities ($v > 200$~km\,s$^{-1}$) are not
reproduced by the model. The intercloud model does not show the HCN(3-2) line in absorption.
 
As the model HCN(3-2) emission, the model HCO$^+$(3-2) emission is more centrally concentrated. 
Otherwise, it shows similar spatial and kinematic structures with somewhat higher brightness temperatures. 

\begin{figure}[!ht]
  \centering
  \resizebox{\hsize}{!}{\includegraphics{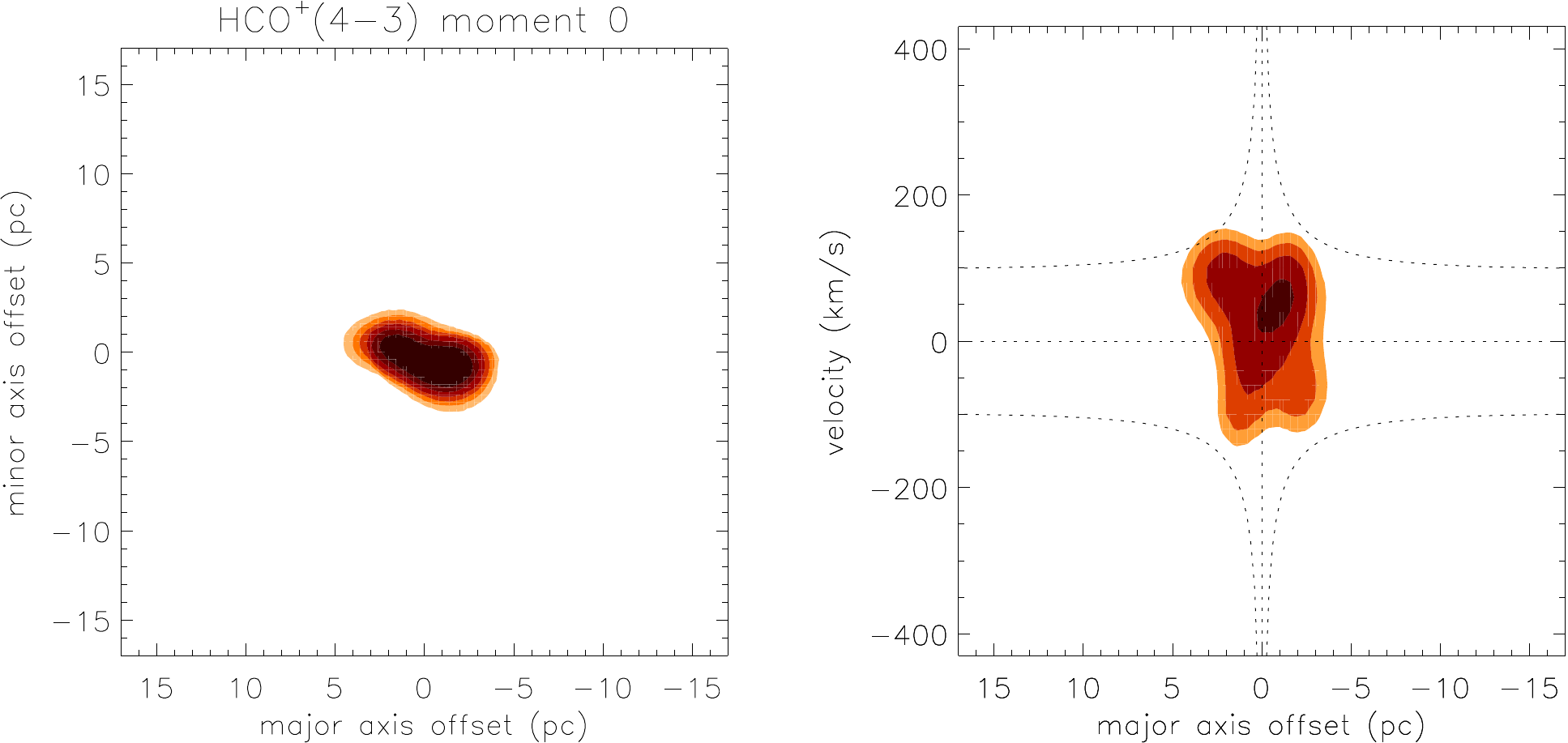}\put(-505,210){\Huge (a)}\put(-205,220){\Huge (b)}}
  \resizebox{\hsize}{!}{\includegraphics{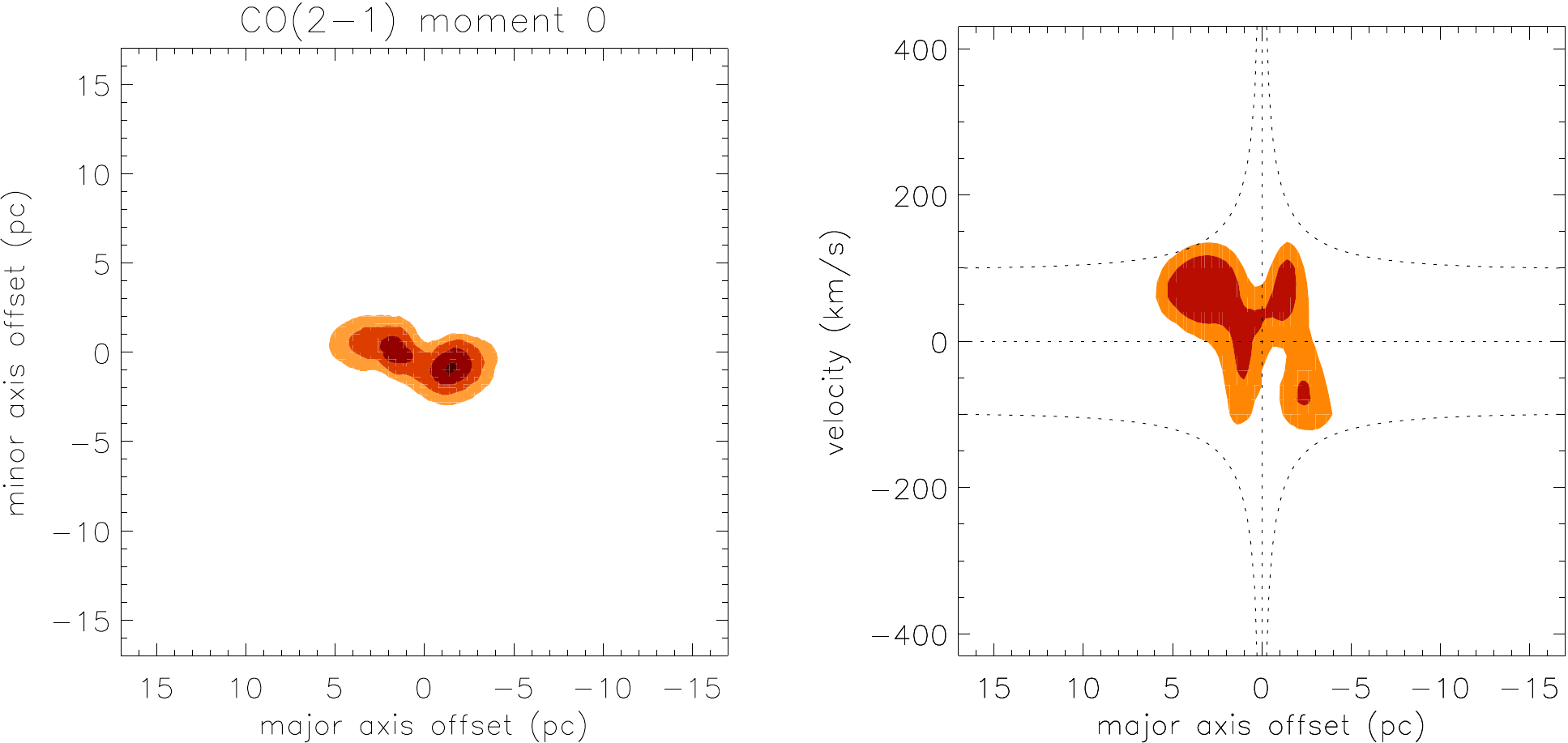}\put(-505,210){\Huge (c)}\put(-205,220){\Huge (d)}}
  \resizebox{\hsize}{!}{\includegraphics{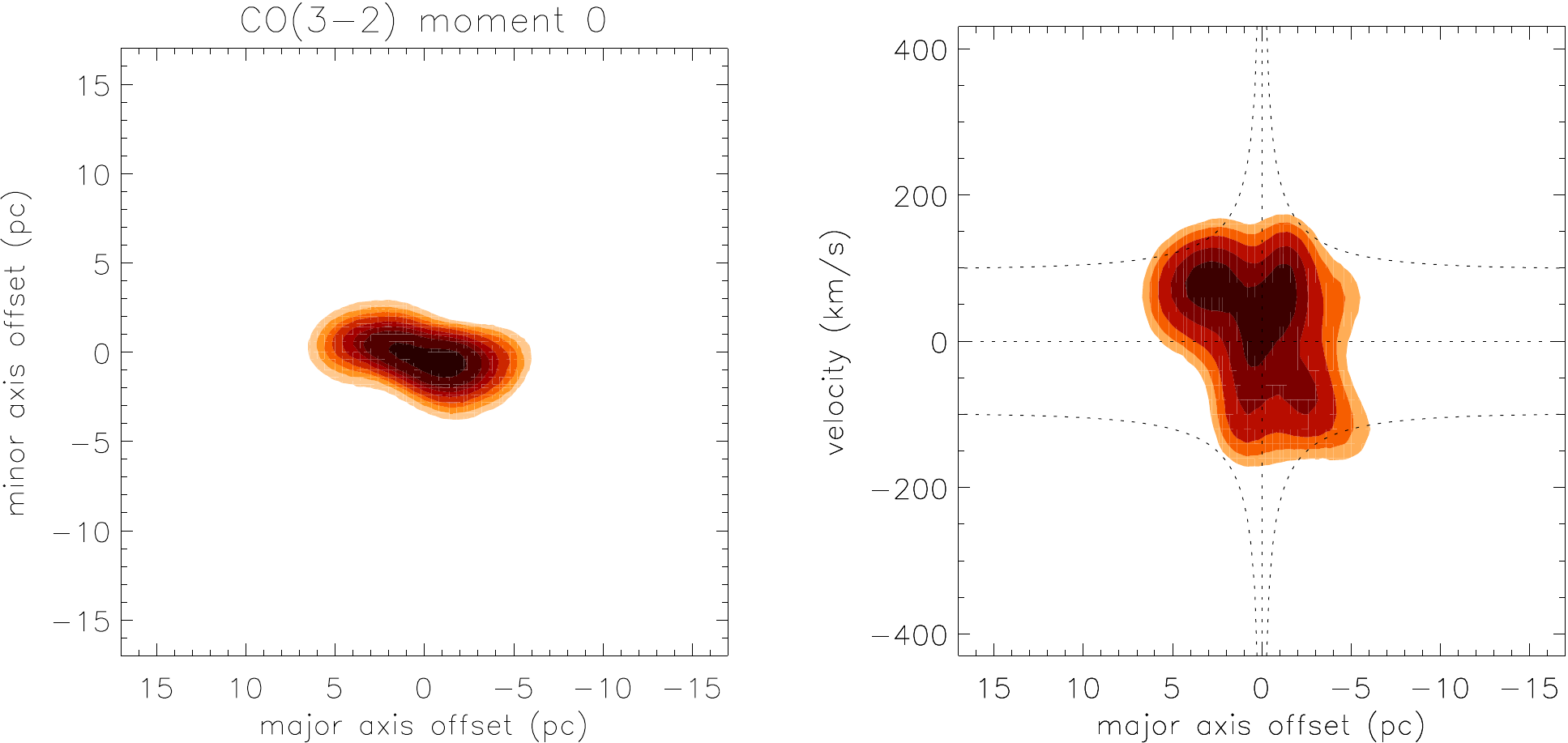}\put(-505,210){\Huge (e)}\put(-205,220){\Huge (f)}}
  \resizebox{\hsize}{!}{\includegraphics{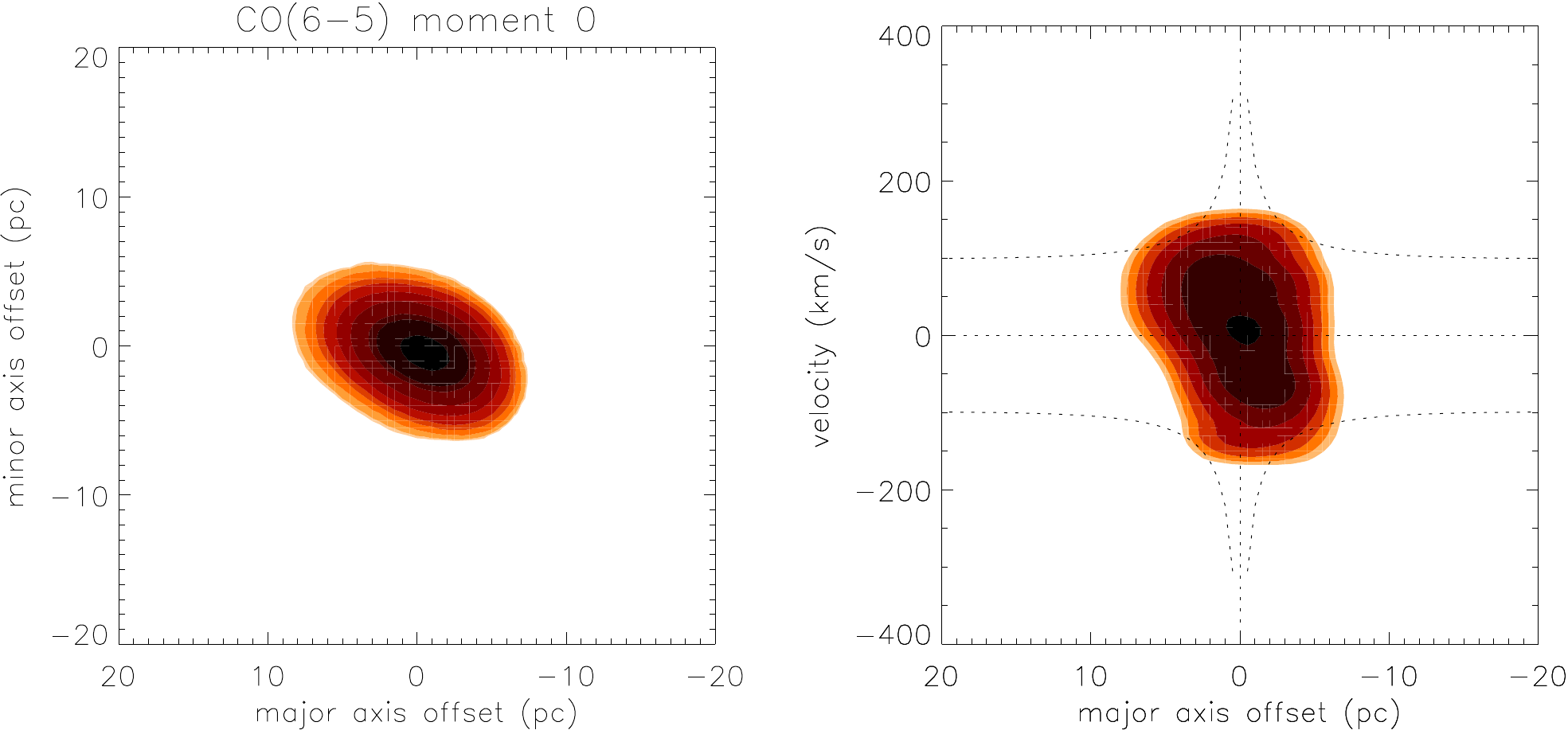}\put(-505,210){\Huge (g)}\put(-205,220){\Huge (h)}}
  \caption{NGC~1068 intercloud gas model. Panel (a) and (b) HCO$^{+}$(4-3) moment~0 maps and pv diagrams, 
    (c) and (d) CO(2-1) moment~0 maps and pv diagrams, (e) and (f) CO(3-2) moment~0 maps and pv diagrams, 
    (g) and (h) CO(6-5) moment~0 maps and pv diagrams. The model levels are the same as 
    those of the observed maps (Figs.~\ref{fig:plottingvollcnd_hcn32radex_new_garciaburillo_co21hr}, \ref{fig:plottingvollcnd_hcn32radex_new_gallimore1},
    and \ref{fig:plottingvollcnd_hcn32radex_new_garciaburillo_hco43hr}).
  \label{fig:plottingvollcnd_hcn32radex_new_garciaburillo_hco43hrQ=30diffvturbfullheat33Q=10_3}}
\end{figure}

The model HCO$^+$(4-3), CO(2-1), CO(3-2), and CO(6-5) are presented in 
Fig.~\ref{fig:plottingvollcnd_hcn32radex_new_garciaburillo_hco43hrQ=30diffvturbfullheat33Q=10_3}.
The corresponding observations are shown in Figs.~\ref{fig:plottingvollcnd_hcn32radex_new_garciaburillo_hco43hr}, 
\ref{fig:plottingvollcnd_hcn32radex_new_garciaburillo_co21hr}, and 
\ref{fig:plottingvollcnd_hcn32radex_new_gallimore1}. The line emission is concentrated within a radius of $\sim 3$~pc for all lines.
In line with the observations, the HCO$^+$(4-3), CO(2-1), CO(3-2), and CO(6-5) lines are not seen in absorption. 

We conclude that the contribution of the intercloud gas to the total emission is expected to be highest at radii $\la 3$~pc. 
The observed absence of line absorption in the HCO$^+$(4-3) lines might be a hint to the existence of an intercloud medium.

\subsubsection{The photodissociation region \label{sec:pdr}}

In Sect.~\ref{sec:XDR} we argue that the X-ray emission is entirely absorbed by dense gas located inside the region
where the gas emits molecular line emission at $R \la 0.3$~pc. This gas is most-probably located within the dust sublimation radius and
is therefore dust-free and optically thin for the UV and optical emission.
All gas surfaces, which are directly illuminated by UV and optical emission from the AGN, give rise to photodissociation regions (PDR).
The models presented in Sects.~\ref{sec:denseclouds} and \ref{sec:intercloudN1068} do not include the effects of optical and UV emission on the gas.
To determine the influence of PDRs on the observed molecular emission distributions, we treat the PDR emission separately and assume
in the following that the gas is only heated by the UV and optical emission, i.e. there is no turbulent mechanical heating.
These PDR strongly emit in molecular lines (e.g., Meijerink \& Spaans 2005, Meijerink et al. 2007).
We assumed a bolometric luminosity of $L \sim 3 \times 10^{44}$~erg\,s$^{-1}$ for NGC~1068 (Vollmer et al. 2018).
The HCN, HCO$^+$, and CO line emission were calculated with the {\tt Meudon PDR code} (Le Petit et al. 2006, Le Bourlot et al. 2012).
The code considers a stationary plane-parallel slab of gas and dust illuminated by a radiation field.
It solves at each point in the cloud, the radiative transfer in the UV taking into account the absorption in the continuum 
by dust and in discrete transitions of H and H2. The model computes the thermal balance taking into account heating processes 
such as the photoelectric effect on dust, chemistry, cosmic rays, etc. and cooling resulting from infrared and millimeter emission of 
the abundant species. Chemistry is solved for any number of species and reactions. 
Once abundances of atoms and molecules and level excitation of the most important species have been computed at each position in the cloud, 
line intensities are deduced by a post-processor code. We assumed $A_{\rm V}=30$ for the plane-parallel slab and a cosmic
ray ionization rate of $\zeta_{\rm CR}=10^{-14}$~s$^{-1}$. Such a high rate was found by Le Petit et al. (2016) who compared their 
PDR models with H$_3^+$ emission from dense clouds in the central molecular zone (CMZ) of the Galaxy.

\begin{figure}[!ht]
  \centering
  \resizebox{\hsize}{!}{\includegraphics{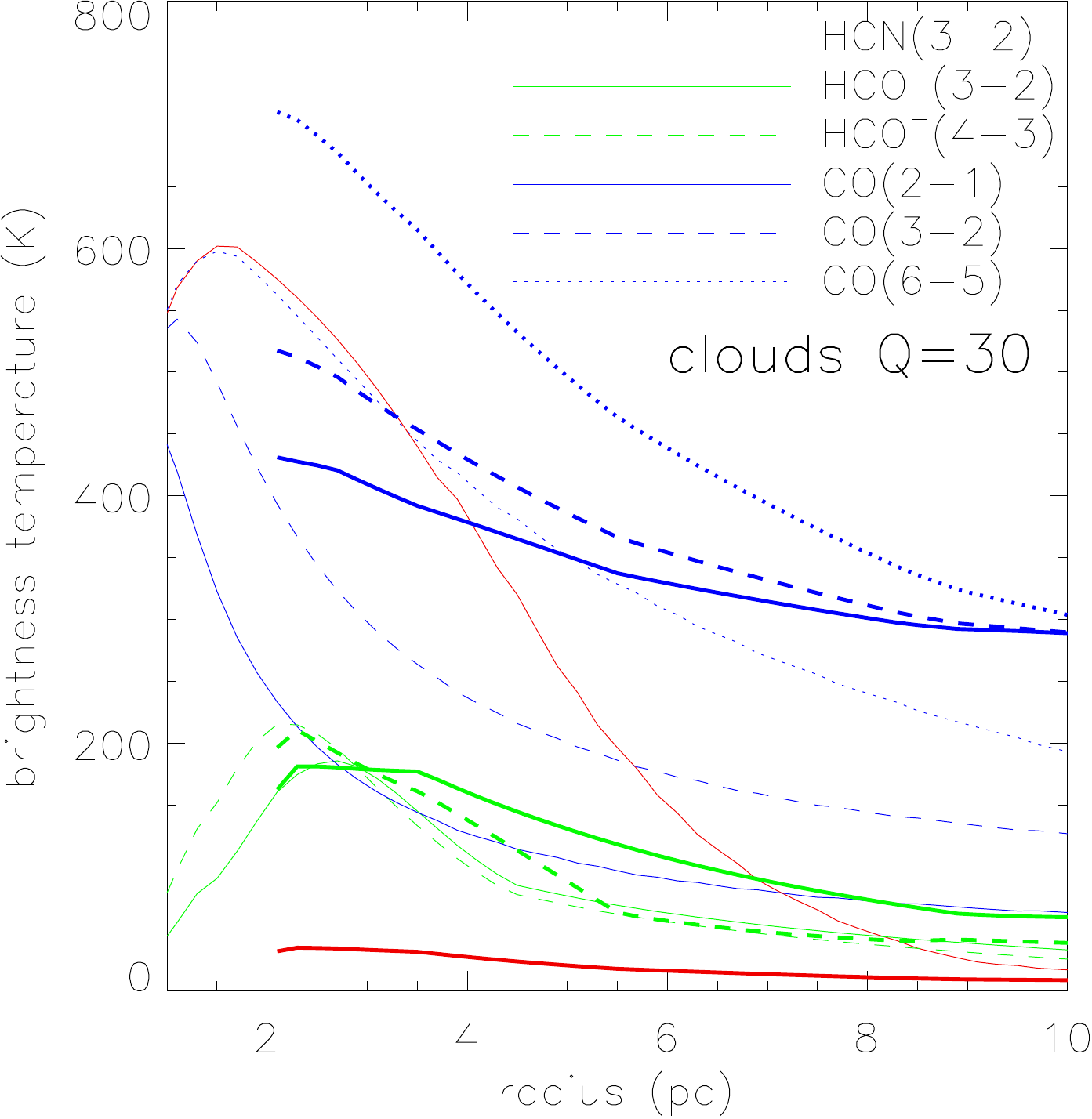}}
  \caption{Radial profiles of the molecular line emission from the dense clouds (thin lines) and the associated PDR regions (thick lines).
    The CR ionization rate is $\zeta_{\rm CR}=10^{-14}$~s$^{-1}$.
  \label{fig:profiles_n1068_plotall_pdr_Q=30}}
\end{figure}

The radial profiles of the molecular line emission for the dense clouds in $Q=30$ gas disk are presented in 
Fig.~\ref{fig:profiles_n1068_plotall_pdr_Q=30}. 
The CO and HCO$^+$(4-3) line emission from the PDR is stronger than, the PDR HCO$^+$(3-2) line emission is comparable to that due to mechanical and 
CR heating for all observed transitions.
On the other hand, the HCN(3-2) line emission from the PDR is much weaker than that due to mechanical and CR heating heating.
The reason for the low HCN(3-2) line emission from the PDR is caused by the temperature gradient of the PDR:
the temperature of the emitting region with an optical depth of $\tau_{\rm HCN} \sim 1$ is much lower in the PDR model than in the model 
with turbulent mechanical heating where all gas is uniformly heated. The high gas temperature at high densities leads to a
strong increase of the HCN abundance via a hot core chemistry.

The 3D configuration of the clouds is a thick ring with an inner wall which is directly illuminated (Vollmer et al. 2018).
Since the observed brightness temperature of the directly illuminated clouds is proportional to the cloud brightness temperature and 
their area filling factor, we expect the contribution of the PDR to the total emission to be dominant close to the
edge of the thick ring at a distance of about $1.5$~pc from the central black hole (Vollmer et al. 2018).
The mean free path of the clouds at the inner edge is about $1$~pc corresponding to the thickness of the PDR-dominated region.

\begin{figure}[!ht]
  \centering
  \resizebox{\hsize}{!}{\includegraphics{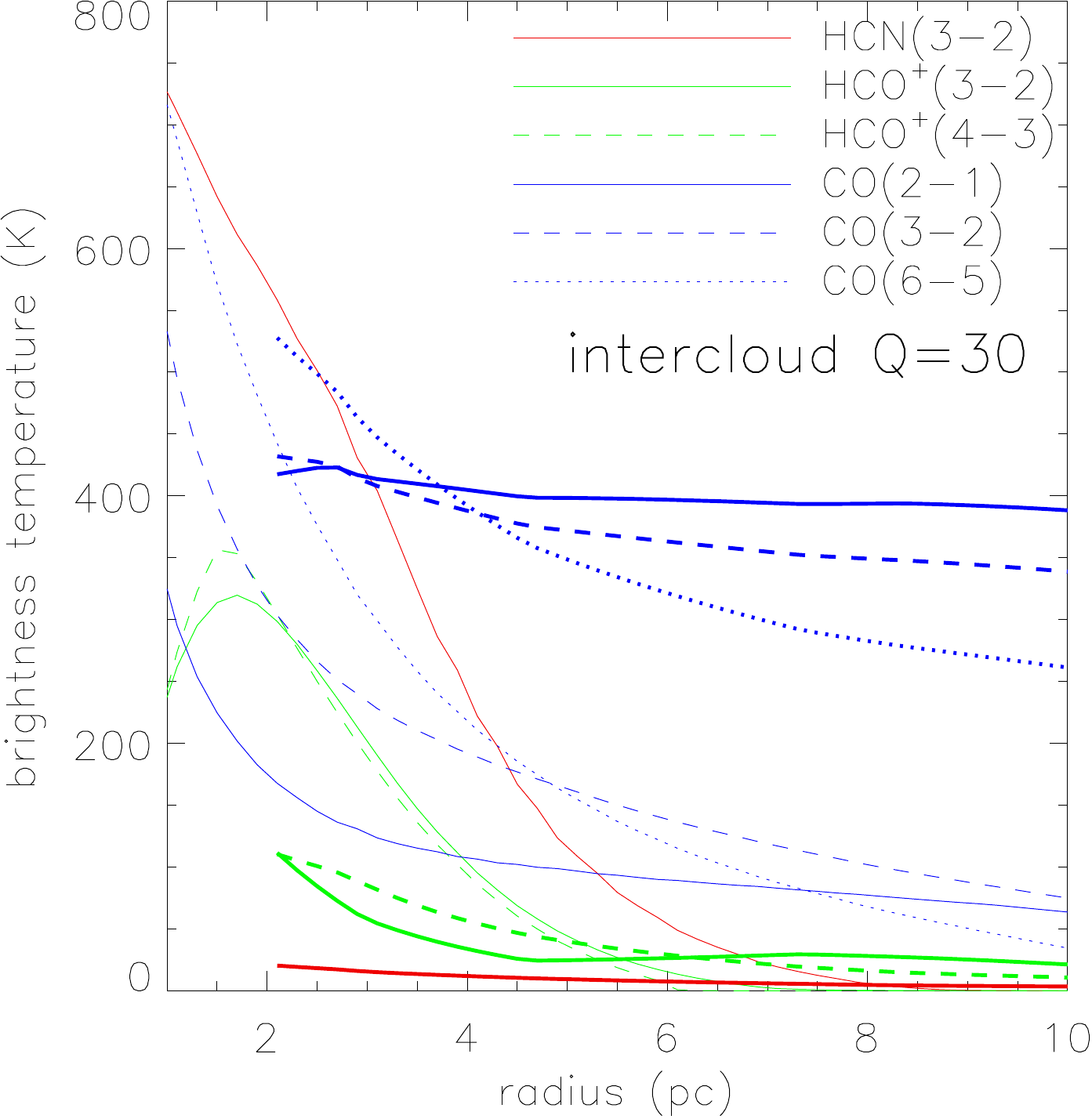}}
  \caption{Radial profiles of the molecular line emission from the intercloud gas (thin lines) and the associated PDR regions (thick lines).
    The CR ionization rate is $\zeta_{\rm CR}=10^{-14}$~s$^{-1}$.
  \label{fig:profiles_n1068_plotall_pdr_higher}}
\end{figure}

As before, we assume that the gas in the ring is turbulent and clumpy. The dense clouds are thus embedded in an intercloud medium.
If the intercloud gas is present, it is more likely that it is directly illuminated than the embedded clouds.
We therefore calculated the PDR molecular line emission for the intercloud medium discussed in Sect.~\ref{sec:intercloudN1068}
(Fig.~\ref{fig:profiles_n1068_plotall_pdr_higher}). 
As for the dense clouds, the PDR CO line emission is stronger, the PDR HCN emission is weaker than that of the mechanically and CR heated 
intercloud medium. For $R < 5$~pc the emission of both PDR HCO$^+$ lines is weaker than, for $R \ge 5$~pc it is comparable to and weaker than
that of the intercloud medium.

We conclude that the central AGN most probably directly illuminates the inner edge of the thick gas ring. Based on our model calculations
we do not expect that the emission from the PDR regions dominates the total molecular line emission at the inner edge of the ring but it
can increase the emission by about a factor of two.

\subsubsection{HCN(4-3), CN(3-2), and CS(7-6)}

Unpublished ALMA Cyle~2 Band~7 HCN(4-3), CN(3-2), and CS(7-6) observations are compared to our models. Details
of the observations are given in Appendix~\ref{sec:almaobs}.

\subsubsection*{Model and observed central spectra}

\begin{figure}[!ht]
  \centering
  \resizebox{\hsize}{!}{\includegraphics{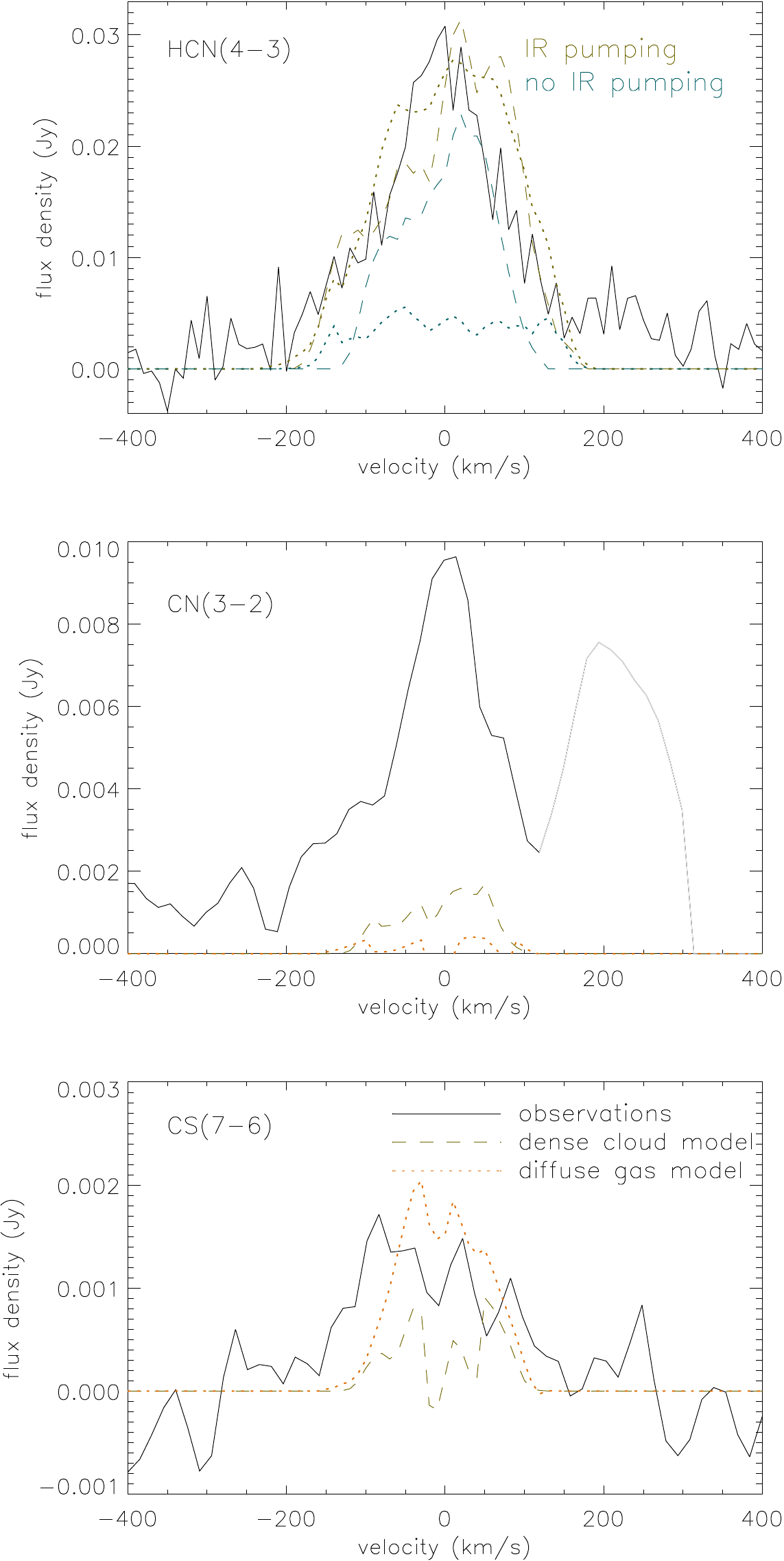}}
  \caption{ALMA integrated spectra of the central source S1 of NGC~1068 (black line). Upper panel: HCN(4-3); middle panel: CN(3-2); lower panel: CS(7-6).
    The gray line corresponds to the second component of the CN(3-2) triplet.
    Colored lines: dashed: S1 spectrum from the dense disk cloud model; dotted: S1 spectrum from the diffuse gas model.
  \label{fig:spectra}}
\end{figure}

Since the HCN(4-3), CN(3-2), and CS(7-6) emission distributions are barely resolved we only focus on the spectrum of the central pixel (i.e., S1).
The observed and model integrated spectra for the dense cloud and diffuse gas models are presented in Fig.~\ref{fig:spectra}.
The height and the width of the ALMA HCN(4-3) spectrum is well reproduced by the dense disk cloud and intercloud models with IR pumping, 
as it is expected from Sect.~\ref{sec:hcn32}.
The integrated intensities of the models without IR pumping are about two times for the dense cloud and four times for the diffuse gas smaller 
than the corresponding models with IR pumping.

For the CN molecule, we focussed our attention to the CN(N=3-2, J=7/2-5/2) transition at $340.248$~GHz, which is strongest line of the CN(3-2) triplet.
The models reproduce the observed linewidth. The model CN emission of the dense disk clouds is about nine times smaller than the observed CN emission.
The CN emission of the diffuse gas model is still smaller.
CN has a high critical density (about half of the HCN critical density) and is primarily formed from photodissociation of HCN and neutral-neutral 
reactions with N, C$_2$, CH$_2$, and CH (Aalto et al. 2002; Boger \& Sternberg 2005; Chapillon et al. 2012). 
Intermediate stages in the reaction pathways involve neutral and ionized carbon (Boger \& Sternberg 2005). CN is thus thought to 
preferentially form in regions illuminated by intense radiation fields (PDRs), whereas it is destroyed in dense gas via CN+H$_2$ $\rightarrow$ HCN+H.
It is thus expected that the PDRs (Sect.~\ref{sec:pdr}) should thus significantly contribute to the CN emission.

The CS(7-6) emission of the dense disk cloud model is about a factor of three smaller than that of the
intercloud model is well comparable to the observed CS(7-6) emission.
Given the uncertainties of the model, we conclude that the model is able to reproduce
the observed CS(7-6) spectrum with the same sulfur abundance as for the CND in the Galactic Center (S/H$=10^{-6}$). 
This abundance is intermediate between that observed in cold dense gas (S/H$ \sim 10^{-7}$; e.g. Quan et al. 2008) and the solar sulfur abundance of 
S/H$=1.5 \times 10^{-5}$ (Jenkins 2009) used in the hot core models of Vidal \& Wakelam (2018). It is consistent with the sulfur abundance found
in the hot core in the star-forming region Orion KL (Esplugues et al. 2014).
We can only speculate that the thermal evaporation of icy 
grain mantles at dust temperatures $>100$~K might be responsible for the enhanced sulfur abundance with respect to cold cores.
In addition, sputtering induced by collisions may be efficient in this turbulent environment to erode the grain mantles and release the sulfuretted 
species to the gas phase as proposed by Fuente et al. (2018) for the Galactic dark cloud Barnard~1.

\section{Discussion \label{sec:discussion}}

The comparison between the moment~0 and moment~1 maps and the pv-diagrams along the major and minor axes
of the different models and the HCN(3-2), HCO$^+$(3-2), HCO$^+$(4-3), CO(2-1), CO(3-2), and CO(6-5) line observations of NGC~1068
showed that the model reproduces the available observations within a factor of about two. In the following we will discuss the 
 stability of counter-rotating disks, the CR ionization 
fractions, elucidate the role of the continuum emission, and discuss the role of outflows or winds. 
Furthermore, the implications of the proposed models are discussed in terms of a twofold outflow and AGN variability.

\subsection{On the stability of counter-rotating disks}

Misalignments between the ionization cones of AGN, and hence the obscuring tori, and the disks of the host galaxies are frequently observed
(e.g., Fischer et al. 2013). Hopkins et al. (2012) stated that these misalignments occur for at least two reasons: first, 
large-scale gas fragmentation can occur (part of a spiral arm or other instability fragments and sinks to the centre),
which can  dramatically  change  the  nuclear  gas  angular  momentum content. Secondly, even in perfectly smooth flows, secondary bars in the presence 
of dissipative gas processes will tend to decouple their angular momentum from the primary bar. Inflow and dissipation lead to a decoupling the inner 
mode angular momentum and orbit plane from that of the outer mode. 
Angular  momentum  exchange  in  the  gas  in  the central regions of galaxies can be 
strongly dominated by supersonic gas shocks surrounding strong torquing regions in the stellar nuclear disc with lopsided/eccentric (m=1) modes
(Hopkins \& Quataert  2011, Baconet al. 2001; Jacobs \& Sellwood 2001; Salow \& Statler 2001; Sambhus \& Sridhar 2002).
These modes can resonantly exchange angular momentum with the pattern at larger radii in the manner of nested bars,
leading (in plane) to possible reversals and counter-rotation of the pattern, which in turn reverses the sense of torques on the gas.
An infall of counter-rotating gas is thus expected, but should be rare.

Quach et al. (2015) and Dyda et al. (2015) studied counter-rotating gas disks analytically and through numerical simulations.
They found that strong supersonic Kelvin-Helmholtz instabilities develop at the interface between the two disks. 
The growth rates are of the order of or larger than the angular rotation rate at the interface. 
Dyda et al. (2015) showed that the onset of the instability leads to the creation of a gap between the two disks (their Fig.11).
The size of the gap is about the vertical size of the outer disk. The gap is created by the mixing of the two counter-rotating components, 
strong heating of the mixed gas, vertical expansion of the gas, and subsequent annihilation of the angular momenta of the two components. 
As a result, the heated gas free-falls towards the center over the surface of the inner disc.
Thereafter, the gap begins to fill as matter from the outer disk moves inwards due to its viscosity. In addition, gas from the inner disk will
also fill the gap from the other side due to viscous diffusion. After a time of $t \sim h/v_{\rm r}$ the instability develops again.
Our $Q=30$ disk model of NGC~1068 has a height of about $0.5$~pc at $R=1.5$~pc and a radial velocity of $v_{\rm r}=12$~km\,s$^{-1}$.
The width of the gap is thus about $0.5$-$1$~pc and the time to fill it is $t=4$-$6 \times 10^4$~yr corresponding to $7$-$11$ dynamical timescales.
This is quite short, but certainly long enough to be observable.
By construction, in our simple sticky particle simulations instabilities cannot develop. The expected gap between the counter-rotating disks 
is thus not present in the simulations. However, the mass accretion rate of the outer thick disk giving rise to $v_{\rm r} \sim 10$-$15$~km\,s$^{-1}$ 
is present.  It is this radial accretion which sets the lifetime of the system.

\subsection{The cosmic ray ionization and gas heating \label{sec:CRion}}

The CR ionization fraction is mainly constrained by the HCO$^+$ emission. For the models where only dense disk clouds are present in
the thick gas disk (Sect.~\ref{sec:denseclouds}) a CR ionization fraction of $\zeta_{\rm CR} \sim 10^{-13}$~s$^{-1}$ yields the best results.
For a significant contribution to the molecular line emission from the intercloud medium, the same CR ionization rate of
is required (Sect.~\ref{sec:intercloudN1068}). The CR ionization rate within the thick gas disk of NGC~1068 is thus 
$100$ times higher than that in the CND ($\zeta_{\rm CR} \sim 10^{-15}$~s$^{-1}$; Sect.~\ref{sec:cnd}).
Within the inner $8$~pc of NGC~1068 the gas heating rate due to cosmic rays is less then $10$\,\% of the turbulent heating rate for the dense gas clouds
and less than $20$\,\% for the intercloud gas.
We thus conclude that within the framework of the model turbulent mechanical heating dominates the heating in the thick gas disk of NGC 1068. 
In the present model turbulence is maintained by the gain of gravitational potential of the accreting gas.

\subsection{The role of the continuum emission}

\begin{figure*}[!ht]
  \centering
  \resizebox{\hsize}{!}{\includegraphics{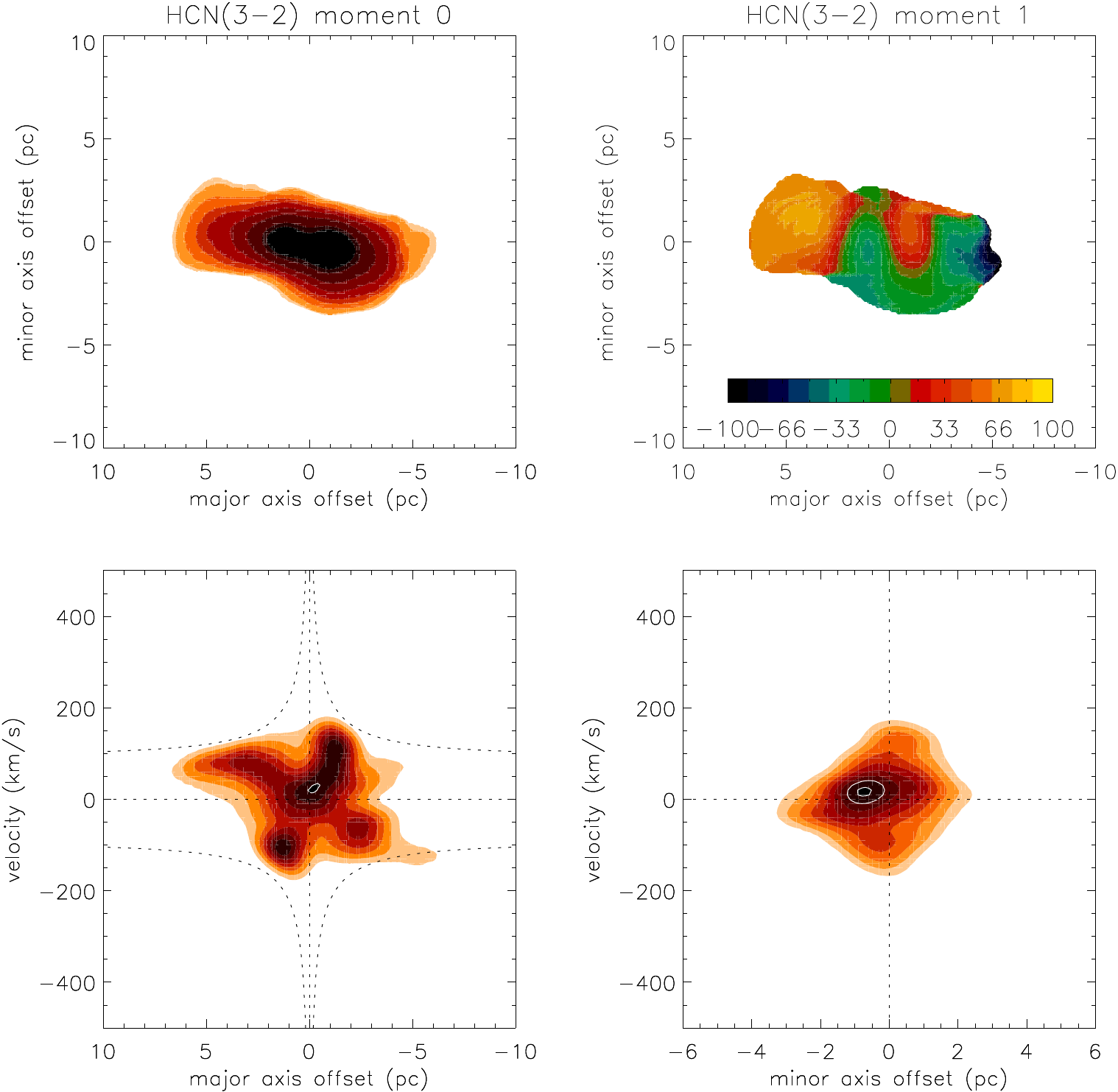}\includegraphics{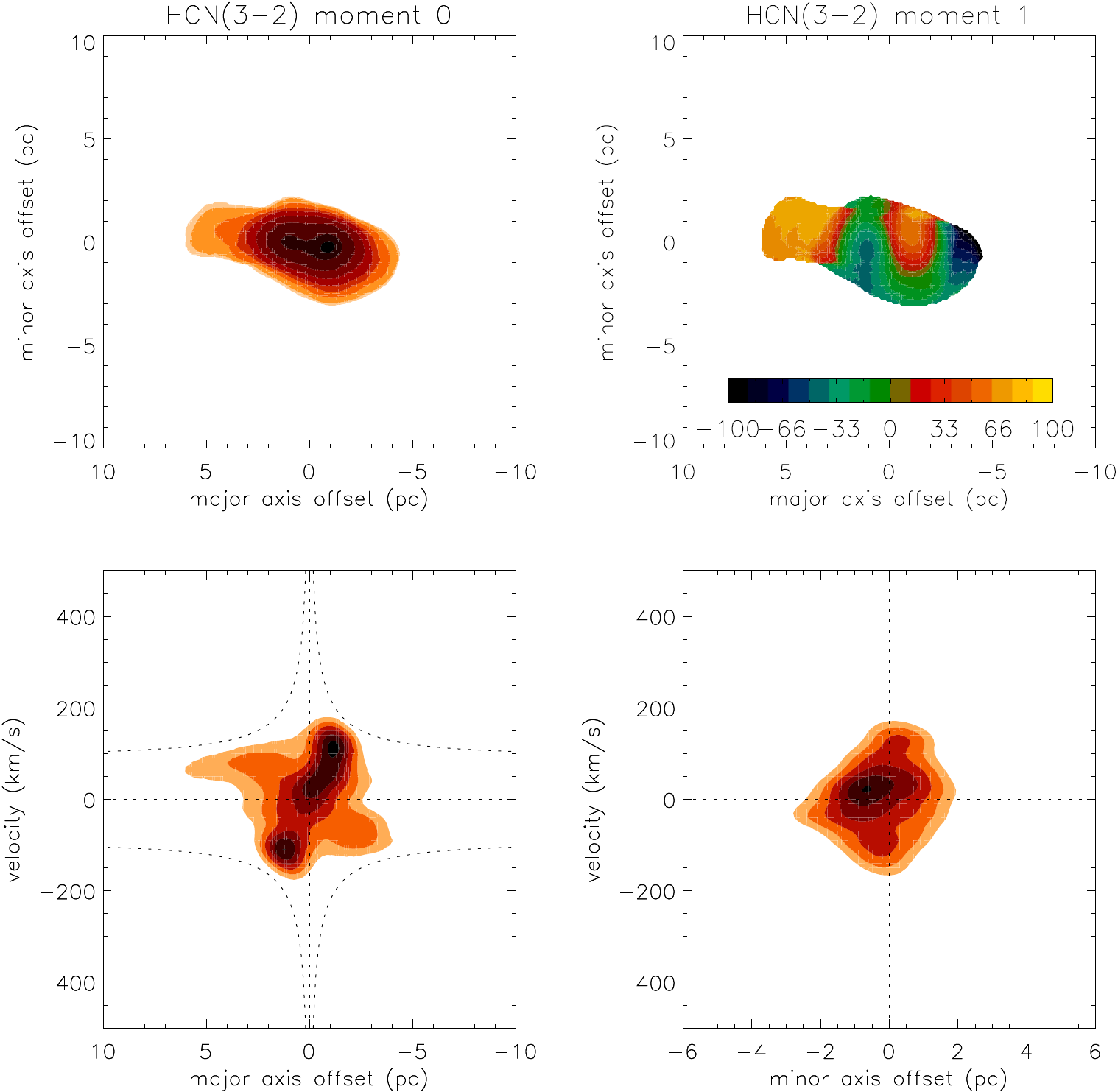}}
  \put(-500,165){\large Q=30  no absorption}
  \put(-240,155){\large Q=30 no absorption}
   \put(-500,155){\large + IR pumping}
   \put(-490,230){\large (a)}\put(-355,230){\large (b)}
   \put(-490,105){\large (c)}\put(-355,105){\large (d)}
  \put(-230,230){\large (e)}\put(-95,230){\large (f)}
  \put(-230,105){\large (g)}\put(-95,105){\large (h)}
  \caption{NGC~1068 dense disk cloud model without absorption. Left panel: $Q=30$ model high resolution HCN(3-2) emission of the
    dense disk cloud model. Right panel: same as left panel but with IR-pumping included in the model.
    Panels (a) and (e) moment~0 maps, (b) and (f) moment~1 maps, (c) and (g) pv diagrams along the major axis,
    (d) and (h) pv diagrams along the minor axis.
    The model levels are the same as those of the observed maps (Fig.~\ref{fig:plottingvollcnd_hcn32radex_newQ=30}).
    The additional white contours levels are $80$ and $90$~K.
  \label{fig:plottingvollcnd_hcn32radex_newQ=30nocont}}
\end{figure*}
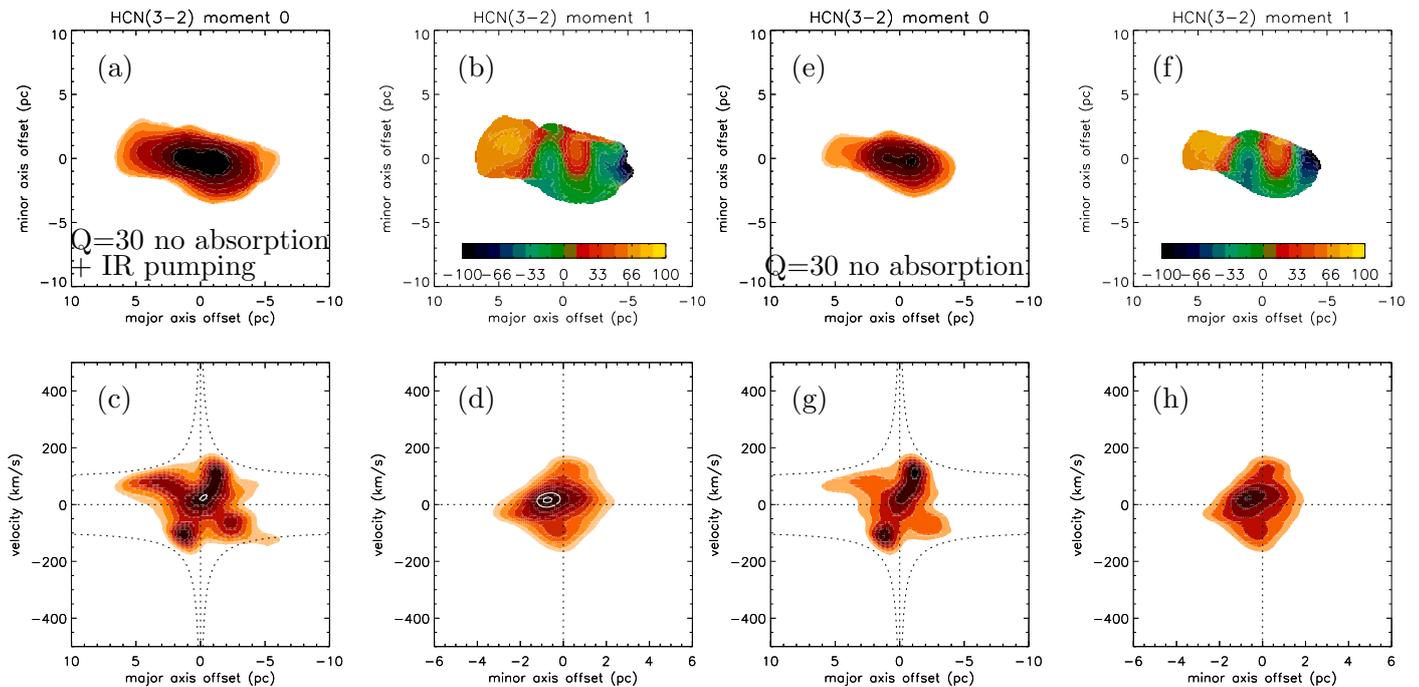

Our modelling allows us to investigate the role of the continuum emission in a simple way: we can remove
the continuum emission, i.e. we can switch off the central AGN.
The resulting moment~0 and moment~1 maps together with the pv diagrams are presented in Fig.~\ref{fig:plottingvollcnd_hcn32radex_newQ=30nocont}
for the $Q=30$ model with and without IR-pumping.
By comparing Fig.~\ref{fig:plottingvollcnd_hcn32radex_newQ=30nocont} to Fig.~\ref{fig:plottingvollcnd_hcn32radex_newQ=30} it becomes clear that 
only the central resolution element is affected by absorption, as expected.
HCN(3-2) emission is detected within a velocity range between $-160$~km\,s$^{-1}$ and $180$~km\,s$^{-1}$.
Whereas the velocity range at negative velocities is compatible with the observed velocity range, HCN(3-2) emission is observed
up to a velocity of $\sim 300$~km\,s$^{-1}$ (panel (b) of Fig.~\ref{fig:plottingvollcnd_hcn32radex_newQ=30}).
HCN(3-2) emission at these high velocities is not present in the model.

We interpret this finding in the following way: 
we adopt the framework of the Vollmer et al. (2018) model, where the continuum emission is produced by the thin maser disk and the
inner edge of the thick gas disk. Whenever a gas cloud of the thick disk is located in front of the continuum emission 
from the point of view of the observer, HCN(3-2) is seen in absorption. We suggest that the observed HCN(3-2)
emission at velocities $> 200$~km\,s$^{-1}$ is emitted by the thin maser disk located within a distance of $\sim 1.5$~pc around 
the central black hole. 
For a black hole mass of $10^7$~M$_{\odot}$ the Keplerian velocity is about $210$~km\,s$^{-1}$ at $1$~pc and $190$~km\,s$^{-1}$ at $1.25$~pc. 
This corresponds to a channel width of $20$~km\,s$^{-1}$. The size of the emitting region is thus about $0.2$~pc whereas the observational beam is $1.4$~pc.
The area filling factor is thus about $(0.25/1.4)^2=0.03$. The observed flux density is about $0.7$~mJy/beam$=30$~K.
Thus, the brightness temperature of the emitting gas is about $30/0.03$~K$=1000$~K. 
We think that such a high brightness temperature can only be reached if we observe at least some HCN maser emission from this region.
The regions of H$_2$O and HCN maser emission are thus coincident.
In addition,  Gallimore et al. (2004; their Fig.~7) showed that a part of the maser spots or maser disk
extend beyond the VLBA $5$~GHz continuum emission distribution in the northwest of the central black hole.
The velocities of these maser spots are indeed higher than $200$~km\,s$^{-1}$.
A significant part of the HCN emission might thus not be absorbed by the central continuum emission.

\subsection{An outer and inner outflow \label{sec:outflow}}

A wind or outflow is observed in NGC~1068 at different scales and in different gas phases:
The line emission of ionized gas at scales of 10 to 100~pc has an hourglass structure and
its kinematics are consistent with an outflow within a hollow-cone structure (May \& Steiner 2017,
Das et al. 2006; Miyauchi \& Kishimoto 2020). At smaller scales, the morphology of the point-source 
subtracted ALMA band~6 continuum  emission (Impellizzeri et al. 2019) and the polar dust emission from 
MIR interferometric observations (Lopez-Gonzaga et al. 2014) are also consistent with an outflow. 
Gallimore et al. (2016), Garcia-Burillo et al. (2019), and Impellizzeri et al. (2019) 
interpreted the high radial velocities and large linewidths of the molecular line emission
along the minor axis of the molecular torus as a molecular outflow in the inner region of 
the massive gas disk  ($R \la 3$~pc). Furthermore, the torus is connected to the 200pc-size gas 
ring through a  network of gas lanes whose kinematics are accounted for by a 3D outflow 
geometry (Garcia-Burillo et al. 2019). We call the latter molecular and dusty wind observed at
scales between a few pc and $\sim 100$~pc the outer wind.

Such an outflow, which is definitely present in NGC~1068, is not included the present model.
The results presented in Sect.~\ref{sec:mollinem} suggest that a good part of the kinematical features of the
molecular line emission, which were interpreted as an outflow, might be due to the perturbed
kinematics caused by the gas-gas collision of two counter-rotating entities. On the other hand,
the present model cannot account for all observed emission: the large linewidth and velocity-asymmetry of the HCN(3-2) and
HCO$^+$(3-2) lines observed in emission in the pv diagram along the minor axis of the molecular torus 
at $R \la 2$~pc (Figs.~\ref{fig:plottingvollcnd_hcn32radex_newQ=30} and 
\ref{fig:plottingvollcnd_hco32radex_newQ=30}) and the absorption wing at negative velocities in the
HCN(3-2) spectrum of the central position (Fig.~\ref{fig:plottingvollcnd_hcn32radex_newQ=30_4}).
At least these two features are most probably caused by the molecular outflow.
It is not excluded that the signatures of the gas-gas collision and the outflow are superimposed in the
inner $R \la 3$~pc of NGC~1068.

The mass inflow rate from the dynamical model at $0.25$~pc (Sect.~\ref{sec:n1068}), although rather variable, is around $0.5$~M$_{\odot}$yr$^{-1}$ 
at the selected timestep. This matches the Vollmer et al. (2018) model quite well for the inflow rate of $0.7$~M$_{\odot}$yr$^{-1}$ in the thin disk.
Most of the inflowing gas must be blown out rather than accreted onto the central black hole.
Within the scenario of Vollmer et al. (2018) a magnetocentrifugal (outer) wind aided by radiation pressure is launched 
from the thick gas disk at a radius of $R \sim 1.5$~pc. This wind is naturally dusty and is assumed to be responsible for
the observed polar infrared emission at the ten parsec scale (Lopez-Gonzaga et al. 2014). The model of Vollmer et al. (2018)
yielded a mass outflow rate of $\dot{M}_{\rm wind} \sim 0.9$~M$_{\odot}$yr$^{-1}$ of the magnetocentrifugal wind compared to
a mass accretion rate of $\dot{M}_{\rm thick\ disk} \sim 1.6$~M$_{\odot}$yr$^{-1}$ of the thick gas disk.
The thin gas disk has still a high mass accretion rate\footnote{With a black hole mass
of $M_{\rm BH} = 10^7$~M$_{\odot}$ the assumed luminosity corresponds to $0.23$ times the Eddington luminosity.}
$\dot{M}_{\rm thin\ disk} \sim 0.7$~M$_{\odot}$yr$^{-1}$. 
This has to be compared to the estimate of the mass accretion rate onto the central black hole of 
$\dot{M}_{\rm BH} \sim L_{\rm bol}/(0.1\,c^2) \sim 0.05$~M$_{\odot}$yr$^{-1}$
given the bolometric luminosity of $L \sim 3 \times 10^{44}$~erg\,s$^{-1}$ (Vollmer et al. 2018). 
It becomes clear that the radial gas inflow within the thin maser disk cannot and is not maintained at radii
$\la 0.1$~pc where the optically emitting hot accretion disk is located. A natural way to explain the 
ratio $\dot{M}_{\rm BH}/\dot{M}_{\rm thin\ disk} \sim 0.1$ is to assume a second mass outflow inside the
dust sublimation radius of $0.25$~pc (Gravity Collaboration 2020). This massive second outflow with a mass
outflow rate of $\dot{M}_{\rm outflow} \sim 0.6$~M$_{\odot}$yr$^{-1}$ is naturally consistent with the suggestion
that the Compton-thick gas layer is located within the dust sublimation radius (Sect.~\ref{sec:XDR}).

\subsection{Sub-Keplerian rotation \label{sec:keplerian}}

It is quite surprising that the observed radial velocities of the HCN, HCO$^+$, and CO lines do not reach
the values expected from the rotation curves at radii $R \ga 2$~pc, especially at positive velocities (Garcia-Burillo et al. 2019, Imanishi et al. 2020,
see also panels (b) and (f) of Fig.~\ref{fig:plottingvollcnd_hcn32radex_new_garciaburillo_co21hr} and panel (b) of 
Fig.~\ref{fig:plottingvollcnd_hcn32radex_new_gallimore1}). 
There are three possible explanations for this finding: (i) a relatively low inclination angle of the outer gas disk,
(ii) non-circular motions of or within the outer disk,
(iii) a significant radial pressure support, and (iv) different emission conditions in the regions of high radial
velocities. (i) A relatively low inclination angle of the outer disk is excluded by the observed moment~0 maps.
Garcia-Burillo et al. (2019) give an inclination angle $i \geq 60^{\circ}$, which is consistent with our results (Table~\ref{tab:bestfit}).
(iii) According to Burkert et al. (2010) the rotation velocity in the presence of a radial gas pressure gradient is
\begin{equation}
v_{\rm rot}^2=v_0^2+\frac{R}{\rho} \frac{{\rm d}p}{{\rm d}R} = v_0^2+\frac{R}{\rho} \frac{{\rm d}}{{\rm d}R} (\rho\,v_{\rm turb}^2)\ ,
\end{equation}
where $v_0$ is the zero-pressure velocity curve and the thermal pressure gradient neglected.
For a model with constant Toomre parameter $Q$ and vertical hydrostatic pressure equilibrium the velocity dispersion is constant and 
the density is given by $\rho = \Omega^2/(\pi\,G\,Q)$. In the case of a constant rotation curve the pressure gradient is
maximum:
\begin{equation}
v_{\rm rot}^2=v_0^2 + v_{\rm turb}^2 \frac{{\rm d}\ln \rho}{{\rm d}\ln R}=v_0^2-2\,v_{\rm turb}^2 \ .
\end{equation}
With $v_0 \sim 100$~km\,s$^{-1}$ and radial velocity dispersion of $v_{\rm turb} \sim 50$~km\,$^{-1}$ 
assuming isotropic turbulence, one obtains $v_{\rm rot}=70$~km\,s$^{-1}$ or a velocity difference of $\Delta v=30$~km\,s$^{-1}$.
Within the radius of influence of the black hole, where $v_0$ corresponds to the Keplerian velocity, the velocity difference is smaller.
This velocity difference can in principle explain the observed velocity difference.
(iv) An intriguing possibility is that the emission of the high radial velocity gas is much weaker than that of the
low radial velocity gas. This can be the case if the molecular emission is dominated by the PDR even in the outer parts of the disk 
because the high radial velocity gas is located in the plane of the sky and we therefore observe these illuminated surfaces
almost edge-on. This orientation leads to a suppressed molecular line emission. This scenario implies that the
parts of the outer gas disk ($R > 2$~pc), for which molecular line emission is detected, are directly illuminated
by the AGN. In this case it is expected that deeper observations reveal gas of higher radial velocities.

We conclude that explanation (ii) - (iv) can explain the difference between the observed radial velocities and
those expected from the mass profile. Given that the distribution of the molecular line emission outside $R=2$~pc
is asymmetric, we think that non-circular motions together with different emission conditions in the regions of high radial
velocities might be the preferred explanations.

\subsection{Plausibility of the proposed scenario}

As stated in Sect.~\ref{sec:introduction}, our models are not intended to reproduce the NGC~1068 observations in detail,
but to serve as a guideline to interpret the available observations. Is our proposed accretion scenario where 
gas clouds fall onto a pre-existing gas disk or ring plausible? It is certainly the simplest model configuration that can qualitatively, 
and to a certain degree quantitatively, explain the available molecular line observations.
The mass accretion rates of the dynamical model are quite high ($\dot{M} \sim 0.1$-$0.5$~M$_{\odot}$yr$^{-1}$; Fig.~\ref{fig:massaccretionrates}).
The inspection of the simulation without infalling gas cloud showed that the outer edge of the ring
moves from a radius of $7$~pc to $3$~pc within $\sim 0.5$~Myr.
This means that the gas cloud has to hit the pre-existing gas disk or ring at the right moment. If the pre-existing gas disk or ring had a larger
size, the time window would be larger. Nevertheless, there has to be a certain timing between the accreting
inner gas disk or ring and the infalling gas cloud. 

We have a scenario in mind where there is permanent chaotic infall
of gas clouds of different angular momentum into the central $20$~pc. If there is no pre-existing gas disk or
ring, such structure will form through a dispersion ring (Sanders 1998). If a second gas clouds falls into the
central $10$~pc within the accretion time of the dispersion ring ($\sim 1$~Myr), we will find a NGC~1068-like scenario.
The accretion time of the CND in the Galactic Center is a few Myr (Vollmer \& Davies 2013). The time window
for the infall of an external gas cloud can thus be about $4$ times larger than that of NGC~1068.
It is quite natural that the infall frequency of external gas clouds is significantly higher in AGNs compared to quiescent galactic centers. 

The center of the Circinus galaxy also harbors an AGN with a maser disk around it (Greenhill et al. 2003).
The ALMA CO, HCN, and HCO$^+$ observations of Izumi et al. (2018) and Kawamuro et al. (2019) showed that the gas located within the
inner $20$~pc rotates in the same sense as the maser disk. The expected rotation velocity of the obscuring thick gas disk is clearly detected (Izumi et al. 2018).
This disk is fed by a gaseous bar structure of a size of $\sim 100$~pc. Thus, contrary to NGC~1068, there seems to be continuous 
infall over more than a dynamical timescale of $t_{\rm dyn} \sim 100$~pc/$100$~km\,s$^{-1} \sim 1$~Myr into the central $20$~pc of the Circinus galaxy.

\subsection{AGN variability}

Infalling gas clouds thus seem to feed the gas disk found at distances of $\sim 10$~pc around the central black hole.
Due to these infall the mass accretion rate into the central $\sim 0.1$~pc will be highly variable.
As seen before, a mass accretion rate of $\dot{M}_{\rm BH} \la 0.1$~M$_{\odot}$yr$^{-1}$ into the central $0.1$~pc is sufficient to make
a galactic nucleus active. Without the infall of an external gas cloud, this mass accretion rate can be maintained over
$\sim 1$0~Myr. If a mass accretion higher than $\ga 0.3$~M$_{\odot}$yr$^{-1}$ is needed to make the galaxy nucleus active,
the timescale becomes $\sim 0.2$~Myr (see Fig.~\ref{fig:massaccretionrates}). The latter timescale is consistent with the flickering timescale found 
by Schawinski et al. (2015). However, we caution the reader not to overinterpret these findings because the interaction
between the outer and inner disks and the outer and inner outflows, which set the accretion rate onto
the central black hole giving rise to the AGN luminosity, is not clear at all.

\section{Summary and conclusions \label{sec:conclusions}}

Recent ALMA high-resolution CO, HCO$^+$, and HCN line observations of the nucleus of NGC~1068 found compact emission at a scale
of $\sim 10$~pc (Imanishi et al. 2018, Garc{\'\i}a-Burillo et al. 2019, Impellizzeri et al. 2019, Imanishi et al. 2020). 
The emission distribution is elongated and shows signs of rotation. The interferometric observations of the
highest angular resolutions revealed a counterrotating gas disk at distances $\la 1.5$~pc from the central black hole.
The observed linewidths are large indicating that the gas disk is geometrically thick.

The gas disk in the nucleus of NGC~1068 has several properties in common with the Circumnuclear Disk (CND) in the Galactic Center:
both are thick rotating gas rings, which are kinematically perturbed. The gas in the CND is resolved into clouds
with sizes of $\sim 0.1$~pc (Montero-Casta{\~n}o et al. 2009) and densities between $10^5$ and $10^6$~cm$^{-3}$ (Harada et al. 2015).

A simple dynamical model consisting of gas clouds that can have partially inelastic collisions is used to reproduce the
main kinematical properties of the available observations (Sect.~\ref{sec:dynmodel}). 
The simplest configuration to explain the kinematics of the CND and the thick gas disk 
in NGC~1068 is gas clouds, which fall onto a pre-existing gas ring.
We made 162 simulations for the CND and 63 simulations for NGC~1068 and compared all meaningful projections to existing molecular line 
observations. The best-fit cloud-ring collision is prograde for the CND and retrograde for NGC~1068.

We applied the analytical model of a turbulent clumpy gas disk of Vollmer \& Davies (2013) calibrated by the
dynamical model and observations to the CND and the thick gas disk with the inner
$20$~pc of NGC~1068 (Sect.~\ref{sec:anamodel}). Within the model framework the energy driving turbulence is supplied
by external infall or the gain of potential energy by radial gas accretion within the disk.
The external mass accretion rate is assumed to be close to the 
mass accretion rate within the disk (the external mass accretion rate feeds the disk at its outer edge).
Within the model, the disk is characterized by the disk mass accretion rate $\dot{M}$ and the Toomre $Q$ parameter, which is
used as a measure of the gas content of the disk. Both quantities are assumed to be constant within the disk.
The analytical model yields the average disk properties and those of the largest turbulent disk clouds
whose size is given by the thickness of C-shocks within the partially ionized gas permeated by a relatively strong magnetic field.

The gas temperature is calculated through the local equilibrium between mechanical and cosmic ray heating on the one hand, and
molecular line cooling (H$_2$O, H$_2$, CO) in the other hand (Sect.~\ref{sec:cloudtemp}). The time-dependent molecule abundances are derived by the
{\tt Nautilus} gas-grain code (Sect.~\ref{sec:network}). Radial profiles of multi-transition CO, HCO$^+$, and HCN brightness temperature 
were calculated.

To each disk gas cloud of the dynamical model the brightness temperature from the analytical model was assigned according to its radius.
In this way datacubes of the molecular lines were produced, which are directly compared to the available observations
via the moment~0 maps, pv diagrams, and spectra along the major axis. For simplicity we divided the turbulent clumpy gas into
an intercloud gas and dense clouds. Furthermore, we evaluated the role of infrared pumping for
the HCN emission and the role of X-ray dominated and photodissociation regions. No wind our outflow component is included in the model.

Our models are not expected to exactly reproduce the available observations, but to serve as a guideline
to interpret these observations. Whereas traditional models derive the gas properties from molecular line ratios, our forward modelling tries to 
reproduce qualitatively and, to a certain extent, quantitatively the observed molecular brightness temperatures. 

Based on our modelling effort we found the following results for the CND in the Galactic Center:
\begin{enumerate}
\item
in the Galactic Center we suggest collisions of two gas clouds with a pre-existing gas ring. This is certainly not a unique solution.
The CND is observed $\sim 0.57$~Myr and $\sim 0.15$~Myr after the impacts (Sect.~\ref{sec:cnd}).
\item
The CND-like model yields $Q=300$ and has a gas mass of $M_{\rm gas} \sim 10^4$~M$_{\odot}$ between $1~{\rm pc} \la R \la 4$~pc.
The mass accretion rate of such a gas ring is a few $10^{-3}$~M$_{\odot}$yr$^{-1}$.
\item
Our model containing only dense disk clouds reproduces within a factor of two the available CO(6-5) (Requena-Torres et al.2012), single dish
HCN(3-2), HCN(4-3), HCO$^+$(3-2), HCO$^+$(4-3), H$^{13}$CN(3-2), H$^{13}$CN(4-3), H$^{13}$CO$^+$(3-2), H$^{13}$CO$^+$(4-3) (Mills et al. 2013), and 
interferometric HCN(4-3) and CS(7-6) (Montero-Casta{\~n}o et al. 2009) observations of the CND (Sect.~\ref{sec:cnd}).
The cosmic ray ionization rate within the CND is $\zeta_{\rm CR}=2 \times 10^{-15}$~s$^{-1}$.
\item  
The HCN abundance is significantly enhanced in the inner $2$~pc due to hot core physics.
\item
To avoid self-absorption of the intercloud medium in the CND, the turbulent heating at the largest scales, which are comparable to the disk height,
has to be decreased by a factor of $\ga 50$ (Sect.~\ref{sec:intercloudCND}). This might be achieved by intermittent turbulent heating and potentially by a strong uniform
magnetic field.
\end{enumerate}

The following results were found for NGC~1068:
\begin{enumerate}
\setcounter{enumi}{4}
\item
the infalling cloud collides with the gas ring on a retrograde orbit. 
The thick gas disk ($Q=30$) is observed $\sim 0.14$~Myr after the impact (Sect.~\ref{sec:n1068}).
\item
The model gas disk at $1.5~{\rm pc} \la R \la 7$~pc has a mass of $M_{\rm gas} \sim 1.4  \times 10^5$~M$_{\odot}$.
The mass accretion rate of such a thick gas disk is $\sim 0.5$~M$_{\odot}$yr$^{-1}$.
\item
Our model reproduces within a factor of $\sim 2$ the available CO(2-1), CO(3-2), CO(6-5), HCN(3-2), HCN(4-3), HCO$^+$(3-2), and HCO$^+$(4-3), and 
CS(7-6) observations of the gas distribution in the inner $20$~pc of NGC~1068 (Sect.~\ref{sec:n1068}).
The cosmic ray ionization rate within the dense disk clouds in the thick gas disk is $\zeta_{\rm CR}=2 \times 10^{-13}$~s$^{-1}$.
\item
As for the CND in the Galactic Center, the HCN abundance is greatly enhanced due to hot core physics.
The CS(7-6) observations of NGC~1068 can be reproduced with a sulfur abundance of S/H$=10^{-6}$.
\item
Based on the available observations it cannot be excluded that the intercloud gas dominates the molecular line emission in NGC~1068 
at radii $\la 2$~pc (Sect.~\ref{sec:intercloudN1068}). 
In order to obtain an optically thin intercloud gas, the Toomre $Q$ parameter has to be set to $Q=100$ and the dissipation 
rate at the driving length must be decreased by at least a factor of $1/200$ compared to Eq.~\ref{eq:turbheat}. 
This is four times less than the factor required for the CND. 
On the other hand, for an optically thick intercloud gas with brightness temperatures high enough to be detected by the available observation setups,
we had to assume a $Q=30$ disk with full heating (Eq.~\ref{eq:turbheat}) and diffuse clouds of size $H/2$ and density of three times the mean density. 
\item
Within the framework of the model turbulent mechanical energy input is the dominant gas heating mechanism within the thick gas disk 
(Sect.~\ref{sec:CRion}). Turbulence is maintained by the gain of potential energy of the accreting gas.
\item
A good part of the kinematical features of the molecular line emission in NGC~1068, which were interpreted 
as an outflow, might be due to the perturbed kinematics caused by the gas-gas collision of two counter-rotating 
entities. It is not excluded that the signatures of the gas-gas collision and the outflow are superimposed in the inner $R \la 3$~pc of NGC~1068.
\item
Since HCN abundances in excess of $x_{\rm HCN}=10^{-8}$ are absent for ionization rates that are higher than a few $10^{-12}$~s$^{-1}$,
the observed high HCN abundances in NGC~1068 cannot be reproduced with models including an
ionization rate which is expected if the gas is directly illuminated by the X-ray emission of the central engine (Sect.~\ref{sec:XDR}). 
We suggest that the bulk of the central X-ray emission is absorbed in a gas layer of Compton-thick gas inside the dust sublimation radius,
maybe by an outflow of dense fully ionized gas. In this case, NGC~1068 would harbor two outflows, an inner ionized and an outer dusty outflow
(see Vollmer et al. 2018).
\item
Infrared pumping increases the HCN(3-2) line emission of the whole gas disk or ring by up to a factor of two (Sect.~\ref{sec:pumping}).
\item 
It is expected that the central engine directly illuminates inner edge of the thick gas disk. Based on our model calculations
we concluded that the molecular line emission can be enhanced by the PDR emission by at most a factor of two at radii between $1$ and $2$~pc 
(Sect.~\ref{sec:pdr}). We do not think that the PDR emission dominates the total emission at these radii.
\end{enumerate}

The role of the intercloud gas in the thick disk and its turbulent heating has to be further investigated.
The determination of the optical depths and brightness temperatures at all scales of the turbulent clumpy ISM as attempted by Vollmer et al. (2017) 
is certainly desirable. However, such a treatment requires a full 3D radiative transfer model for the molecular line emission.
In the present model the radiative transfer is only calculated within the gas clouds and optically thick clouds can hide the emission
of other clouds located behind them.
We conclude that the scenario of infalling gas clouds onto pre-existing gas rings in galactic centers is viable and consistent
with available observation of the CND in the Galactic Center and the dense gas distribution within the inner $20$~pc of NGC~1068.
We suggest that the observed AGN flickering (Schawinski et al. 2015) might be linked to the proposed scenario of 
permanent chaotic infall of gas clouds of different angular momentum into the central $20$~pc of galactic centers.


\appendix

\section{The analytical model \label{sec:amod}}

The analytical model is fully described in Vollmer \& Davies (2013).
Here we only give an overview. We assume a turbulent clumpy gas disks where the energy to drive turbulence is supplied
by external infall or the gain of potential energy by radial gas accretion within the disk. The size of the largest 
turbulent gas clouds is determined by the size
of a continuous (C-type) shock propagating in dense molecular clouds with a low ionization fraction at a given velocity dispersion. 
We use the expressions derived by Vollmer \& Davies (2013) for the expected volume and
area filling factors, mass, density, column density, and velocity dispersion of the clouds. The latter is based on 
scaling relations of intermittent turbulence whose open parameters are estimated for the Circumnuclear Disk
in the Galactic Center. The meaning of all model parameters is given in Table~\ref{tab:model}.

\begin{table*}
      \caption{Model parameters and their meaning}
         \label{tab:model}
      \[
       \begin{array}{|l|l|l|l|}
        \hline
	{\rm large\ scale\ disk} & {\rm CND} & {\rm NGC~1068} & \\
	\hline
	\hline
	v_{\rm rot}(20~{\rm pc}) & 100~{\rm km\,s}^{-1} & 100~{\rm km\,s}^{-1} & {\rm rotation\ velocity\ (Guesten\ et\ al.\ 1987, \ Hicks\ et\ al.\ 2009)} \\
	v_{\rm A,0} & 1.0~{\rm km\,s}^{-1} & 1.5~{\rm km\,s}^{-1} & {\rm Alfven\ velocity\ (Vollmer\ \&\ Davies\ 2013)} \\
	M_{\rm dyn}(20~{\rm pc}) & 4.4 \times 10^7~{\rm M}_{\odot} & 4.4 \times 10^7~{\rm M}_{\odot} & {\rm total\ enclosed\ (dynamical)\ mass\ (Genzel\ et\ al.\ 2010,} \\
         & & & {\rm Hicks\ et\ al.\ 2009)} \\
	M_{\rm gas} & 4.5 \times 10^4~{\rm M}_{\odot} & 3.6 \times 10^5~{\rm M}_{\odot}  & {\rm total\ gas\ mass\ (this\ work)} \\
	v_{\rm turb} & 20~{\rm km\,s}^{-1} & 50~{\rm km\,s}^{-1} & {\rm gas\ turbulent\ velocity\ dispersion\ (Guesten\ et\ al.\ 1987,} \\
          & & & {\rm Hicks\ et\ al.\ 2009)} \\
	Q & 300 & 30 & {\rm disk\ Toomre\ parameter\ (Eq.}~\ref{eq:toomq}) \\
        \dot{M} & 3.8~{\rm M}_{\odot}{\rm yr}^{-1}  & 2 \times 10^{-3}~{\rm M}_{\odot}{\rm yr}^{-1} & {\rm disk\ mass\ accretion\ rate\ (this\ work)} \\
        \eta & 0.1 & 0.1 & {\rm dissipated\ energy\ fraction\ per\ cloud-cloud\ collision\ (Sect.~\ref{sec:dynmodel})} \\
	\alpha & 3.7 \times 10^{13}~{\rm cm}^{3}{\rm s}^{-1}{\rm g}^{-1} & 3.7 \times 10^{13}~{\rm cm}^{3}{\rm s}^{-1}{\rm g}^{-1} & {\rm gas\ particle\ collision\ coefficient\ (Draine\ et\ al.\ 1983)} \\
	\gamma & 600~{\rm cm}^{-\frac{3}{2}}{\rm s}^{\frac{1}{2}} & 300~{\rm cm}^{-\frac{3}{2}}{\rm s}^{\frac{1}{2}} & {\rm ion-neutral\ coupling\ (Williams \ et\ al.\ 1998,\ McKee\ et\ al.\ 2012)} \\
	\zeta_{\rm CR} & 2 \times 10^{-15}~{\rm s}^{-1} & 2 \times 10^{-13}~{\rm s}^{-1} & {\rm cosmic\ ray\ ionization\ rate\ (Yusef-Zadeh\ et\ al.\ 2013,\ this\ work)} \\ 
        L_{\rm X} & 0 & 5 \times 10^{43}~{\rm erg\,s}^{-1} & {\rm X-ray\ luminosity\ (Bauer\ et\ al.\ 2015)} \\
        M_{\rm BH} & 4 \times 10^6~{\rm M}_{\odot} & 1 \times 10^7~{\rm M}_{\odot} & {\rm black\ hole\ mass\ (Genzel\ et\ al.\ 2010,\ Greenhill\ et\ al.\ 1996)} \\
	R & & & {\rm galactic\ radius} \\
        \Omega &  &  & {\rm angular\ velocity} \\
	\rho &  &  & {\rm disk\ midplane\ density} \\
	\Sigma &  &  & {\rm disk\ surface\ density} \\
	H &  &  & {\rm disk\ height} \\	
	l_{\rm driv} &  &  & {\rm turbulent\ driving\ lengthscale} \\
	\Phi_{\rm V} &  &  & {\rm cloud\ volume\ filling\ factor}\ ({\rm Eq.}~\ref{eq:phivturb}) \\
	\Phi_{\rm A} &  & & {\rm cloud\ area\ filling\ factor}\ ({\rm Eq.}~\ref{eq:vturbcl}) \\
        x_{i} &  &  & {\rm ionization\ fraction}\ ({\rm Eq.}~\ref{eq:xion}) \\
        \zeta_{\rm X} &  &  & {\rm X-ray\ ionization\ rate}\ ({\rm Eq.}~\ref{eq:zetax}) \\
        \Gamma_{\rm g} &  &  & {\rm gas\ heating\ rate}\ ({\rm Eq.}~\ref{eq:heating}) \\
        \Gamma_{\rm turb} &  &  & {\rm turbulent\ heating\ rate}\ ({\rm Eq.}~\ref{eq:turbheat}) \\
        \Gamma_{\rm CR} &  &  & {\rm cosmic\ ray\ heating\ rate}\ ({\rm Eq.}~\ref{eq:crheat}) \\
        \Gamma_{\rm X} &  &  & {\rm X-ray\ heating\ rate}\ ({\rm Eq.}~\ref{eq:xheat}) \\
        H_{\rm X} &  &  & {\rm X-ray\ energy\ deposition\ rate\ per\ particle}\ ({\rm Eq.}~\ref{eq:hx}) \\
	\hline
	\hline
	{\rm small\ scale\ clouds} & & & \\
	\hline
	\hline 
	r_{\rm sh} & & & {\rm compression\ rate\ of \ the \ C-shock}\ ({\rm Eq.}~\ref{eq:comprate}) \\
	M_{\rm cl} & & & {\rm cloud\ mass} \\
	r_{\rm cl} &  &  & {\rm cloud\ radius}\ ({\rm Eq.}~\ref{eq:lll}) \\
	l_{\rm cl} & & & {\rm cloud\ size} \\
	\rho_{\rm cl} &  & & {\rm cloud\ density} \\
	N_{\rm cl} & & & {\rm cloud\ surface\ density} \\
	c_{\rm s} & & & {\rm local\ sound\ speed}\\
	t_{\rm ff}^{\rm cl} & & & {\rm cloud\ free\ fall\ time}\\
	T_{\rm g} & & & {\rm gas\ temperature} \\
        B & & & {\rm magnetic\ field} \\
	\hline
        \end{array}
      \]
\end{table*}

The averaged overall disk is treated as an accretion disk in hydrostatic equilibrium (e.g., Pringle 1981) with a given
mass accretion rate $\dot{M}$ and Toomre parameter 
\begin{equation}
\label{eq:toomq}
Q=\frac{v_{\rm turb} v_{\rm rot}}{\pi G \Sigma R}=\frac{v_{\rm turb}}{v_{\rm rot}} \frac{M_{\rm dyn}}{M_{\rm gas}}\ ,
\end{equation}
where $v_{\rm turb}$ and $v_{\rm rot}$ are the turbulent and rotation velocities, and $M_{\rm dyn}$ and $M_{\rm gas}$ are the total enclosed mass
and the gas mass. The Toomre $Q$ parameter is used as a measure of the gas content of the disk 
with $Q=1$ for the maximum disk gas mass. If the disk is in hydrostatic equilibrium the midplane density is given by $\rho=\Omega^2/\pi/G/Q$, where
$\Omega= v_{\rm rot}/R$ is the angular velocity.

Thermal instabilities triggered by small-scale gas compression
lead to the condensation of dense cold clouds (Parravano 1987).
Within the strong shock approximation the gas compression rate in the shock is given by
\begin{equation}
\label{eq:comprate}
r_{\rm sh}=\frac{\rho_{\rm cl}}{\rho}=\sqrt{2}\frac{v_{\rm turb}}{v_{\rm A,0}}\ ,
\end{equation}
where the $v_{\rm A,0}=B/\sqrt{4 \pi \rho}$ is the neutral Alfv\'{e}n speed in the diffuse medium with
magnetic field $B$ (see, e.g. Chen \& Ostriker 2012). The neutral Alfv\'{e}n speed of the ISM clouds of different 
densities is almost constant $v_{\rm A,0} \sim 1$~km\,s$^{-1}$ (Crutcher 1999). 
Assuming recombination-ionization equilibrium, the shock thickness is given by Eq. 43 of Chen \& Ostriker (2012):
\begin{equation}
l=2^{\frac{7}{4}}\frac{\sqrt{v_{\rm A,0} v_{\rm turb}}}{\alpha x_{i} (\mu_{i}/\mu_{n}) \rho}\ ,
\label{eq:lll}
\end{equation}
where $\alpha=3.7 \times 10^{13}$~cm$^{3}$s$^{-1}$g$^{-1}$ (Draine et al. 1983) is the collision coefficient,
$x_{i}$ the ionization fraction, and $(\mu_{i}/\mu_{n})=(30/2.3)=13$ the fraction between the mean ion and neutral molecular weights.
We assume a cloud radius $r_{\rm cl}=l$.

The ionization fraction is given by 
\begin{equation}
\label{eq:xion}
x_{i}= \gamma \big(\frac{\zeta_{\rm CR}}{n_{\rm H}}\big)^{\frac{1}{2}}\ ,
\end{equation}
where $\gamma=600$-$2000$~cm$^{-\frac{3}{2}}$s$^{\frac{1}{2}}$ (Williams et al. 1998; McKee et al. 2010)
and $n_{\rm H}=\rho/(2.3 \times m_{\rm p})$. We adopted $\gamma \sim 600$~cm$^{-\frac{3}{2}}$s$^{\frac{1}{2}}$.
If the neutral Alfv\'{e}n speed is higher, $\gamma$ and $\zeta_{\rm CR}$ have to be adapted such that
$(v_{\rm A,0}/1~{\rm km}\,{\rm s}^{-1}) = (\gamma/600~{\rm cm}^{-\frac{3}{2}}{\rm s}^{\frac{1}{2}})/\sqrt{(\zeta_{\rm CR}/\zeta_{\rm CR,0})}$,
where $\zeta_{\rm CR,0}$ is the adopted cosmic ray ionization rate.

The disk mean density is linked to the cloud density via the volume filling factor $\rho=\rho_{\rm cl} \Phi_{\rm V}$.
An approximation for the volume filling factor is 
\begin{multline}
\Phi_{\rm V}=\frac{1}{\sqrt{2}}\frac{v_{\rm A,0}}{v_{\rm turb}}= \\
0.02 (v_{\rm A,0}/1~{\rm km\,s}^{-1})/(v_{\rm turb}/30~{\rm km\,s}^{-1})\ .
\label{eq:phivturb}
\end{multline}

In the case of compressible turbulence the mean volume rate of energy transfer becomes $\epsilon_{\rm V}= \rho_{l} \epsilon = \rho_{l} v_{{\rm turb,} l}^{3}/l$
(e.g., Fleck 1996). If the density scales with $\rho_{l} \propto l^{-3 \alpha}$, the turbulent velocity dispersion obeys the relation
$v_{l} \propto l^{1/3+\alpha}$ (Fleck 1996, Kritsuk et al. 2007). Extending the $\beta$-model to compressible turbulence leads to 
\begin{equation}
\bar{\epsilon}_{\rm V} = \rho_{l} (l_{\rm driv}/l)^{D-3} v_{{\rm turb,} l}^{3}/l = const\ .
\label{eq:scaling}
\end{equation}
With $\rho_{l} \propto l^{-3 \alpha}$, the turbulent velocity dispersion obeys the relation $v_{l} \propto l^{\frac{1}{3}(3\alpha+D-2)}$.
For $D=2$ one obtains $v_{l} \propto l^{\alpha}$ and the energy flux 
$\Delta E/(\Delta A \Delta t) = \rho_{l} v_{{\rm turb,} l}^{3}$ is conserved. 
In this case the scaling relation (Eq.~\ref{eq:scaling}) becomes
\begin{equation}
\big(\frac{l_{\rm cl}}{l_{\rm driv}}\big) \rho_{\rm cl} \frac{v_{\rm cl}^{3}}{l_{\rm cl}} = \rho \frac{v_{\rm turb}^{3}}{l_{\rm driv}}\ .
\end{equation}
With the area filling factor $\Phi_{\rm A}= \Phi_{\rm V} H/r_{\rm cl} = \rho/\rho_{\rm cl} H/r_{\rm cl}$ one obtains
\begin{equation}
\frac{v_{\rm cl}^{3}}{l_{\rm cl}} = \Phi_{\rm A} \frac{v_{\rm turb}^{3}}{l_{\rm driv}}\ .
\label{eq:vturbcl}
\end{equation}

\begin{figure}[!ht]
  \centering
  \resizebox{\hsize}{!}{\includegraphics{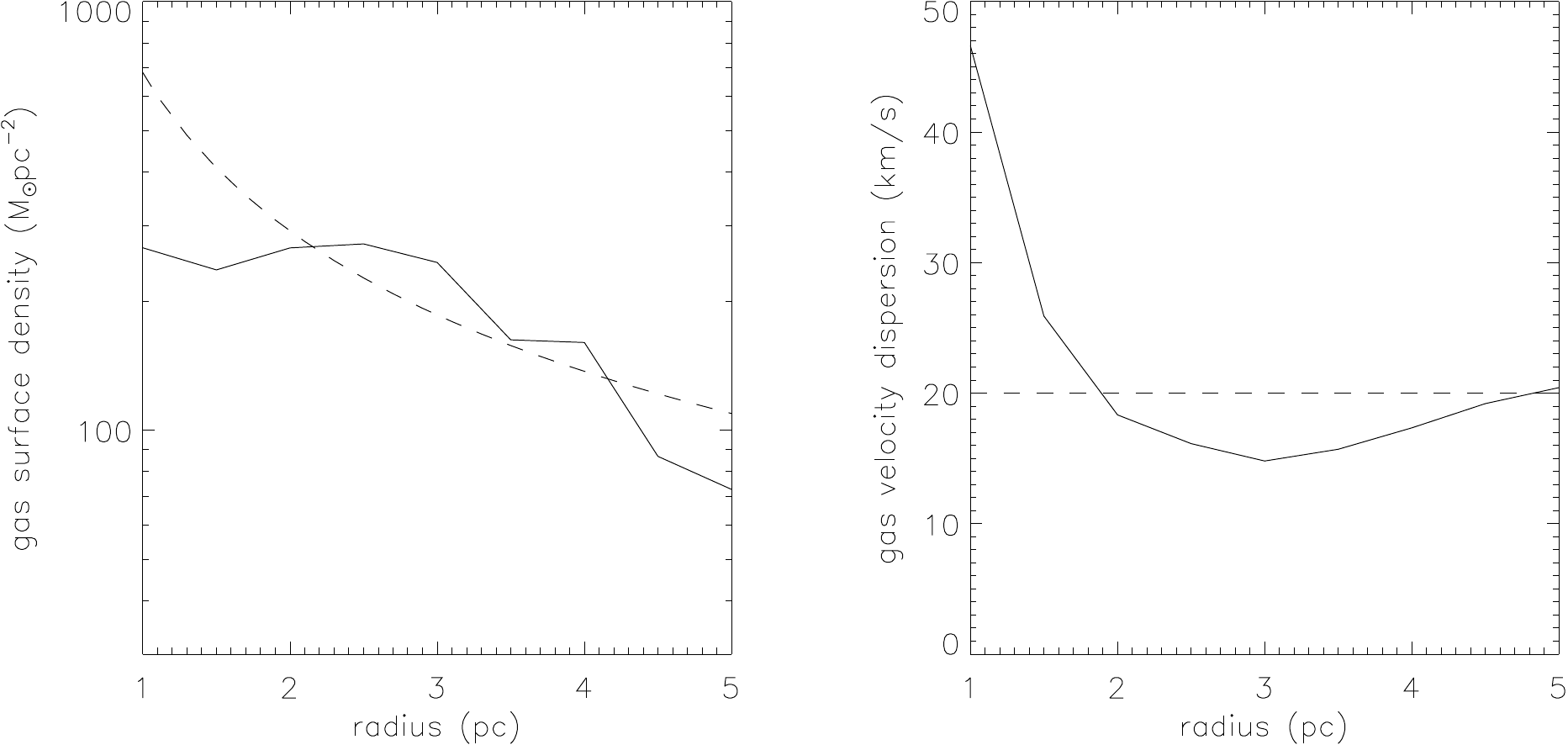}\put(-490,230){\Huge (a) CND model}\put(-200,230){\Huge (b)}}
  \resizebox{\hsize}{!}{\includegraphics{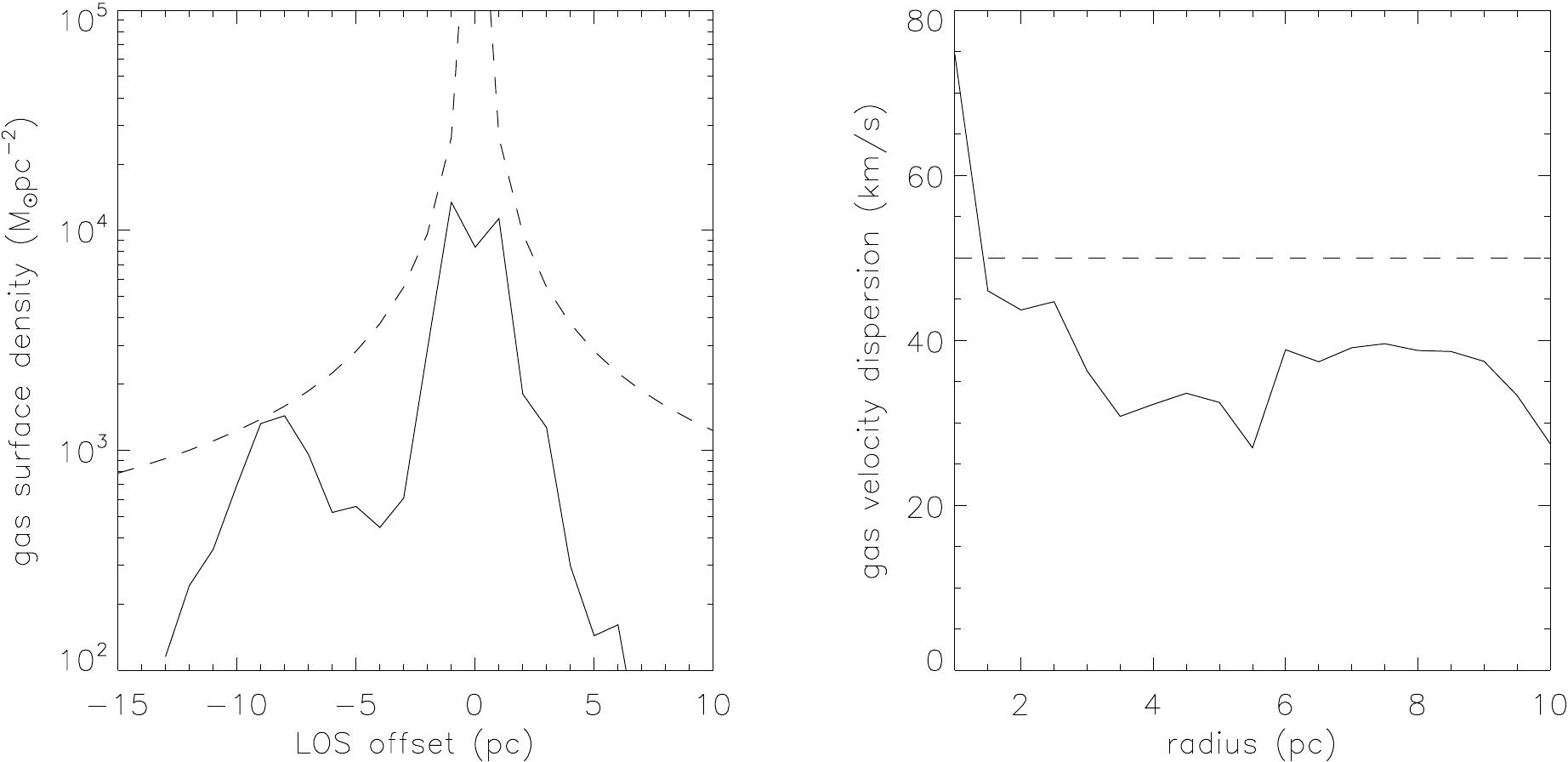}\put(-490,230){\Huge (c) N1068 model}\put(-200,230){\Huge (d)}}
  \caption{Calibration of the analytical model (dashed lines) by the dynamical model (solid lines) at the time of interest.
    Panel (a) radial profile of the gas surface density of the CND-like model.
    Panel (b) radial profile of the gas velocity dispersion of the CND-like model.
    Panel (c) gas surface density cut along the line-of-sight of the NGC~1068-like model.
    Panel (d) radial profile of the gas velocity dispersion of the NGC~1068-like model.
  \label{fig:sigmadisk}}
\end{figure}

For each distance from the central black hole the size, density, and velocity dispersion of the turbulent gas clouds
are determined using Eqs.~\ref{eq:lll}, \ref{eq:phivturb}, and \ref{eq:vturbcl}, respectively.
Each model is characterized by two input parameters: the Toomre $Q$ parameter or the gas surface density and the turbulent velocity dispersion 
of the gas disk $v_{\rm turb}$. These parameters where chosen such that the analytical models are consistent
with the dynamical model at the time of interest. The comparison between the gas surface densities
and gas velocity dispersions of the CND and NGC~10168 models are presented in Fig.~\ref{fig:sigmadisk}.
Since the NGC~1068 model has an asymmetric mass distribution, we show surface density cuts along the line-of-sight;
radial profiles are shown otherwise. 
The gas surface densities of the analytical models are on the higher end of but well comparable to the model surface densities.
We decided to set the gas velocity dispersions of the analytical models to those derived from observations
(G\"usten et al. 1987, Hicks et al. 2009), which are about $25$\,\% higher than those of the dynamical model.

\section{Molecular line emission \label{sec:molli}}

We employ the emission line modelling used by Vollmer et al. (2017).
A molecular line source is usually observed by chopping the telescope's beam between
on- and off-source positions and measuring the difference in antenna temperatures.  
In general, the difference in brightness temperatures is
\begin{equation}
\Delta T^*_{\rm A}=\big(1-{\rm e}^{-\tau}\big)\frac{h\nu}{k}\big(\frac{1}{{\rm e}^{h\nu/kT_{\rm ex}}-1}-
\frac{1}{{\rm e}^{h\nu/kT_{\rm bg}}-1}\big)\ ,
\end{equation}
where $\tau$ is the optical depth of the line, $\nu$ the frequency of the observations, $h$ and $k$ the
Planck and Boltzmann constants, and $T_{\rm ex}$ and $T_{\rm bg}$ the excitation and background brightness
temperatures, respectively. 

Considering only a single collider (H$_2$) for simplicity, the excitation temperature is
\begin{equation}
\frac{1}{T_{\rm ex}}=\big(\frac{1}{T_{\rm g}}+(\frac{A_{ul}}{n q_{ul}}\frac{T_{\rm bg}}{T_*})\frac{1}{T_{\rm bg}}\big)/(1+\frac{A_{ul}}{nq_{ul}}\frac{T_{\rm bg}}{T_*})\ ,
\end{equation}
where $T_*=h \nu_{ul}/k$, $n$ is the gas density, $nq_{ul}$ the collisional de-excitation rate, and $A_{ul}$ the Einstein coefficients of 
the transition $ul$. The background brightness temperature $T_{\rm bg}$ is the sum
of the effective emission temperatures of the galaxy's dust $T_{\rm eff\ dust}$ and the cosmic background at the 
galaxy redshift $T_{\rm CMB}$ (see Eq.~17 of da Cunha et al. 2013).

Following Vollmer et al. (2017), we consider two-level molecular systems in which the level populations are determined by
a balance of collisions with H$_2$, spontaneous decay and line photon absorption, and stimulated emission with
$\tau > 1$. 
The molecular abundances were calculated using the chemical network (see Sect.~\ref{sec:network}).
For simplicity, we neglected the hyperfine structure of HCN. 

The rotation constants, Einstein coefficients, and collision rates were taken from the Leiden Atomic and 
Molecular Database (LAMDA; Sch\"{o}ier et al. 2005). The CO collision rates were provided by Yang et al. (2010).
The HCN collision rates were taken from the He--HCN rate coefficients calculated by Dumouchel et al. (2010), 
scaled by a factor of $1.36$ to go to HCN--H$_2$ (see Green \& Thaddeus 1976). The HCO$^+$ collision rates were taken from Flower (1999).

The comparison of our results with those of RADEX (van der Tak et al. 2007) showed that our calculations tend to overestimate the CO optical depths 
by up to a factor of two and those of HCN and HCO$^+$ by a factor of two to three. This is caused by the approximate treatment of the line emission,
which is more accurate for the CO optical depth than for the HCN and HCO$^+$ optical depths. This approximation was used because it gave us a
high flexibility for the model calculations.
Whereas the HCN and HCO$^+$ brightness temperatures are underestimated by up to a factor of two, the CO brightness temperatures agree with those of RADEX
with typical differences of about $10$ to $30$\,\%. 
Given the overall uncertainties of the analytical and chemical models, we think that these uncertainties are acceptable. 
We estimate the overall uncertainty of the brightness temperatures caused by the uncertainties of the analytical and chemical models and our approximate treatment 
of the molecular line emission to be about a factor of two. To be consistent with RADEX, we divided all model optical optical depths by a factor of two.
We used RADEX for the calculations of the CS and CN optical depths and brightness temperatures.

\section{HCN, HCO$^+$, H$^{13}$CN, and H$^{13}$CO$^+$ spectra along the major axis \label{sec:spectramaj}}

\begin{figure*}[!ht]
   \resizebox{\hsize}{10cm}{\includegraphics{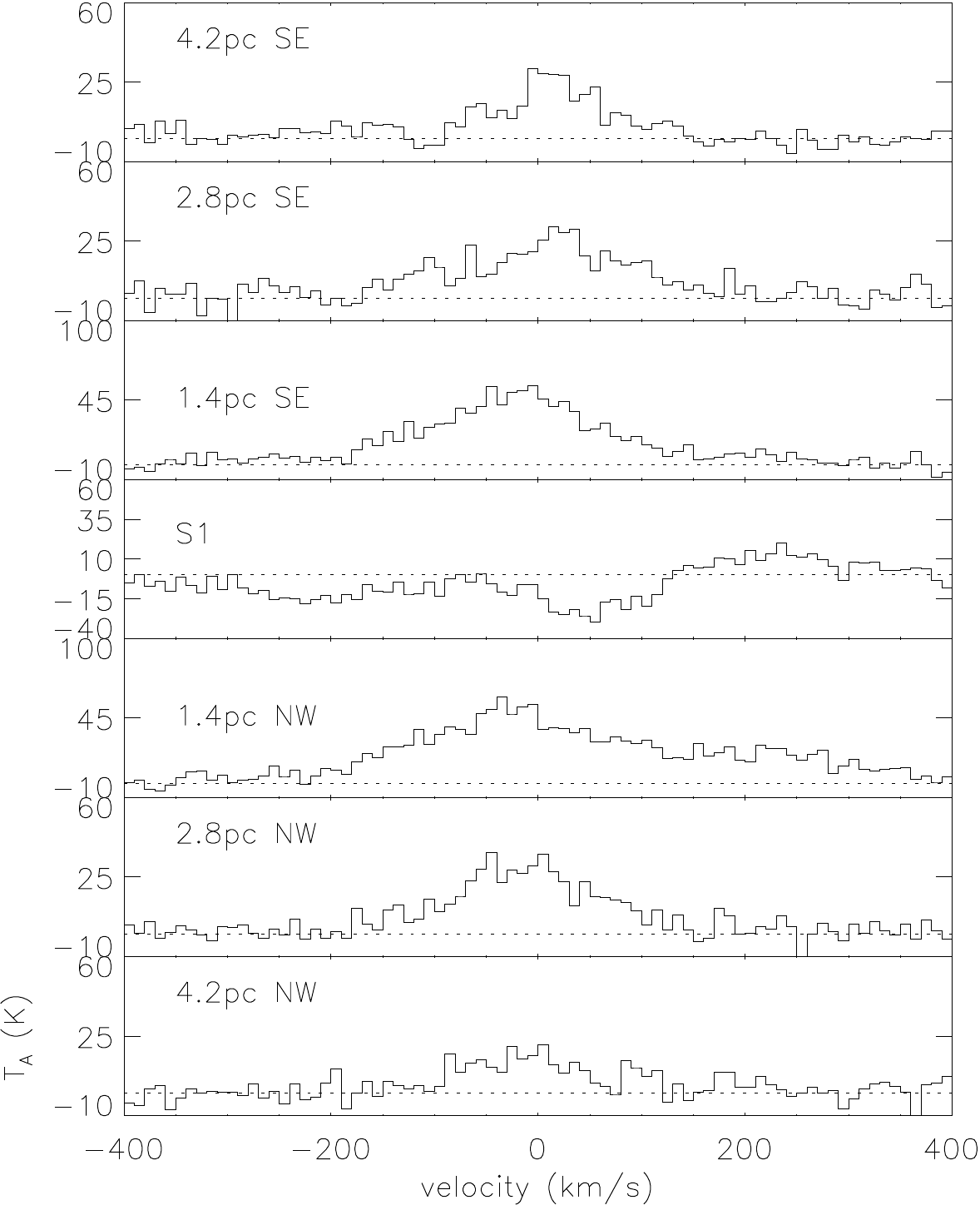}\includegraphics{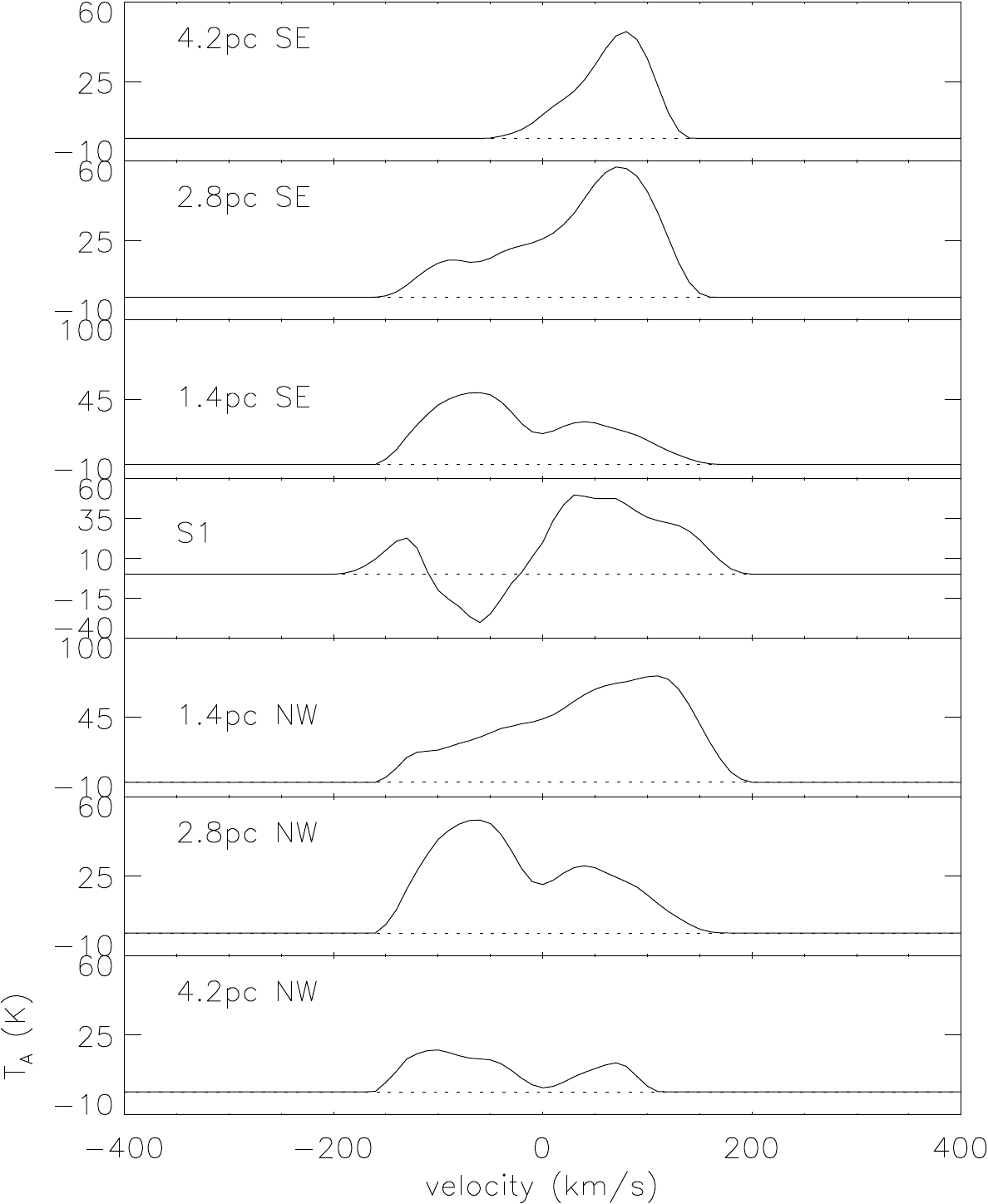}}
   \put(-360,270){\Large Observations}
   \put(-60,270){\Large model}
  \caption{NGC~1068. Left panel: HCN(3-2) high resolution spectra along the major axis (from Impellizzeri et al. 2019 data). Right panel:
    high-resolution HCN(3-2) spectra along the major axis of the dense disk cloud model with IR pumping.
  \label{fig:plottingvollcnd_hcn32radex_newQ=30_4}}
\end{figure*}

We present the observed HCN(3-2) spectra from
Impellizzeri et al. (2019) from resolution elements along the major axis in Fig.~\ref{fig:plottingvollcnd_hcn32radex_newQ=30_4}.
It is remarkable that the observed spectra outside the central resolution element have linewidths in excess of $150$~km\,s$^{-1}$ and the velocities
of their maxima are close to the systemic velocity, i.e. rotation is barely visible in these spectra.
Moreover, the central resolution element shows the HCN(3-2) line in absorption over a velocity range between $-400$ and $150$~km\,s$^{-1}$.
The line is seen in emission around a radial velocity of $\sim 250$~km\,s$^{-1}$.

The corresponding $Q=30$ model HCN(3-2) spectra are presented in Fig.~\ref{fig:plottingvollcnd_hcn32radex_newQ=30_4}.
Whereas the emission peaks and the covered velocity range of the models are higher but comparable to the observed emission peaks, the shapes of 
the model lines are different from the observed line shapes: the model spectra have their maximum at $\sim 80$~km\,s$^{-1}$ and show a 
secondary maximum at $\sim -70$~km\,s$^{-1}$, whereas the observed spectra have their maximum much closer to the systemic velocity
($|\Delta v| \la 50$~km\,s$^{-1}$). The model HCN(3-2) line is seen in absorption between $-100$ and $0$~km\,s$^{-1}$ in the central spectrum. 
The model absorption is thus much narrower and less deep than the observed absorption.
Moreover, the model emission at positive velocities has no observed counterpart and there is no model emission at a 
velocity of $250$~km\,s$^{-1}$ in the central spectrum. The signatures of rotation are thus more visible in the model than in the observations.

\begin{figure*}[!ht]
  \resizebox{\hsize}{!}{\includegraphics{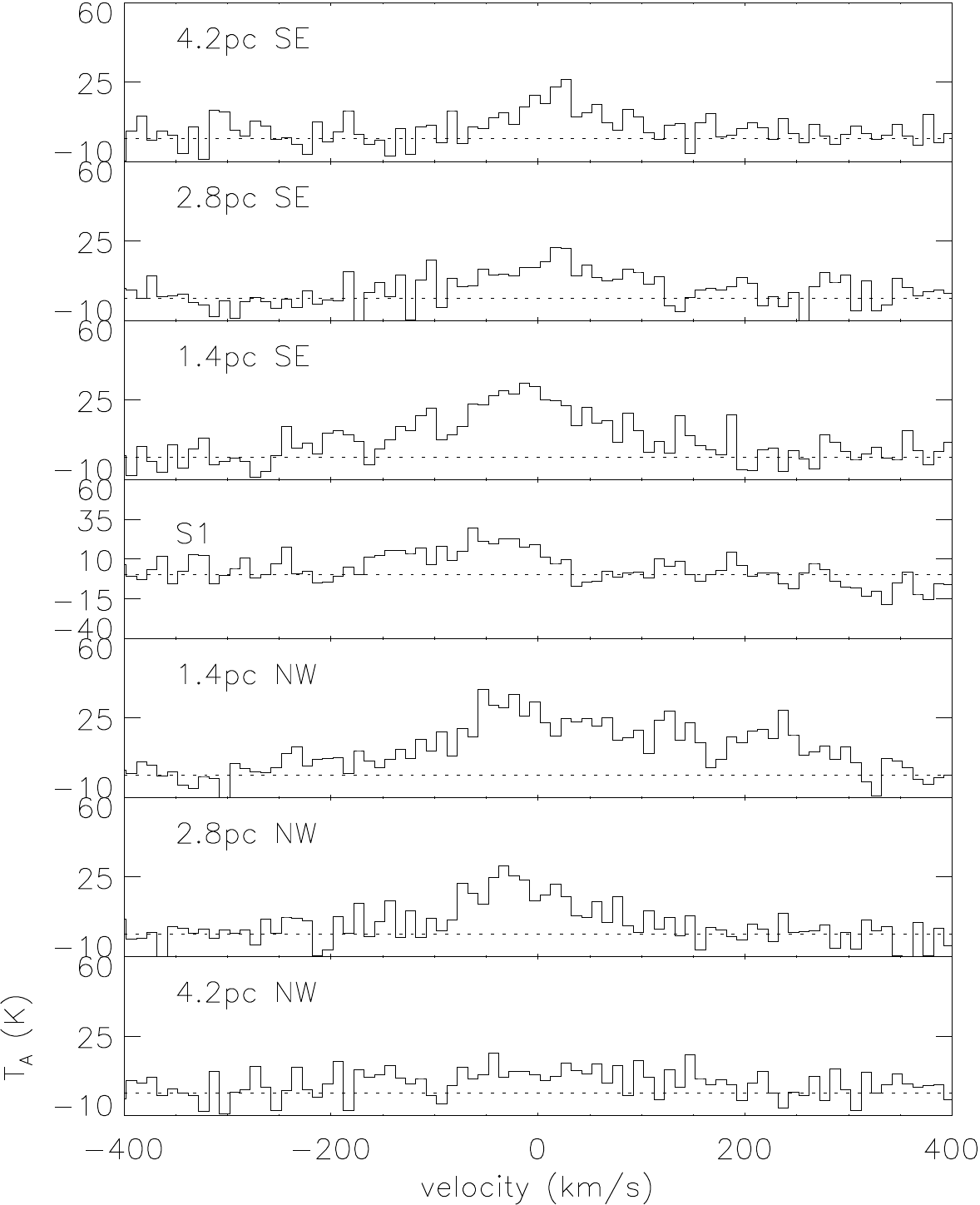}\includegraphics{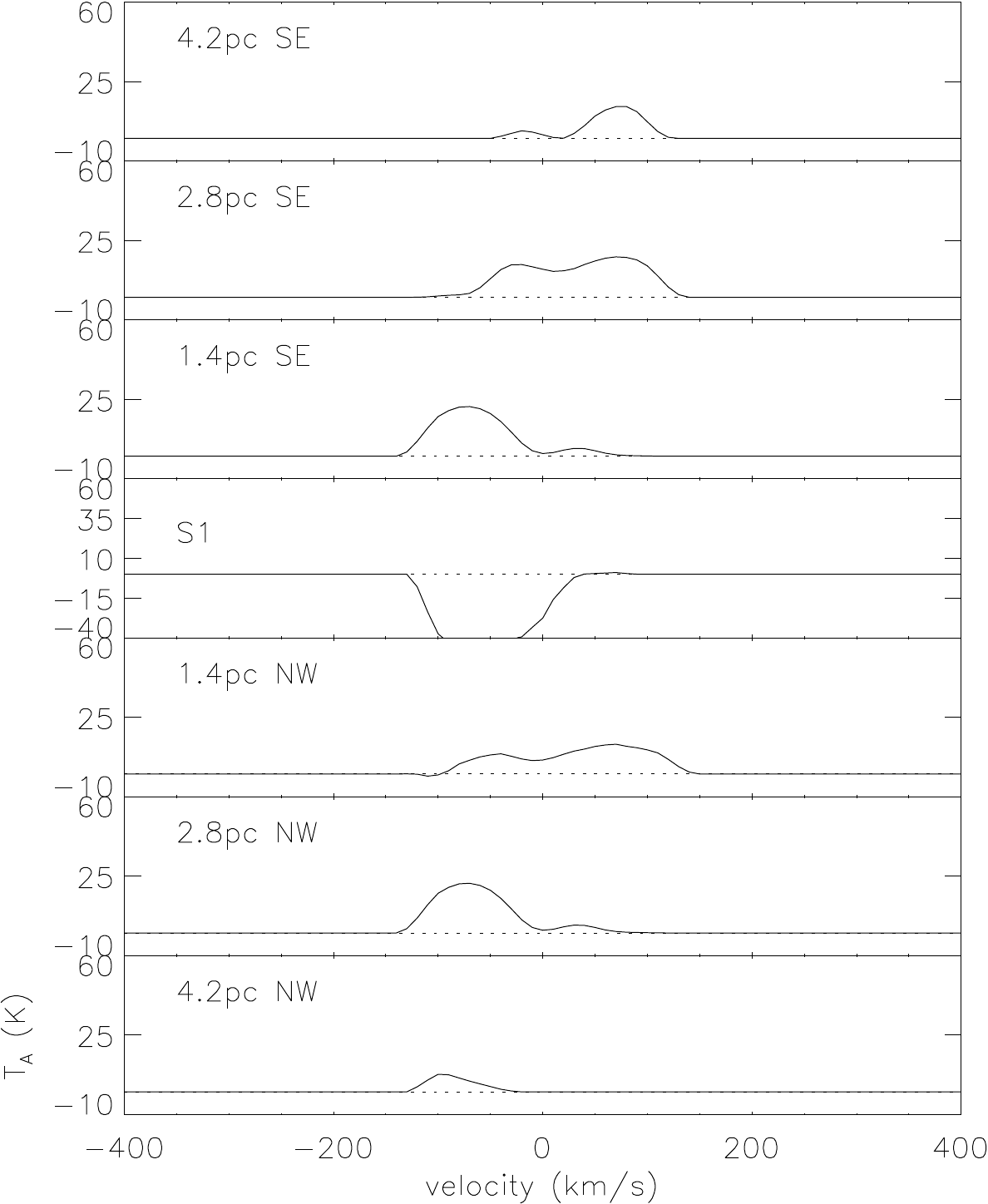}}
    \put(-360,305){\Large Observations}
    \put(-60,305){\Large model}
  \caption{NGC~1068.  Left panel: HCO$^+$(3-2) high resolution spectra along the major axis (from Imanishi et al. 2020 data).
    Right panel: dense disk cloud model high-resolution HCO$^{+}$(3-2) spectra along the major axis.
  \label{fig:plottingvollcnd_hco32radex_newQ=30_4}}
\end{figure*}

The high-resolution HCO$^+$(3-2) spectra corresponding to Fig.~\ref{fig:plottingvollcnd_hcn32radex_newQ=30_4} are presented
in Fig.~\ref{fig:plottingvollcnd_hco32radex_newQ=30_4}.
The main difference between the HCO$^+$(3-2) and HCN(3-2) spectra of the model are $\la 1.5$ times lower flux densities of the 
HCO$^+$(3-2) emission and a broader and deeper absorption around $200$~km\,s$^{-1}$. 
The model HCO$^+$(3-2)/HCN(3-2) ratio is significantly smaller for the peaks at $\sim 80$~km\,s$^{-1}$  than for the rest of the
spectra.

Imanishi et al. (2020) presented high-resolution H$^{13}$CN(3-2) and H$^{13}$CO$^{+}$(3-2) observations of NGC~1068.
At their sensitivity limit (see Table~\ref{tab:obstab}), no emission of the two lines was detected within the inner $20$~pc of NGC~1068. 
Only the H$^{13}$CN(3-2) line was seen in absorption within the central resolution element (Fig.~10 of Imanishi et al. 2020).
The depth of the absorption is about $-20$~K and the velocity width is about $\Delta v \sim 250$~km\,s$^{-1}$.
We included the H$^{13}$CN(3-2) line emission into our model by assuming an isotope ratio of $^{12}C/^{13}C=30$ (see Fig.~2 of Halfen et al. 2017).
As expected, the model H$^{13}$CN(3-2) emission is much fainter than the HCN(3-2) emission.
It is only seen in emission ($\sim 6\sigma$) in the resolution element at a distance of $1.4$~pc northwest of the central black hole.
The H$^{13}$CN(3-2) line is seen in absorption in the central resolution element, as it is observed.
The model absorption depth is narrower ($\sim 130$~km\,s$^{-1}$) and about two times deeper than the observed absorption depth.
\begin{figure}[!ht]
  \resizebox{\hsize}{!}{\includegraphics{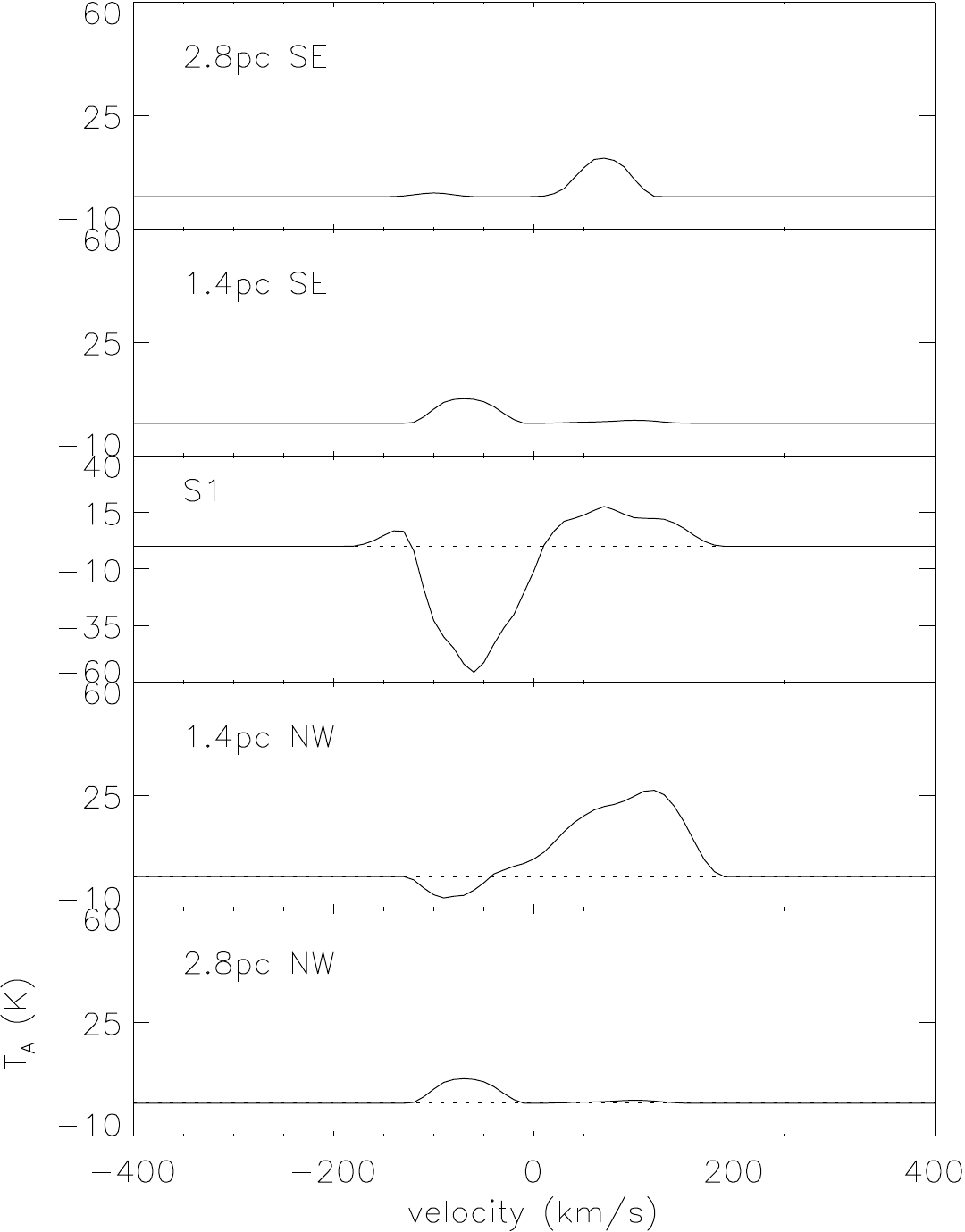}}
  \put(-60,300){\Large model}
  \caption{NGC~1068. Dense disk cloud model high-resolution H$^{13}$CN(3-2) spectra along the major axis. 
  \label{fig:plottingvollcnd_h13cn32radex_newQ=30_4}}
\end{figure}

\begin{figure}[!ht]
  \resizebox{\hsize}{!}{\includegraphics{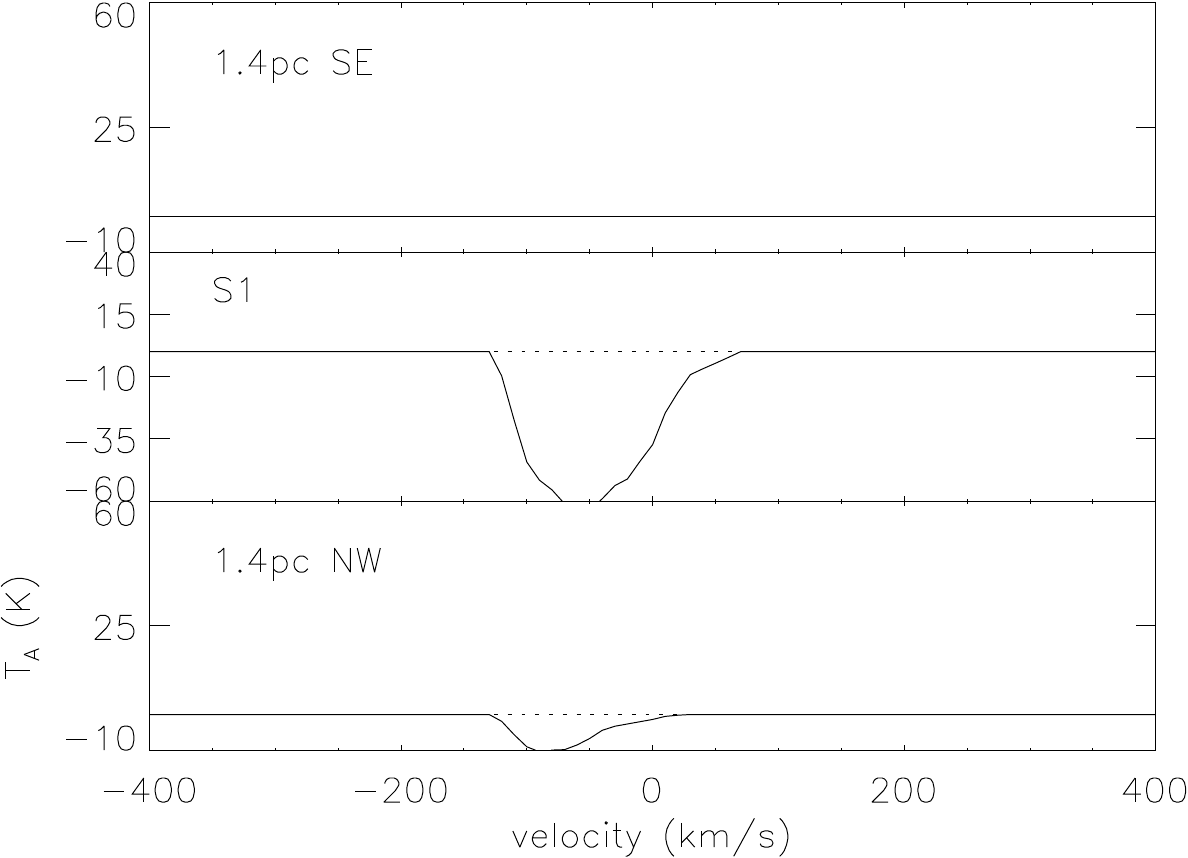}}
  \put(-60,170){\Large model}
  \caption{NGC~1068. Dense disk cloud model high-resolution H$^{13}$CO$^{+}$(3-2) spectra along the major axis. 
  \label{fig:plottingvollcnd_h13co32radex_newQ=30_4}}
\end{figure}

The model H$^{13}$CO$^{+}$(3-2) spectra (Fig.~\ref{fig:plottingvollcnd_h13co32radex_newQ=30_4}) do not
show any emission, as it is observed. Contrary to the observations, the H$^{13}$CO$^{+}$(3-2) line is seen in absorption in the model.
We conclude that the model only qualitatively resembles the observed HCN(3-2) and HCO$^+$(3-2) spectra.  
Given the tiny amount of H$^{13}$CN(3-2) emission and the absence of H$^{13}$CO$^{+}$(3-2) emission in the model spectra, 
the assumed isotope ratio $^{12}C/^{13}C = 30$ is justified.

\section{ALMA observations \label{sec:almaobs}}

During ALMA Cycle 2 (project code 2013.1.00014.S), NGC~1068 was observed
in Band~7 centered at $\lambda \sim 866$~$\mu$m. We used the longest baseline
configurations available. The baseline range was 15 - 2270~m. Structures larger than
$7.2''$ are filtered out. Total time on source was $18.5$~minutes. Source
observations were interleaved with short scans of nearby bright
calibrator sources. The phase calibrators was J0239$+$0416,
located $4.3^{\circ}$ from NGC~1068. The spectral windows were tuned to observe three continuum
windows, 2~GHz total bandwidth each, and one tuned to observe HCN ($J = 4 \to 3$; $\nu_0 = 354.50547$~GHz) at barycentric redshift 1150~km\,s$^{-1}$  
(Brinks et al. 1997). The channel-width of the HCN line
measurements is $\sim 0.4$~km\,s$^{-1}$, spanning about $\pm 800$~km\,s$^{-1}$ total
bandwidth. Calibration and data reduction followed standard
procedures for ALMA observations in CASA. NGC~1068 proved bright enough for
self-calibration over scan intervals. Two rounds of phase-only self-calibration
reduced sidelobe artifacts and improved the background rms.
The photometric accuracy,  $\sim 10$\,\%, is limited by the photometric
uncertainties of the flux calibrators (3C454.3, J0224+0659, J0238+166,
and J2253+1608). The absolute astrometric accuracy is limited by the transfer of calibration solutions from
the complex gain calibrators to NGC~1068. The expected accuracy for the observed phase calibrators is about 5\% of the 
beam width (Asayama et al. 2017), corresponding to 7~mas (0.5~pc).

Images and data cubes were produced using multiscale CLEAN (Cornwell 2008) and Briggs weighting (Briggs 1995). 
Spectral line channels were combined to create cubes with $10$~km\,s$^{-1}$ channel-width. The background rms noise levels
are 0.078~mJy~beam$^{-1}$ and 1.1~mJy~beam$^{-1}$ for the continuum and HCN channel maps, respectively. The synthetic restoring beam is $0.15''
\times 0.11''$, PA~$78^{\circ}$. The primary beams is about $20''$. As emission is
detected only in the inner $3''$, and our focus is on the inner arcsecond, the data are not corrected for primary beam
attenuation. The expected attenuation at $1.5''$ from the pointing center (i.e., S1) is less than 1\% at either band.

We discovered additional line detections during the Band~7 continuum
processing. Both CS(7-6) and the two strongest lines of the CN(3-2) triplet appeared in bands tuned for continuum. These data were processed following the
same procedure as used for the CO and HCN observations. The final data cubes were binned into $15$~km\,s$^{-1}$ channels during imaging and CLEAN
deconvolution.

Table~\ref{tab:S1lines} summarizes the integrated emission line properties, which were derived from a Gaussian fit to the emission line profile. 
For the purposes of the summary, we fitted the two strongest lines of the CN(3-2) triplet.
\begin{table*}[!ht]
\caption{Line Properties for Nuclear Radio Source S1}
\label{tab:S1lines}
\begin{tabular}{lllll}
  \hline
Transition          & Rest  & Integrated  & Radial & FWHM    \\
          & Frequency &  Line Flux &  Velocity &     \\
 & (GHz) & (Jy\,km\,s$^{-1}$) & (km\,s$^{-1}$) & (km\,s$^{-1}$) \\
\hline
HCN 4-3             & 354.50547      & 5.3 (0.3)            & 1139 (3)        & 210 (20) \\
CN N=3-2, J=7/2-5/2 & 340.24787      & 1.60 (0.04)          & 1147 (5)        & 166 (5) \\
CN N=3-2, J=5/2-3/2 & 340.03157      & 1.22 (0.04)          & 1147 (5)        & 166 (5) \\
CN N=3-2 (total)    &                & 2.82 (0.06)          & 1147 (5)        & 166 (5) \\
CS 7-6              & 342.88287      & 0.35 (0.03)          & 1090 (10)       & 260 (30) \\
\end{tabular}
\end{table*}


\begin{thebibliography}{}

\bibitem[Aalto et al.(2002)]{2002A&A...381..783A} Aalto, S., Polatidis, A.~G., H{\"u}ttemeister, S., et al.\ 2002, A\&A, 381, 783

\bibitem[ALMA Partnership(2017)]{TechnicalManual}  Asayama, S.,  Biggs, A., de Gregorio, I., Dent, B., Di Francesco, J., Fomalont, E., 
Hales, A., Hibbard, J., Marconi, G., Kameno, S., Vila Vilaro, B., Villard, E., Stoehr, F. 2017, ALMA Cycle 5 Technical Handbook

\bibitem[Bacon et al.(2001)]{2001A&A...371..409B} Bacon, R., Emsellem, E., Combes, F., et al.\ 2001, A\&A, 371, 409

\bibitem[Bauer et al.(2015)]{2015ApJ...812..116B} Bauer, F.~E., Ar{\'e}valo, P., Walton, D.~J., et al.\ 2015, ApJ, 812, 116

\bibitem[Boger \& Sternberg(2005)]{2005ApJ...632..302B} Boger, G.~I. \& Sternberg, A.\ 2005, ApJ, 632, 302

\bibitem[Bonning et al.(2007)]{2007ApJ...659..211B} Bonning, E.~W., Cheng, L., Shields, G.~A., et al.\ 2007, ApJ, 659, 211

\bibitem[Boonman et al.(2001)]{2001ApJ...553L..63B} Boonman, A.~M.~S., Stark, R., van der Tak, F.~F.~S., et al.\ 2001, ApJL, 553, L63

\bibitem[Briggs(1995)]{none} Briggs, D. S. 1995, PhD thesis, The New Mexico Institute of Min-ing and Technology

\bibitem[Brinks et al.(1997)]{1997Ap&SS.248...23B} Brinks, E., Skillman, E.~D., Terlevich, R.~J., et al.\ 1997, Ap\&SS, 248, 23

\bibitem[Burtscher et al.(2016)]{2016A&A...586A..28B} Burtscher, L., Davies, R.~I., Graci{\'a}-Carpio, J., et al.\ 2016, A\&A, 586, A28

\bibitem[Carroll \& Goldsmith(1981)]{1981ApJ...245..891C} Carroll, T.~J. \& Goldsmith, P.~F.\ 1981, ApJ, 245, 891

\bibitem[Chapillon et al.(2012)]{2012A&A...537A..60C} Chapillon, E., Guilloteau, S., Dutrey, A., et al.\ 2012, A\&A, 537, A60

\bibitem[Chen \& Ostriker(2012)]{2012ApJ...744..124C} Chen, C.-Y. \& Ostriker, E.~C.\ 2012, ApJ, 744, 124

\bibitem[Christopher et al.(2005)]{2005ApJ...622..346C} Christopher, M.~H., Scoville, N.~Z., Stolovy, S.~R., et al.\ 2005, ApJ, 622, 346

\bibitem[Combes et al.(2019)]{2019A&A...623A..79C} Combes, F., Garc{\'\i}a-Burillo, S., Audibert, A., et al.\ 2019, A\&A, 623, A79

\bibitem[Cornwell(2008)]{2008ISTSP...2..793C} Cornwell, T.~J.\ 2008, IEEE Journal of Selected Topics in Signal Processing, 2, 793

\bibitem[Crutcher(1999)]{1999ApJ...520..706C} Crutcher, R.~M.\ 1999, ApJ, 520, 706

\bibitem[da Cunha et al.(2013)]{2013ApJ...766...13D} da Cunha, E., Groves, B., Walter, F., et al.\ 2013, ApJ, 766, 13

\bibitem[Das et al.(2006)]{2006AJ....132..620D} Das, V., Crenshaw, D.~M., Kraemer, S.~B., et al.\ 2006, AJ, 132, 620

\bibitem[Davies et al.(2007)]{2007ApJ...671.1388D} Davies, R.~I., M{\"u}ller S{\'a}nchez, F., Genzel, R., et al.\ 2007, ApJ, 671, 1388

\bibitem[Draine(1983)]{1983ApJ...270..519D} Draine, B.~T.\ 1983, ApJ, 270, 519

\bibitem[Dumouchel et al.(2010)]{2010MNRAS.406.2488D} Dumouchel, F., Faure, A., \& Lique, F.\ 2010, MNRAS, 406, 2488

\bibitem[Dyda et al.(2015)]{2015MNRAS.446..613D} Dyda, S., Lovelace, R.~V.~E., Ustyugova, G.~V., et al.\ 2015, \mnras, 446, 613

\bibitem[Esplugues et al.(2014)]{2014A&A...567A..95E} Esplugues, G.~B., Viti, S., Goicoechea, J.~R., et al.\ 2014, A\&A, 567, A95

\bibitem[Etxaluze et al.(2011)]{2011AJ....142..134E} Etxaluze, M., Smith, H.~A., Tolls, V., et al.\ 2011, AJ, 142, 134

\bibitem[Fischer et al.(2013)]{2013ApJS..209....1F} Fischer, T.~C., Crenshaw, D.~M., Kraemer, S.~B., et al.\ 2013, ApJS, 209, 1

\bibitem[Fleck(1996)]{1996ApJ...458..739F} Fleck, R.~C.\ 1996, ApJ, 458, 739

\bibitem[Flower(1999)]{1999MNRAS.305..651F} Flower, D.~R.\ 1999, MNRAS, 305, 651

\bibitem[Gallimore et al.(1996)]{1996ApJ...462..740G} Gallimore, J.~F., Baum, S.~A., O'Dea, C.~P., et al.\ 1996, ApJ, 462, 740

\bibitem[Gallimore et al.(2004)]{2004ApJ...613..794G} Gallimore, J.~F., Baum, S.~A., \& O'Dea, C.~P.\ 2004, ApJ, 613, 794

\bibitem[Gallimore et al.(2016)]{2016ApJ...829L...7G} Gallimore, J.~F., Elitzur, M., Maiolino, R., et al.\ 2016, ApJL, 829, L7

\bibitem[Garc{\'\i}a-Burillo et al.(2016)]{2016ApJ...823L..12G} Garc{\'\i}a-Burillo, S., Combes, F., Ramos Almeida, C., et al.\ 2016, ApJL, 823, L12

\bibitem[Garc{\'\i}a-Burillo et al.(2019)]{2019A&A...632A..61G} Garc{\'\i}a-Burillo, S., Combes, F., Ramos Almeida, C., et al.\ 2019, A\&A, 632, A61

\bibitem[Garc{\'\i}a-Burillo et al.(2021)]{2021A&A...652A..98G} Garc{\'\i}a-Burillo, S., Alonso-Herrero, A., Ramos Almeida, C., et al.\ 2021, A\&A, 652, A98

\bibitem[Genzel et al.(2010)]{2010RvMP...82.3121G} Genzel, R., Eisenhauer, F., \& Gillessen, S.\ 2010, Reviews of Modern Physics, 82, 3121

\bibitem[Gillessen et al.(2009)]{2009ApJ...692.1075G} Gillessen, S., Eisenhauer, F., Trippe, S., et al.\ 2009, ApJ, 692, 1075.

\bibitem[Gravity Collaboration et al.(2019)]{2019A&A...625L..10G} Gravity Collaboration, Abuter, R., Amorim, A., et al.\ 2019, A\&A, 625, L10

\bibitem[Gravity Collaboration et al.(2020)]{2020A&A...634A...1G} Gravity Collaboration, Pfuhl, O., Davies, R., et al.\ 2020, \aap, 634, A1

\bibitem[Green \& Thaddeus(1976)]{1976ApJ...205..766G} Green, S. \& Thaddeus, P.\ 1976, ApJ, 205, 766

\bibitem[Greenhill et al.(1996)]{1996ApJ...472L..21G} Greenhill, L.~J., Gwinn, C.~R., Antonucci, R., et al.\ 1996, ApJL, 472, L21

\bibitem[Greenhill et al.(2003)]{2003ApJ...582L..11G} Greenhill, L.~J., Kondratko, P.~T., Lovell, J.~E.~J., et al.\ 2003, ApJL, 582, L11

\bibitem[Guesten et al.(1987)]{1987ApJ...318..124G} Guesten, R., Genzel, R., Wright, M.~C.~H., et al.\ 1987, ApJ, 318, 124

\bibitem[Halfen et al.(2017)]{2017ApJ...845..158H} Halfen, D.~T., Woolf, N.~J., \& Ziurys, L.~M.\ 2017, ApJ, 845, 158

\bibitem[Harada et al.(2015)]{2015A&A...584A.102H} Harada, N., Riquelme, D., Viti, S., et al.\ 2015, A\&A, 584, A102

\bibitem[Hasegawa \& Herbst(1993)]{1993MNRAS.261...83H} Hasegawa, T.~I. \& Herbst, E.\ 1993, MNRAS, 261, 83

\bibitem[Hersant et al.(2009)]{2009A&A...493L..49H} Hersant, F., Wakelam, V., Dutrey, A., et al.\ 2009, A\&A, 493, L49


\bibitem[Hicks et al.(2009)]{2009ApJ...696..448H} Hicks, E.~K.~S., Davies, R.~I., Malkan, M.~A., et al.\ 2009, ApJ, 696, 448

\bibitem[Hicks et al.(2013)]{2013ApJ...768..107H} Hicks, E.~K.~S., Davies, R.~I., Maciejewski, W., et al.\ 2013, ApJ, 768, 107

\bibitem[Hopkins \& Quataert(2011)]{2011MNRAS.415.1027H} Hopkins, P.~F. \& Quataert, E.\ 2011, MNRAS, 415, 1027

\bibitem[Hopkins et al.(2012)]{2012MNRAS.425.1121H} Hopkins, P.~F., Hernquist, L., Hayward, C.~C., et al.\ 2012, MNRAS, 425, 1121

\bibitem[Hirota et al.(1998)]{1998ApJ...503..717H} Hirota, T., Yamamoto, S., Mikami, H., et al.\ 1998, ApJ, 503, 717

\bibitem[Hsieh et al.(2018)]{2018ApJ...862..150H} Hsieh, P.-Y., Koch, P.~M., Kim, W.-T., et al.\ 2018, ApJ, 862, 150

\bibitem[Hsieh et al.(2021)]{2021ApJ...913...94H} Hsieh, P.-Y., Koch, P.~M., Kim, W.-T., et al.\ 2021, ApJ, 913, 94

\bibitem[Imanishi et al.(2016)]{2016ApJ...822L..10I} Imanishi, M., Nakanishi, K., \& Izumi, T.\ 2016, ApJL, 822, L10

\bibitem[Imanishi et al.(2018)]{2018ApJ...853L..25I} Imanishi, M., Nakanishi, K., Izumi, T., et al.\ 2018, ApJL 853, L25

\bibitem[Imanishi et al.(2020)]{2020ApJ...902...99I} Imanishi, M., Nguyen, D.~D., Wada, K., et al.\ 2020, ApJ, 902, 99

\bibitem[Impellizzeri et al.(2019)]{2019ApJ...884L..28I} Impellizzeri, C.~M.~V., Gallimore, J.~F., Baum, S.~A., et al.\ 2019, ApJL, 884, L28


\bibitem[Izumi et al.(2018)]{2018ApJ...867...48I} Izumi, T., Wada, K., Fukushige, R., et al.\ 2018, ApJ, 867, 48.

\bibitem[Jackson et al.(1993)]{1993ApJ...402..173J} Jackson, J.~M., Geis, N., Genzel, R., et al.\ 1993, ApJ, 402, 173

\bibitem[Jacobs \& Sellwood(2001)]{2001ApJ...555L..25J} Jacobs, V. \& Sellwood, J.~A.\ 2001, ApJL, 555, L25

\bibitem[Jenkins(2009)]{2009ApJ...700.1299J} Jenkins, E.~B.\ 2009, ApJ, 700, 1299

\bibitem[Lacy et al.(1994)]{1994ApJ...428L..69L} Lacy, J.~H., Knacke, R., Geballe, T.~R., et al.\ 1994, ApJL, 428, L69

\bibitem[Lazarian(2006)]{2006IJMPD..15.1099L} Lazarian, A.\ 2006, International Journal of Modern Physics D, 15, 1099

\bibitem[Le Bourlot et al.(2012)]{2012A&A...541A..76L} Le Bourlot, J., Le Petit, F., Pinto, C., et al.\ 2012, \aap, 541, A76

\bibitem[Le Petit et al.(2006)]{2006ApJS..164..506L} Le Petit, F., Nehm{\'e}, C., Le Bourlot, J., et al.\ 2006, ApJS, 164, 506

\bibitem[Le Petit et al.(2016)]{2016A&A...585A.105L} Le Petit, F., Ruaud, M., Bron, E., et al.\ 2016, A\&A, 585, A105

\bibitem[Li et al.(2012)]{2012ApJ...760...33L} Li, P.~S., Myers, A., \& McKee, C.~F.\ 2012, ApJ, 760, 33

\bibitem[Liu et al.(2012)]{2012ApJ...756..195L} Liu, H.~B., Hsieh, P.-Y., Ho, P.~T.~P., et al.\ 2012, ApJ, 756, 195

\bibitem[Lodato \& Bertin(2003)]{2003A&A...398..517L} Lodato, G. \& Bertin, G.\ 2003, A\&A, 398, 517

\bibitem[L{\'o}pez-Gonzaga et al.(2014)]{2014A&A...565A..71L} L{\'o}pez-Gonzaga, N., Jaffe, W., Burtscher, L., et al.\ 2014, A\&A, 565, A71

\bibitem[Lopez-Rodriguez et al.(2020)]{2020ApJ...893...33L} Lopez-Rodriguez, E., Alonso-Herrero, A., Garc{\'\i}a-Burillo, S., et al.\ 2020, ApJ, 893, 33

\bibitem[Kokubo(2018)]{2018PASJ...70...97K} Kokubo, M.\ 2018, PASJ, 70, 97

\bibitem[Kritsuk et al.(2007)]{2007ApJ...665..416K} Kritsuk, A.~G., Norman, M.~L., Padoan, P., et al.\ 2007, ApJ, 665, 416

\bibitem[Mac Low(1999)]{1999ApJ...524..169M} Mac Low, M.-M.\ 1999, ApJ, 524, 169

\bibitem[Maloney et al.(1996)]{1996ApJ...466..561M} Maloney, P.~R., Hollenbach, D.~J., \& Tielens, A.~G.~G.~M.\ 1996, ApJ, 466, 561

\bibitem[Marinucci et al.(2016)]{2016MNRAS.456L..94M} Marinucci, A., Bianchi, S., Matt, G., et al.\ 2016, MNRAS, 456, L94

\bibitem[May \& Steiner(2017)]{2017MNRAS.469..994M} May, D. \& Steiner, J.~E.\ 2017, MNRAS, 469, 994

\bibitem{a39} McKee C.~F., Li P.~S., Klein R.~I., 2010, ApJ, 720, 1612

\bibitem[Meijerink \& Spaans(2005)]{2005A&A...436..397M} Meijerink, R. \& Spaans, M.\ 2005, A\&A, 436, 397

\bibitem[Meijerink et al.(2007)]{2007A&A...461..793M} Meijerink, R., Spaans, M., \& Israel, F.~P.\ 2007, A\&A, 461, 793

\bibitem[Mills et al.(2013)]{2013ApJ...779...47M} Mills, E.~A.~C., G{\"u}sten, R., Requena-Torres, M.~A., et al.\ 2013, ApJ, 779, 47

\bibitem[Miyauchi \& Kishimoto(2020)]{2020ApJ...904..149M} Miyauchi, R. \& Kishimoto, M.\ 2020, ApJ, 904, 149

\bibitem[Momferratos et al.(2014)]{2014MNRAS.443...86M} Momferratos, G., Lesaffre, P., Falgarone, E., et al.\ 2014, MNRAS, 443, 86

\bibitem[Montero-Casta{\~n}o et al.(2009)]{2009ApJ...695.1477M} Montero-Casta{\~n}o, M., Herrnstein, R.~M., \& Ho, P.~T.~P.\ 2009, ApJ, 695, 1477

\bibitem[Nelson \& Langer(1997)]{1997ApJ...482..796N} Nelson, R.~P. \& Langer, W.~D.\ 1997, ApJ, 482, 796

\bibitem[Neufeld \& Kaufman(1993)]{1993ApJ...418..263N} Neufeld, D.~A. \& Kaufman, M.~J.\ 1993, ApJ, 418, 263

\bibitem[Neufeld et al.(1995)]{1995ApJS..100..132N} Neufeld, D.~A., Lepp, S., \& Melnick, G.~J.\ 1995, ApJS, 100, 132

\bibitem[Padoan et al.(1997)]{1997ApJ...474..730P} Padoan, P., Jones, B.~J.~T., \& Nordlund, {\r{A}}. P.\ 1997, ApJ, 474, 730

\bibitem[Padovani et al.(2009)]{2009A&A...501..619P} Padovani, M., Galli, D., \& Glassgold, A.~E.\ 2009, A\&A, 501, 619

\bibitem[Pan \& Padoan(2009)]{2009ApJ...692..594P} Pan, L. \& Padoan, P.\ 2009, ApJ, 692, 594

\bibitem[Parravano(1987)]{1987A&A...172..280P} Parravano, A.\ 1987, A\&A, 172, 280

\bibitem[Pringle(1981)]{1981ARA&A..19..137P} Pringle, J.~E.\ 1981, ARA\&A, 19, 137

\bibitem[Quach et al.(2015)]{2015MNRAS.446..622Q} Quach, D., Dyda, S., \& Lovelace, R.~V.~E.\ 2015, MNRAS, 446, 622

\bibitem[Quan et al.(2008)]{2008ApJ...681.1318Q} Quan, D., Herbst, E., Millar, T.~J., et al.\ 2008, ApJ, 681, 1318

\bibitem[Requena-Torres et al.(2012)]{2012A&A...542L..21R} Requena-Torres, M.~A., G{\"u}sten, R., Wei{\ss}, A., et al.\ 2012, A\&A, 542, L21


\bibitem[Rolffs et al.(2011)]{2011A&A...536A..33R} Rolffs, R., Schilke, P., Zhang, Q., et al.\ 2011, A\&A, 536, A33

\bibitem[Ruaud et al.(2015)]{2015MNRAS.447.4004R} Ruaud, M., Loison, J.~C., Hickson, K.~M., et al.\ 2015, MNRAS, 447, 4004

\bibitem[Sakamoto et al.(2010)]{2010ApJ...725L.228S} Sakamoto, K., Aalto, S., Evans, A.~S., et al.\ 2010, ApJL, 725, L228

\bibitem[Salow \& Statler(2001)]{2001ApJ...551L..49S} Salow, R.~M. \& Statler, T.~S.\ 2001, ApJL, 551, L49

\bibitem[Sambhus \& Sridhar(2002)]{2002A&A...388..766S} Sambhus, N. \& Sridhar, S.\ 2002, A\&A, 388, 766

\bibitem[Sanders(1998)]{1998MNRAS.294...35S} Sanders, R.~H.\ 1998, MNRAS, 294

\bibitem[Sani et al.(2012)]{2012MNRAS.424.1963S} Sani, E., Davies, R.~I., Sternberg, A., et al.\ 2012, MNRAS, 424, 1963

\bibitem[Schawinski et al.(2015)]{2015MNRAS.451.2517S} Schawinski, K., Koss, M., Berney, S., et al.\ 2015, MNRAS, 451, 2517

\bibitem[Sch{\"o}ier et al.(2005)]{2005A&A...432..369S} Sch{\"o}ier, F.~L., van der Tak, F.~F.~S., van Dishoeck, E.~F., et al.\ 2005, A\&A, 432, 369

\bibitem[Semenov et al.(2010)]{2010A&A...522A..42S} Semenov, D., Hersant, F., Wakelam, V., et al.\ 2010, A\&A, 522, A42

\bibitem[Stoer]{Stoer} Stoer and R. Bulirsch, Introduction to numerical analysis, 3rd edition, Springer, New York, 2002. ISBN 978-0-387-95452-3

\bibitem[Stone et al.(1998)]{1998ApJ...508L..99S} Stone, J.~M., Ostriker, E.~C., \& Gammie, C.~F.\ 1998, ApJL, 508, L99

\bibitem[Kawamuro et al.(2019)]{2019PASJ...71...68K} Kawamuro, T., Izumi, T., \& Imanishi, M.\ 2019, PASJ, 71, 68

\bibitem[Tielens \& Hollenbach(1985)]{1985ApJ...291..722T} Tielens, A.~G.~G.~M. \& Hollenbach, D.\ 1985, ApJ, 291, 722

\bibitem[Tsuboi et al.(2018)]{2018PASJ...70...85T} Tsuboi, M., Kitamura, Y., Uehara, K., et al.\ 2018, PASJ, 70, 85

\bibitem[van der Tak et al.(2007)]{2007A&A...468..627V} van der Tak, F.~F.~S., Black, J.~H., Sch{\"o}ier, F.~L., et al.\ 2007, A\&A, 468, 627

\bibitem[Vidal \& Wakelam(2018)]{2018MNRAS.474.5575V} Vidal, T.~H.~G. \& Wakelam, V.\ 2018, MNRAS, 474, 5575

\bibitem[Vollmer \& Duschl(2001)]{2001A&A...367...72V} Vollmer, B. \& Duschl, W.~J.\ 2001a, A\&a, 367, 72

\bibitem[Vollmer \& Duschl(2001)]{2001A&A...377.1016V} Vollmer, B. \& Duschl, W.~J.\ 2001b, A\&A, 377, 1016

\bibitem[Vollmer \& Duschl(2002)]{2002A&A...388..128V} Vollmer, B. \& Duschl, W.~J.\ 2002, A\&A, 388, 128

\bibitem[Vollmer et al.(2003)]{2003A&A...407..515V} Vollmer, B., Zylka, R., \& Duschl, W.~J.\ 2003, A\&A, 407, 515

\bibitem[Vollmer et al.(2004)]{2004A&A...413..949V} Vollmer, B., Beckert, T., \& Duschl, W.~J.\ 2004, A\&A, 413, 949

\bibitem[Vollmer et al.(2008)]{2008A&A...491..441V} Vollmer, B., Beckert, T., \& Davies, R.~I.\ 2008, A\&A, 491, 441

\bibitem[Vollmer \& Davies(2013)]{2013A&A...556A..31V} Vollmer, B. \& Davies, R.~I.\ 2013, A\&A, 556, A31

\bibitem[Vollmer et al.(2017)]{2017A&A...602A..51V} Vollmer, B., Gratier, P., Braine, J., et al.\ 2017, A\&A, 602, A51

\bibitem[Vollmer et al.(2018)]{2018A&A...615A.164V} Vollmer, B., Schartmann, M., Burtscher, L., et al.\ 2018, A\&A, 615, A164

\bibitem[Wakelam et al.(2015)]{2015ApJS..217...20W} Wakelam, V., Loison, J.-C., Herbst, E., et al.\ 2015, ApJS, 217, 20

\bibitem[Wiegel 1994]{} Wiegel, W. 1994, Diploma thesis, Univ. Heidelberg

\bibitem{a61} Williams J.~P., Bergin E.~A., Caselli P., Myers P.~C., Plume R., 1998, ApJ, 503, 689 

\bibitem[Yang et al.(2010)]{2010ApJ...718.1062Y} Yang, B., Stancil, P.~C., Balakrishnan, N., et al.\ 2010, ApJ, 718, 1062

\bibitem[Yusef-Zadeh et al.(2013)]{2013ApJ...762...33Y} Yusef-Zadeh, F., Hewitt, J.~W., Wardle, M., et al.\ 2013, ApJ, 762, 33

\end{thebibliography}
\end{document}